\documentclass[11pt]{thesis}

\usepackage[german,british]{babel}
\usepackage[latin1]{inputenc}
\usepackage{type1cm}

\usepackage{units}
\usepackage{amsmath,amssymb,amsbsy,amssymb}
\usepackage{aas_macros}
\usepackage{graphicx}
\usepackage[font=small,bf]{caption}
\usepackage[authoryear]{natbib}

\usepackage{ifpdf}
\ifpdf
  \usepackage[pdftex]{hyperref}
  \hypersetup{
    plainpages=false,%
    colorlinks,%
    citecolor=black,%
    filecolor=black,%
    linkcolor=black,%
    urlcolor=black}
\else
  \newcommand\texorpdfstring[2]{#1}
  \newcommand\url[1]{\texttt{#1}}
\fi

\defcitealias{1993A&A...269..581R}{RK93}
\defcitealias{1998A&A...335..488F}{F98}    
\defcitealias{1999ApJ...514L..99K}{KBSTN99}
\defcitealias{2005A&A...436..585D}{AB05}
\defcitealias{2006ApJ...653.1266J}{JML06}
\defcitealias{2008AN....329..619G}{Gressel et al., 2008a}
\defcitealias{2008A&A...486L..35G}{Gressel et al., 2008b}

\usepackage{relsize}
\newcommand\HI{H{\smaller I}}
\newcommand\OVI{O{\smaller VI}}

\newcommand\cm{\,\rm cm}
\newcommand\pccm{\,\rm cm^{-3}}
\newcommand\s{\,\rm s}
\newcommand\g{\,\rm g}
\newcommand\erg{\,\rm erg}
\newcommand\K{\,\rm K}

\newcommand\Myr{\,\rm Myr}
\newcommand\Gyr{\,\rm Gyr}
\newcommand\muG{\, \mu{\rm G}}

\newcommand\kms{\,\rm km\,s^{-1}}
\newcommand\kmss{\,\rm km\,s^{{\text -}\!1}}  
\newcommand\ms{\,\rm m\,s^{-1}}
\newcommand\pc{\,\rm pc}
\newcommand\kpc{\,\rm kpc}
\newcommand\vis{\,\rm cm^{2}s^{-1}}
\newcommand\Msun{\,\rm\,M_\odot}
\newcommand\kB{k_{\,\rm B}}

\newcommand\same[1]{\hspace{1.0em} --- #1}
\newcommand\yes{$\bullet$}
\newcommand\no{$\circ$}

\newcommand\Mn{\!-\!}
\newcommand\Pl{\!+\!}
\newcommand\Eq{\!=\!}
\newcommand\tms{\!\times\!}
\newcommand\cdt{\!\cdot\!}
\newcommand\lft{\left(\!}
\newcommand\rgt{\!\right)}

\newcommand\degr{^{\circ}}
\newcommand{\simgt}{\!\stackrel{>}{_{\sim}}\!}
\newcommand{\simlt}{\!\stackrel{<}{_{\sim}}\!}

\newcommand{\itl}{\tilde{\imath}}
\newcommand{\jtl}{\tilde{\jmath}}

\newcommand\hlf{\frac{1}{2}}
\newcommand\phlf{\!+\!\frac{1}{2}}
\newcommand\mhlf{\!-\!\frac{1}{2}}

\newcommand\xx{\hat{\mathbf{x}}}
\newcommand\yy{\hat{\mathbf{y}}}
\newcommand\zz{\hat{\mathbf{z}}}
\newcommand\ee{\hat{\mathbf{e}}}
\newcommand\rr{\mathbf{r}}

\newcommand\V{\mathbf{v}}
\newcommand\vrms{v_{\rm rms}}
\newcommand\U{\mathbf{u}}
\newcommand\mU{\bar{\mathbf{u}}}
\newcommand\B{\mathbf{B}}
\newcommand\mB{\bar{\mathbf{B}}}
\newcommand\Tf{\mathcal{B}}
\newcommand\mTf{\bar{\mathcal{B}}}

\newcommand{\fv}[2]{\left<#1_#2\right>}
\newcommand{\mn}[1]{\overline{#1}}
\newcommand{\rms}[1]{\left<\right.\!#1\!\left.\right>}

\newcommand\visfl{\boldsymbol{\tau}}
\newcommand\talph{\boldsymbol{\alpha}}
\newcommand\teta{\tilde{\boldsymbol{\eta}}}

\newcommand\EMF{\mathcal E}
\newcommand\Reyn{\rm Re}
\newcommand\Rm{\rm Rm}
\newcommand\Pran{\rm Pr}
\newcommand\Pm{\rm Pm}
\newcommand\St{\rm St}
\newcommand\Co{\Omega_{*}}

\renewcommand\rho{\varrho}

\DeclareMathAlphabet{\mathbf}{OML}{cmm}{b}{it}

\newcommand\thesistitle{%
  Supernova-driven~Turbulence~and
  Magnetic~Field~Amplification
  in~Disk~Galaxies
}

\begin{document}

\title{%
  {\large
    Astrophysikalisches Institut Potsdam\\ 
    Magnetohydrodynamik \\
  }
  \vfill
  {\Huge{\scshape{\thesistitle}}}
  \vfill
}
\author{
  Dissertation\\ zur Erlangung des akademischen Grades\\
  Doktor der Naturwissenschaften (Dr. rer. nat.)\\
  in der Wissenschaftsdisziplin Astrophysik\\ \\ \\ \\
  eingereicht an der\\
  Mathematisch-Naturwissenschaftlichen Fakultät\\
  der Universität Potsdam \\ \\ \\ \\
  von\\  {\Large \scshape{Oliver Gressel}}\\  aus Mosbach
}
\date{\vfill Babelsberg, im August 2008}



\ifpdf\hypersetup{pageanchor=false}\fi

\thispagestyle{empty}
\begin{center}
{\large 
  \hrulefill~ Astrophysikalisches Institut Potsdam ~\hrulefill\\ 
  Magnetohydrodynamik \\
}
\vfill
\begin{doublespace}
  \begin{center}
    {\Huge \scshape{\thesistitle}}
  \end{center}
\end{doublespace}
\vfill
\includegraphics[width=0.8\columnwidth]{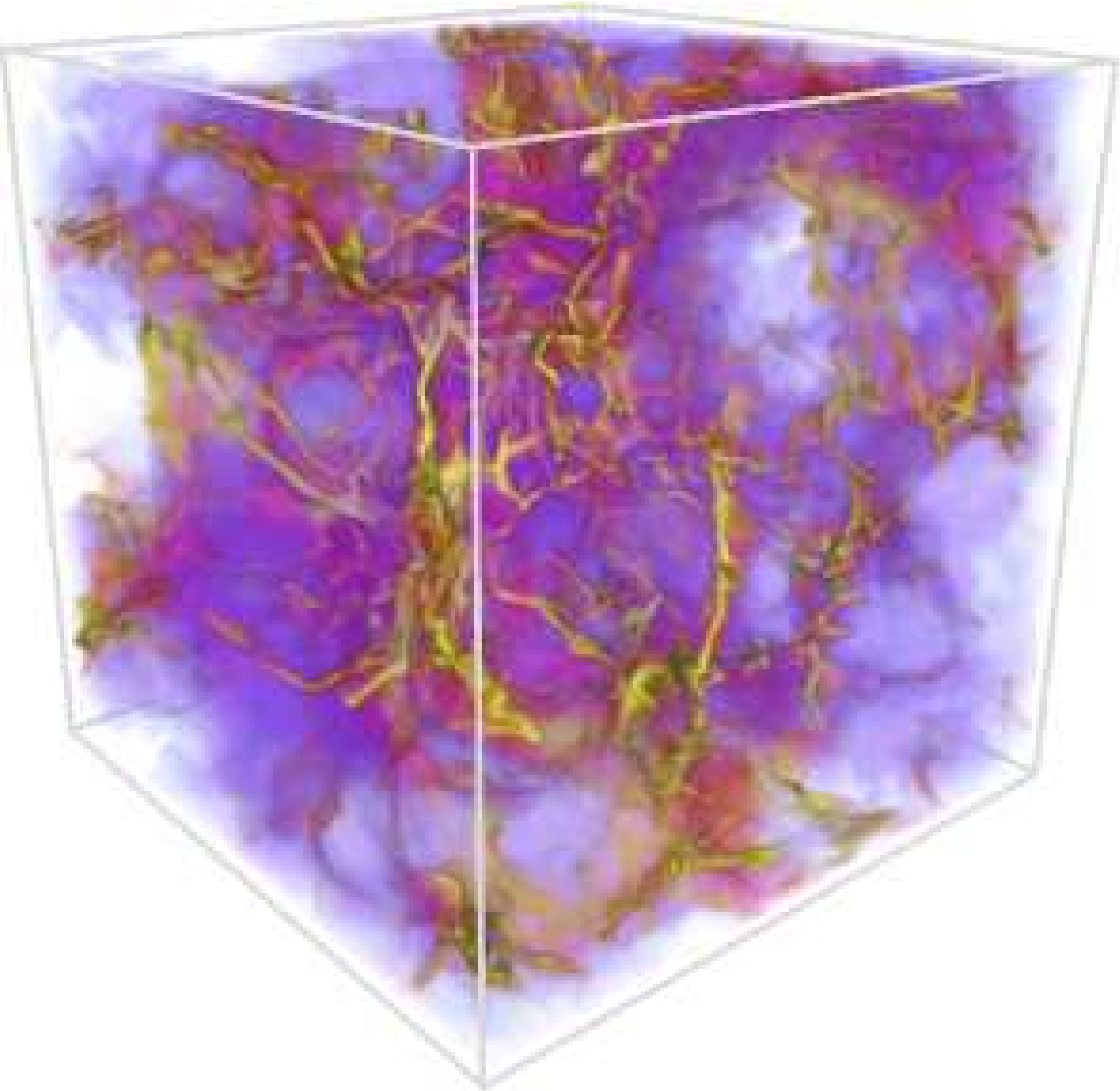}
{\LARGE 
  \scshape{\hskip2em
    \includegraphics[width=0.2\columnwidth]{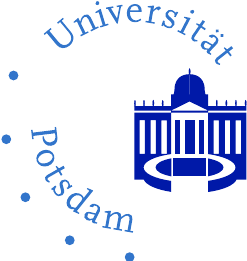}%
    \hfill Oliver Gressel \hskip2em\nobreak}
}
\end{center}
\clearpage


\thispagestyle{empty}

\parbox[t]{0.85\columnwidth}{

\textbf{Abstract:~}%
Supernovae are known to be the dominant energy source for driving turbulence
in the interstellar medium. Yet, their effect on magnetic field amplification
in spiral galaxies is still poorly understood. Analytical models based on the
uncorrelated-ensemble approach predicted that any created field will be
expelled from the disk before a significant amplification can occur. By means
of direct simulations of supernova-driven turbulence, we demonstrate that this
is not the case. Accounting for vertical stratification and galactic
differential rotation, we find an exponential amplification of the mean field
on timescales of 100$\Myr$. The self-consistent numerical verification of such a
``fast dynamo'' is highly beneficial in explaining the observed strong
magnetic fields in young galaxies. We, furthermore, highlight the importance
of rotation in the generation of helicity by showing that a similar mechanism
based on Cartesian shear does not lead to a sustained amplification of the
mean magnetic field. This finding impressively confirms the classical picture
of a dynamo based on cyclonic turbulence.

\vspace{20pt}

\textbf{Kurzzusammenfassung:~}%
Supernovae sind bekanntermaßen die dominante treibende Energiequelle für
Turbulenz im interstellaren Medium. Dennoch ist ihre Auswirkung auf die
Verstärkung von Magnetfeldern in Spiralgalaxien weitestgehend unverstanden.
Analytische Modelle, die auf der Annahme eines unkorrelierten Ensembles
beruhen, sagen voraus, dass das erzeugte Feld aus der galaktischen Scheibe
herausgedrängt wird bevor eine substantielle Verstärkung erfolgen kann.
Mithilfe numerischer Simulationen supernovagetriebener Turbulenz zeigen wir,
dass dies nicht der Fall ist. Unter Berücksichtigung einer vertikalen
Schichtung und differentieller galaktischer Rotation beobachten wir eine
exponentielle Verstärkung des mittleren Magnetfeldes auf einer Zeitskala von
100 Mio.~Jahren. Diese selbstkonsistente numerische Bestätigung eines
``schnellen Dynamos'' erlaubt es, die beobachteten starken Magnetfelder in
jungen Galaxien zu erklären. Darüberhinaus stellen wir die Wichtigkeit der
Rotation bei der Erzeugung von Helizität heraus, indem wir zeigen, dass ein
ähnlicher Effekt basierend auf kartesischer Scherung nicht zu einer
Verstärkung des mittleren Magnetfeldes führt. Dies bestätigt eindrucksvoll das
klassische Bild zyklonischer Turbulenz.

}

~\par\vfill
\begin{minipage}[b]{0.5\columnwidth}
  \textbf{Cover illustration:} Volume rendering of the mass density for model
  F4--BOX, showing the dense filaments and cavities created by supernova
  explosions. Image produced by VAPOR (\texttt{www.vapor.ucar.edu}).
\end{minipage}


\maketitle
\ifpdf\hypersetup{pageanchor=true}\fi

\tableofcontents

\chapter{Astrophysical Context}
\label{ch:context}

\section{Introduction} 
\label{sec:intro}

The modern world cannot be imagined without electromagnetic fields. Yet, the
common notion of ``magnetism'' is still very much attached to the picture of
small black ferromagnets -- a picture that is completely adverse to the highly
dynamical nature of magnetic fields throughout the cosmos.

In 1877, Werner von Siemens received a patent for his so-called
``dynamo-electric machine'', a generator which worked without such permanent
magnets. Instead, it received its magnetic field from a dynamical
amplification mechanism powered by the very current it produces. Due to the
inherent feedback loop, the machine could be seeded from the residual
magnetisation of its coils, making the expensive permanent magnets redundant.
Only 42 years later, in 1919, Sir Joseph Larmor proposed that a similar
mechanism, based on electromagnetic induction, might be responsible for the
magnetic field of the sun. Today, dynamo theory is successfully applied to a
wide range of celestial bodies \citep{2004maun.book.....R} and can well
explain the ubiquitous magnetic fields in planets, stars, accretion disks, and
even galaxies. Very much like in the electric generator, the fundamental
mechanism in a dynamo is the conversion of mechanical energy into magnetic
energy. In this respect, the galactic dynamo poses a very special
implementation of such a device: since the kinematic structure as well as the
magnetic field topology are directly observable, it embodies an exciting
challenge for dynamo theory.

\subsection{Galactic magnetic fields -- the observer's account} 

It is now 60 years since the first observations of galactic magnetic fields,
which were based on the polarisation of starlight. The effect was
independently detected by \citet{1949Sci...109..165H} and
\citet{1949Sci...109..166H} -- the results were in fact published on
consecutive pages in the same issue of \emph{Science}.
\citet{1949PhRv...75.1605D}, in the same year, explained this polarisation by
the alignment of elongated interstellar dust grains in an external magnetic
field. Apart from the polarisation of background sources, these grains also
emit polarised infrared radiation. Like Zeeman splitting, which requires
stronger fields, these methods are mainly used within the Milky Way, but are
not sensitive enough to obtain good measurements for distant galaxies.

The most powerful method to map galactic fields in external galaxies is the
synchrotron emission from the relativistic electron component within the
interstellar plasma. These so-called cosmic rays (CRs) are believed to be
accelerated in the shocks of supernova remnants and in protostellar jets.
Under the influence of the Lorenz force, the electrons spiral around the
magnetic field lines and produce the characteristic polarised emission that
can be observed with radio telescopes. Because the energetic electrons are
distributed over a wide spectrum, this emission can be found over a wide range
of radio frequencies. Whereas the intensity of the total emission primarily is
a measure for the column density of the cosmic rays, it can be translated to a
magnetic field strength (of the total field) via the assumption of energy
equipartition between the two components.

The degree of polarisation is interpreted as a measure of the field
regularity, i.e., strong polarisation implies large-scale coherent fields.
Field structures on angular scales below the beam size will, in turn, lead to
a depolarisation of the signal -- an indication of dominant small-scale
fields. From the emission process, the orientation of the observed
polarisation vectors is perpendicular to the magnetic field lines, or more
precisely, to their projection onto the plane of the sky. The observed
orientation of the signal is, however, modified by the additional effect of
Faraday rotation. On its passage through the magnetised plasma, the linear
polarisation vector of the electromagnetic wave is subject to a rotation which
is proportional to the line-of-sight component of the magnetic field. The
origin of this effect lies in the different effective refraction index for the
left- and right-handed circularly polarised waves. 

Although this complication, at first, might seem tedious to deconvolve, it can
in fact be used as a further source of information. This is because the
amount of Faraday rotation depends on the wavelength of the radiation as
$\lambda^2$. With multiband observations over a wide range of radio
frequencies, it thus becomes possible to measure the direction and amplitude
of the regular field component along the line-of-sight. Because the
orientation of the polarisation vector alone is always ambiguous by $180\degr$,
the additional effect from Faraday rotation is also helpful to address
questions with respect to the overall parity of the observed regular fields.
Although there is the possibility to develop a ``tomography'' based on this
effect, the sensitivity and resolution of the current instrumentation is still
insufficient to create three-dimensional maps of galactic magnetic fields.
With powerful facilities under construction (Low Frequency Array, LOFAR) and
in the design phase (Square Kilometre Array, SKA), the near future awaits a
drastic leap ahead \citep{2008Ap&SS.314..107B,2008arXiv0804.4594B}.

\begin{figure}
  \begin{minipage}[b]{0.70\columnwidth}
    \includegraphics[width=0.95\columnwidth]%
    {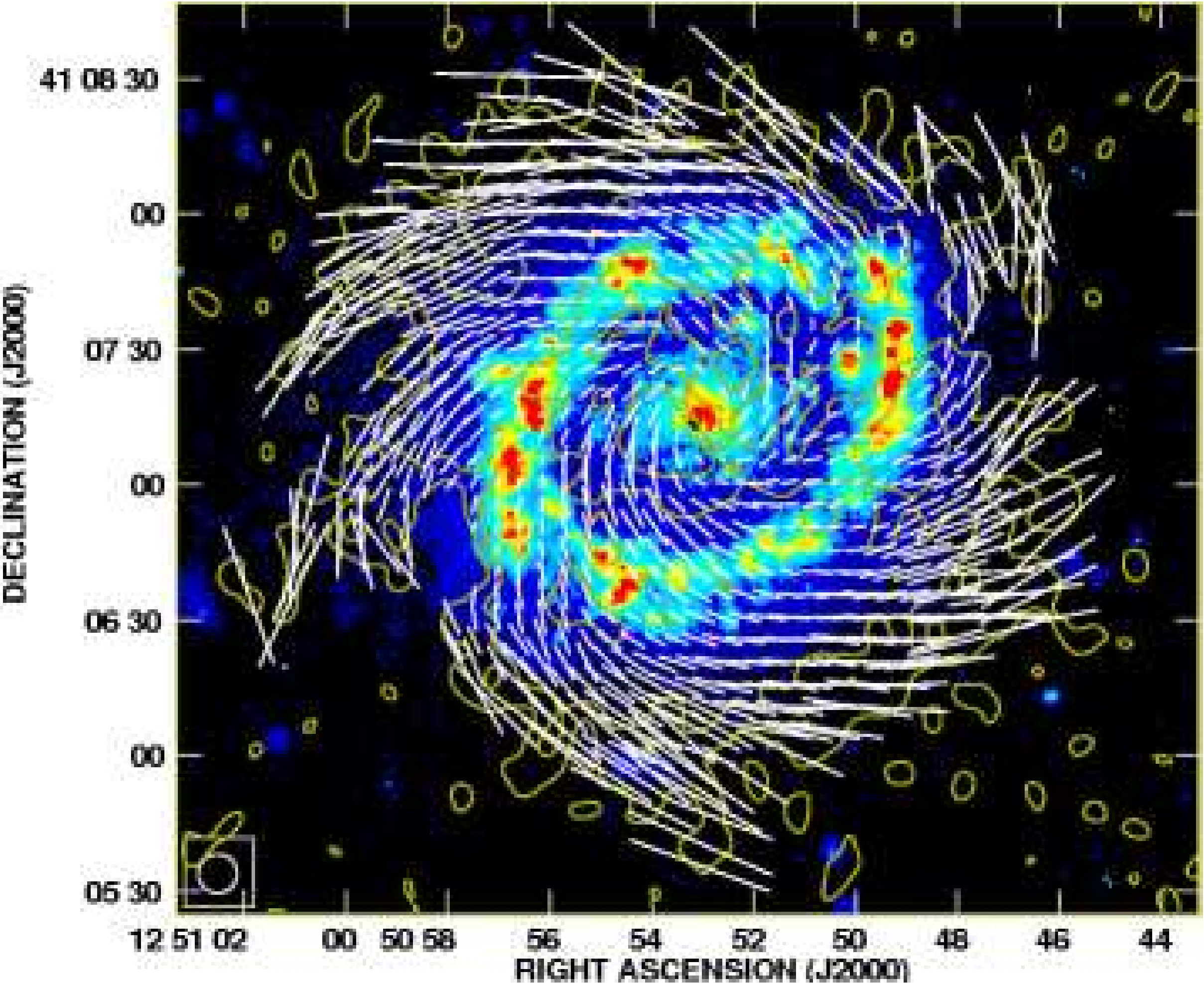}
  \end{minipage}
  \begin{minipage}[b]{0.28\columnwidth}
    \caption[Magnetic spiral arms of the ringed galaxy NGC~4736]{Magnetic
      spiral arms of the ringed galaxy NGC~4736 \citep{2008ApJ...677L..17C}.
      Polarised intensity contour map at 8.46~GHz with observed magnetic field
      vectors of the polarisation degree overlaid on the H$\alpha$ image
      \citep[from][]{2003MNRAS.344..527K}. Figure courtesy of K.T. Chy{\.z}y.}
    \label{fig:ngc4736}
  \end{minipage}
\end{figure}

Figure~\ref{fig:ngc4736} shows a state-of-the-art radio observation of the
magnetic field within the nearby ringed galaxy NGC~4736
\citep{2008ApJ...677L..17C}. This particular case is in so far remarkable, as
the magnetic field shows a distinct spiral pattern emerging at the very centre
of the disk and extending far beyond the ring structure of the gas
distribution. Unlike expected under the assumption of passively advected
magnetic fields, the field lines do not follow the disk structure, but cross
the inner ring at a remarkably high pitch angle of $35\degr$. Together with
the high field strength of $30$ and $13\muG$ in the total and regular magnetic
field, respectively, this lends strong support in favour of the presence of an
efficient dynamo mechanism \citep{2008ApJ...677L..17C}.

In this example we have already encountered some of the central observational
findings dynamo theory has to be confronted with: (i) field amplitudes of
several $10\muG$ -- created within a lifetime of a few $\Gyr$, (ii)
considerable pitch angles of up to $35\degr$, (iii) a characteristic ratio of
the regular field over the turbulent field -- varying as a function of star
formation activity, and (iv) an even, i.e., quadrupolar, symmetry with respect
to the galactic plane. The latter assumption is based on observation of halo
fields in nearby edge-on galaxies \citep[see e.g.][]{1995A&A...302..691D}.

Because the interstellar medium has long been known to be in a highly
turbulent state, it is expected that also the mechanism relevant for the field
amplification is closely related to the turbulence. The very hypothesis of a
large-scale coherent field emerging out of unordered motion poses a compelling
example for the self-organisation of a chaotic system. Before we proceed with
the analysis of the dynamo process itself, we thus want to briefly introduce
the setting of this galactic spectacle.

\subsection{Interstellar turbulence} 

The interstellar medium (ISM) is an extremely diluted, turbulent gas which
fills the otherwise void space between the stars \citep[see][for a
review]{2001RvMP...73.1031F}. Due to the vast multitude of important physical
processes, the ISM has an extremely rich and heterogeneous structure
\citep{1978ppim.book.....S}. As the most prominent feature, a thermal cooling
instability (TI) leads to the formation of compact, cold atomic H- and
molecular H$_2$-clouds ($20$--$100\K$) which are enclosed by a diffuse, hot
plasma ($10^6\K$). Enriched with dust grains and cosmic rays, the interstellar
cocktail is permanently stirred by supernovae (SNe), stellar winds and
protostellar jets.\footnote{%
  For estimations of the expected energy inputs from the various possible
  sources see Section C in \citet{2004RvMP...76..125M}.} %
With an estimated rate of two per century, for our own Galaxy, supernova
explosions are commonly perceived as extremely rare events. Albeit two per
century is a high number on cosmic timescales, we will never observe the
related turbulent motions with our own eyes. Yet, the ``time-lapse''
animations of our simulation runs nicely illustrate the vigorous driving
caused by the ubiquitous explosions and one can gain an intuitive feeling for
the vibrant dynamics of the flow.

Based on the observed light-curves and chemical abundances, supernovae have
originally been classified into type~I and type~II events. Later, this
classification has been refined and subtypes have been introduced for the
first type. With respect to the kinetic feedback from SNe, the exact mechanism
leading to the explosion is unimportant, and we are mainly interested in the
spectral class of the stellar progenitor. If we, in the following, speak of
``type~I'' SNe, we implicitly mean type~Ia, corresponding to, e.g., pair SNe
in common envelope binary systems. Accordingly, ``type~II'' means type~II
including type~Ib/Ic. These events have their common origin in massive OB
stars. The main characteristics of O and B stars is that they form in stellar
associations of 10--100 objects. Due to their high mass, these stars evolve
rapidly and have a short main sequence lifetime during which the OB
association will not disperse significantly. In turn, the occurring type~II
SNe are highly correlated both in space and time and form what is known as a
super-bubble (SB). As we will see later, this morphological difference can
drastically change the way in which the SNe structure the interstellar medium.

The estimated thermal energy input due to a single supernova is $10^{51}\erg$
-- this is only about four orders of magnitude less than the turbulent energy
contained in the interstellar medium of the whole Milky Way. A high fraction
of this energy is directly converted into kinetic energy by the rapid
expansion of the remnant. It is believed that a certain fraction of the SN
energy is deposited in the form of cosmic ray electrons. Based on the high
diffusivity of the electron gas, \citet{1992ApJ...401..137P} suggested that the
cosmic ray pressure can give rise to a buoyancy instability. It has, however,
been argued that even in equipartition the kinetic influence of the CR
component is not necessarily significant \citep{2006MNRAS.373..643S}.

Although the ISM makes up for only about five percent of the baryonic matter
content within the Galaxy, it plays an essential role as the carrier of the
magnetic field, i.e., the magnetic flux within the plasma exceeds the flux
from the stars by many orders of magnitude. Interstellar magnetic fields are
believed to be of some importance for star formation theory -- not in the
classical picture of magneto-static support, but rather as a modifying agent
for the gravoturbulent fragmentation and the subsequent protostellar collapse
\citep[see][for a recent account]{2004RvMP...76..125M}. Magnetic tension
forces can, e.g., efficiently redistribute angular momentum and thus initiate
or enhance local collapse. Due to their assumed importance for the dynamics of
turbulent flows, magnetic fields might also affect the slope of the turbulent
cascade. The self-similar nature of the turbulent inertial range, in turn, is
thought to be a key parameter for the determination of the core mass function
and, ultimately, the initial mass function (IMF) of the stellar population.

\section{Field amplification in spiral galaxies} 

The discussion about the origin of the observed galactic fields is divided
into two major schools of reasoning: While dynamo theory argues that large
scale fields are due to a dynamic process (contemporary field), the opposing
standpoint is based on a frozen-in field stemming from the formation of the
galaxy (primordial field). The main argument against the latter comes from the
observed high values of the turbulent diffusivity ($\sim10^{26}\vis$) which
would lead to a decay of any ordered magnetic field component within about
$0.7\Gyr$ \citep*{1998A&A...329..911R}. To overcome this diffusive process
astrophysicists seek for an adequate source for the production of magnetic
field \citep[see e.g.][for an extensive review]{1996ARA&A..34..155B}.

\subsection{General considerations} 

Undoubtedly, the generation of the azimuthal field can be explained by
differential rotation which is the dominating galactic flow pattern. Any
radial field will be instantly sheared out into the azimuthal direction,
thereby converting kinetic to magnetic energy. Neglecting diffusive processes,
this mechanism would in principle work until the field itself is strong enough
to counteract the differential rotation, i.e., until a substantial amount of
angular momentum is redistributed via the $R\phi$-component of the Maxwell
stress tensor. In reality, the amplification through shear will only be
transient, and decay will occur long before equipartition is reached. This is
because the winding-up in the azimuthal direction inherently increases the
wavenumbers of the radial structures making them susceptible to dissipative
processes. In general, we can say that differential rotation alone does not
lead to a sustained amplification of the galactic magnetic field. Apart from
this, observations indicate large pitch angles for the magnetic field, i.e.,
the direction of the magnetic field lines deviates strongly (up to $35\degr$)
from the direction of the velocity field. To explain these large angles one
clearly needs a robust mechanism to regenerate the radial field
\citep{2005mpge.conf..117E}.

The generation of a mean magnetic field from turbulent fluctuations can be
explained via the so-called $\alpha$~effect \citep{1980mfmd.book.....K}, which
parameterises the correlations of the small-scale turbulent velocity $\U'$ and
the magnetic field $\B'$, giving rise to a mean electromotive force $\EMF=
\overline{\U'\tms\B'}$. In the case of homogeneous, isotropic turbulence, it
can be shown that the EMF is directly linked to the kinetic helicity
$\U'\cdt(\nabla\tms\U')$ of the turbulent flow. Generally, only a flow that
exhibits some sort of asymmetry can produce a non-vanishing helicity and thus
contribute to the induction equation. In the case of the ISM, there are three
characteristics: (i) the axis of rotation, (ii) the galactic shear gradient,
and (iii) the vertical gradient in density and turbulence intensity.

The first contribution is embodied in the Coriolis force, which will give
uprising turbulent eddies a definite skewness, i.e., the turbulence becomes
cyclonic. In connection with the vertical stratification, this leads to a
non-zero mean kinetic helicity, which will, in turn, produce the desired
poloidal magnetic field component. As we will see from our simulations, this
term is essential for the SN-driven galactic dynamo to operate.%
\footnote{Note that rotation is also an essential ingredient for the
  magneto-rotational instability, which depends on the interaction of the
  Alfv{\'e}nic mode with epicyclic oscillations
  \citep[cf.][]{1969A&A.....1..388F}.} %

The effect of the shear is twofold: Although it cannot lead to a dynamo
itself, it couples the radial field to the azimuthal direction, thus preparing
half of the dynamo loop. Moreover, the huge reservoir of kinetic energy stored
in the differential rotation provides a potential additional power source for
the dynamo. The coupling between a (weak) turbulent $\alpha$~mechanism and
(strong) shear is commonly referred to as $\alpha\Omega$~dynamo. Since this
type, however, is characterised by vanishingly small pitch angles, we prefer
to speak of an $\alpha^2\Omega$~dynamo, where both effects give contributions
of the same magnitude.

The role of the third term, which is responsible for the so-called diamagnetic
pumping \citep{1992A&A...260..494K}, has been interpreted in two diametrical
ways -- yet, with rather similar implications. As would be expected from
intuition, the pumping is always directed towards the less turbulent regions
\citep*[cf.][]{1993IAUS..157..321R}. Therefore, if one assumes a profile of
the turbulent velocity which decreases with $z$, the pumping will lead to an
expulsion of the generated field into large galactic heights, counteracting
its production. A profile of the turbulent velocity that increases with
galactic height can, on the other hand, be justified from considerations based
on the observed density stratification \citep{1996A&A...311..451F}. In this
case, the pumping will be directed inward, which implies that the magnetic
field will be compressed in the disk midplane. Although this may seem
beneficial, at first glance, too strong pumping will confine the field to a
smaller and smaller region around the midplane. Similarly to the folding of
radial structures in the case of dominant shear, this implies sharp vertical
gradients in the magnetic field, and hence a considerable source of
dissipation. The admissible strength of pumping that allows for dynamo
solutions has been explored by \citet*{1994A&A...286...72S}.

We conclude that it is crucial to determine the mutual strength of the terms
arising from (i) and (iii) to answer the question whether a galactic dynamo
may operate. Furthermore, the interplay between the effects due to (i) and
(ii) will determine the pitch angle that can be obtained by the dynamo
process. These two questions are also tightly related to the saturation
mechanism. Depending on which of the processes reaches its saturated state
first, the properties of the dynamo might drastically change in the quenched
regime.

\subsection{Mean-field models} 

Irrespective of the underlying physics, one can simply apply a closure to the
mean-field induction equation in the form of an $\alpha$~prescription%
\footnote{Although based on the same concept, this dynamo $\alpha$ shall not
  be confused with the famous viscosity parameter introduced by
  \citet{1973A&A....24..337S}.}. %
Although this approach is inherently limited in its predictive power, it can
already reproduce many of the observed features of the galactic field and its
topology. \citet{1998A&A...333...27R}, for example, study three-dimensional
dynamos with a prescribed, vertically stratified velocity dispersion. Apart
from the stratification, the turbulence intensity is assumed to increase in
the spiral arms. The antisymmetric off-diagonal part of the $\alpha$~tensor
constitutes a diamagnetic pumping term $\gamma_{\rm dia} \propto -\tau_{\rm
  c}\nabla\rms{u'^2}$, as well as a buoyancy term $\gamma_{\rm buo} \propto
\tau_{\rm c}\rms{u'^2}\,\nabla\log\rho$. While these pumping terms, like the
turbulent diffusivity $\eta_T \propto \tau_{\rm c}\rms{u'^2}$, scale linearly
with the correlation time $\tau_{\rm c}$, the diagonal components, which
comprise the dynamo effect, scale with $\Omega\tau_{\rm c}^2$. This implies
that, at a sufficiently high correlation time, the dynamo mechanism becomes
strong enough to overcome the diffusive and advective loss terms. With a
powerful $\alpha$~effect, one can thus easily explain the observed high pitch
angles. In addition, the lower level of diffusivity in the inter-arm regions
leads to stronger regular fields there.

Albeit this finding is in agreement with radio observations, the increased
velocity dispersion in the spiral arms is not \citep{1996ARA&A..34..155B}.
Therefore, in a refined approach, \cite*{1999A&A...350..423R} apply a uniform
velocity dispersion in connection with a correlation time $\tau_{\rm c}$,
which is now assumed to vary with the spiral pattern. The new model does again
explain the stronger inter-arm fields, although now based on assumptions
compatible with observations. This example, to some extent, demonstrates what
is meant with ``limited predictive power'': a conclusion is only as good as
the weakest assumption it is based on. Moreover, a model relying on an
unjustified premise can still lead to a correct prediction -- but what is the
value of this prediction, then?

The dynamo models presented so far are largely based on the analytical
description derived within the quasilinear framework
\citep[e.g.][]{1990GApFD..50...53R}. As a complement to this paradigm (which
generally does not depend on SNe as a driving mechanism) the mean-field models
of \citet{2000A&A...358..125F} are founded on a simplified treatment of single
supernova explosions as developed by \citeauthor{1992ApJ...391..188F}. These
two very different approaches to the derivation of closure parameters by means
of analytical considerations shall be briefly introduced in the following
sections.

\section{Analytical derivations} 

To strengthen the predictive power of large eddy simulations (LES) based on a
turbulence model one usually aims to derive the free parameters of the theory
from empirical data. Unlike in classical turbulence modelling, in the case of
mean-field MHD it is, however, not possible to directly obtain the closure
parameters from laboratory experiments. In the absence of empirical grounds,
and before simulations became affordable to replace experiments, the
coefficients had to be derived analytically on the basis of plausible
assumptions; the only aid to this being crude observations of column densities
and turbulent velocity dispersions.

\subsection{Second order theory} 

Although the so-called second order correlation%
\footnote{sometimes also referred to as first order smoothing approximation
  (FOSA)} %
approximation (SOCA) has its origins in the analytical treatment of the solar
dynamo \citep*{1966ZNatA..21..369S}, it is sufficiently general to be applied
to any kind of cyclonic turbulence. The basic idea behind SOCA is to obtain
$\EMF=\overline{\U'\tms\B'}$ for a prescribed velocity field $\U'$ from the
time integrated induction equation for $\B'$ (cf. Sec.~\ref{sec:SOCA}). This
only becomes possible by neglecting the terms quadratic in the fluctuations
\citep[cf. Ch.~3 in][]{1980mfmd.book.....K} -- hence the name. Although the
formal assumptions on the underlying turbulence are very restrictive, it turns
out that SOCA expressions give reasonable predictions, even when some of the
prerequisites are not fulfilled. This applies, e.g., to the required limit of
small Strouhal numbers, $\St=\tau_{\rm c}\,u'/l_{\rm c} \ll 1$, with a
correlation time and length $\tau_{\rm c}$ and $l_{\rm c}$, respectively.

Since SOCA theory only depends on the underlying small-scale flow structure,
its predictions can be directly checked from a comparison with simulations. In
general, the scheme is derived in the spectral decomposition of the velocity
field $\U'$; however, for the sake of giving explicit expressions, authors
usually assume turbulence with a single scale and apply mixing-length theory.
In this case, the only free parameter, when comparing the results to
simulations, is the coherence time $\tau_{\rm c}$ of the turbulent flow field.

General SOCA results for homogeneous, isotropic turbulence demonstrate that
the dynamo effect is strongly related to the kinetic helicity
$\U'\cdot\nabla\tms\U'$, and the turbulent diffusion scales with the velocity
$\U'^2$. Another basic result, which can be understood intuitively, is the
fact that the turbulent transport (the so-called diamagnetic pumping) follows
$-\nabla\U'$. Taking into account the effects of stratification in the density
and turbulent velocity, \citet{1993A&A...269..581R} derived an $\alpha$~effect
depending on the combined gradient of the two quantities. While for the
antisymmetric part of the tensor the gradients can be combined into a common
gradient $\nabla\log(\rho\,u')$, for the diagonal elements it takes the form
of a weighted product $\nabla\log(\rho^s\,u')$. The weighting factor $s$ is
found to depend on the rotation rate and approaches the value
$s\rightarrow3/2$ in the limit of slow rotation.

\subsection{The uncorrelated-ensemble approach} 
\label{sec:isolated}

Disregarding the mutual interaction of the supernova remnants,
\citet{1992ApJ...391..188F} analytically derived the dynamo effect for
isolated, spherically symmetric explosions embedded in a homogeneous ambient
medium. In a series of papers the model was incrementally refined. The dynamo
profiles $\alpha(z)$ computed from this method had amplitudes of a few hundred
meters per second and extended across the asymptotic diameter ($\sim 200\pc$)
of a typical remnant. Since no vertical stratification was included in the
early models, the profiles were perfectly antisymmetric with respect to the
midplane.

The model relied crucially on the central assumption that the net effect of
the SNe can be seen as an ``ensemble of uncorrelated events''. That is, the
effects of the explosions at various galactic heights $z$ were convolved with
an assumed vertical SN distribution (with scale heights of $90\pc$ for the
type~II and $325\pc$ for the type~I SNe) to yield the effect of an ensemble of
explosions. Because of the odd symmetry of the profiles, the positive
contribution from the upper half of the remnant would almost cancel out with
the negative contribution from the lower half.\footnote{%
  This effect could only be avoided by assuming an unrealistic
  $\delta$-distribution for the SNe, which means that all explosions are
  assumed to occur in the galactic midplane.} %
Due to the larger scale height of the type~I SNe, their contribution was
diminished to negligible values. The net contribution of the type~II SNe was
found to be on the order of a few ten meters per second.

To make matters worse, these models found a vertical transport velocity on the
order of $\kms$, which was more than a factor of $50$ higher than the
$\alpha$~effect. This means that the field was basically blown away before it
could be amplified by the turbulence. Under these conditions the operability
of a galactic dynamo seemed highly improbable
\citep*[cf.][]{1994A&A...286...72S}.

Because single SNe turned out to be much too weak, also the effects due to
several correlated supernovae were considered, albeit still as isolated
shells. The idea behind this was that the correlation between the explosions
would lead to a higher coherence time $\tau_{\rm c}$ of the generated
turbulence. From general considerations, it could be assumed that the
$\alpha$~effect would grow with $\tau_{\rm c}^2$, whereas the pumping would
only be affected linearly. In consequence, the ratio $\hat{\gamma}$ between
the two effects would decrease like $\tau_{\rm c}^{-1}$. The inclusion of SBs
gave amplitudes of the $\alpha$~effect of about $400\ms$ but the pumping
effect still remained dominant by a factor of $15$.

To describe the evolution of a single remnant more closely and include
non-axisymmetric effects like the galactic shear gradient, the models were
later extended to a semi-analytical approach which involved numerical
simulations for the expansion of the remnant. Yet, the overall effect was
still computed via a convolution assuming an ensemble of explosions. First
two-dimensional simulations to compute the azimuthal dynamo parameter for a
single supernova have been performed by \citet*{1993A&A...274..757K}. This
model has been generalised to 3D by \citet*{1996A&A...305..114Z}, who computed
the full dynamo $\alpha$~tensor for a single remnant. Still, the key issue of
a dominating turbulent pumping remained. The numerical treatment of the full
expansion phase of the remnant also made it possible to study the non-linear
aspects of magnetic quenching \citep{1996A&A...313..448Z}. Like in the
kinematic case, turbulent buoyancy overwhelms any $\alpha$~effect, and the
ratio $\hat{\gamma}$ was even found to increase in the saturated regime.

Because the asymptotic radius of the remnant critically depends on the
pressure of the ambient medium, one expects pear- or peanut-shaped envelopes
for super bubbles breaking out of the galactic disk
\citep{1988ApJ...324..776M}. In this scenario, the profiles caused by the
single explosions are no longer assumed to be independent of their position.
Due to the pressure stratification of the interstellar medium, the part of the
remnant pointing away from the midplane inflates to a larger radius, thus
contributing stronger to the $\alpha$~effect. With the new asymmetry in the
profiles, the cancellation effect, inherent in the convolution, was
drastically reduced. Accordingly, \citet{1998A&A...335..488F} computed values
of $6.0\kms$ and $2.6\kms$ for the radial and azimuthal $\alpha$~effect. Also
the ratio $\hat{\gamma}$ could be further reduced to a value of $\simeq 6$,
which does not exclude a dynamo solution per se, but still might constitute a
burden for the field amplification mechanism.

To conclude our review on the treatment of isolated remnants (which was, in a
way, paradigmatic for the research on galactic magnetic fields for over a
decade) we want to note that all models based on the described approach found
an \emph{outward} transport of the mean magnetic field. While this is indeed
correct for a single remnant, the hypothesis that this feature will carry over
to the ensemble of explosions is not.

\section{Object of investigation} 

As will become obvious, the major limitation to the uncorrelated-ensemble
approach is directly related to its central assumption: It considers
non-interacting, isolated events taking place on a uniform, smooth background.
This is definitely not the case for the ISM, which due to the thermal
instability \citep{1965ApJ...142..531F} is a highly clumpy medium. Within the
ISM, most of the mass is concentrated in cold molecular clouds, and most of
the volume is occupied by warm and hot neutral and ionised gas. Collisions of
expanding supernova remnants lead to strong fragmentation and subsequently to
the formation of over-dense filaments and clumps (see cover illustration).
These structures, in turn, will rain back into the gravitational potential
(mainly of the stellar component) comprising the so-called Galactic Fountain.
Because of the high conductivity of the ionised ISM, the magnetic flux is, to
a good approximation, frozen-in with respect to the fluid. If we further
assume that the regions of strong field are correlated with the high density
clumps, this opens the possibility to efficiently hinder the magnetic flux
from escaping the galaxy. This intricate effect, that can properly be captured
only in fully dynamical MHD simulations, might drastically reduce the value of
$\hat{\gamma}$.

The central aim of the current investigation is to perform direct simulations
of interstellar turbulence to gain knowledge about the $\alpha$~effect which
is thought to be responsible for the creation of the strong observed magnetic
fields within spiral galaxies. Taking into account the immense complexity of
the non-linear system (given by the magnetohydrodynamic equations of motion at
high Reynolds numbers) this $\alpha$~effect cannot be derived analytically --
at least not from first principles: As has been argued in the preceeding
sections, available theoretical derivations are either based on doubtful
assumptions (uncorrelated-ensemble approach), or depend on properties of the
turbulence (i.e., the velocity dispersion as a function of $z$ within the
quasilinear approach) that cannot be inferred directly from the interstellar
medium. With the aid of direct simulations, it becomes possible to challenge
these predictions and define a new foundation for mean-field modelling.

\subsubsection{Organisation of the thesis}

All simulations have been performed on massively parallel computers applying
the new version 3 of the {NIRVANA} MHD fluid code \citep{2004JCoPh.196..393Z}.
To account for the differential background rotation of spiral galaxies, the
code has been extended to the framework of the local shearing box
approximation \citep{2007CoPhC.176..652G}. For a realistic representation of
the multi-phase interstellar medium a radiative cooling term was included --
see Appendix~\ref{ch:validation} for implementation details and tests
performed. In Chapter~\ref{sec:model}, we describe the geometric setup, the
initial- and boundary-conditions and introduce the various effects which enter
our box model of the interstellar medium. The results of the direct simulation
runs are presented in Chapter~\ref{ch:results}. This part has a rather wide
scope and presents very general results, which might have implications for
star formation theory (as an external boundary condition) and possibly even
for cosmological simulations (as a subgrid scale model). From the obtained
kinematic quantities, we can already estimate the $\alpha$~effect on the basis
of SOCA expressions. Moreover, in the case of differential rotation, we
observe an exponentially growing mean field dynamo. In Chapter~\ref{ch:dynamo}
the origin of this dynamo (and the question why no dynamo is observed in the
case of Cartesian shear) is studied in the framework of mean-field theory.

To be able to compare the simulation results with the analytical description,
we infer the $\alpha$~effect by means of passive tracer fields -- this
so-called test-field method has only recently been suggested
\citep{2005AN....326..245S} and allows to simultaneously measure all relevant
tensor components. The particular advantage of the method over previous
statistical approaches lies in the fact that the inversion problem is
well-posed. This, however, requires that test-fields are separately evolved by
an additional set of (passive) induction equations. The implementation of the
method is documented in Appendix~\ref{ch:test_fields}.

\subsubsection{Restrictions}

For the sake of simplicity, we restrict our simulations to the case of a
vanishing vertical net flux, i.e., $\bar{B}_z=0$. This is justified by
observations of edge-on galaxies, which suggest that the vertical component of
the regular magnetic field will only significantly contribute in the far halo
and the fields are generally aligned with the disk
\citep{1995A&A...302..691D}. Consequently, we do not consider a possible
contribution of $\alpha_{zz}$. We want to point out that this is merely a
practical choice and there is no fundamental limitation, which would per se
exclude an investigation of $\alpha_{zz}$ by means of the test-fields method.

In our approach, the dynamo coefficients are assumed to be regular functions
depending on the vertical coordinate $z$ only. Following a more general
paradigm (where non-local effects are captured via integration kernels)
\citet*{2008A&A...482..739B} have recently observed a scale dependence for the
dynamo effect. While this work marks a valuable extension of the test-field
method, the corresponding advances are not reflected in our current
implementation. In addition, we do not consider a possible temporal evolution
of the $\alpha$~parameters, nor a dependence on the magnetic field strength.
Particularly, all dynamo coefficients are measured in the weak field regime,
i.e, neglecting the effects of $\alpha$~quenching. Due to the additional
magnetic contribution in the total pressure, simulations at equipartition
field strength will presumably require higher numerical resolution.
Nevertheless, the saturation mechanism will, of course, be a subject of
subsequent investigations.

For the key input parameters of our model,\footnote{%
  This particularly includes the supernova rate $\sigma$, the rotation
  frequency $\Omega$, the shear parameter $q$, and the midplane pressure
  $p_0$, which are all more ore less constrained from observations
  \citep{2001RvMP...73.1031F}.} %
detailed dependencies are known as functions of the galactocentric radius $R$,
and the galactic height $z$ \citep[see e.g.][]{1998A&A...335..488F}. Although
it seems obvious, and is indeed tempting, to follow a 1+1 approach and derive
dynamo parameters depending on both $R$ and $z$, this endeavour is currently
infeasible due to the high demands in computational resources. It has,
furthermore, been pointed out that long-term variations in the external
parameters (gravity, supernova rate, etc.) should be accounted for on
timescales of a few hundred million years. While this is certainly true, we
currently refrain from unnecessary complications of the model. Because we are
interested in the very fundamental mechanisms of field creation, such a
complication would merely imply an obscuration of the relevant processes. As a
concluding remark, we want to note that more specific constraints and
restrictions, e.g. with respect to particular physical effects, will be
discussed where applicable.



\cleardoublepage
\chapter{Modelling the Interstellar Medium}
\label{sec:model}

\section{General setup} 

\begin{figure}
  \center\includegraphics[width=0.85\columnwidth]{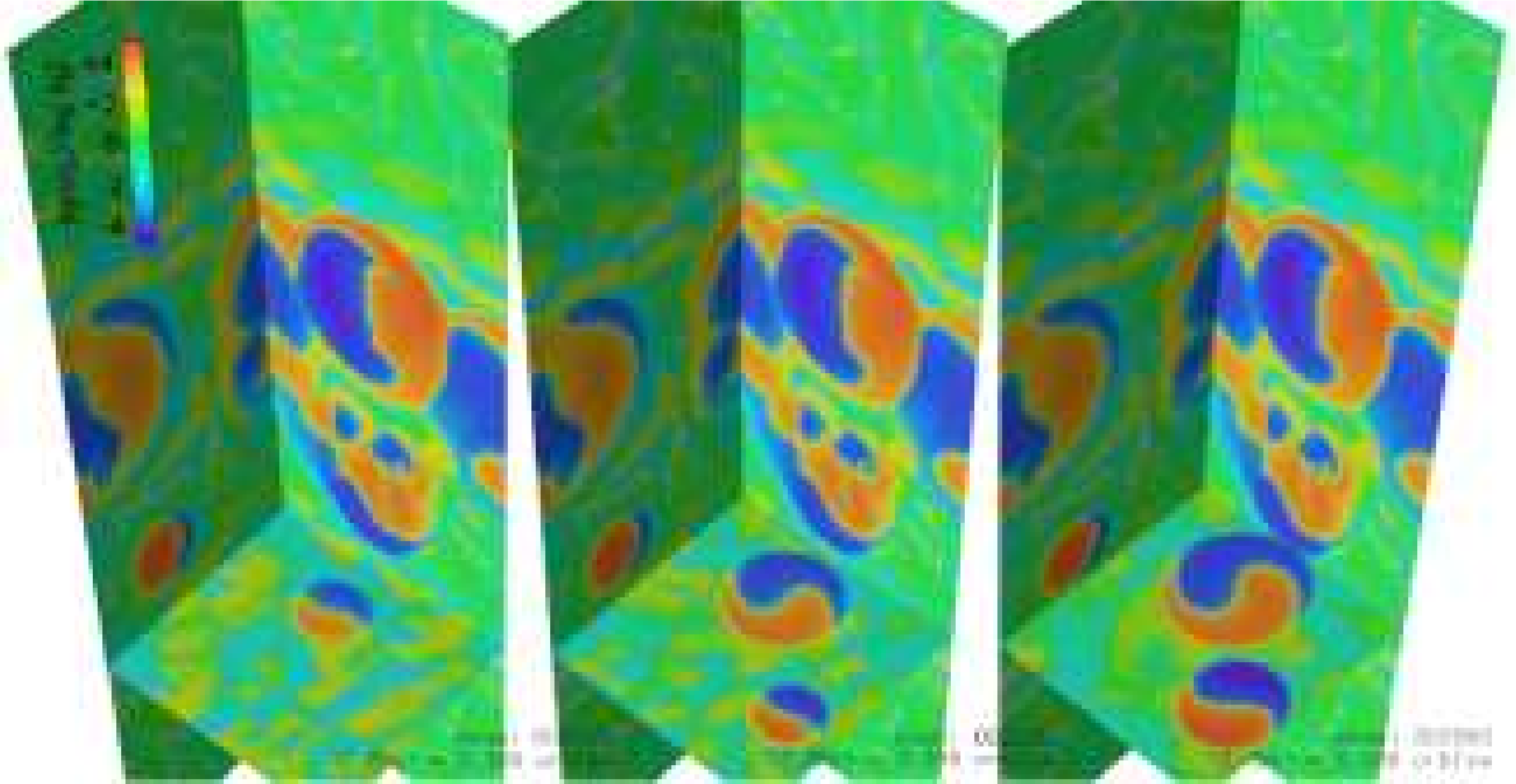}
  \caption[Renderings of the simulation box at an early time]{%
    Renderings of the simulation box at an early time of the evolution. The
    colour coding indicates the signed logarithm of the radial velocity $v_R$.
    The three representations show horizontal slices at different heights $z$.
    At this early stage the single remnants are still visible, highlighting
    the differential twist within the expanding cavities.}
  \label{fig:twist}
\end{figure}

In our endeavour to understand the creation of galactic magnetic fields, we
aim to implement a self-consistent dynamo based on first principles. Only the
direct verification of such a process can serve as a reliable gauge for
mean-field modelling. To study the field amplification by SN feedback, we
simulate the dynamic evolution of the differentially rotating, vertically
stratified, turbulent interstellar medium. Our numerical approach utilises a
three-dimensional setup in the framework of magnetohydrodynamics. For our
simulations we use the NIRVANA%
\footnote{freely available at \url{http://nirvana-code.aip.de}} %
MHD fluid code \citep{2004JCoPh.196..393Z,2005CoPhC.170..153Z}, which employs
state-of-the-art numerical algorithms. The particular discretisation is based
on the Godunov-type central scheme for 2D conservation laws developed by
\citet*{2005SIAM....23..707K}. This scheme has been extended to
three-dimensional MHD and combined with constrained transport
\citep{1988ApJ...332..659E} to solve the induction equation ensuring a
divergence-free evolution of the magnetic field. Within the NIRVANA code, a
semi-discrete approach is applied, where the spatially discretised equations
in the flux conservation form are integrated in time via a third-order
Runge-Kutta scheme.

The physical effects considered cover viscous and resistive terms as well as
thermal conduction and optically thin radiative heating and cooling -- the
latter is particularly important to grasp the heterogeneous nature of the
multi-phase ISM. The adopted computational domain covers a box of $0.8\times
0.8\times 4.0\kpc^3$, vertically centred around the midplane and representing
a local patch of the galactic disk (see Fig.~\ref{fig:twist} for a general
impression). Parameters like the midplane density $\rho_0$, the rotation rate
$\Omega$, the stellar gravitational potential, and the shear parameter $q$ can
be adjusted to suit conditions at varying distance from the galactic centre.

In the scope of the current work, we do not consider effects related to
radiative transport, non-equilibrium chemistry, photo-ionisation, feedback
from protostellar jets, nor anisotropic heat conduction. Compared to the
violent energy input from the SNe, these phenomena are believed to be of
secondary importance for the overall dynamics of the ISM, and thus for the
dynamo process. Moreover, we do not include the self-gravity of the
interstellar plasma, which is thought to only become important on scales that
are well below the current numerical resolution. At the present level of
modelling, we also chose to neglect effects due to a cosmic ray component and
focus on the kinetic and thermal energy input from the SNe. This is not,
because we think cosmic rays do not play a significant role within the ISM,
but rather because we want to understand the responsible effects in a
bottom-up approach. This means that we aim to consider models of increasing
complexity and only proceed to the inclusion of the next relevant effect after
we have substantially understood the contribution of the last.

Combining radiative cooling, differential rotation, and vertical
stratification already seems to blow this philosophy out of proportions.
These effects, however, mark the bare minimum configuration necessary for a SN
driven dynamo to operate. As will become clear, the latter two constituents
serve in generating the needed anisotropy of the turbulence, whereas the role
of the cooling is more subsidiary. Besides its implications for the Galactic
Fountain, the radiation coupling is indispensable to balance the thermal
energy input from the SNe. Simulations based on the injection of kinetic
energy alone \citep{2006MNRAS.370..415M} indeed show that there is no halfway
approach in modelling SN turbulence.

\citet{2004ApJ...605L..33H} treat the cosmic ray gas in the diffusion
approximation and have demonstrated that the associated buoyant instability,
in conjunction with shear, can exponentially amplify the mean magnetic field.
Besides the direct input of kinetic energy via SN explosions, this Parker-type
instability is the most promising dynamo mechanism that has been proposed for
galaxies so far. In regions of lower stellar activity, the magneto-rotational
instability (MRI) has to be considered as a field amplification mechanism
\citep{2004A&A...423L..29D,2007ApJ...663..183P}, too. Due to the limited
resolution of the current models the unstable MRI modes remain only marginally
resolved in the early evolution of the simulations. Whenever applicable we
will try to compare the outcome of our runs with results from MRI- and
CR-simulations.

\subsection{Geometric considerations} 

Unlike in MRI simulations, where the box dimension can be chosen freely by
making the equations non-dimensional, the inclusion of radiative cooling
defines a meaningful physical length scale to our problem. Because we
simultaneously have to resolve the diameter of the supernova remnants along
with the thickness of their thin shell, this already sets considerable
measures on the dynamic range to be covered. While, on one hand, the domain
has to be sufficiently small, such that the assumptions for the local
expansion of the equations of motion are justified, the box, on the other
hand, has to cover the integral scale of the turbulence and the large-scale
flow. In the thick disk, the driving scale of the turbulence is around
$100\pc$. Away from the midplane, where the ambient pressure is comparatively
low, supernova remnants can, however, easily grow to diameters of several
hundred parsecs. Due to the assumed periodicity in the horizontal plane, this
can lead to self-interactions of remnants if the box size is chosen too small.

Per se, such an interaction is not impossible, but because the supernova rate
declines exponentially with height, it is reasonably improbable that a
corresponding ghost remnant appears in the neighbouring domain within the
lifetime of the remnant. In the early models with a horizontal dimension of
$0.5\kpc$, this effect was clearly visible; at the current box size of
$0.8\kpc$ it is much less pronounced. The larger diameter of the box
particularly becomes important when we include differential rotation. This is
because, within the shearing box formalism, the opposite radial boundaries
have an offset velocity $q\Omega L_x$ with respect to each other. This offset
would lead to an additional correlation in the mean electromotive force of an
self-interacting remnant. To exclude this artificial effect as good as
possible we decided to accept the disadvantages of a coarser grid resolution.

Concerning the vertical extent of the domain \citet{2007EAS....23...87D}
stress the importance to cover the so-called disk-halo interaction. In what
the authors term the ``duty-cycle'' of the Galaxy, hot gas can escape the
gravitational potential similarly to an exhaust valve. Colliding shells in the
halo will then lead to density enhancements and, due to the radiative cooling,
to the formation of dense cores; these will rain back towards the midplane,
forming the so-called Galactic Fountain \citep{1980ApJ...236..577B}. Since
vertical transport processes put strong constraints on the proposed SN-driven
dynamo, it is vital to include this disk-halo circulation. The chosen vertical
extent of our model can only be seen as a minimum requirement for this. Due to
the limitations in computing power, a further extension of the vertical
dimension will require adaptive mesh techniques. Albeit we have successfully
tested this in the case without differential rotation, the mesh refinement
leads to prohibitive complications when considering shearing periodicity as
necessary for the case of true galactic rotation.

\subsection{Boundary conditions} 
\label{sec:BCs}

As already mentioned, our model assumes sheared periodic boundary conditions
in the horizontal directions. Although this is the natural choice when one is
interested in the local behaviour of a given flow, one has to keep in mind the
intrinsic limitations. To illustrate this, let us consider an arbitrary
horizontal slice through the computational domain. If we want to compute the
line integral of the electromotive force along two opposite edges, inherently,
the values at two corresponding points are identical. The orientation of the
line elements, however, is opposite. This means, the contributions will
exactly cancel out and, consequently, the line integral over the closed loop
vanishes. Applying Stokes' theorem, this implies that the vertical component
of the magnetic flux is ideally conserved \citep*{1995ApJ...440..742H}.
Because observations indicate that the vertical component of the regular field
will only significantly contribute in the far halo
\citep{1996ARA&A..34..155B}, we here focus on the case with zero vertical flux
and only consider the radial and azimuthal components of the regular magnetic
field.

The boundary conditions in the vertical direction allow the gas to flow out
but will inhibit inflow. Hydrodynamic variables are extrapolated with
vanishing gradients. With the vertical boundaries at $2\kpc$, the issue of
losing matter due to strong outflows is drastically reduced compared to the
original model by %
\citet*[][hereafter \citetalias{1999ApJ...514L..99K}]{1999ApJ...514L..99K},
where this posed a major limitation to an extended temporal evolution.
Depending on the supernova rate, and whether we apply clustering for type~II
SNe (cf. Sec.~\ref{sec:driving}), we only lose about five percent of the
matter per $\Gyr$, indicating that we in fact grasp the essential part of the
Galactic Fountain.

For the magnetic field, we apply vertical boundary conditions where the
transversal components are extrapolated with zero gradient, and the normal
component is reconstructed from the solenoidal constraint. In contrast to the
commonly applied pseudo-vacuum conditions (where the transverse components are
set to zero), this allows magnetic stresses to be exerted on the surface,
i.e., we locally tolerate non-vanishing Poynting fluxes, but avoid artificial
changes in the field topology. Unlike in MRI-simulations, where the vertical
boundaries are a source of complication, we do not observe numerical
difficulties in our present setup.

\subsection{Model equations} 

The equations of resistive MHD are solved in the local shearing box approach.
We apply a co-rotating Cartesian coordinate system with $\xx$, $\yy$, and
$\zz$ being the unit vectors along the radial, azimuthal, and vertical
direction. The conserved quantities $\rho$, $\rho\V$, $e$, and $\B$, i.e.,
density, momentum, total energy, and magnetic field are evolved according to
the following set of non-linear equations:
\begin{eqnarray}
      \partial_t\rho +\nabla\cdt(\rho \V) & = & 0\,, \nonumber\\
      \partial_t(\rho\V) +\nabla\cdt
          [\rho\mathbf{vv}+p^{\star}-\mathbf{BB}] & = &
           - 2 \rho\Omega\zz\tms\V
           + 2\rho\Omega^{2}qx\,\xx \nonumber\\ & & 
           + \rho g(z) \zz +\nabla\cdt\visfl\,, \nonumber\\
      \partial_t e + \nabla\cdt
          [(e\Pl p^{\star})\V\Mn(\V\cdt\B)\B] & = &
          + 2\rho\Omega^{2}qx\xx\cdt\V
          + \rho g(z) \zz\cdt\V \nonumber\\ & &
          + \nabla\cdt[\visfl\V + \eta\B\tms(\nabla\tms\B) + \kappa \nabla T] 
          \nonumber\\ & &
          + \Gamma_{\rm SN} - \rho^2 \Lambda(T) + \rho \Gamma(z)\,, \nonumber\\
      \partial_t \B -\nabla\tms(\V\tms\B -\eta \nabla\tms\B)&=& 0\,,
\end{eqnarray}
with the supplemental solenoidal constraint $\nabla\cdt\B=0$. We adopt units
where the magnetic vacuum permeability $\mu_0$ is set to unity and define the
total pressure $p^{\star}=p+\nicefrac{1}{2}\B^2$, assuming an adiabatic
equation of state, $p = (\gamma-1)\epsilon$, with $\gamma=\nicefrac{5}{3}$.
The thermal energy density $\epsilon$ is computed from the total energy as
\begin{equation}
  \epsilon=e-\nicefrac{1}{2}\rho\V^2 -\nicefrac{1}{2}\B^2\,,
\end{equation}
with the exception of regions with $\epsilon<0.07e$, where we apply a ``dual
energy'' formalism to avoid numerical inaccuracies in the above equation. The
non-ideal fluxes comprise the gradient in the temperature $T$ and the viscous
stress tensor
\begin{equation}
  \visfl =\tilde{\nu}\left(\nabla\V +(\nabla\V)^{\top}
         -\nicefrac{2}{3}(\nabla\cdt\V)\right),
\end{equation}
with $\tilde{\nu}$ the dynamic viscosity parameter. Furthermore, $\eta$
denotes the magnetic diffusivity and $\kappa$ the coefficient of (isotropic)
thermal heat conduction.

The background shear of the flow is characterised by the parameter $q=d \ln
\Omega/d \ln R$, where $R$ is the radius in a cylindrical coordinate system
rooted at the galactic centre. The case $q=-1$ corresponds to a flat rotation
profile as approximately applicable at the solar circle. The terms $2\rho
\Omega^{2}qx\xx\cdot\V\,$ and $\rho g(z) \zz\cdot\V\,$ represent work done
against the tidal and gravitational potential, respectively. Former stems from
the tidal force $2\rho\Omega^{2}qx\,\xx$ arising in the local expansion of the
equations of motion \citep*{1995ApJ...440..742H}. Independent of the shear
rate, we include the effect of the Coriolis force $-2\rho\Omega\zz\times\V$.
To isolate the effects of the shear, we also perform fiducial runs with a
value of $q=0$, representing solid body rotation, and runs with $q=-1$ and
Coriolis forces disabled to study the case of plain shear.

Following \citetalias{1999ApJ...514L..99K}, we use the gravitational potential
of \citet{1989MNRAS.239..605K}, which includes contributions from a stellar
disk and a central halo. The corresponding vertical gravitational acceleration
is
\begin{equation}
  g(z) = -\frac{a_1 z}{\sqrt{z^2+z_0^2}}\ -\ a_2 z, \label{eq:grav}
\end{equation}
with constants $a_1$=$1.42\tms10^{-3}\kpc\Myr^{-2}$,
$a_2$=$5.49\tms10^{-4}\Myr^{-2}$, and $z_0$=$180\pc$.

The additional source terms $\Gamma_{\rm SN}-\rho^2\Lambda(T)+\rho\Gamma(z)$ 
in the total-energy equation represent the thermal energy input due to
supernovae and optically thin radiative cooling/heating and will be described
in section~\ref{sec:energy_sources}.

\subsection{Dissipative terms} 
\label{sec:visc_terms}

As mentioned above, we include a full treatment of the non-ideal terms. The
motivation for this is two-fold: While the main goal is to provide distinct
conditions at the dissipation scale, yielding well defined Reynolds numbers
$\Reyn$ and magnetic Reynolds numbers $\Rm$, we do not want to conceal that
including viscous terms also helps to stabilise the numerical treatment in
regions with high density contrast. To restrict the impact of the viscous
mixing to regions where it is needed and to prevent additional over-cooling
at the cloud interfaces, we scale the dynamic viscosity coefficient
$\tilde{\nu}$ with the density and apply a constant kinematic viscosity of
$\nu=0.5\tms10^{25}\vis$. This approach, which has also been used by
\citet*{2007ApJ...654..945B}, allows the definition of a Prandtl number
$\Pran=\nu/\kappa\,\rho\,c_p$ (with $c_p$ the specific heat capacity at
constant pressure) and magnetic Prandtl number $\Pm=\nu/\eta$ that are
independent of density, and thus constant throughout the domain.

The main reason to introduce thermal conduction is related to the development
of a thermal instability \citep[TI][]{1965ApJ...142..531F} below temperatures
of $\simeq 6000\K$ (cf. Sec.~\ref{sec:cooling} for the definition of the
cooling function used). Since the Field instability, in the inviscid case, has
a finite growth rate in the limit of high wavenumbers, numerical modelling of
this instability is inherently prone to artificial growth of unstable modes at
grid scale. On the other hand, when considering a finite value for the thermal
conduction coefficient $\kappa$, unstable modes are substantially damped below
the so-called Field length
\begin{equation}
  \lambda_{\rm F} = 2\pi \left[ 
    \frac{\rho^2\Lambda}{\kappa T} (1-\beta) \right]^{-1/2}\,,
\end{equation}
in the case of the cooling function $\Lambda$ only depending on the
temperature $T$, and with $\beta=d\ln\Lambda/d\ln T$. To prevent unphysical
growth of the instability and to guarantee a converged solution,
\citet{2004ApJ...602L..25K} have introduced the so-called Field condition,
which states that $\lambda_{\rm F}$ has to be resolved with at least three
grid cells; accordingly, we chose a value of $\kappa_0=4.08\tms 10^8 \erg
\cm^{-1} \K^{-1}\s^{-1}$. To avoid a further suppression of the numerical
timestep in regions of high temperature gradient, we scale this coefficient
with the mass density\footnote{%
  This is, of course, contrary to the Spitzer scaling $\sim T^{5/2}$.} %
and prescribe $\kappa=\kappa_0\,\rho/\rho_0$, yielding a constant Prandtl
number of $\Pran\simeq 4$.

To date, the condition introduced by \citet{2004ApJ...602L..25K} is widely
disregarded by many authors. Notable exceptions are the MRI simulations by
\citet{2004ApJ...601..905P} and a TI study by \citet*{2007ApJ...654..945B}.
The opposite standpoint, represented by a number of authors \citep[see
e.g.][]{2005ApJ...630..911G,2006ApJ...653.1266J}, is to neglect thermal
conduction. The general argumentation is that numerical diffusion defines a
``numerical Field-length'' that is thought to sufficiently damp small-scale
modes of the instability. \citet{2004A&A...425..899D} argue that molecular
heat conduction is suppressed perpendicular to the magnetic field lines (and
thus also isotropically for sufficiently tangled fields) and that turbulent
transport takes an important role.

Unlike in laboratory plasmas or within the sun, the magnetic Prandtl number
$\Pm$ of the ISM is thought to be very high -- \citet{2005PhR...417....1B}
estimate a value of $\Pm\simeq 4\tms10^{11}$. In numerical simulations with
limited dynamic range, one is, however, usually restricted to values close to
unity. This is because both the viscous and resistive length scales have to be
resolved on the numerical grid. Since many magnetic phenomena depend
critically on this number, it is important to prescribe a distinct value for
$\Pm$. Neglecting viscous terms, $\Pm$ is determined by the intrinsic
properties of the numerical scheme, which are hardly traceable. The
implications of this issue on turbulence caused by MRI have recently been
studied by \citet{2007A&A...476.1113F,2007A&A...476.1123F}.

For practical purposes, we choose $\Pm=2.5$ equivalent $\eta= 0.2\tms
10^{25}\vis$, which is still two orders of magnitude smaller than the expected
turbulent diffusivities $\eta_{\rm t}$. \citet*{2007ApJ...668..110O} apply a
similar value of $\eta=0.3\tms10^{25}\vis$ in their reference model of the CR
driven buoyant instability. In fact, their dynamo crucially relies on the
presence of a molecular diffusivity, and the authors find the efficiency of
the field amplification to depend on this parameter. With a reference rotation
frequency of $25\kms\kpc^{-1}$ and a box dimension of $L_x=0.8\kpc$, we yield
Reynolds numbers $\Reyn=L_x^2\Omega/\nu \simeq 1000$ and $\Rm\simeq2500$, for
our standard run.

The role of the microscopic diffusivity in defining $\Rm$ in numerical
simulations poses a very subtle question. The diffusive time scale associated
with the low values of the magnetic diffusivity within the ISM by far exceeds
the Hubble time. This means that any efficient mean-field dynamo will have to
operate on a time scale different than the diffusive one -- this is usually
termed ``fast dynamo'', as opposed to a slow dynamo, which depends on the
reconnection on microscale. In view of the limited magnetic Reynolds number of
numerical simulations, this implies that a necessary criterion for the
robustness of the field amplification mechanism at realistic $\Rm$ is the
persistence of the effect for low values of $\eta$.

\section{Energy source terms} 
\label{sec:energy_sources}

\begin{figure}
  \center\includegraphics[width=\columnwidth]{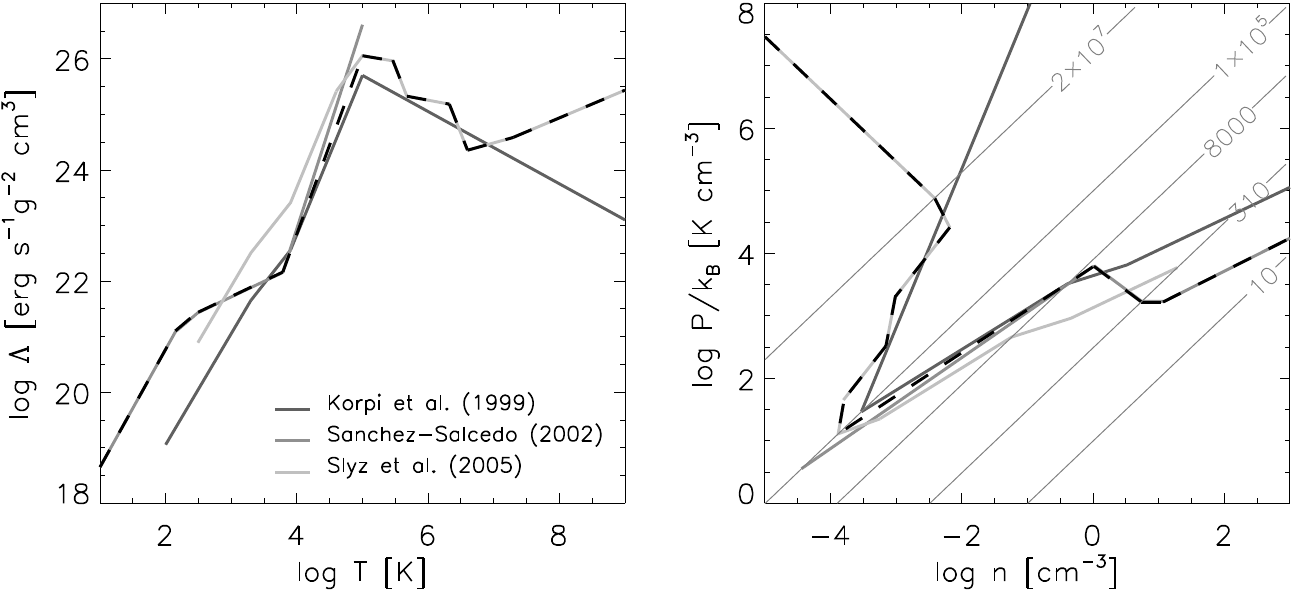}
  \caption[Optically thin cooling functions and corresponding
           equilibrium curves]{%
      Optically thin cooling functions (left panel) and corresponding
      equilibrium curves, for $\Gamma$=$\Gamma_0$, in phase space (right
      panel). Our model is indicated by a dashed line and combines branches of
      the curves used by \citet{2002ApJ...577..768S} for ${T\le6102\K}$ and
      \citet{2005MNRAS.356..737S} for $T\ge10^5\K$.}
    \label{fig:cooling}
\end{figure}

Since one of the main goals of this work is to predict the vertical structure
of the velocity dispersion and mean flow, artificial forcing naturally has to
be excluded as a driving mechanism. In an attempt to model interstellar
turbulence without the need for a complex thermodynamical treatment
\citet{2006MNRAS.370..415M} apply a forcing that is based on localised
expansion waves. This approach, however, is limited to subsonic flows and is
found to not produce any vorticity and hence helicity. This shows that there
is no intermediate level of modelling and one has to consider driving via the
injection of thermal energy -- which, in turn, makes it mandatory to include
radiative cooling.

\subsection{Radiative cooling and ambient heating} 
\label{sec:cooling}

\begin{table}
  \begin{minipage}[c]{0.66\columnwidth} \small
    \begin{tabular}{ccc}\heading{$T_{i}\ [\K]$&
        $\Lambda_{i}\ [\erg\,\s^{-1}\g^{-2}\cm^{3}\K^{-\beta_i}]$&$\beta_{i}$}
      10&3.420$\times10^{16}$&~2.12\\ 
      141&9.100$\times10^{18}$&~1.00\\ 
      313&1.110$\times10^{20}$&~0.56\\ 
      6102&1.064$\times10^{10}$&~3.21\\ 
      $10^{5}$&1.147$\times10^{27}$&-0.20\\ 
      2.88$\times10^{5}$&2.290$\times10^{42}$&-3.00\\ 
      4.73$\times10^{5}$&3.800$\times10^{26}$&-0.22\\ 
      2.11$\times10^{6}$&1.445$\times10^{44}$&-3.00\\ 
      3.98$\times10^{6}$&1.513$\times10^{22}$&~0.33\\ 
      2.00$\times10^{7}$&8.706$\times10^{20}$&~0.50\\\hline
    \end{tabular}
  \end{minipage}
  \begin{minipage}[t]{0.33\columnwidth}
    \caption[Parameters for the prescribed cooling function]{%
      Parameters for the prescribed cooling function as defined by 
      Eq.~(\ref{cooling}) and illustrated in Fig.~\ref{fig:cooling}.}
    \label{tab:cooling}
  \end{minipage}
\end{table}

We treat the interstellar medium as an optically thin plasma and prescribe the
coupling to the radiation field via a piecewise power law of the form:
\begin{equation}
  \Lambda(T) = \Lambda_i\,T^{\beta_i},\quad\text{for } T_i \le T < T_{i+1}\,.
  \label{cooling}
\end{equation}
The used parameters are documented in Table~\ref{tab:cooling} and are
essentially a combination of the cooling curves used by
\citet*{2002ApJ...577..768S} for $T \le 6102\K$ and
\citet{2005MNRAS.356..737S} for $T \ge 10^5\K$. In the latter work, the branch
for the high temperature range is adopted from \citet{1987ApJ...320...32S},
while in the former the neutral phase is based on equilibrium models by
\citet{1995ApJ...443..152W}. We want to mention that, improving over previous
work of \citetalias{1999ApJ...514L..99K}, we include the thermally unstable
range between $141\K$ and $6102\K$, which, via a thermal instability, leads to
the formation of a cold ISM phase \citep{1965ApJ...142..531F}. This comprises
an important step towards the proper inclusion of the Galactic Fountain, which
might also have implications on the vertical magnetic field transport.

The equilibrium curve in the right panel of Figure~\ref{fig:cooling} divides
the plane of the phase space into two regions: Above the curve, the plasma is
cooled, while below the curve, it is subject to net heating. If we now
consider a parcel of fluid sitting on the equilibrium curve and displace it
by slightly increasing (decreasing) its density while keeping the pressure
constant, we will accordingly reduce (increase) its temperature. In regions
with positive slope $\beta-1$ of the equilibrium curve, this implies that we
arrive at a region of net heating (cooling) and the disturbed parcel will
return to its stable equilibrium state. In the temperature interval between
$141\K$ and $6102\K$, however, the slope of the curve is negative, which means
that the fluid that is displaced towards the cold (warm) side of the curve
will experience net cooling (heating) and will be further driven away from
its equilibrium state until it reaches an adjacent stable branch. This
argumentation can be similarly repeated for isochoric and adiabatic
disturbances \citep*{2002ApJ...577..768S}, yielding the following conditions
for TI:

\begin{equation}
  \beta=d\ln\Lambda/d\ln T <\ \begin{cases}
    \quad 1 & \text{isobaric} \\
    \quad 0 & \text{isochoric} \\
    \ (1-\gamma)^{-1} & \text{adiabatic.}
  \end{cases}
\end{equation}

There are various effects that contribute to the diffuse heating of the
interstellar gas. The most important of these are thought to be photoelectric
heating and ionising radiation from OB stars. As the stellar component is
exponentially distributed in the galactic disk, and the photons can escape the
disk through a diffusive process, one has to assume some form of vertical
dependence for the background heating rate. Since we want to focus on the
dynamical rather than the thermal evolution of the ISM, we refer the reader to
section 2.2 of \citet{2006ApJ...653.1266J}, where the issue is discussed in
some detail. For practical purposes we use a prescription of the form:
\begin{equation}
  \Gamma(z) = \Gamma_0 \times 
  \begin{cases}
    e^{-\frac{z^2}{2z_0\,H_\Gamma}} & \text{if}\ |z| \le z_0 \\
    e^{ \frac{z_0}{2H_\Gamma}}\,
    (e^{-\frac{z}{H_\Gamma}} + 10^{-5}) & \text{otherwise},
  \end{cases}
\end{equation}
where we adopt $H_\Gamma=300\pc$ and choose $z_0=H_\Gamma/5$. The midplane
heating rate is set to $\Gamma_0=0.015\erg\,\s^{-1}$. Spatial variations of
the heating rate (due to the inhomogeneous character of the ISM) have to be
neglected since the required radiative transfer methods are beyond the scope
of this work. Because of the dominance of the SN driving, this does not pose a
serious limitation to our model.

\subsection{Supernova driving} 
\label{sec:driving}

In our simulations, the driving of the turbulence is accomplished via
localised injections of thermal energy, modelling supernova explosions. The
events are discrete in time and spatially confined -- the details of the
injection process are described in Appendix~\ref{sec:isolated_SNR}. In the
basic model, the remnants are exponentially distributed in the vertical
direction with a scale height of $325\pc$ for type~I and $90\pc$ for type~II
SNe. The reference galactic frequencies are $\sigma_{\rm I} = 4\Myr^{-1}
\kpc^{-2}$ and $\sigma_{\rm II} = 30\Myr^{-1} \kpc^{-2}$, respectively. The
associated explosion energies are $10^{51}$ and $1.14\tms10^{51}\erg$
\citep{2001RvMP...73.1031F}. The higher energy for type~II SNe accounts for
the contribution of a stellar wind of the massive progenitor.

In early simulation runs, we found that a static vertical distribution of SNe
gave rise to a swing amplification of the basic vertical oscillation mode
within the external gravitational potential. This was due to the fact that
whenever the density peak, and with it the centre of mass (CM), was deflected
from $z=0$, the peak of the SN driving would remain in the midplane and thus
further drive away the peak. The amplified base mode had an oscillation period
of roughly $100\Myr$. In a first approach, we inhibited this undesired effect
by distributing the SNe around the vertical centre of mass, rather than
statically around $z=0$. This cured the amplification of the CM mode, but, at
a later time, introduced an artificial split-up of the disk. In this
situation, two distinct portions of the disk would oscillate at opposite phase
leaving the centre of mass, and thus the centre of the driving, unaffected.

In an alternative approach, we introduced an artifical damping force
$f_c=2\,v_z^{\rm CM}(a_1/z_0+a_2)^{1/2}$, with constants taken from
Equation~(\ref{eq:grav}), to critically damp the harmonic part of the
oscillation. This also removed the swing amplification, but did not inhibit
the fragmentation of the thick disk, either. Finally, we decided to drop the
concept of a static SN distribution and adopted a prescription where the
vertical distribution of the type~II SNe was determined by the gas density
profile. This is justified by the assumption that the star formation rate is
proportional to the local gas density and by the short lifetimes of massive
stars compared to the evolution time of our model. To account for a finite
lifetime, we applied a $10\Myr$ running average for the density profile. With
the newly defined SN distribution, both the artificial oscillation and the
disruption of the thick disk can be successfully avoided. Due to the
additional kinetic pressure from the SNe, and depending on the supernova rate,
the disk now smoothly disperses (cf. Sec.~\ref{sec:gas_prof}).

Within our model, we make an important distinction between type~I and type~II
SNe: The latter are spatially clustered by the (artificial) constraint that
the density at the explosion site be above average (with respect to a
horizontal slab) while the former are spatially uncorrelated
\citepalias{1999ApJ...514L..99K}. Although this may seem very crude, it gives
a fraction of clustered events comparable with observations
\citep{2001RvMP...73.1031F}. The reference simulation with clustering switched
off reveals that the general morphology is affected quite strongly by the
clustering, which indicates the importance of this effect. This in mind, a
possible extension for the current model will be a more realistic prescription
for the representation of clustered events. \citet{2005A&A...436..585D}, for
example, use a rather sophisticated approach, where a number of stellar
particles is formed whenever a certain local star-forming criterion is
fulfilled. The created stellar particles are distributed according to an
assumed initial mass function and then dispersed with random velocities of
the order of $5\kms$, modelling an OB association. The position of the
particles is evolved over a typical stellar lifetime to determine the position
of the new SNRs.

\section{The initial model} 
\label{sec:strat}

Previous stratified ISM models including SNe
\citepalias{1999ApJ...514L..99K,2005A&A...436..585D,2006ApJ...653.1266J} and
MRI \citep{2007ApJ...663..183P} all start from an isothermal initial state at
a prescribed temperature. The main drawback of this approach is that the
isothermal stratification is not in balance with respect to the cooling. If
the gas is allowed to cool to its equilibrium state, the hydrostatic condition
is violated, and the disk will instantaneously collapse. In the subsequent
evolution, the dynamic pressure from SNe or MRI can, of course, balance this
process, but the unphysical initial disturbance remains.

The gravitational energy released in the early phase of the collapse is on the
order of the energy deposited by a single supernova. The disturbance, however,
has a low wavenumber and will thus be damped very inefficiently. This means
that the model has to be evolved long enough to erase all the traces from the
collapse. In order to keep the initial fingerprint as gentle as possible, we
propose a more sophisticated initial model, where the vertical profiles of the
density and pressure are numerically integrated to be in combined hydrostatic
\emph{and} radiative equilibrium. To obtain such a dual solution, we deploy
the slope of the radiative equilibrium curve (shown in the right panel of
Fig.~\ref{fig:cooling}) as an effective equation of state.

Balancing the pressure gradient across a length ${\rm d}z$ with the weight of
the corresponding fluid parcel, we arrive at the following differential
equation, defining the vertical density stratification:
\begin{eqnarray}
  \frac{{\rm d}\rho}{{\rm d}z} & = &  
  \left( \rho\,g(z) - \frac{\partial p}{\partial z} \right)
  \left( \frac{\partial p}{\partial \rho} \right)^{-1}\,.
  \label{eq:drhodz}
\end{eqnarray}
In addition to the explicit dependence on the gravitational potential $g(z)$,
we have to consider the implicit dependencies, hidden in the assumed equation
of state, given by $p_{\rm eq}(\rho,z)$. This pressure is derived from the
equilibrium temperature $T_{\rm eq}(\rho,z)$, which we obtain from the balance
of heating and cooling:
\begin{equation}
  \rho\,\Lambda_i\,T^{\beta(T)}=\Gamma(z)\,.
\end{equation}
This equation is solved by means of an iterative root finder and, via
$\Gamma(z)$, explicitly depends on the vertical coordinate $z$. The actual
partial derivatives, which further depend on the logarithmic slope $\beta$ of
the cooling function, are then given by
\begin{equation}
  \frac{\partial p}{\partial \rho} =  
     (1-\beta^{-1})\,\frac{p}{\rho} \qquad\textrm{and}\qquad
  \frac{\partial p}{\partial z}    = 
     \frac{p}{\beta\,\Gamma(z)}
     \frac{{\rm d}\Gamma(z)}{{\rm d}z}\label{eq:dp}\,.
\end{equation}
Substituting (\ref{eq:dp}) into Equation~(\ref{eq:drhodz}) results in an
ordinary differential equation, which we numerically integrate with a
second-order Runge-Kutta method.%
\footnote{ To avoid the discontinuities in the piecewise constant slope
  $\beta(T)$, we smear out the corresponding steps via transfer functions and
  sample the resulting curve to a cubic lookup table.}%

\begin{figure}
  \begin{minipage}[b]{0.66\columnwidth}
    \includegraphics[width=0.9\columnwidth]{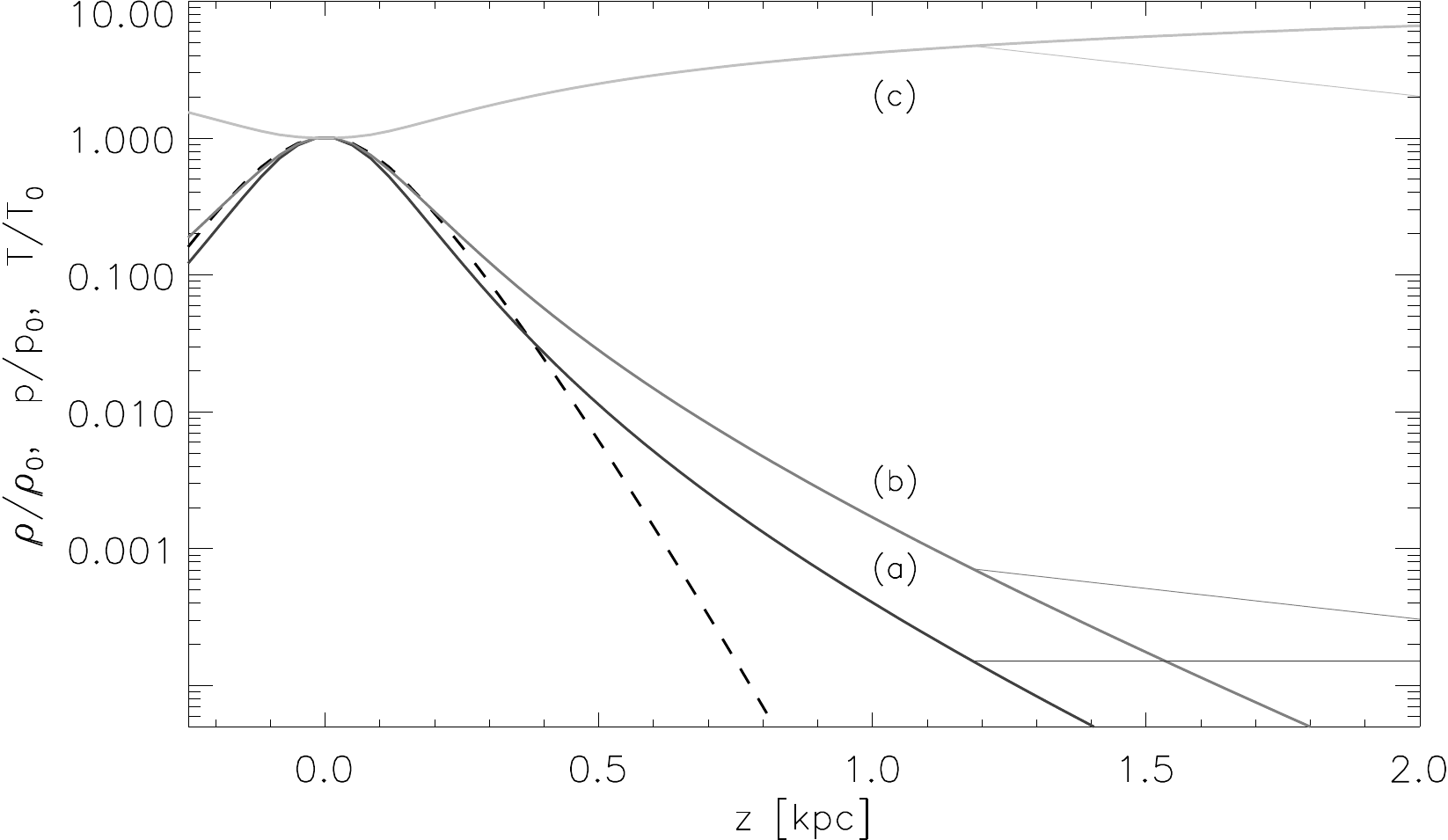}
  \end{minipage}
  \begin{minipage}[b]{0.33\columnwidth}
    \vskip-4ex%
    \caption[Vertical profiles of initial density, pressure, and temperature]{%
      Vertical profiles of initial density (a), pressure (b), and temperature
      (c) compared to the isothermal solution (dashed line). For $z>1.2\kpc$
      we also plot the curves for the truncated outer density profile.
      \vspace{10pt}}
  \end{minipage}
  \label{fig:strat}
\end{figure}

The computed profiles are shown in Figure~\ref{fig:strat}, and one can see
that, beyond $z\simeq 0.5\kpc$, the radiatively stable profiles are
considerably flatter than the isothermal solution, whereas for the inner disk,
the density profile is somewhat steeper. Due to these differences, the
temperature varies by a factor of about five. To avoid unphysically high
density contrasts we limit the density to a typical intergalactic value. The
exact number is taken from the lower left tip of the equilibrium curve in the
right panel of Figure~\ref{fig:cooling}. We chose this value since there
cannot be a stable stratification beyond this point. We want to remark that
this ambient density is only set during the initial setup, i.e., we do not
restrict any of the fluid variables during the evolution of our model. The
variation of the pressure in the flat part of the density profile is due to
the implicit dependence on the heating rate.

Without setting a lower limit on the density, the profiles can be evolved
absolutely stable, i.e., the kinetic energy remains five orders of magnitude
smaller than the thermal energy in the box over many dynamical time scales.
This is due to the fact that the cooling times are extremely long in the low
density regions, where we would actually leave the stable branch of the
cooling function. Since we apply an ambient density, we essentially violate
hydrostatic equilibrium for the outer part of our disk, which results in
material falling inwards. This effect is, however, by far not as dramatic as
the complete collapse of the disk in the isothermal case. Therefore we are
confident to reach a steady state with the initial signatures erased
considerably earlier than in comparable models.



\cleardoublepage
\chapter{Simulation Results}
\label{ch:results}

\section{General evolution} 

In the following, we will discuss the outcome of the simulations conducted
during the last three years. All the presented computations have been
performed on the \texttt{sanssouci} (256 CPUs) and \texttt{babel} (560 CPUs)
Beowulf-clusters at the Astrophysical Institute Potsdam (AIP). The total
amount of computing power that went into the simulations is on the order of
$10^6$ CPU hours.

An overview of the various models can be found in Table~\ref{tab:models},
where we also introduce the basic naming scheme according to the most
fundamental parameters of our model: the rotation rate $\Omega$ and the
supernova frequency $\sigma$. While we measure the former in units of
$\Omega_0=25\kms\kpc^{-1}$, the latter is a multiple of the reciprocal sum
$\sigma_0 = 1/\left( 1/\sigma_{\rm I} + 1/\sigma_{\rm II} \right)$ of the
corresponding frequencies $\sigma_{\rm I} = 4\Myr^{-1} \kpc^{-2}$ and
$\sigma_{\rm II}=30\Myr^{-1} \kpc^{-2}$.

The listed models can be subdivided into distinct groups, which serve the
investigation of certain aspects of our setup. The subset Q4, H4, F4, for
example, allows to study the dependence on the supernova frequency, whereas we
will use the sequence F1--F8 to explore the dependence on the rotation rate.
The T4-series of models marks a very early stage of the project and is
affected by some deficiencies, which have been resolved in later runs. The
main focus of these simulations lies in the examination of morphological
features with respect to the clustering (T4--NCL), the exclusion of type~I SNe
(T4--SNII), and the effect of small-scale magnetic fields (T4--KIN, which
implements the kinematic case $\beta_{\rm P}=\infty$).

For the ``standard'' model F4, we conducted a series of comparison runs to
study fundamentally different geometrical situations. Model F4--SHR is
identical to model F4, but has the Coriolis forces disabled.\footnote{%
  Since $\Omega$ still enters the definition of the shear rate $q\Omega$, we
  keep the nomenclature ``F4'', in this case.} %
This marks the transition from differential rotation to Cartesian shear.
Because the curvature terms vanish in the shearing box approximation, the
linear profile of the background shear is identical for both geometries, i.e.,
the only difference lies in the dynamics of the flow. In a complementary
setup, labelled F4--ROT, we study the case of solid body rotation.

In addition, we investigate the case without thermal instability (T4--noTI),
and models with stronger and weaker external gravitational forces,
respectively. To compute spectra at a higher resolution and to demonstrate
that the stratification of the disk is in fact relevant for the mean-field
dynamo, we also performed a non-stratified box model F4--BOX
\citep[cf.][]{2004ApJ...617..339B} at a resolution of $256^3$ grid cells.

\begin{table}\begin{center}
\begin{tabular}{lcccccccccc}\heading{
& domain $[\kpc\,]$& grid& $\Delta\ [\pc\,]$& SNe& cl.& $g_z$& 
$\!\sigma/\sigma_0\!$ & $\!\Omega/\Omega_0\!$ & $q$ & $\beta_{\rm P}$}\\[-8pt]
Q4       & $0.8^2,\pm 2.133$& $96^2\tms512$& 8.3& 
         I+II& \yes& \yes& 0.25& 4.0& -1& $2\tms10^7$\\
H4       & $0.8^2,\pm 2.133$& $96^2\tms512$& 8.3& 
         I+II& \yes& \yes& 0.50& 4.0& -1& $2\tms10^7$\\
T4       & $0.8^2,\pm 2.000$& $96^2\tms480$& 8.3& 
         I+II& \yes& \yes& 0.75& 4.0&  0& $2000$\\[8pt]
T4--NCL  & $0.8^2,\pm 2.000$& $96^2\tms480$& 8.3& 
         I+II& \no & \yes& 0.75& 4.0&  0& $2000$\\
T4--SNII & $0.8^2,\pm 2.000$& $96^2\tms480$& 8.3& 
         II&   \yes& \yes& 0.75& 4.0&  0& $2000$\\
T4--KIN  & $0.8^2,\pm 2.000$& $96^2\tms480$& 8.3& 
         I+II& \yes& \yes& 0.75& 4.0&  0& $\infty$\\[8pt]
F1       & $0.8^2,\pm 2.133$& $96^2\tms512$& 8.3& 
         I+II& \yes& \yes& 1.00& 1.0& -1& $2\tms10^7$\\
F2       & $0.8^2,\pm 2.133$& $96^2\tms512$& 8.3& 
         I+II& \yes& \yes& 1.00& 2.0& -1& $2\tms10^7$\\
F4       & $0.8^2,\pm 2.133$& $96^2\tms512$& 8.3& 
         I+II& \yes& \yes& 1.00& 4.0& -1& $2\tms10^7$\\
F8       & $0.8^2,\pm 2.133$& $96^2\tms512$& 8.3& 
         I+II& \yes& \yes& 1.00& 8.0&-1& $2\tms10^7$\\[8pt]
F4--ROT  & $0.8^2,\pm 2.000$& $96^2\tms480$& 8.3& 
         I+II& \yes& \yes& 1.00& 4.0&  0& $2\tms10^7$\\
F4--SHR  & $0.8^2,\pm 2.133$& $96^2\tms512$& 8.3& 
         I+II& \yes& \yes& 1.00& -- & -1& $2\tms10^7$\\
F4--BOX  & $0.4^3$          & $256^3$      & 1.6& 
         I+II& \yes& \no & 1.00& 4.0&  0& $10^5$\\[4pt]
\hline
\end{tabular}
\end{center}
\caption[Overview of conducted models]{%
  Overview of conducted models. The letters 'Q', 'H', 'T', and 'F'
  indicate quarter, half, three-quarter, and full SN rate $\sigma_0$,
  respectively, whereas numbers give the rotation rate in units of $\Omega_0$. 
  Clustering (column 'cl.') applies to type~II SNe only. To avoid artificial
  anisotropies, the grid spacing $\Delta$ is kept constant in all directions.}
\label{tab:models}
\end{table}

\subsection{Buildup of turbulence} \label{sec:general_results} 

\begin{figure}
  \center\includegraphics[width=\columnwidth]{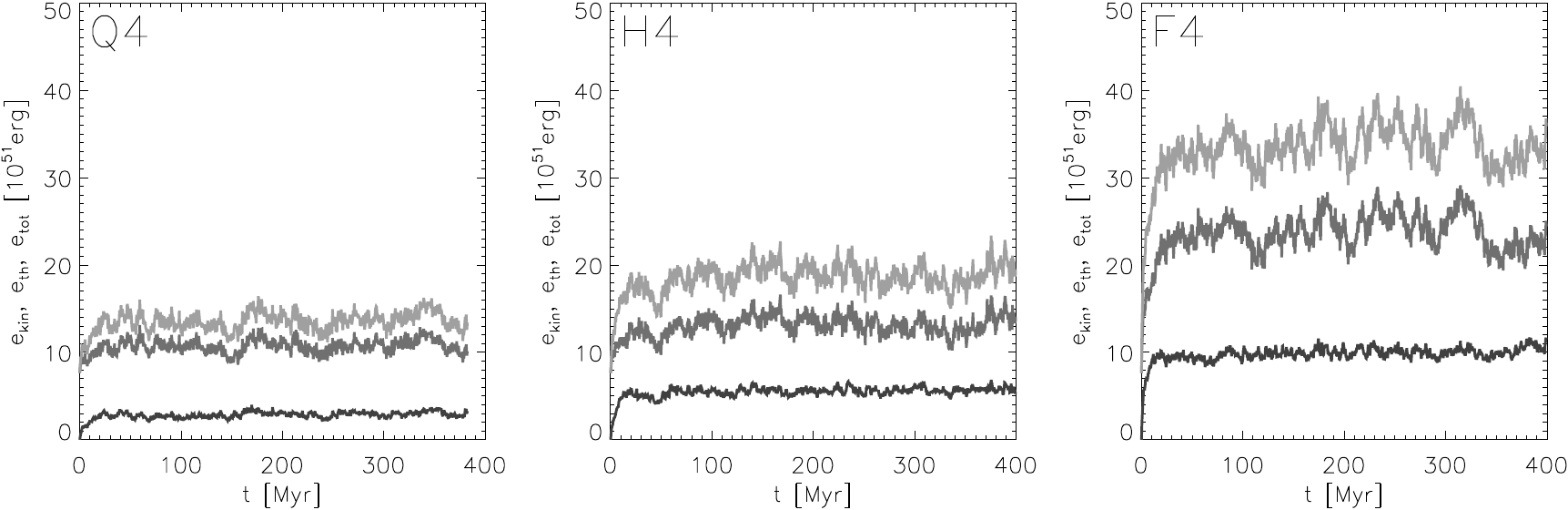}
  \caption[Evolution of kinetic, thermal, and total energy for models
    Q4, H4, and F4]{%
      Temporal evolution of kinetic, thermal, and total energy for models Q4,
      H4, and F4. A steady state is reached after $\simeq 50\Myr$ showing
      fluctuations on a $10\%$ level.}
  \label{fig:ene_QHF4}
\end{figure}

In the models of %
\citet[][hereafter \citetalias{2006ApJ...653.1266J}]{2006ApJ...653.1266J} and
\citet[][hereafter \citetalias{2005A&A...436..585D}]{2005A&A...436..585D}, the
disk is initially out of equilibrium with respect to the cooling. The reason
for this is that the authors computed the hydrostatic equilibrium for an
isothermal configuration. However, from the right panel of
Figure~\ref{fig:cooling} one can see that the isothermal contours are
considerably steeper than the equilibrium curve. This means that the gas,
especially in the dense region near the midplane, will rapidly relax towards
its thermal equilibrium state, consequently violating the hydrostatic balance.
The initiated collapse, in turn, will induce strong shock waves bouncing back
from the midplane. Since our initial model is in thermal equilibrium for
heights up to $1.2\kpc$ (cf. Sec.~\ref{sec:strat}), we do not observe any
initial collapse, but only slight accretion of low density material at the
outer boundaries. Turbulence builds up smoothly, and a large fraction of the
volume has been reached by at least one explosion after about $20\Myr$. After
$50\Myr$, the turbulence reaches a quasi stationary state, and the kinetic
energy contained within the box reaches values of $2.9 (\pm0.3)$, $5.6
(\pm0.4)$, and $10.0 (\pm0.5) \tms 10^{51}\erg$ for the models Q4, H4, and F4
(see Fig.~\ref{fig:ene_QHF4}). The thermal energy settles at a level of $10.6
(\pm0.8)$, $13.4 (\pm1.0)$, and $24.2 (\pm1.8)$, respectively. For the models
H4 and F4, the ratio between the kinetic and thermal energy has the same value
of $\simeq 0.4$, while the ratio is somewhat lower for model Q4. The kinetic
energy deposited by the SNe scales linearly with the SN rate.

\subsection{Disk morphology} 
\label{sec:morphology}

\begin{figure}
  \center\includegraphics[width=0.95\columnwidth]{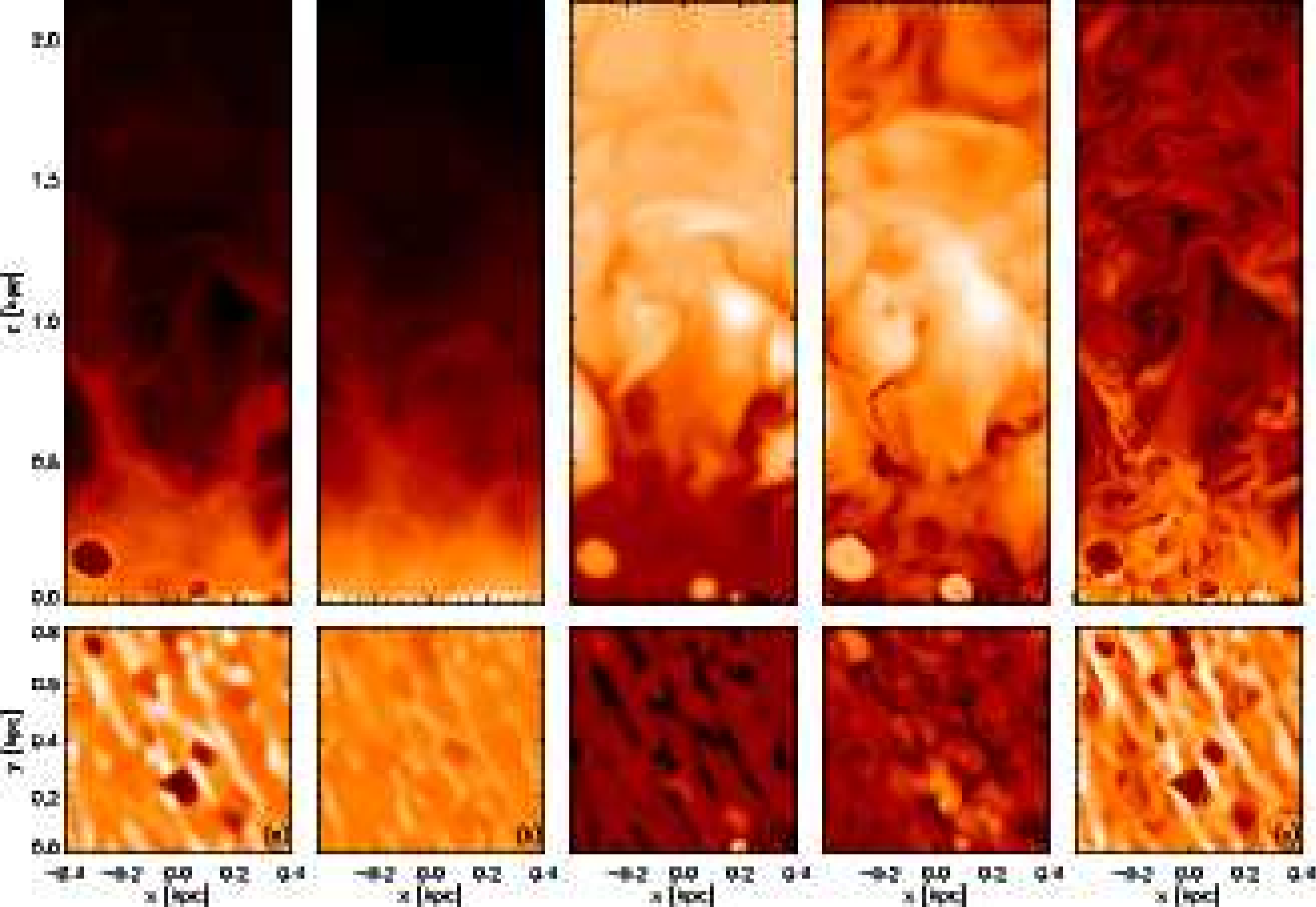}
  \caption[Renderings of the simulation box for model Q4]{%
    Vertical slices of the top half of model Q4 (upper panels) and horizontal
    slices through the midplane (lower panels), at $t=161\Myr$. Quantities
    shown are: (a) number density $[\cm^{-3}]$, (b) column density
    $[\cm^{-2}]$, (c) temperature $[\K]$, (d) velocity $[\kms]$, and (e)
    magnetic field $[\muG]$. The logarithmic colour coding extends over ranges
    $[-4.38, 1.20]$, $[17.6,21.6]$, $[ 2.02, 6.98]$, $[-0.96, 2.36]$, and
    $[-5.21,-0.86]$, respectively.}
  \label{fig:slc_Q4}
  \center\includegraphics[width=0.95\columnwidth]{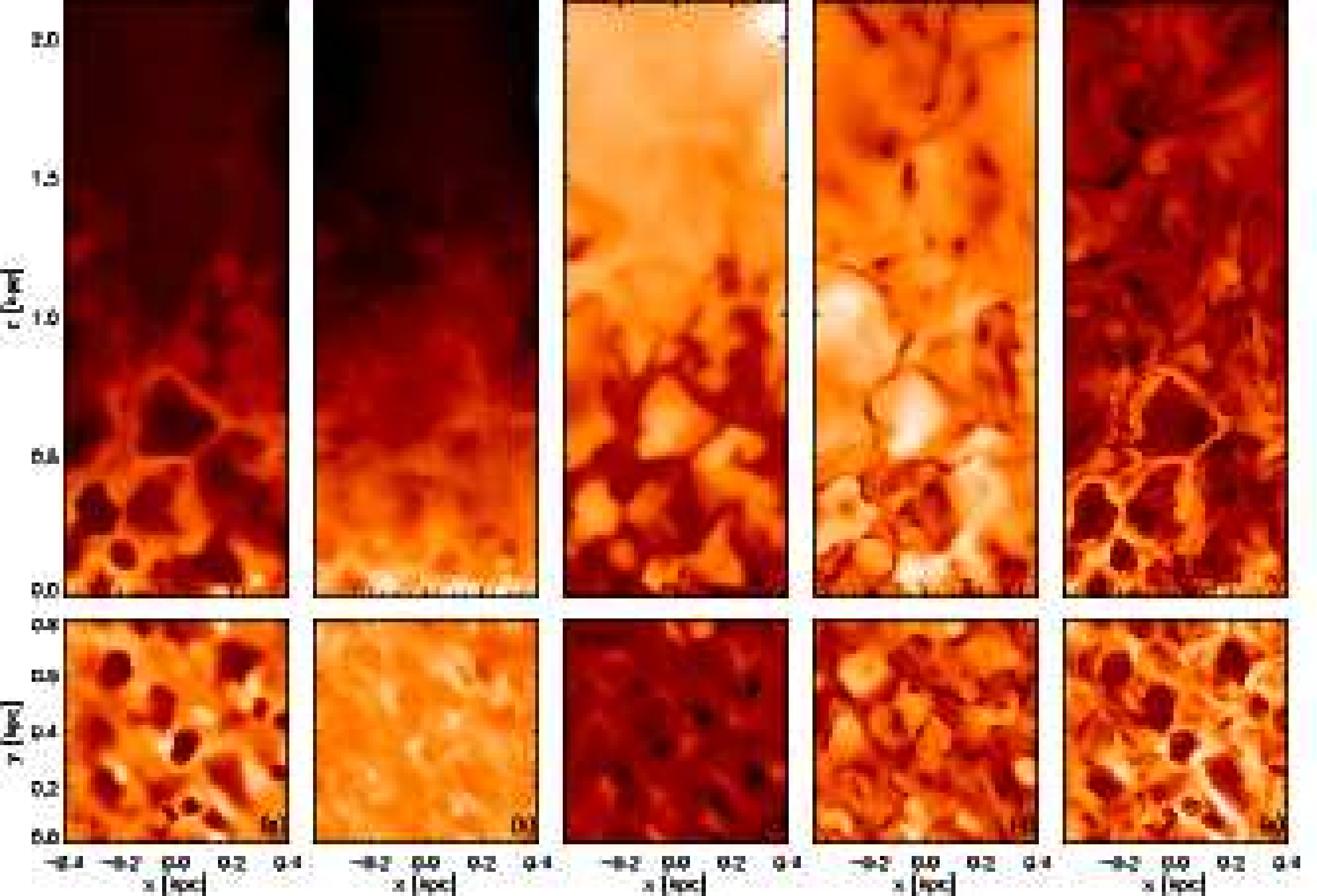}
  \caption[\same{ for model F4}]{%
    Same as Fig.~\ref{fig:slc_Q4}, but for model F4 at a time 
    $t=164\Myr$. The logarithmic colour coding extends over ranges $[-4.45,
    1.10]$, $[17.20,21.83]$, $[ 2.07, 6.66]$, $[-0.85, 2.39]$, $[-5.47,-0.91]$.}
  \label{fig:slc_F4}
\end{figure}

Figure~\ref{fig:slc_Q4} shows vertical and horizontal slices through the
simulation box of model Q4 at a time $t=161\Myr$. Most of the material is
contained in cold clumps forming a $150$--$200\pc$ wide disk. Close to the
midplane, the network of clumps and filaments is permeated by strong shocks
from the SNe, which are blowing hot cavities. There is a number of interesting
ideas in conjunction with the formation and lifetime of molecular clouds, and
whether turbulence within those can be driven by external shocks. We currently
cannot aid these discussions because we are lacking resolution to properly
capture the small-scale structures. The dense clumps in our simulations,
however, seem to be rather short lived, transient entities. If these cloudlets
are formed away from the midplane, they are gravitationally accelerated and
develop a head-tail configuration. Similar structures have been observed in
connection with so-called high velocity clouds \citep[HVCs, see e.g.][for a
recent review]{2006RvMA...19...31R}, and it indeed seems alluring to identify
the cloudlets in our simulations with HVCs.

From the horizontal slices, we observe that newly created structures are
perpetually sheared out by the differential rotation, resulting in rather
elongated filaments. This is not the case for model F4 (see
Fig.~\ref{fig:slc_F4}), where SNe occur at a four times higher rate, and
structures are destroyed by new SN events before they become significantly
sheared out. This morphological difference easily explains why the regular
component of the magnetic field is stronger in regions of low activity (cf.
Sec.~\ref{sec:reg_tur}).

\begin{figure}
  \center\includegraphics[width=0.95\columnwidth]{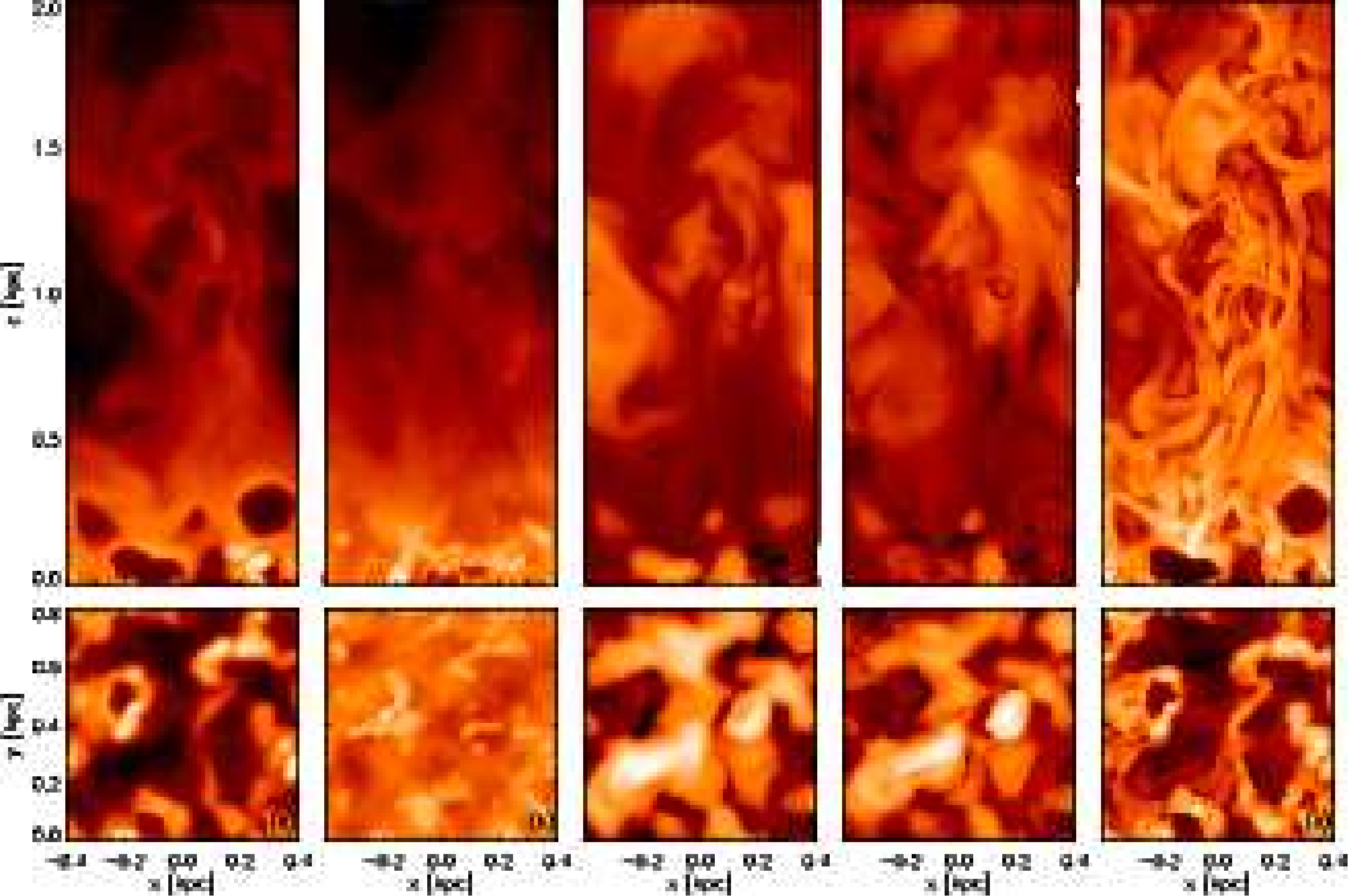}
  \caption[\same{ for model T4--NCL}]{%
    Same as Fig.~\ref{fig:slc_Q4}, but for model T4--NCL at a time 
    $t=170\Myr$. The colour coding extends over ranges $[-4.56, 1.25]$,
    $[17.96,21.78]$, $[ 1.97, 8.09]$, $[-0.10, 3.14]$, $[-4.69, 0.30]$.}
  \label{fig:slc_T4_NCL}
\end{figure}

\begin{figure}
  \center\includegraphics[width=\columnwidth]{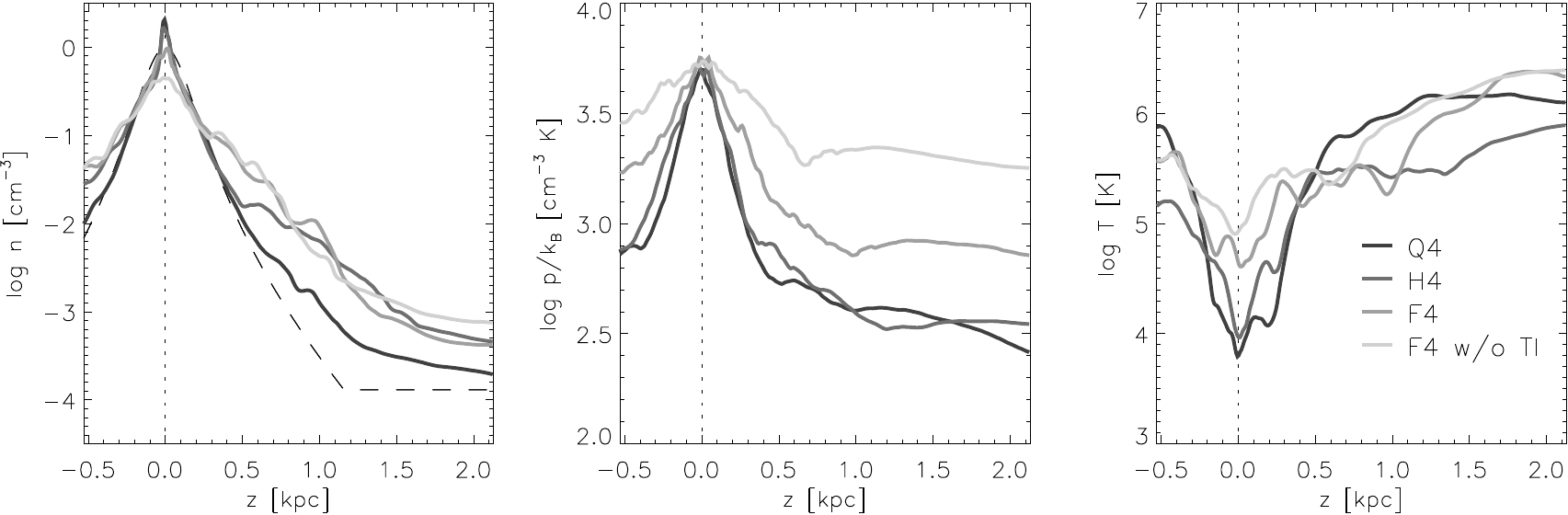}
  \caption[Vertical profiles of number density, pressure, and temperature]{%
    Time averaged vertical profiles of number density $n$, pressure $p/\kB$,
    and temperature $T$ for the models Q4, H4, and F4. For the density we
    also show the initial profile (dashed).}
  \label{fig:prof_QHF4}
\end{figure}

Looking at the lower panels (a) and (e) of Figure~\ref{fig:slc_Q4}, one can
already see that, at least for the region around the midplane, there exists a
significant correlation between the density and the magnetic field amplitude
-- this will be discussed in detail in Section~\ref{sec:rhoB}. While single
SNRs are largely confined to the midplane, super bubbles break out of the
central disk and drive moderate vertical flows. Their dense shells, that are
further compressed by shocks, will form clouds, which can efficiently cool,
and will, in turn, rain back into the gravitational potential, thus forming
what is termed the Galactic Fountain
\citep{1980ApJ...236..577B}.\footnote{%
  This effect is illustrated in Fig.~\ref{fig:rho_uz_1kpc}, where the vertical
  velocity distribution is shown.} %
Despite the limited domain of $z=\pm2\kpc$, we only lose $\simeq 5\%$ of the
total mass per $\Gyr$ through the top and bottom boundaries of our box. If we
turn off the clustering (model T4--NCL), the morphology changes quite
drastically. Instead of well confined supper bubbles we see more disrupted
features and chimney-like structures that channel strong vertical outflows
(see Fig.~\ref{fig:slc_T4_NCL}). The velocity dispersion in the hot phase is
twice as high as in the clustered case; also about five times more mass is
lost through the top and bottom boundaries. These differences demonstrate the
importance of a proper modelling of clustered explosions.

\section{Vertical disk structure} 

\begin{figure}
  \center\includegraphics[width=\columnwidth]{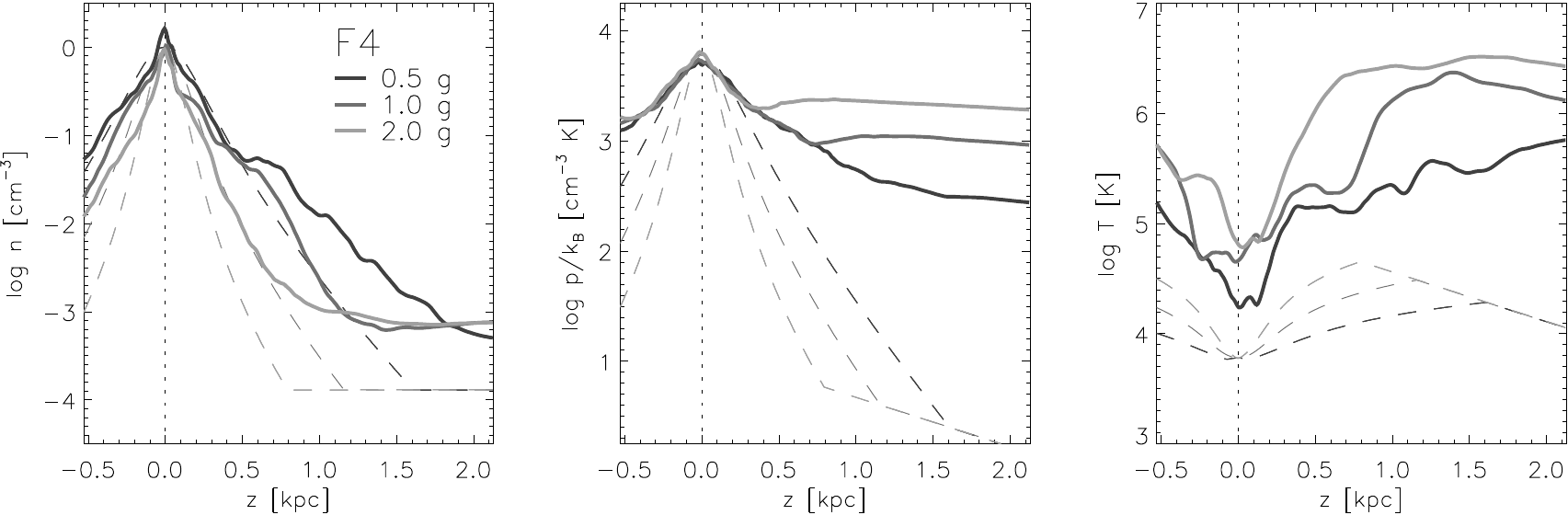}
  \caption[\same{for model F4 at varying gravity}]{%
    Same as Fig.~\ref{fig:prof_QHF4}, but averaged over $t=80$--$100\Myr$ for
    model F4 at normal gravity and with the gravity force enhanced/reduced by
    a factor of two. Dashed lines show the initial profiles.}
  \label{fig:prof_F4g}
\end{figure}

The vertical stratification constitutes an important ingredient of our model
of the galactic disk. We now want to analyse the disk structure resulting from
the quasi stationary balance of the various contributions. Starting from the
hydrostatic equilibrium of the initial model, we add the kinematic and thermal
pressure form the supernova explosions. To understand this new situation, the
hydrodynamic equilibrium has to be replaced by a more intricate dynamical
equilibrium. How will this affect the structure and thickness of the disk?

In view of the expected vertical transport processes, which will play an
important role for the operability of the galactic dynamo, we also want to
study the vertical distribution of the turbulent velocity and the galactic
wind launched by the SNe. How are these quantities affected by the Galactic
Fountain resulting from TI? How do they depend on the depth of the
gravitational well? With the high computational demands of our simulations, we
can only begin to address these questions. In the prospect of global models,
where parameters will strongly vary (e.g. with the galactocentric radius), it
will be highly desirable to understand the underlying relationships on basis
of a semi-analytical description.

\subsection{Thermodynamic structure} \label{sec:gas_prof} 

Within the range of parameters studied, the thermodynamic structure of the
disk moderately depends on the supernova rate. This is illustrated in
Figure~\ref{fig:prof_QHF4}, where we plot time averaged vertical profiles of
the number density, pressure, and temperature for our models Q4, H4, and F4.
For the central region of the disk, we determine a scale height of $\simeq
100\pc$. Between $0.5$ and $1.5\kpc$ the profiles become flatter and roughly
follow an exponential with scale height of $\simeq 400\pc$ \citep[cf. Sec.~3.1
in][and references therein]{2006ApJ...653.1266J}. With an increase in the
supernova rate, the inner disk will more and more disperse resulting in a
flatter profile and a lower midplane density, respectively a higher midplane
pressure.

To study the influence of the thermal instability on the vertical structure
(and ultimately on the field amplification process), we repeat our setup F4
with all parameters unchanged except the coefficients of the cooling function,
which we now take from \citet{1999A&A...350..230K}. As we can see in
Figure~\ref{fig:cooling}, the cooling curve of
\citeauthor{1999A&A...350..230K} is similar to ours but does not include the
characteristic S-shape below $6000\K$. Moreover, the curves strongly diverge
for temperatures above $10^7\K$, resulting in a discrepancy of two orders of
magnitude at $10^8\K$.

The main effect of the neglect of the thermal instability, unsurprisingly, is
the absence of a cold inner disk, as can be seen in
Figure~\ref{fig:prof_QHF4}. While the profile near the midplane is less
peaked, the warm thick disk is not changed much compared to model F4. This is
consistent with the fact that the two cooling curves are very similar in the
temperature domain prevailing at these galactic heights. The pressure
stratification without TI is somewhat shallower and already flattens out at
$z\simeq0.6\kpc$.

Although most visible in the case without TI, the characteristic kink in the
pressure profile is present, more or less pronounced, in all of our models and
also appears in the simulations of \citetalias{2006ApJ...653.1266J}. This kink
marks the transition into the hot halo above the disk and seems to be tightly
related to the kinetic structure of the turbulent stratification (see
Sec.~\ref{sec:velocity_prof} below). While this transition seems to be largely
independent of the applied supernova rate, its position depends on the
gravitational potential. This is illustrated in Figure~\ref{fig:prof_F4g},
where we plot the same profiles for model F4 with half and double the
gravitational acceleration. Since our potential $\Phi(z)$ consists of two
components reflecting a galactic halo and a stellar population (with an
assumed vertical distribution), one may think of various other possible
modifications to the external forces.

Rather surprisingly, the pressure profiles coincide for the inner part on the
disk and the individual curves fork to an ambient value at distinct points
(see centre panel of Fig.~\ref{fig:prof_F4g}). At the same time, the density
curves flatten out and reach an average floor density of $\simeq0.001\pccm$.
As expected, the scale height of the disk decreases with the steepness of the
gravitational potential. It is, however, not obvious why the ambient pressure
scales with the external force. This might be related to the static vertical
distribution of the type~I SNe causing more explosions in the low density
halo, where the gas is less efficiently cooled.

\subsection{Dynamical equilibrium and wind} 
\label{sec:velocity_prof}

\begin{figure}
  \center\includegraphics[width=\columnwidth]{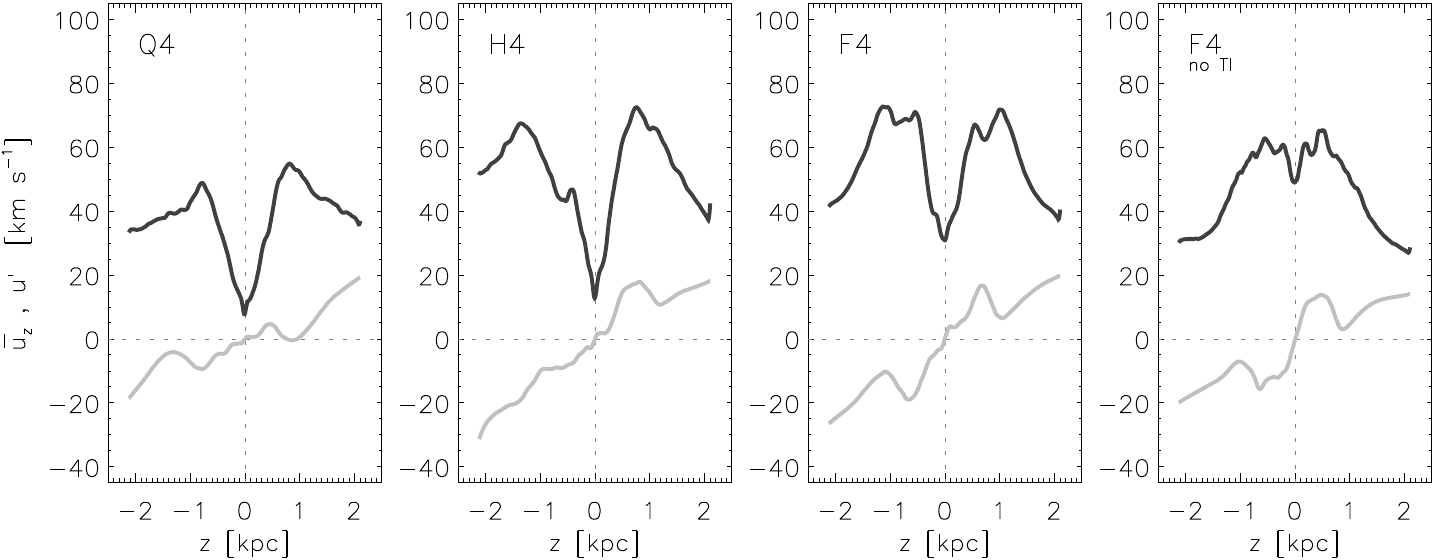}
  \caption[Profiles of the mean flow and the turbulent velocity]{%
    Vertical profiles of the mean flow $\mn{u}_z$ (light) and the turbulent
    velocity $u'$ (dark) for models Q4, H4, and F4 averaged over
    $t=90$--$160\Myr$. The random component shows a distinct double-peaked
    shape, which is also reflected in the wind.} 
  \label{fig:zdisp}
\end{figure}

The time averaged turbulent and mean velocity profiles for the three models
Q4, H4, and F4 (latter with and without TI) are depicted in
Figure~\ref{fig:zdisp}. The vertical structure of the turbulent velocity shows
a distinct M-shape, which peaks at $\pm1\kpc$. The positive gradient of $u'$
in the inner part of the disk strongly suggests an inward transport of the
mean magnetic field component, as will be discussed in more detail in
Section~\ref{sec:transport}.

The strong dip near the midplane is less pronounced in the case without TI,
where no cold inner disk forms. The inner part of the profiles is similarly
shaped as the ones obtained from MRI turbulence
\citep*{2004A&A...423L..29D,2007ApJ...663..183P}, but considerably steeper.
Crudely extrapolating the fall-off in turbulence intensity, we estimate that
MRI might become important in maintaining the observed velocity dispersions at
galactic heights of $|z| \simgt 3\kpc$. While the overall amplitude of the
turbulence increases with the SN intensity, its gradient seems to be less
affected. The same holds for the wind, which reaches an amplitude of $\simeq
20\kms$ at the upper end of the box -- irrespective of the SN rate. For the
case without TI, the vertical gradient in the turbulent velocity is reduced
(see rightmost panel in Fig.~\ref{fig:zdisp}).

\begin{figure}
  \center\includegraphics[width=\columnwidth]{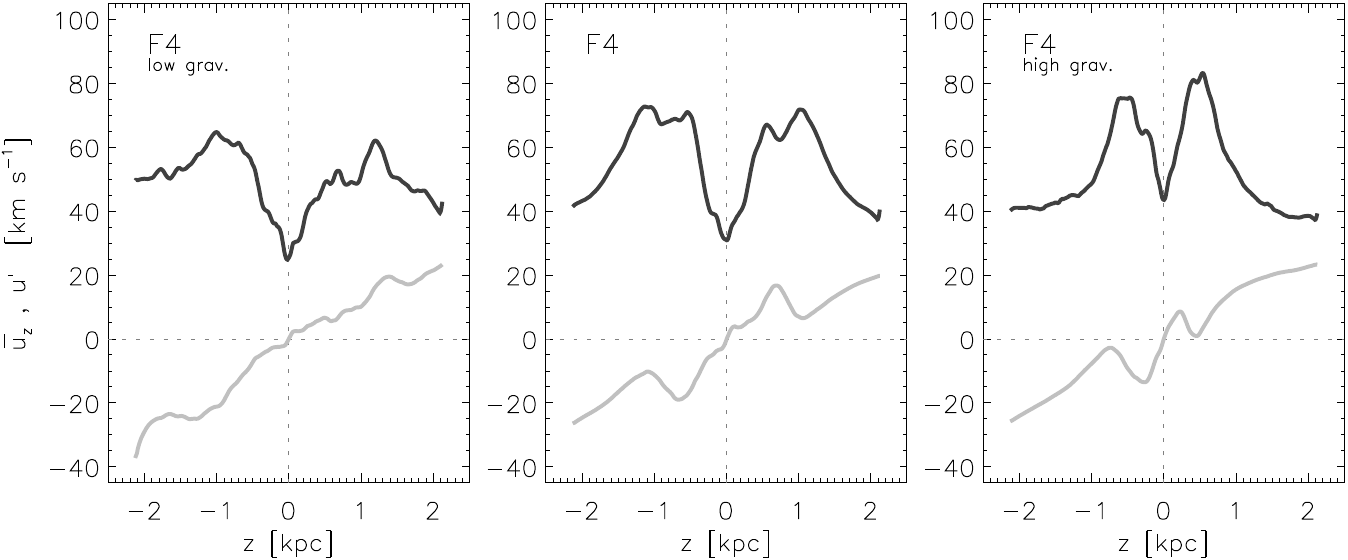}
  \caption[\same{for model F4 at varying gravity}]{%
    Same as Fig.~\ref{fig:zdisp}, but for model F4 at varying external
    gravitational potential. The M-shaped profile of $u'$ becomes narrower for
    stronger gravity.}
  \label{fig:zdisp_grav}
\end{figure}

The wind shows a distinct modulation around $z\simeq 0.8\kpc$, i.e., the
material ejected out of the thick disk decelerates towards a ``stagnation''
point and is accelerated again. Unlike the velocity dispersion, the structure
of the wind is less affected by the neglect of thermal instability; the
amplitude of its modulation is diminished by about $50\%$, however. 

The effect of gravity on the vertical structure of the turbulent velocity and
wind can be seen from Figure~\ref{fig:zdisp_grav}. Again, irrespective of the
applied potential, the wind climbs to a value of $20\kms$ at the boundary of
our box. Unlike the overall slope, the modulation in the wind changes with the
strength of the external force, and the deviations from the linear wind
profile are shifted towards the midplane when gravity is increased.
Accordingly, the M-shaped profile of the turbulent velocity becomes narrower.
This indicates that the turbulence and the mean flow are naturally linked with
respect to their vertical structure.

The characteristic modulation of the galactic wind can be understood in terms
of a dynamical equilibrium equation. If we write down the 1D Reynolds-averaged
Euler equation,\footnote{%
  For simplicity, we do not include the contributions from the cosmic ray
  pressure (not present in our simulations) and the magnetic pressure
  (negligible in the high $\beta_{\rm P}$ regime we consider).}
\begin{equation}
  \partial_t\,(\rho \mn{u}_z) +\partial_z\,\left[
  \rho \mn{u}_z^2 + \frac{1}{3}\rho\mn{u_z'^2}+ p \right] 
  = - \rho\,\partial_z\phi(z)\,,
\end{equation}
we see how the different effects are linked. In contrast to the original Euler
equation, we obtain an additional term, known as the Reynolds stress
$\rho\mn{u'_i u'_j}$, which describes the back-reaction of the turbulence on
the mean flow. The isotropic part of this tensor can be identified with the
kinetic pressure $\nicefrac{1}{3}\,\rho\mn{u_z'^2}$. Assuming a stationary
solution $\partial_t\,(\rho \mn{u}_z)=0$, one can derive a condition
\begin{equation}
  \mn{u}_z^2 + \frac{1}{3}\mn{{u'}_{\!\!z}^2}+ p/\rho + \phi(z) = {\rm const}\,.
  \label{eq:dyn_eq}
\end{equation}
for dynamical equilibrium. Neglecting the self-gravity of the interstellar gas,
we apply a static external potential $\phi(z)$, but this still leaves us with
three independent quantities. For a static equilibrium (${\mn{u}_z=0}$), one
can simply obtain the turbulent velocity stratification, e.g., from the
observed density profile \citep{1996A&A...311..451F}. Since we observe a wind
in our simulations, we have to consider all terms in
Equation~(\ref{eq:dyn_eq}), however.

\begin{figure}
  \center\includegraphics[width=\columnwidth]{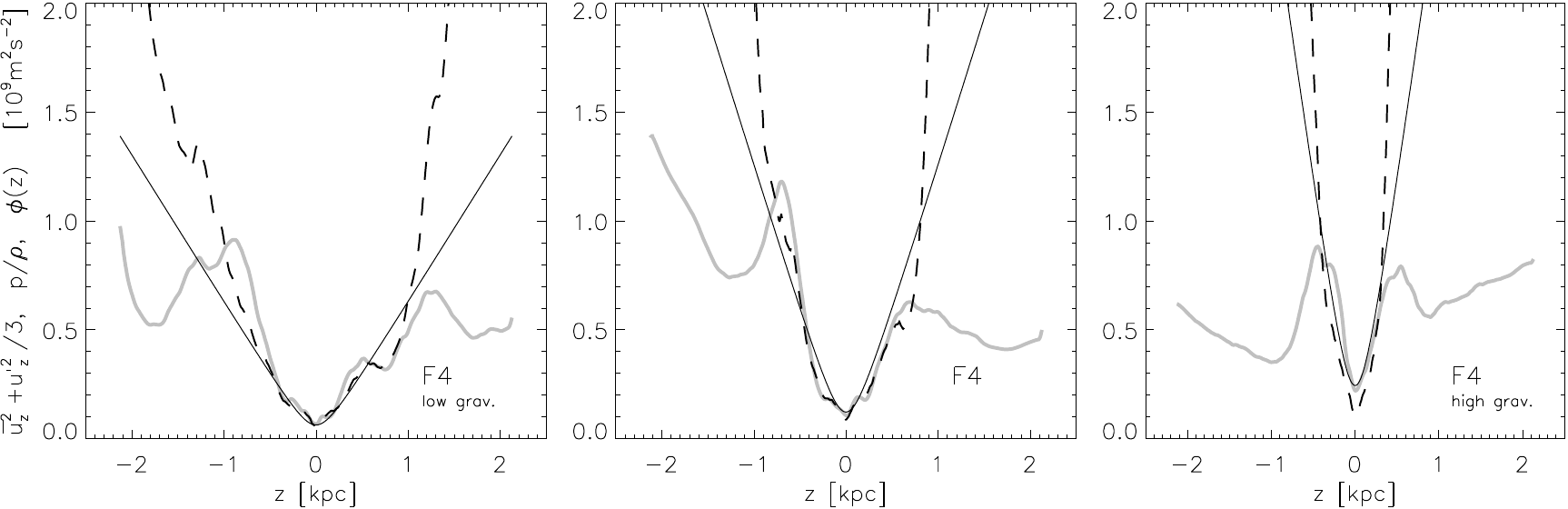}
  \caption[Energy distribution for model F4 at varying gravity]{%
    Vertical profiles of the total kinetic energy (grey line), thermal energy
    (dashed line), and gravitational potential (solid line) for model F4 at
    varying external gravity.}
  \label{fig:dyn_eq}
\end{figure}

If we assume energy equipartition, we presume that the sum of the kinetic
terms should be equal to the external potential and the thermal temperature as
illustrated in Figure~\ref{fig:dyn_eq}, where we plot the corresponding terms
of Equation~(\ref{eq:dyn_eq}). We see that, in the central region around the
midplane, the different energy forms are indeed balanced. The point where the
kinetic energy drops from its equipartition value coincides with the peak in
the turbulent velocity, the local maximum in the wind profile, and the point
where the pressure curve becomes flat. This can be understood as follows:
while inside this characteristic point the pressure force and the turbulence
gradient oppose each other (leading to the equipartition), outside this point
they act together. Since the pressure profile flattens and the profile in $u'$
steepens, the resulting wind reaches a ``stagnation'' point, before it is
accelerated again by the combined force of the kinetic and thermal pressure.

\section{Thermal and kinetic distribution} 

For a brief look at various distribution functions, we separate the
contributing phases into four distinct temperature classes, which we select
according to the branches of our cooling function. To only capture the bulk of
the unstable regime and not cut through the stable populations around
$T=6102\K$ and $T=141\K$ (see Table~\ref{tab:cooling}), we define the cold
neutral medium (CNM) to include material below $T=200\K$ and attribute the
temperature range from $200\K$ to $4400\K$ to the unstable phase. This range
is joined with the warm ionised medium (WNM) up to $10^5\K$ followed by the
hot ionised medium (HIM) above $10^5\K$. This classification is to some point
arbitrary and other authors indeed chose different intervals.\footnote{%
  \citetalias{2005A&A...436..585D}, e.g., select five intervals divided at
  temperatures of $200\K$, $7900\K$, $16{\rm kK}$, and $10^{5.5}\K$;
  \citetalias{2006ApJ...653.1266J} only distinguish three states divided at
  $200\K$ and $17{\rm kK}$.}%

\subsection{Occupation fractions} 

\begin{figure}
  \center\includegraphics[width=0.9\columnwidth]{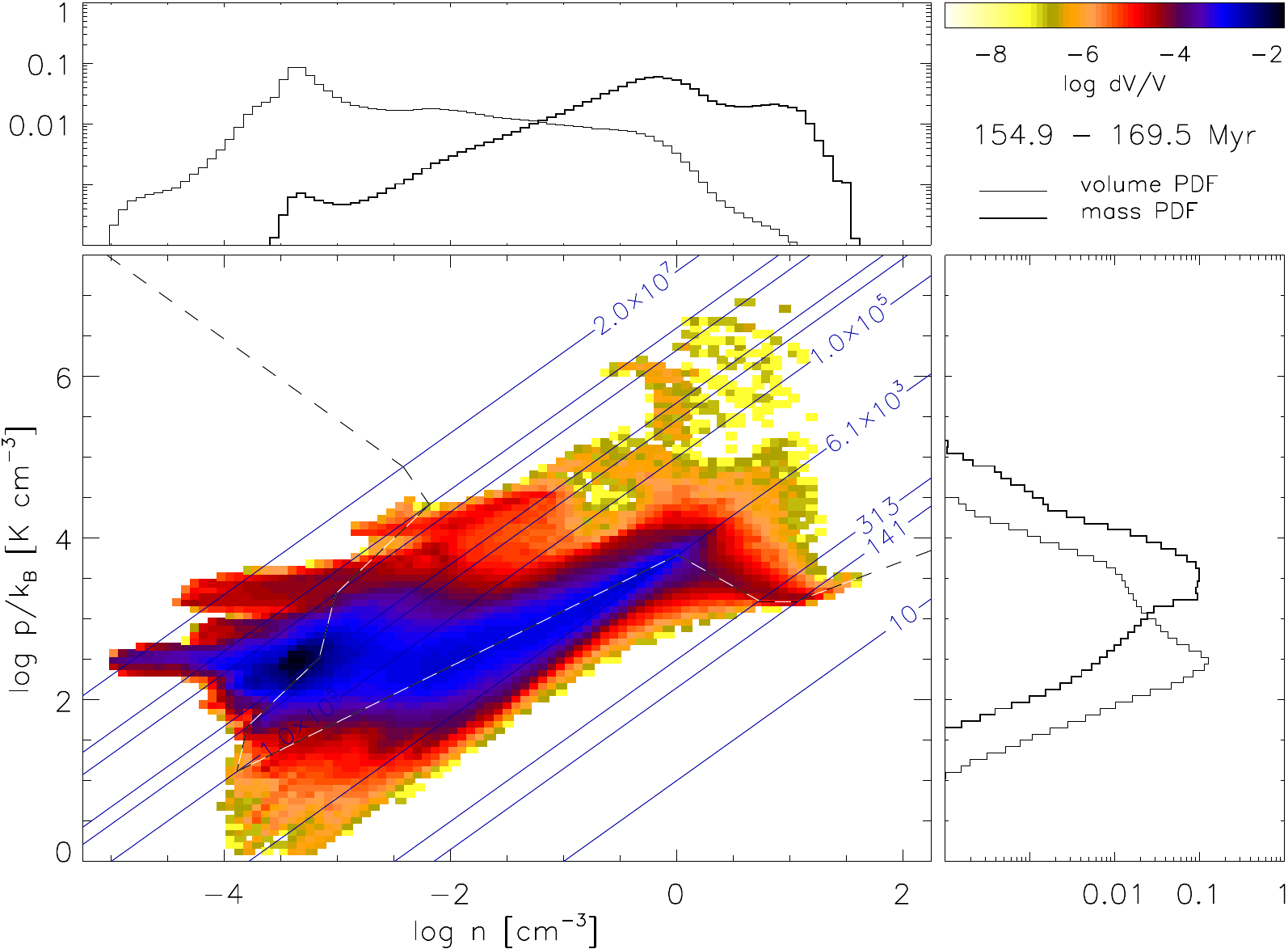}
\caption[Phase diagram of the supernova heated plasma]{%
  Phase diagram of the supernova heated plasma in the midplane of model H4 for
  $t=155$--$170\Myr$. Overplotted are isothermal contours (labelled in $\K$)
  and the equilibrium cooling curve (dashed line). Adjacent plots show volume-
  (thin line) and mass- (thick line) weighted PDFs.}
\label{fig:phase}
\end{figure}

In Figure~\ref{fig:phase}, we present the phase space distribution of the disk
gas for our model H4 at half the galactic supernova rate. The colour coding
shows the logarithmic volume fraction ${\rm d}V/V$. Overplotted are the
equilibrium cooling curve (dashed line) and contours of constant temperature
(labelled in $\K$). The stable branches of the equilibrium curve are richly
populated, but unlike predicted by the classical two-phase model of
\citet*{1969ApJ...158..173G}, there also exists gas in the radiatively
unstable regime.

In phase space, the hot material at a range of medium densities (located in
the upper part of Fig.~\ref{fig:phase}) corresponds to recent explosions.
While the hot interior of the supernova remnant cools slowly and effectively
moves to lower pressures on isothermal lines, the dense shell of the SNR cools
more efficiently and takes a much steeper path in the phase space diagram. The
almost isobaric tails at the left side of the plot can be identified as old
remnants in approximate pressure equilibrium with their surroundings. The
molecular clumps make up the very right end of the distribution and extend
down to temperatures of a few ten Kelvin. The vast amount of material is
restricted to a pressure range covering 1--2 orders of magnitude which is much
less than the range in density that extends over more than six orders.
Compared to the considerably higher resolved models of
\citetalias{2006ApJ...653.1266J} and \citetalias{2005A&A...436..585D}, our
simulations to some extent suffer from insufficient resolution to properly
grasp the cold, dense phase. It is claimed that a spatial resolution of $\sim
1\pc$ is necessary for converged results \citep[cf.  Sec.~3.4
in][]{2004A&A...425..899D}. The limited resolution of our model is also
reflected in the mass- and volume-fractions, which we want to discuss in the
following.

For model H4 we find $0.1$, $8$, $70$, and $22\%$ of the volume occupied by
the cold, unstable, warm, and hot gas. At the full supernova rate (model F4)
this is slightly shifted towards $52\%$, and $40\%$ for the WNM, and HIM,
respectively. The corresponding mass fractions are $4$, $30$, $62$, and $4\%$
(CNM, unstable, WNM, and HIM) for model H4, and $3$, $28$, $60$, and $9\%$ for
model F4. The volume fractions for the warm and hot medium at full supernova
rate agree well with the values of $\sim 50\%$ and $41$--$51\%$ given by
\citetalias{2006ApJ...653.1266J}, who apply a very similar cooling function
and heating rate. As already reported in \citetalias{1999ApJ...514L..99K}, we
find the filling factor of the hot phase to depend critically on whether
clustering is applied. With fully uncorrelated explosions, we register values
up to $f_h \simeq 0.7$ in agreement with the analytical prediction for this
case by \citet{1977ApJ...218..148M}. In contrast,
\citetalias{2005A&A...436..585D}, who restrict their explosion sites to
regions of low temperature and high density, find a much lower value of $f_h
\simeq 0.2$. Observationally, this parameter is also very poorly constrained;
\citep{1992FCPh...15..143D} estimates an upper limit of $0.5$.

As would be expected from the turbulent nature of the simulations, a
considerable amount of the WNM resides in the thermally unstable regime. For
model F4 we find equal parts by mass in the stable respectively unstable WNM
below $5000\K$, for model H4 this reduces to $45\%$ by mass. Although somewhat
lower than the reported $60$--$70\%$ in \citetalias{2006ApJ...653.1266J} and
\citetalias{2005A&A...436..585D}, this is still consistent with the
observational constraints by \citet{2003ApJ...586.1067H}.

\subsection{Distribution functions} 

\begin{figure}
  \center\includegraphics[width=\columnwidth]{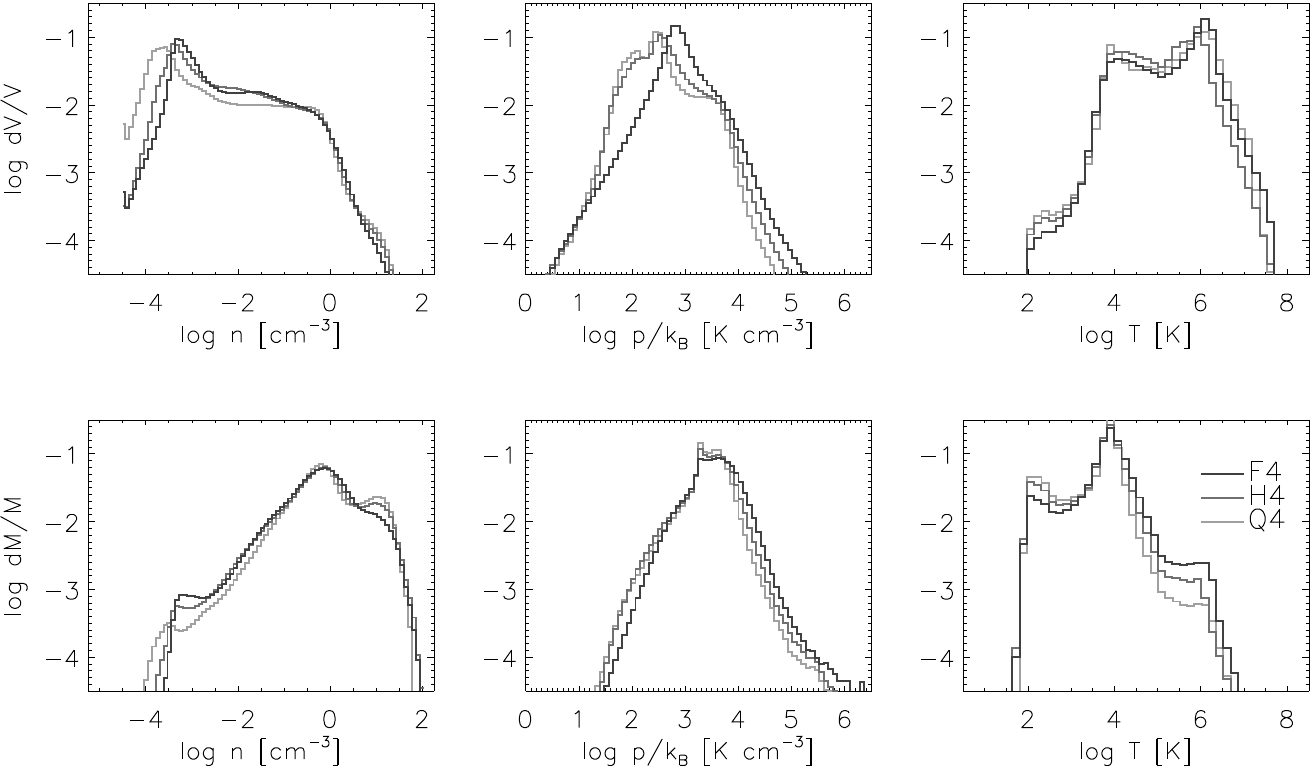}
\caption[Comparison of distribution functions for models Q4, H4, and F4]{%
  Comparison of distribution functions for models Q4, H4, and F4 averaged over
  $t=50\Myr$. We show volume- (upper) and mass-weighted (lower row) PDFs of
  number density $n$ (left), gas pressure $p/\kB$ (centre), and temperature
  $T$ (right panel).}
\label{fig:dist}
\end{figure}

To aid the classification of the thermodynamic states of the various phases,
it is useful to compute volume- and mass-weighted histograms of the
thermodynamic quantities. As can already be seen from Figure~\ref{fig:phase},
the distribution of the gas is very broad. While the classical two-phase model
assuming pressure equilibrium between the phases \citep*{1969ApJ...158..173G}
predicts a bimodal density distribution together with a simple
Dirac-distribution for the pressure, this is certainly not the case for such a
turbulent environment. Nevertheless, our PDFs, which we plot in
Figure~\ref{fig:dist}, bear some resemblance with the ones proposed by the
three-phase model of \citet{1977ApJ...218..148M}: The mass-weighted density
PDF (lower left panel of Fig.~\ref{fig:dist}) shows three rather distinct
peaks corresponding to a HIM at $n=10^{-3.5}\cm^{-3}$ and $T=10^{6.5}\K$, a
warm phase around $10^4\K$, and a component consisting of thermally unstable
and cold gas at densities of $\simeq 10\cm^{-3}$.

There has been some discussion on whether the distinct peaks of the
distribution can be fitted with log-normal distributions and to what extent
the pressure histogram exhibits power-law tails \citep[see
e.g.][]{2005ApJ...630..911G}. With the limited resolution of our models, it
does, however, not seem advisable to enter this discourse. While the lower end
of our pressure histogram is largely insensitive to the applied supernova
rate, it will extend to higher pressures for increasing supernova activity.
Quite noticeably, the density peak at $\log n \simeq 1$ disappears in our
model F4, indicating that higher resolved simulations are indispensable to
study turbulence at higher SN rates.

\begin{figure}
  \begin{minipage}[b]{0.7\columnwidth}
    \includegraphics[width=0.97\columnwidth]{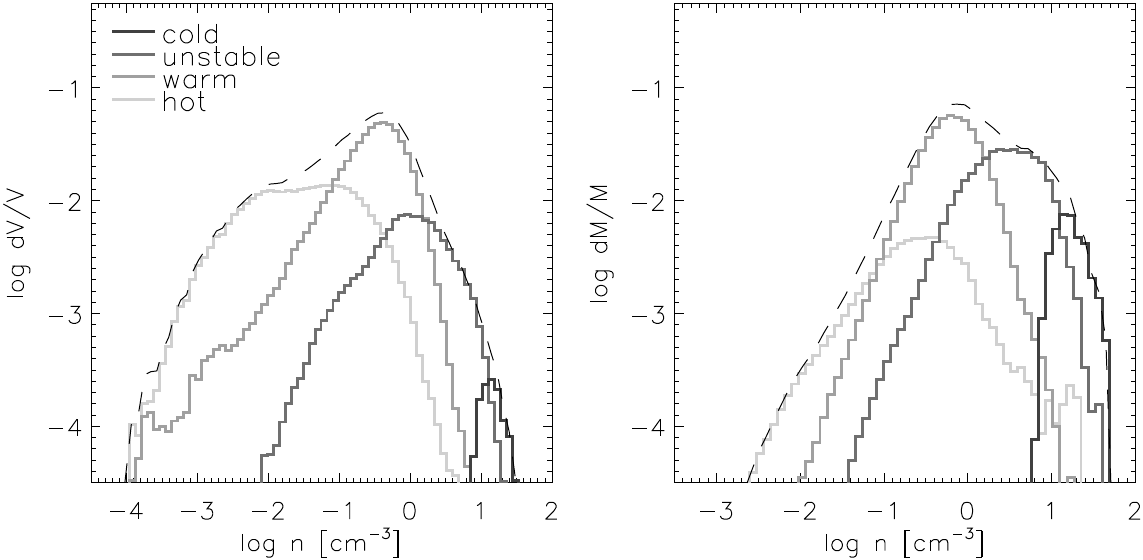}
  \end{minipage}
  \begin{minipage}[b]{0.29\columnwidth}
    \caption[Density distribution functions according to different phases]{%
      Density distribution functions for the gas contained within $|z|<133\pc$
      of model F4. The plots illustrate how the various phases contribute to
      the total PDF (dashed envelope).\vspace{10pt}}
    \label{fig:dist_ph}
  \end{minipage}
\end{figure}

To conclude the discussion on thermodynamic properties,
Figure~\ref{fig:dist_ph} demonstrates how the different phases contribute to
the total density distribution. The bulk of the PDF is determined by the warm,
unstable, and hot phases who show broad distributions in density. In contrast
to this, the contribution of the cold phase is strongly limited to the high
density regime. This is contrary to the simulations of
\citetalias{2005A&A...436..585D}, who find cold gas down to densities of
$n\simeq 0.01\cm^{-3}$ for the HD run and even as low as $n\simeq
10^{-3}\cm^{-3}$ for the MHD run. Since our simulations are generally in the
weak field regime, a comparison with their HD run seems more appropriate.
Still, there remains a discrepancy of three orders of magnitude which can only
be attributed to the inclusion of thermal conduction in our models.

\subsection{Velocity dispersions} \label{sec:dispersion} 

Long before the era of the space telescopes and their highly resolved images
of supernova remnants and giant molecular clouds, it was already well
established that the interstellar medium is in a highly turbulent state. This
knowledge was derived from observations of broadened spectral lines, both in
emission and absorption. When atoms and molecules absorb or emit photons, the
associated spectral lines are Doppler shifted with the relative velocity
between the observer and the source. Within a turbulent plasma, the atoms and
molecules move at random velocities. Combining the intensity of all photons
the profile of a certain spectral feature is convolved with the velocity
distribution function, yielding a Doppler broadened spectral line. The width
of such a line is commonly referred to as velocity dispersion, indicating the
amplitude of the turbulent velocity field.

\begin{figure}
  \begin{minipage}[b]{0.70\columnwidth}
    \includegraphics[width=0.95\columnwidth]{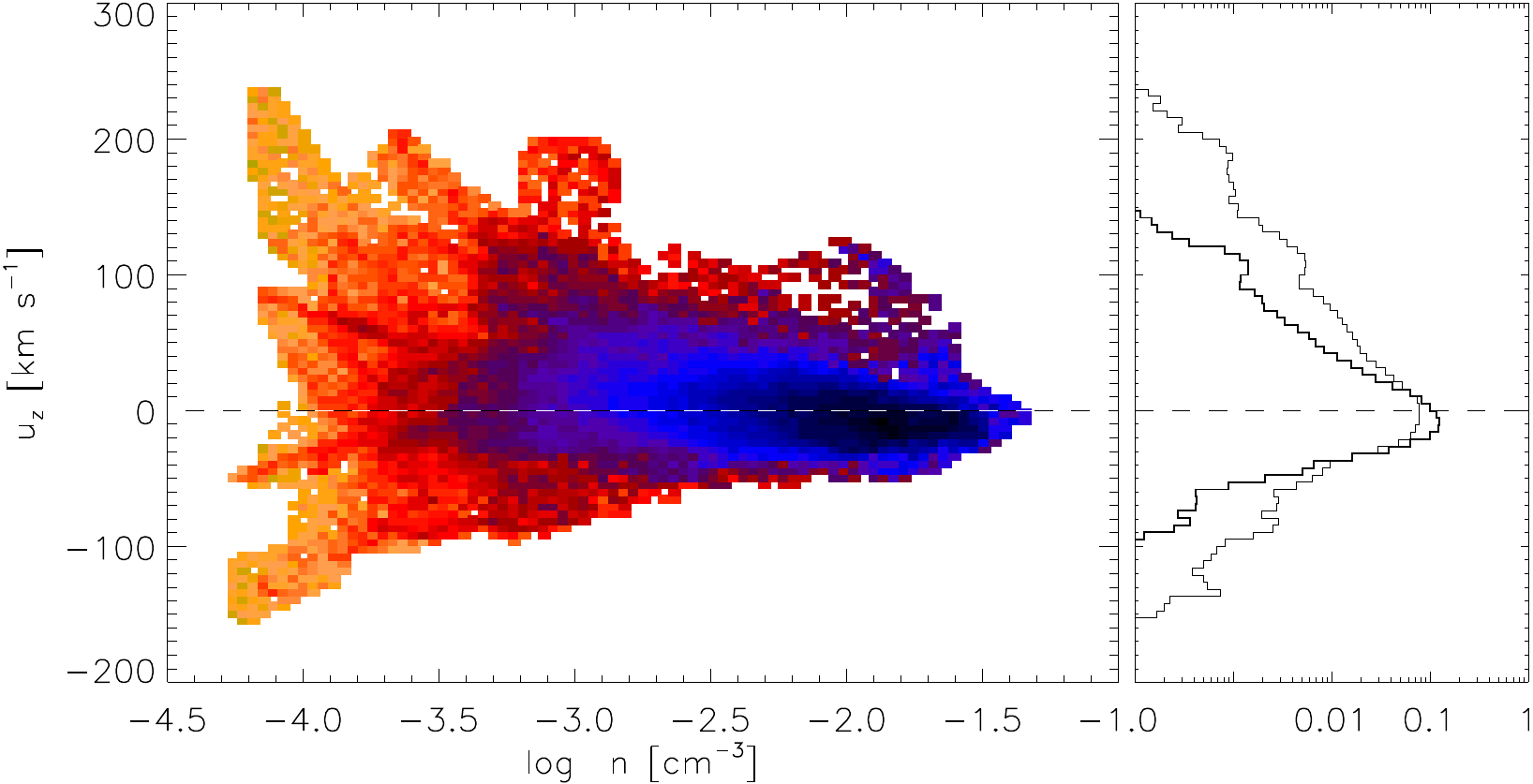}
  \end{minipage}
  \begin{minipage}[b]{0.29\columnwidth}
    \caption[2D histogram over vertical velocity and number density]{%
      2D histogram of the velocity $u_z$ and number density $n$. Colour coding
      indicates the logarithmic mass fraction within a horizontal slice at
      $z=1\kpc$. The adjacent plot shows the mass- (thick line) and
      volume- (thin line) integrated PDFs.\vspace{10pt}}
    \label{fig:rho_uz_1kpc}
  \end{minipage}
\end{figure}

An example for a turbulent velocity distribution can be seen from
Figure~\ref{fig:rho_uz_1kpc}, where we plot the 2D histogram of the vertical
velocity versus density. One can see that the distribution is biased towards
negative velocities, i.e. inflow, for higher densities. Aside from the
systematic offset due to the mean flow, we observe a dispersion that
considerably broadens towards low densities, i.e. high temperatures. Also note
that due to this trend the width of the mass weighted histogram is somewhat
smaller than that of the volume weighted histogram. Taking into account the
contribution from the galactic wind, which especially becomes important in the
outer parts of the disk, we consequently remove the mean component of the flow
from the velocity field before computing rms~values.

\subsubsection{Line of sight velocity in \HI}

For a direct comparison with observations, we also compute line-of-sight (LoS)
velocity dispersions. Generally, these 1D dispersions are smaller compared to
3D by a factor of $1.5$--$2.0$, reflecting projection effects. For
homogeneous, isotropic turbulence, the projection factor for the rms~velocity
is given by
\begin{equation}
  \left( 
    \frac{1}{4\pi} \iint (\sin(\theta)\sin(\phi))^2\,{\rm d}\phi\,{\rm d}\theta
  \right)^{-1/2} = \sqrt{\frac{8}{\pi}} \simeq 1.6\,.
\end{equation}
From an observers point of view, there exists a number of ions whose spectral
lines are used as tracers at different temperature regimes and wavelengths.
Assuming equilibrium ionisation, one could in principle compute ion densities
directly from the thermodynamic properties. This allows to determine mass
weighted LoS column densities and velocity dispersions as seen in observations
\citep[see e.g.][for a study of \OVI]{2005ApJ...634L..65D}. To mimic the
velocity profile of the \HI\ emission line, \citet*{2006ApJ...638..797D} select
regions with temperature $T\le12{\rm kK}$ and density $n\ge 0.25 \cm^{-3}$ to
define the mass weighted LoS dispersion $\sigma_{\rm H_{^I}}$.

For our basic models Q4, H4, and F4, we compute velocity dispersions from six,
eight, respectively nine snapshots between $t=100$--$200\Myr$ and find values
of $3.1$, $4.3$, and $5.9\kms$ for $\sigma_{\rm H_{^I}}$. This is somewhat
smaller than the observed dispersion of $7$--$9\kms$ for the warm \HI\ 
intercloud medium \citep{1985ApJ...289..792K}. Contrary to the simulations of
\citet*{2006ApJ...638..797D}, who report a constant value of $\sim 3\kms$
irrespective of the applied SN rate, we find $\sigma_{\rm H_{^I}}$ to be
increasing with the supernova rate. Besides assuming a much higher galactic SN
frequency of $2.58\tms 10^2\Myr^{-1}\kpc^{-2}$, their models neglected
stratification.

A possible explanation for the constant dispersion in \HI\ might be the limited
resolution ($7.8\pc$) of their simulations, which -- albeit comparable to ours
($8.3\pc$) -- might have more dramatic influences at higher SN rates. Apart
from the different parameter regime, the authors for their simulations use the
ZEUS code which implements a non-conservative formulation of the energy
equation. In this formulation, the kinetic energy dissipated due to the
numerical truncation error at grid scale is not recovered in the thermal
energy. This feed back mechanism, which is naturally present in the
conservative formulation (and can only partly be covered by artificial
viscosity within a non-conservative scheme), is particularly important when
using a radiative cooling function with a strong temperature dependence.
Without this ``viscous'' heating process at the cloud inter-cloud interfaces,
there is less cold material being returned to the thermally unstable regime
and TI cannot be tapped efficiently to mediate the turbulent energy, injected
via the SNe, towards higher densities. It, however, remains to be checked
whether this can explain the saturated velocity dispersions.

\subsubsection{Comparison of ISM phases}

\begin{table}
 \begin{minipage}[c]{0.70\columnwidth}
  \begin{tabular}{lrlrlrl}\heading{&Q4 &&H4 &&F4 &}
   cold & $  2.4$ & $(\pm 0.2)$& $  3.1$ & $(\pm 0.3)$& $  4.8$ & $(\pm 0.6)$\\
   unst.& $  9.4$ & $(\pm 2.4)$& $ 12.1$ & $(\pm 2.3)$& $ 12.7$ & $(\pm 1.2)$\\
   warm & $ 12.7$ & $(\pm 0.7)$& $ 16.8$ & $(\pm 1.3)$& $ 18.7$ & $(\pm 0.7)$\\
   hot  & $ 48.1$ & $(\pm13.5)$& $ 58.1$ & $(\pm15.6)$& $ 62.7$ & $(\pm 8.0)$\\
   \hline
  \end{tabular}
 \end{minipage}
 \begin{minipage}[c]{0.29\columnwidth}
  \caption[Turbulent velocities for models Q4, H4, and F4]{%
    Turbulent velocities in $\kms$, averaged over 6,8, and 9 snapshots within
    $t=100$--$200\Myr$ for models Q4, H4, and F4.}
  \label{tab:vrms}
 \end{minipage}
\end{table}

In Table~\ref{tab:vrms}, we report volume weighted turbulent rms~velocities
for three runs with varying SN rate. The values for the cold phase show the
same trend as the mass weighted LoS velocities discussed above. Compared to
\citetalias{2005A&A...436..585D}, who find $\vrms\simeq 7\kms$ for the
$T<200\K$ gas in their HD run, we fall short of this by $\sim 30\%$. Recalling
the moderate resolution of our models, we do not consider this a dramatic
difference, however. The trend of increased velocity dispersion with higher
supernova activity is less pronounced in the unstable gas. This implies that
the cooling instability plays an important role in maintaining the velocity
dispersion in the warm neutral medium. Additionally, while most of the given
numbers are only fluctuating on a $10\%$ level, the values for the unstable
phase show strong temporal fluctuations, reflected in standard deviations of
up to $25\%$ for model Q4. This might indicate that this temperature regime,
particularly in regions of low SN activity, is prone to intermittent dynamics.
As a note of caution, it shall be remarked that, as mentioned above, the
choice of the temperature interval defining the unstable phase is rather
delicate. Finally, the velocity dispersions in the warm ($19\kms$) and hot
($63\kms$) phases of our run F4 are consistent with the ones by
\citetalias{2005A&A...436..585D} and the observational references cited
therein.

\section{Spectral Properties} 
\label{sec:spec}

Turbulence inherently displays structures on a wide range of spatial scales.
The concept of spectral analysis of the turbulent flow dates back to
Richardson, who initiated the paradigm of the turbulent cascade
\citep[see][for an introduction]{1995tlnk.book.....F}. Based on the assumption
that the driving forces (dominant at the injection scale) as well as
dissipative terms (dominant at small scales) have little influence on the
dynamics of the flow within an intermediate wavenumber regime,
\citet{1941DoSSR..30..301K}, hereafter K41, derived his famous $k^{-5/3}$ law
for the spectral energy within the so-called inertial range. In this range,
the flow is dominated by the non-linear terms of the Navier-Stokes equations
justifying the central assumption of a constant spectral transfer rate of the
turbulent kinetic energy.

Kolmogorov's theory, which has been independently derived by
\citet{1948ZPhy..124..614W}, strictly only applies to homogeneous, isotropic,
incompressible turbulence in the limit of infinite Reynolds numbers. Although
there have been many attempts to generalise the concept to the more complex
cases of compressible, anisotropic, and particularly, magnetohydrodynamic
turbulence \citep[see, e.g.,][for a recent review]{2003LNP...614...56C}, there
is no coherent picture of the turbulent inertial range in the general case.
Lacking laboratory experiments at sufficiently high Reynolds numbers, the
current research on MHD turbulence is largely based on numerical simulations
with increasing spectral resolution, and spacecraft measurements of the solar
wind. Only recently, turbulence data from infrared observations in OMC1 has
been analysed by \citet{2006A&A...445..601G}.

\subsection{Energy spectra} 

To suite the demand of higher spatial resolution, we have performed an
additional run F4--BOX with a smaller, cubic domain of $(400\pc)^3$. At a
resolution of $1.6\pc$, this model is very similar to the one of
\citet{2004ApJ...617..339B}, which neglects vertical gravity and the
dissipative terms in the MHD equations. Compared to the aforementioned model,
that applied a box length of $200\pc$ at a resolution of $0.8\pc$, we choose a
coarser grid to be able to better capture the large-scale structure of the
flow. This is because, with an expected integral scale of the supernova
remnants of $80\pc$, the smaller box will likely exhibit periodicity effects.

\begin{figure}
  \center\includegraphics[width=\columnwidth]{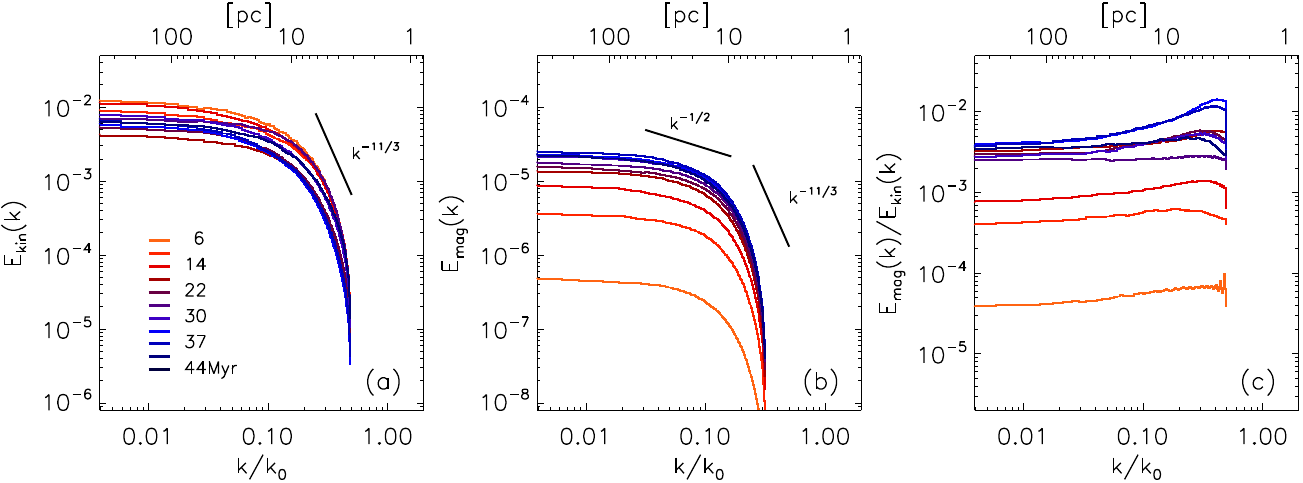}
  \caption[One-dimensional power spectra for model F4--BOX]{%
    One-dimensional power spectra of the kinetic (a) and magnetic (b) energy
    and the corresponding ratio (c), for six snapshots of model F4--BOX at the
    indicated times. Wavenumbers are in grid units $k_0$, the corresponding
    spatial scale can be inferred from the upper axis. In panels (a) and (b),
    the energy is measured in the same arbitrary units.}
  \label{fig:pwsp}
\end{figure}

In Figure~\ref{fig:pwsp}, we plot the one-dimensional power spectra of the
kinetic and magnetic energy for several snapshots of our model F4--BOX. The
predicted slope of $\nicefrac{-5}{3}$ of the Kolmogorov theory for the
inertial range of three-dimensional spectra has to be divided by the surface
element $k^2$ of a shell in k-space to yield the corresponding value for the
1D spectrum. The resulting slope of $\nicefrac{-11}{3}$ for one-dimensional
spectra has been overplotted in the figure. Due to the limited resolution of
our simulations we, however, do not observe an inertial range in our
simulations. According to the Kazantsev theory for the small-scale dynamo
\citep[see section 5.2 in][for a recent account]{2005PhR...417....1B}, the
three-dimensional magnetic spectrum should exhibit a power-law dependence of
slope $\nicefrac{3}{2}$, which (translated to our 1D spectrum) corresponds to
a slope of $\nicefrac{-1}{2}$, as indicated in Figure~\ref{fig:pwsp}.

Surprisingly enough, unlike in forced turbulence simulations, the driving
scale of the turbulence is not clearly visible in the energy spectra.
Similarly to \citet{2004ApJ...617..339B}, we find the kinetic spectrum to
fluctuate around a constant value, while the magnetic spectrum grows in time.
The impression that the magnetic energy saturates at late times is somewhat
misleading since the kinetic energy is particularly low at this instant in
time. If one compensates for this dependence and plots the magnetic energy
spectrum normalised by the kinetic energy $E_{\rm kin}(0)$, instead, one
observes that the magnetic energy only grows slowly with respect to the
kinetic energy at the late stage.

Further evidence that the small-scale dynamo is indeed quenched at this time
comes from the fact that the shapes of the magnetic and kinetic energy spectra
become very similar (see panel (c) in Fig.~\ref{fig:pwsp}). We see that the
magnetic component is strongest in the small scales at $\sim 5\pc$ with the
peak slightly moving towards smaller scales at later times
\citep*[cf.][]{2002ApJ...567..828S}. At $t=44\Myr$, the profile becomes almost
flat. However, considering that the magnetic flux is ideally conserved for
triple periodic boundary conditions, this setup is, to some extent, of
academic interest only.

\subsection{Velocity structure functions} 
\label{sec:velo_struc}

\begin{figure}
  \center\includegraphics[width=\columnwidth]{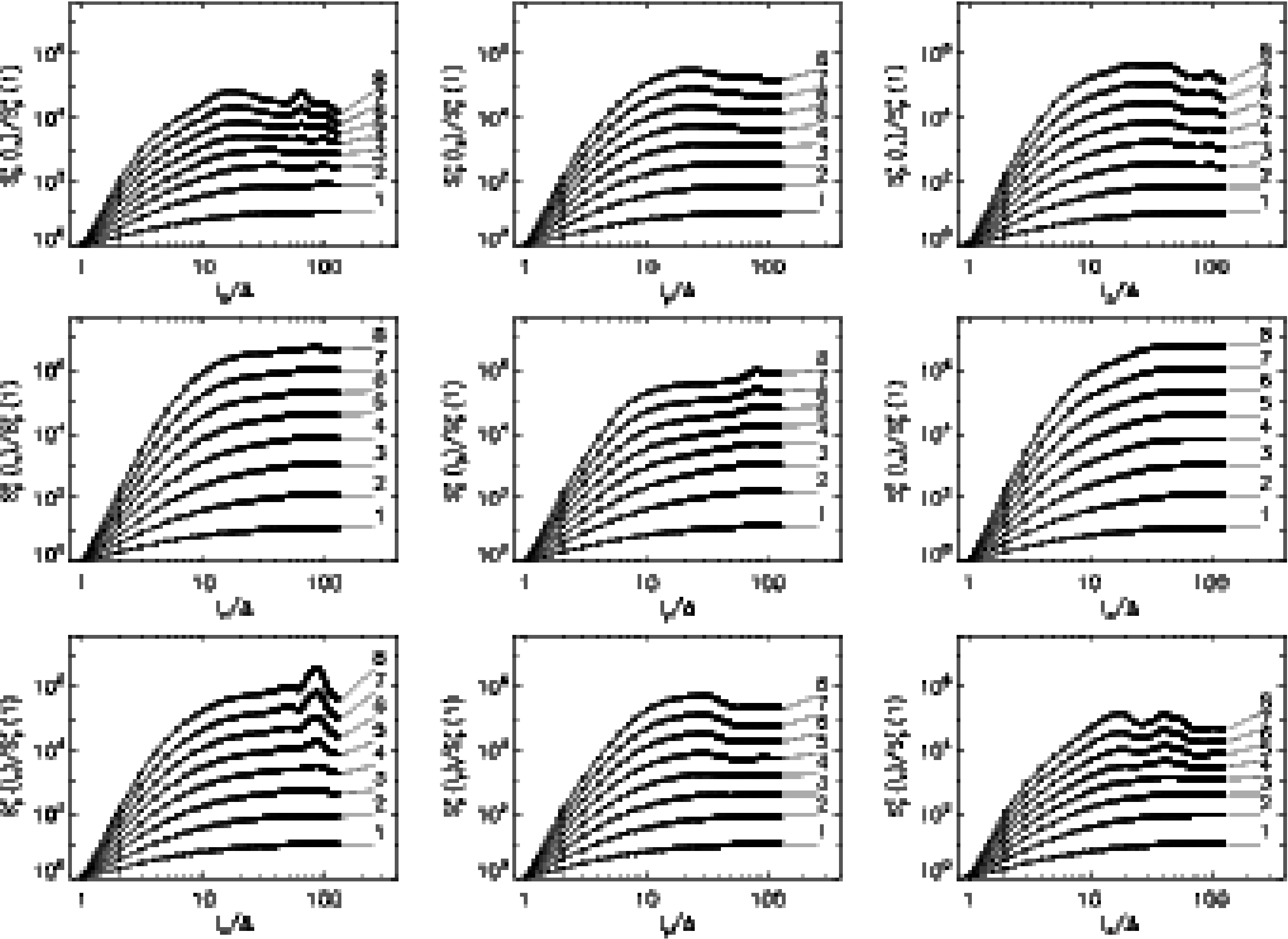}
  \caption[Velocity structure functions]{%
    Velocity structure functions of order $p=1$--$8$ for model F4--BOX at
    $t=20\Myr$. The diagonal and off-diagonal panels represent the
    longitudinal and transversal structure functions.}
  \label{fig:stf_v}
\end{figure}

To better understand the spectral structure of the supernova driven turbulence
in our simulations, we will now briefly introduce the concept of structure
functions and their associated scaling exponents. The velocity structure
function ${\mathbf S}_p(l)$ of order $p$ along $\xx$ is
\begin{equation}
  {\mathbf S}_p(l)=\rms{|{\mathbf v}({\mathbf r}+l\xx)
                        -{\mathbf v}({\mathbf r})|^p}\quad
                        \propto\ l^{\zeta_{p}}\,
\end{equation}
and accordingly for the other space dimensions. The idea behind this framework
is that intermittent structures, i.e., structures that depart from strict
self-similarity, will have different effects on the exponents $\zeta_{p}$
which describe the scaling behaviour of the structure functions.

In Figure~\ref{fig:stf_v}, we plot structure functions computed from a single
snapshot of model F4--BOX at $t=20\Myr$. The plots are arranged in a grid
representing the components of the velocity vector (rows) and the spatial
direction of the argument $l$ (columns), i.e., the diagonal and off-diagonal
panels show the longitudinal and transversal structure functions,
respectively. Most notably, the highest order moments of the upper and lower
left plots exhibit distinct features at a scale of $90$--$100\pc$ which
coincides with the asymptotic size of a single supernova remnant near the
midplane. Note that the features are not present in the lower order moments of
the structure functions and, hence, do not show up in the power spectra
(Fig.~\ref{fig:pwsp}), either.

\begin{figure}
  \center\includegraphics[width=\columnwidth]{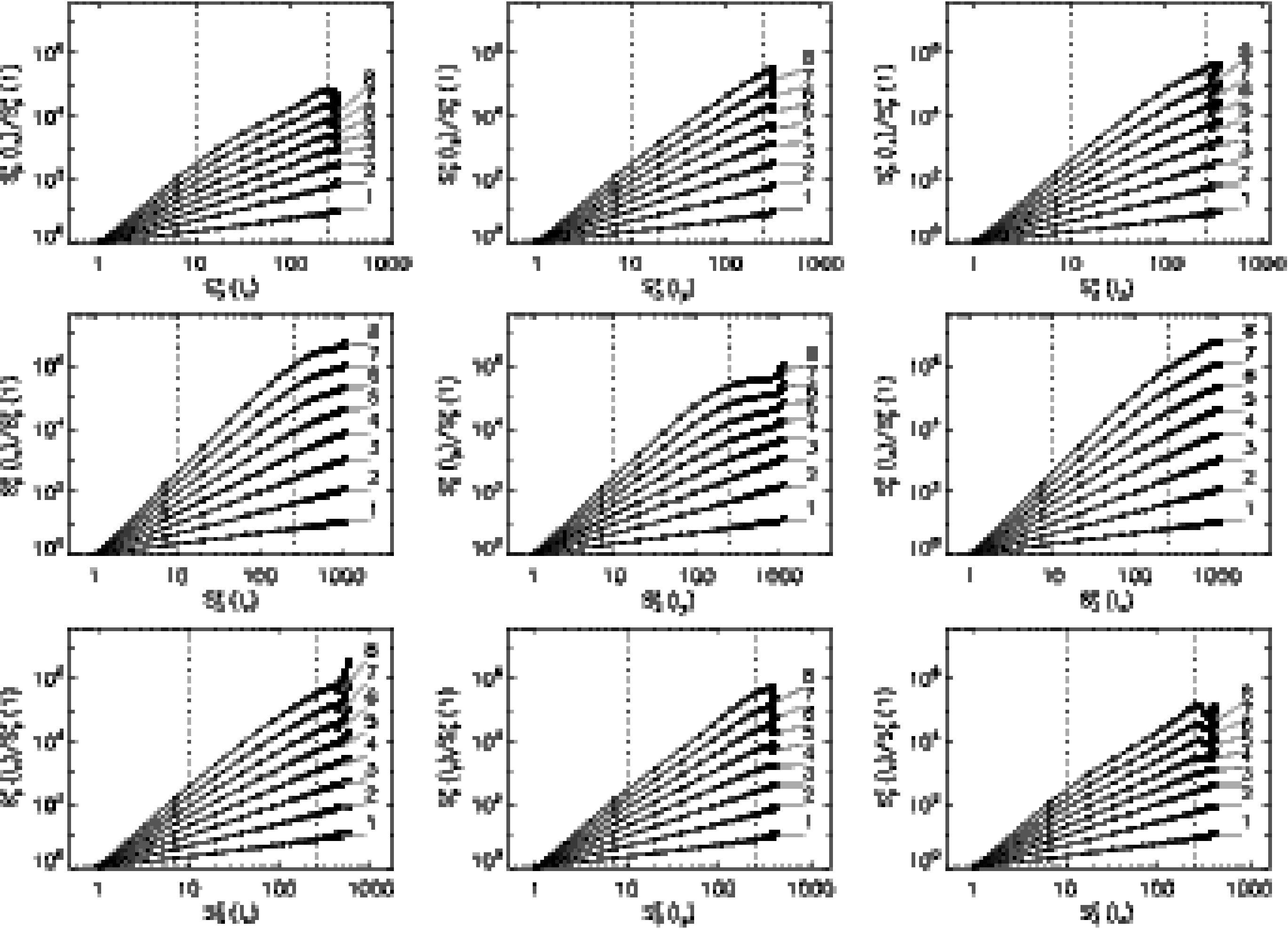}
  \caption[\same{assuming extended self similarity}]{%
    Same as Fig.~\ref{fig:stf_v}, but plotted versus $S_3(l)$, i.e., assuming
    extended self similarity. Dotted lines indicate the range used for fitting
    the logarithmic slopes $\zeta_p$.}
  \label{fig:stf_v_ess}
\end{figure}

The universality of Kolmogorov's four-fifths law for $S_3(l)$ has led
\citet{1993PhRvE..48...29B} to the discovery that the power-law character of
the structure functions emerges more clearly when plotted against the third
order one. The authors also showed that the range of scales which exhibit self
similar scaling can be extended towards the dissipation scale and hence termed
the new technique extended self similarity (ESS). The reason for this is that
the dissipation seems to affect all moments of the structure function in the
same manner, which means that the particular choice of $S_3(l)$ is to some
extent arbitrary. To obtain the scaling exponents $\zeta_p$ from our
simulations, we apply ESS as depicted in Figure~\ref{fig:stf_v_ess} and fit
the resulting slopes. Before discussing the outcome of our simulations we now
want to give a short account on recent phenomenological theories related to
intermittent turbulence.

\subsubsection{Intermittency corrections}

While the Kolmogorov picture for incompressible turbulence predicts a linear
dependence of the form $\zeta_{p}=h\,p$ with $h=1/3$, experimental results
show deviations from this behaviour for $p>3$ \citep[see][and references
therein]{1994PhRvL..72..336S}. One idea related to this discrepancies is that
dissipation does not occur in the whole volume but only in a fractal subset
related to, e.g., vortex filaments (in the case of incompressible turbulence)
or current sheets (in the case of MHD turbulence). The geometric approach of
the analytical $\beta$~model \citep[cf.][]{1995tlnk.book.....F} explicitly
takes into account the fractal co-dimension $C=3-D$ of the assumed dissipative
structures. The subset of the turbulent flow that is available for the
turbulent cascade is then effectively reduced by a factor $f_l\propto l^C$.
The resulting scaling exponents are simple linear functions $\zeta_p =
\frac{p}{3}+C(1-\frac{p}{3})$. A turbulent flow may, however, exhibit a range
of scalings, e.g., near shock structures the scaling parameter $h$ may be
reduced. Since structure functions of higher order are more strongly
influenced by singularities, the lowest value of $h$ will dominate at a given
order $p$. The scaling exponent $\zeta_p=\inf [ h\,p+3-D(h)]_h$ is then
defined as the infimum of the given fractal model of scaling $h$. This means
that the functions $\zeta(p)$ and $D(h)$ are related via the Legendre
transform \citep[cf. Sec.~5.3 in][]{1995tlnk.book.....F}. These so-called
multi-fractal or multi-scaling models can be expressed in a closed formula
\begin{equation}
  \zeta_{p}^{\rm SL}=\frac{p}{9}+2\,
  \left(1-\left(\frac{2}{3}\right)^{p/3}\right)\,,\label{eq:zeta_SL}
\end{equation}
as derived by \citet{1994PhRvL..72..336S}, who assumed a value $C=2$
corresponding to one-dimensional vortex filaments. The formula has been
further generalised by \citet{1995PhRvE..52..636P}, who propose a relation
\begin{equation}
  \zeta_{p}^{\rm PP}=
  \frac{p}{g}(1-x)+C\,\left(1-(1-\frac{x}{C})^{p/g}\right)\,,\label{eq:zeta_PP}
\end{equation}
where the coefficients $g$ and $x$ depend on the (assumed) scaling of the eddy
velocity $v_{l}\sim l^{1/g}$ and turnover time $\tau_{l}\propto l^{x}$. In
particular, these relations can be chosen to suite the conditions of MHD
turbulence \citep*[see e.g.][for a more recent account]{2003PhRvE..67f6302M}.

A corresponding specialisation relevant for interstellar turbulence has been
proposed by \citet{2002ApJ...569..841B}, who takes the paradigm of so-called
strong Alfv{\'e}nic turbulence \citep{1995ApJ...438..763G} as a starting
point. This phenomenological description of magnetic turbulence is based on
the assumption of a critical balance between the Alfv{\'e}nic timescale of
magnetic perturbations along the field lines and the hydrodynamic turnover
time of eddies perpendicular to this direction. The central prediction of the
phenomenology are a scale dependent anisotropy $k_{\parallel}\propto
k_{\perp}^{2/3}$ and scaling relations $v_{l}\propto l^{1/3}$ (i.e.,
Kolmogorov) and $\tau_{\rm GS} \propto l^{2/3}$, respectively. Based on these
relations together with an assumed co-dimension of $C=1$ (corresponding to
dissipation in current-sheets), \citet{2002ApJ...569..841B} arrives at a
scaling
\begin{equation}
  \zeta_{p}^{\rm B}=\frac{p}{9}+1-3^{-p/3}\,. \label{eq:zeta_B}
\end{equation}
Most notably, the respective energy-spectrum $E(k_{\perp})$ is found to
exhibit a scaling exponent $-(1+\zeta_{2})\sim-1.74$, which is compatible to
the observational Larson law
\begin{equation}
  \left\langle u_l^2 \right\rangle \sim l^{\,0.74 \dots 0.76}
\end{equation}
based on molecular cloud data \citep{1981MNRAS.194..809L}.

\begin{figure}
  \center\includegraphics[width=0.96\columnwidth]{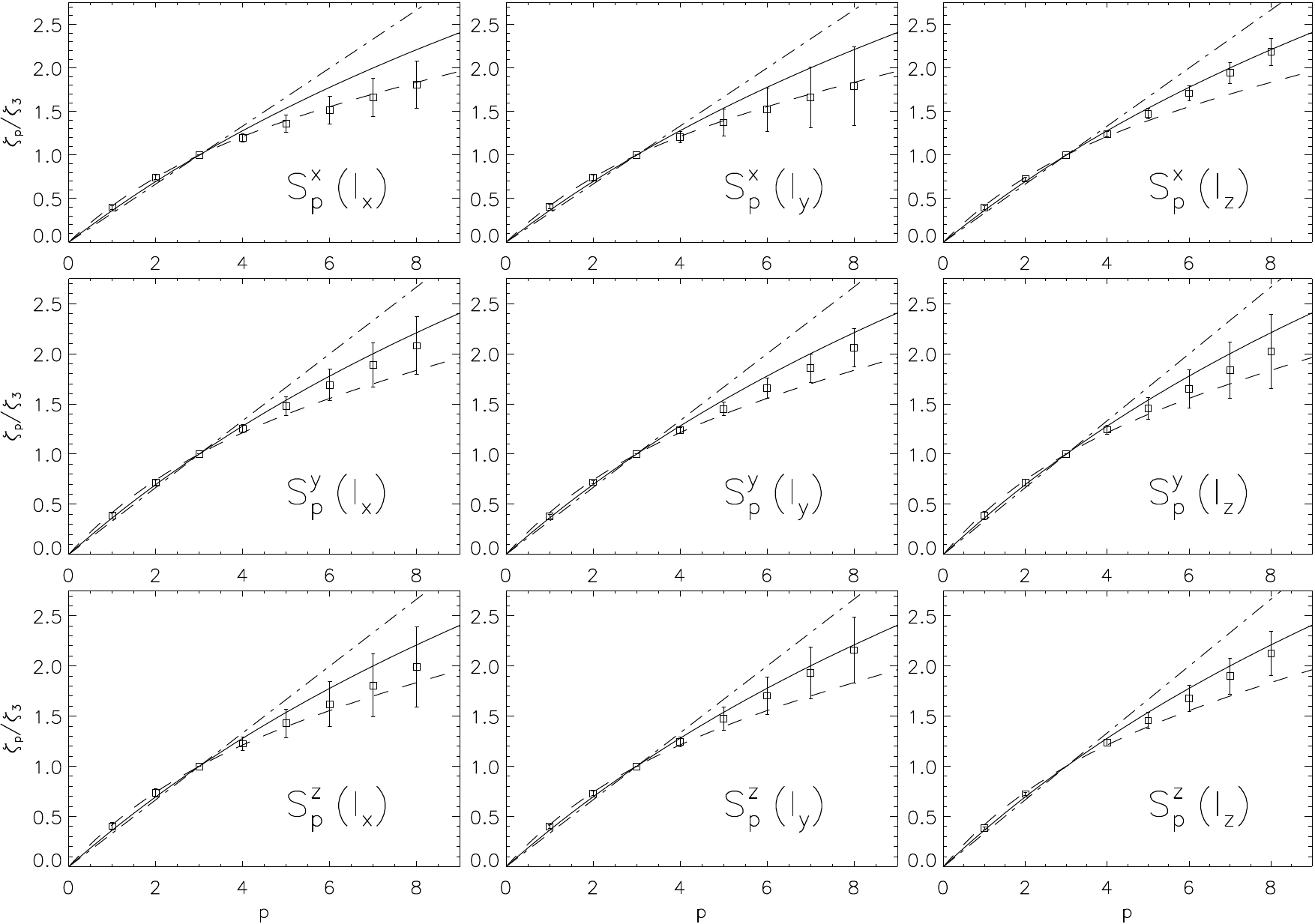}
  \caption[Scaling exponents of the velocity structure functions]{%
    Scaling exponents $\zeta_p$ of the velocity structure functions $S_p(l)$
    averaged over seven snapshots between $t=20$ and $31.5\Myr$. The
    overplotted curves indicate (from top to bottom) K41 scaling
    $\zeta_p=p/3$, the She-L{\'e}v{\^e}que formula $\zeta_p^{\rm SL}$,
    Eq.~(\ref{eq:zeta_SL}), and the Boldyrev formula $\zeta_p^{\rm B}$,
    Eq.~(\ref{eq:zeta_B}).}
    \label{fig:stf_scl_v}
\end{figure}

In Figure~\ref{fig:stf_scl_v}, we present the scaling exponents $\zeta_p$ as
obtained from model F4--BOX. The error-bars correspond to the standard
deviation with respect to the time averaging, the uncertainty from the fit
procedure itself is considerably smaller. The scalings are consistent with
both the She-L{\'e}v{\^e}que and Boldyrev formula. This basically means that
interstellar turbulence does not fit a single scheme of ``vortex filaments''
versus ``current sheets'' but rather comprises a multitude of fractal
dimensions. Yet, the obtained curves nicely match the shape of the predicted
log-Poisson model, lending support to this particular mathematical approach
for describing the intermittency corrections. In contrast to our results,
\citet{2007ApJ...665L..35D}, without exception, find scalings consistent with
the Boldyrev formula. For the case of their MHD run, the authors, however,
observe a certain deviation from $D=2$, which they interprete as a
modification due to the back reaction of the (dynamically significant)
magnetic fields.

From our simulations, we do not observe a clear trend between the longitudinal
and transversal structure functions, which is in agreement with theory and
simulations by \citetalias{2006ApJ...653.1266J}, who actually find a trend but
rationalise it away by stating that the results are very sensitive to the
range of data points used for fitting. We find a similar behaviour for the
higher order moments: Going back to Figure~\ref{fig:stf_v_ess}, we see that,
e.g. in the middle left panel, the lower order structure functions are nearly
perfect power laws whereas the moments with $p>5$ flatten out for higher
values of $S_3(l)$. If the third order moment already shows non-monotonic
behaviour, the situation becomes even worse. Since the fits are strongly
affected by these data points, we have chosen to restrict the fit procedure to
the subinterval where the ESS curves exhibit obvious power laws (as indicated
by the dotted lines in Fig.~\ref{fig:stf_v_ess}). In conclusion, we can say
that a decisive interpretation (and comparison) of the inferred scalings will
require an improvement of the current methods.

\subsection{Magnetic structure functions} 

In the case of incompressible MHD turbulence, the velocity and the magnetic
field can be combined into the Els{\"a}sser variables ${\mathbf z}^{\pm}=
{\mathbf v}\pm {\mathbf v_{\rm A}}$ with the Alfv{\'e}n speed ${\mathbf
  v}_{\rm A}$ representing the magnetic field. The symmetry between magnetic
field and velocity is reflected in the analogy between the equation of
vorticity conservation and the induction equation. In consequence, one expects
that the turbulent properties of the two fields are tightly related -- but
does this still hold in the case of compressible MHD turbulence? To make
matters even more intricate, our simulations of the turbulent ISM comprise
an extremely large density contrast. The role of density fluctuations in the
interstellar turbulence is far from being understood, and recent numerical
investigations of \citet*{2007ApJ...658..423K} point at the fact that the
intermittency of the density and velocity field are indeed different.

\begin{figure}
  \center\includegraphics[width=\columnwidth]{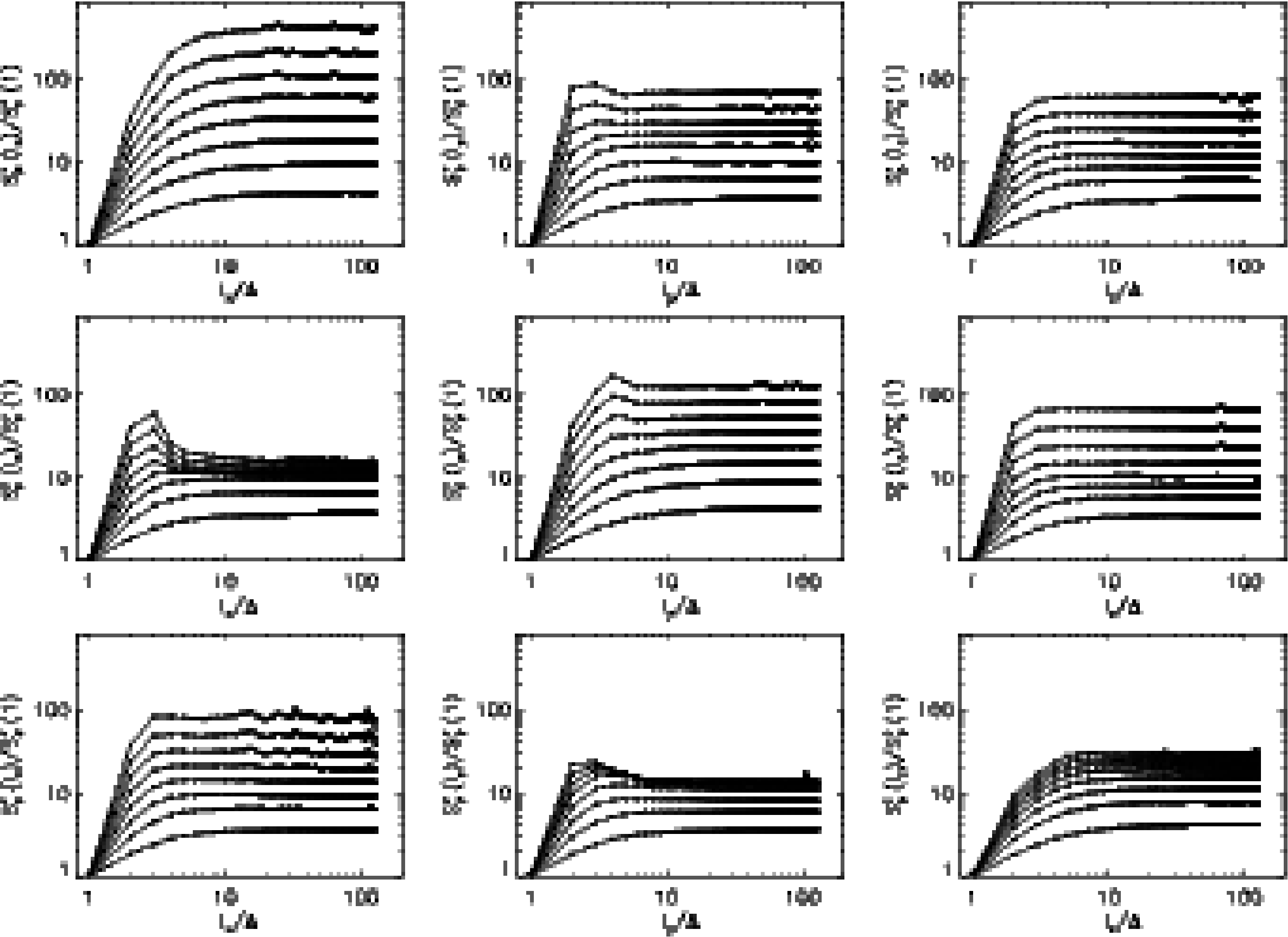}
  \caption[Magnetic structure functions]{%
    Same as Fig.~\ref{fig:stf_v}, but for the magnetic structure functions of
    order $p=1$--$8$ for model F4--BOX at time $t=20\Myr$.}
  \label{fig:stf_B}
\end{figure}

The velocity structure functions that we presented in the previous section
were mainly dominated by the hot and warm ISM components. In contrast, the
magnetic field is strongest in regions of high density (cf.
Sec.~\ref{sec:rhoB}). This means that the emerging magnetic structures are
largely confined to the cold ISM phase which in turn is very clumpy. In
consequence, the magnetic structure functions which we plot in
Figure~\ref{fig:stf_B} are almost flat on scales larger than a few parsec,
implying uncorrelated fields in-between the clumps. With only a few points
representing the inertial range, it becomes difficult to obtain reliable
scaling exponents for the magnetic field. Where this is possible, e.g. for
$S_p^x(l_x)$ (see upper left panel of Fig.~\ref{fig:stf_B}), we find a scaling
that is similar to the one for the velocity.

In conclusion, we can say that higher-oder two-point statistics are a viable
tool to study the intermittency corrections for the turbulent cascade. We have
demonstrated that they can proof useful even in the very complex scenario of
compressible interstellar turbulence. The high density contrast due to
condensations from thermal instability and the violent driving from SNe,
however, will make it necessary to consider density weighted quantities as a
modification to the current method. It is also perceived that an understanding
of the underlying theory is more easily gained by means of artificially forced
simulations. This is because simple setups are easier to restrict to certain
parameter regimes, and hence the dependence on a particular parameter
\citep[e.g. the Mach number of the flow, ][]{2004PhRvL..92s1102P} can be
extracted more clearly.

\section{The magnetised interstellar plasma} 

Apart from the mean-field dynamo, which is the main interest of this thesis,
the amplification of the small-scale, unordered field by driven turbulence
constitutes an intriguing question in its own right \citep[see e.g.][and
references therein]{2004ApJ...612..276S}. The fundamental difference between
the two processes is that the small-scale dynamo is always present in flows
with sufficiently high $\Rm$ (even in the absence of net helicity), whereas
the large-scale mechanism depends crucially on existing gradients to produce
an inverse cascade, i.e., to transfer magnetic energy to larger scales.

\subsection{Small-scale dynamo} 
\label{sec:rhoB}

For the amplification of the small-scale magnetic field within the
interstellar medium there are two fundamental mechanisms: (i) compression in
the shells of supernova remnants and (ii) shear from turbulent motions.%
\footnote{As a third effect, one might argue that the field can also be
  enhanced in condensations arising from the cooling instability. To the
  knowledge of the author, this has not been studied, yet.} %
The basic difference between field compression and shear is that in the former
magnetic flux (with respect to a Lagrangian fluid element) is conserved,
whereas in the latter magnetic flux can be created or destroyed.

While compressible amplification is probably dominant in the disk midplane,
where SNRs and SBs are more strongly confined to existing cavities, shearing
motions become more important with galactic height, where the explosions more
easily break up into unordered turbulence. This effect might also exhibit some
resolution dependence in the sense that the break up of the shells is probably
enhanced at higher numerical resolution because the nonlinear thin shell
instability \citep[NTSI, see e.g.][]{2007ApJ...665..445H} is more adequately
resolved.

Irrespective of the strong compressional driving via SNe,
\citet{2004ApJ...617..339B} and also \citet{2005MNRAS.356..737S} find the
solenoidal component of the velocity field to be dominating by more than one
order of magnitude. Only at the driving scale the two components contribute
equally. Similarly, \citet{1999A&A...350..230K} report $60$--$90\%$ of the
kinetic energy to be in vortical motions, and partly attribute this to the
so-called baroclinic effect. Conclusively, the high level of solenoidal
motions strongly supports the presence of a small-scale dynamo in the ISM.

If we assume that in the limit of ideal MHD the magnetic flux is frozen into
the fluid, i.e., bound to a Lagrangian fluid element, we can derive a simple
scaling relation. If we compress such an element, the density will scale
inversely with the volume, i.e., with the third power of the associated length
$l$. Because the magnetic flux through the surface is conserved, this means
that the magnetic field strength will scale with $l^{-2}$ implying a relation
$|B|\propto\rho^{\nicefrac{2}{3}}$.

A different approach, which goes back to \citet{1953ApJ...118..113C}, assumes
that the Alfv{\'e}n speed scales with the turbulent velocity $\delta v$,
yielding a relation of the form $|B|\propto\rho^{\nicefrac{1}{2}}\,\delta v$.
Since the flow in our simulations is supersonic and also super-Alfv{\'e}nic,
it is not clear in how far this scaling is applicable to our case. Radio
observations by \citet{1997A&A...320...54N} support a similar exponent of
$0.48\,\pm0.05$ for length scales down to about $100\pc$.

\subsection{Correlation with density} 

Numerical simulations of SN-driven turbulence generally produce a large
scatter in the $\rho|B|$ relation. \citetalias{2005A&A...436..585D} do not
even consider a fit and claim the magnetic field to be uncorrelated with
density. However, these models are already saturated with respect to the
magnetic pressure. In our simulations, when we try to correlate the magnetic
field strength with the density, we find a very broad distribution with
considerable scatter. If we, nevertheless, fit a power law relation we infer a
slope of about two-fifths, consistent with the results of
\citet{2005ApJ...634..390B}, who found a best-fit value of $0.386$. It is
worthwhile remarking that this is rather similar to the relation of our
initial model, where we compute a slope of $0.348$. Since our initial
stratification is based on a radiatively stable solution and assumes a
constant plasma parameter $\beta_{\rm P}$, it comprises the various effects
related to the radiative cooling and heating, but also the assumption of
equipartition of the magnetic field strength with the thermal energy. This
points at the possibility that the correlation between density and magnetic
field strength might be determined from equipartition with thermal energy
rather than equipartition with kinetic energy as assumed in
\citet{1953ApJ...118..113C}.

\begin{figure}
  \includegraphics[width=0.32\columnwidth]{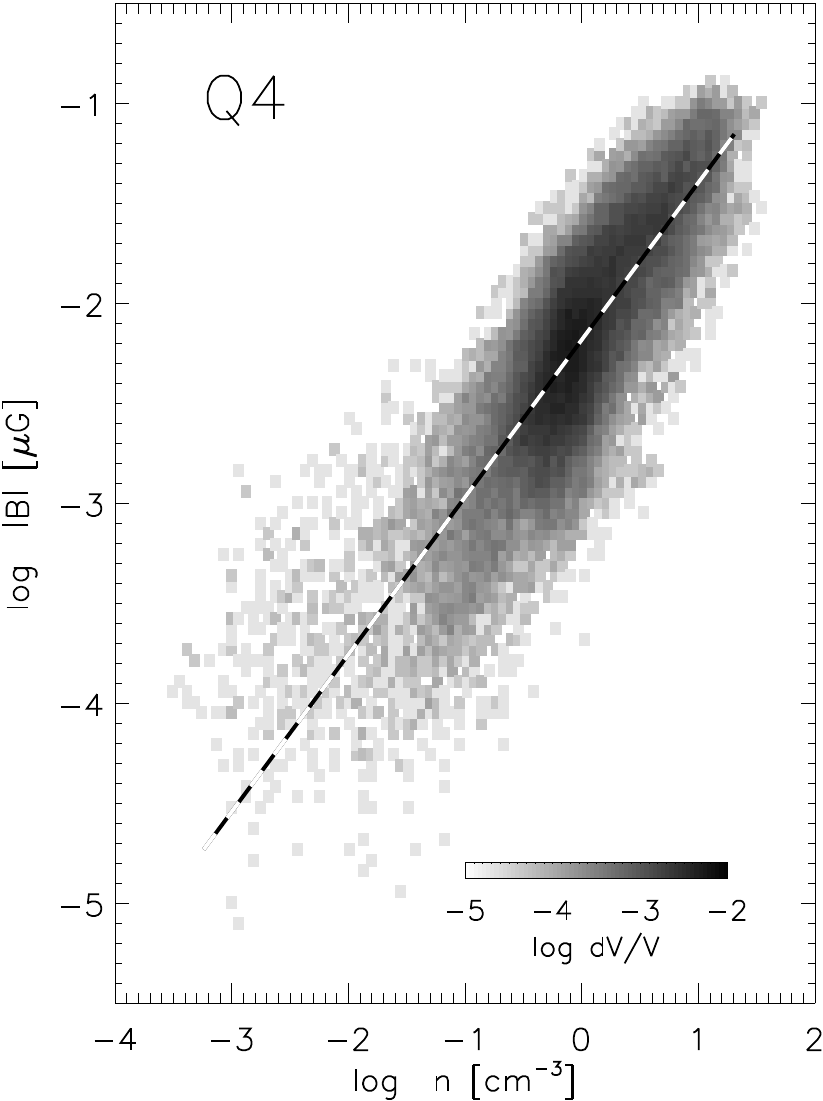}\hfill%
  \includegraphics[width=0.32\columnwidth]{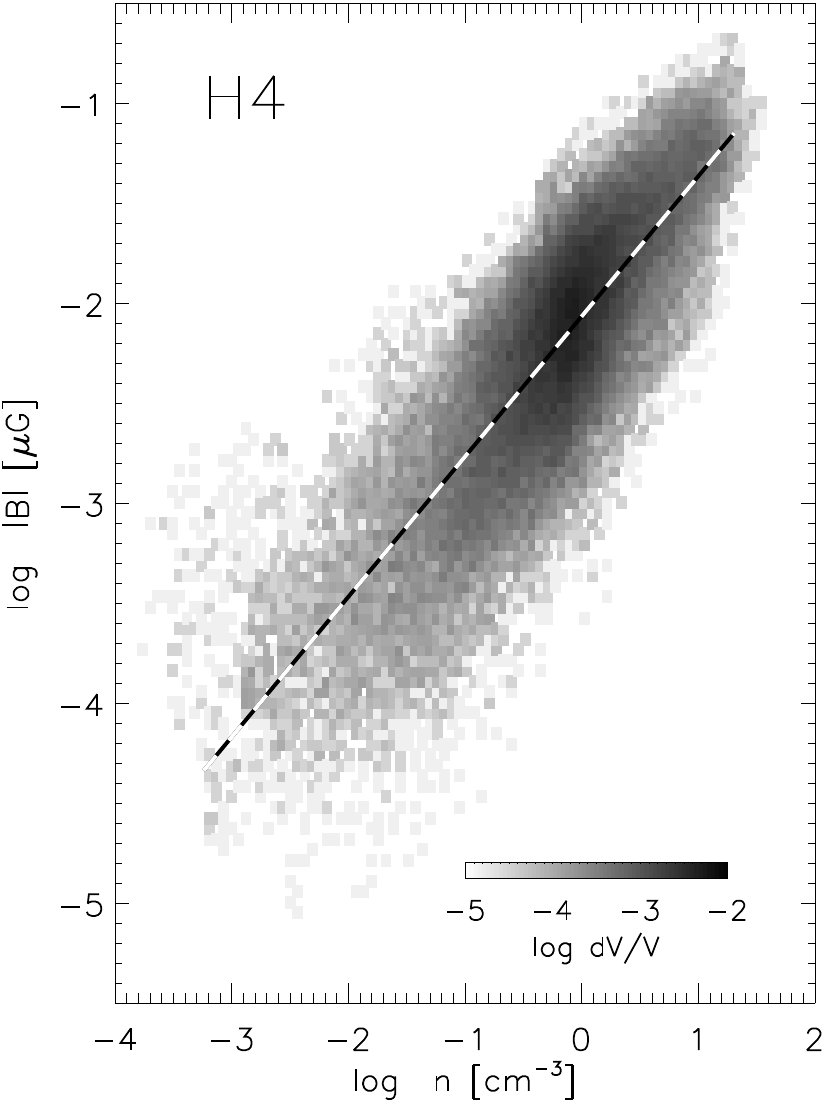}\hfill%
  \includegraphics[width=0.32\columnwidth]{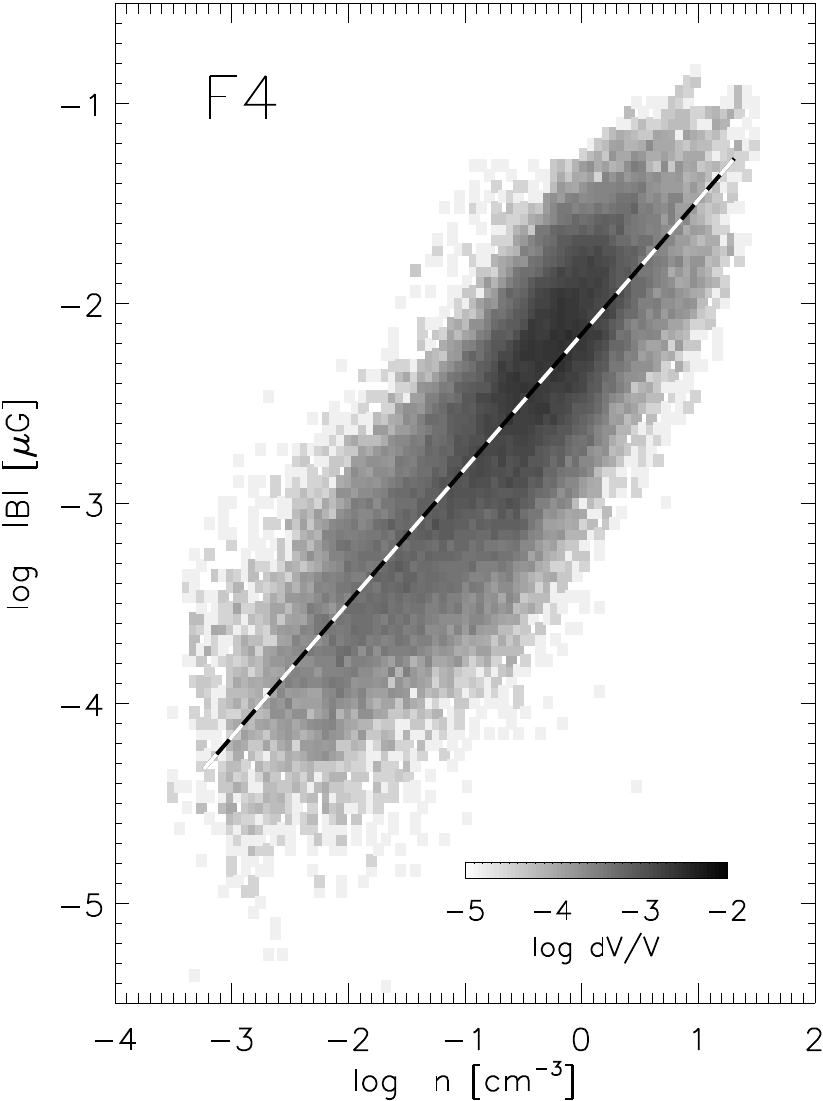}
  \caption[Correlation between density and magnetic field]{%
    Magnetic distribution within the midplane of models Q4, H4, and F4 between
    $t=120$ and $160\Myr$. Greyscales indicate the logarithmic volume fraction
    ${\rm d}V/V$ as a function of logarithmic density and magnetic field
    strength. The overplotted linear regression shows slopes of
    $0.787\,\pm0.005$ for model Q4, $0.701\,\pm0.005$ for model H4, and
    $0.671\,\pm0.005$ for model F4, respectively.}
\label{fig:rhoB_QHF}
\end{figure}

The distinctness of the correlation considerably improves if we restrict
ourselves to the disk midplane. In Figure~\ref{fig:rhoB_QHF}, we show scatter
plots of density versus magnetic field amplitude for the midplane gas of our
models Q4, H4, and F4. By means of a linear regression we fit slopes of
$0.787$ for model Q4, $0.701$ for model H4, and $0.671$ for model F4,
respectively. This is even a bit steeper than would be expected from the
simple picture of adiabatic field compression. The additional enhancement of
the magnetic field in high density regions might be explained by the modified
effective equation of state due to the radiative cooling.

\subsection{Vertical field structure} 

\begin{figure}
  \center\includegraphics[width=\columnwidth]{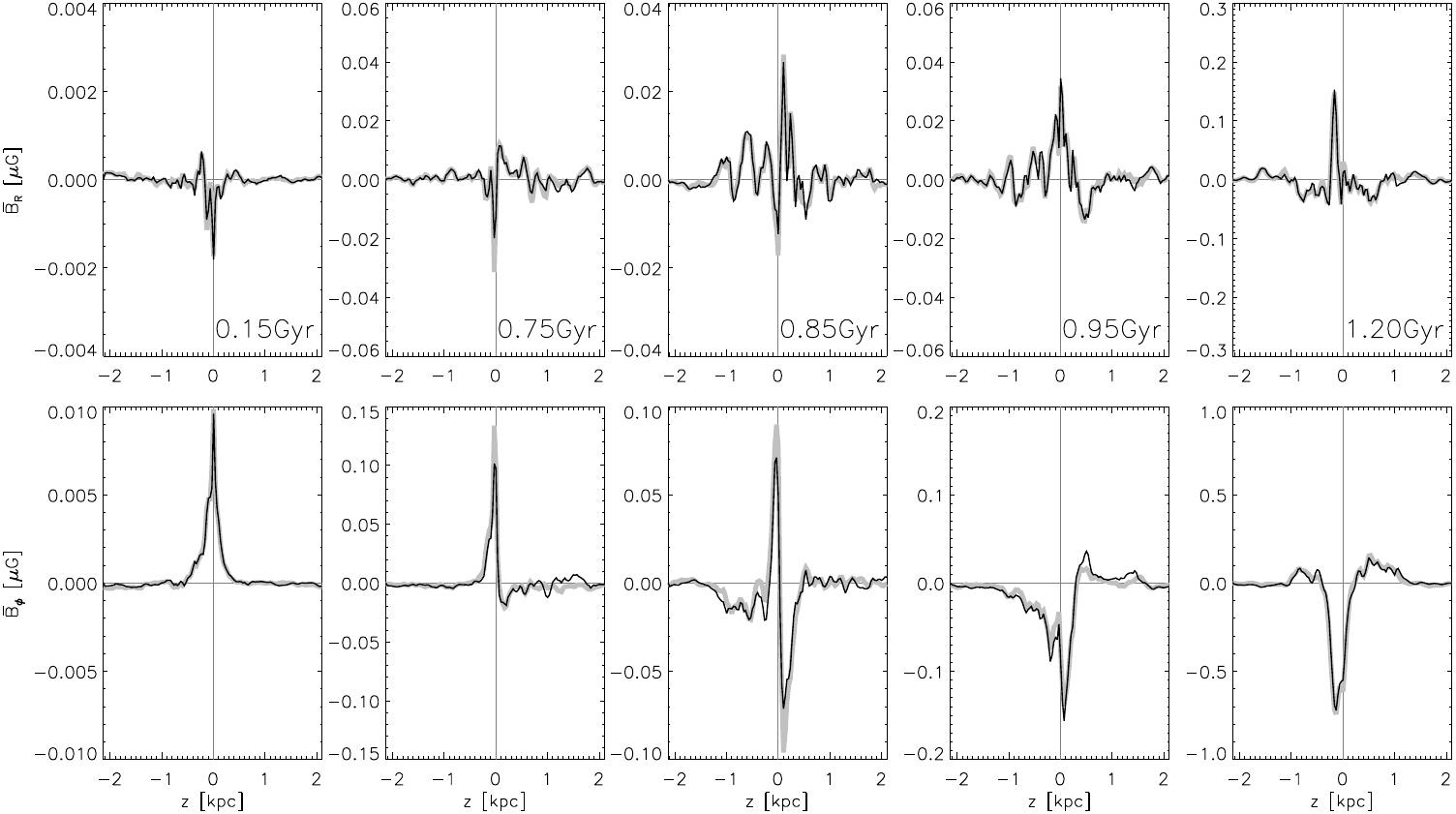}
  \caption[Profiles of the regular radial and azimuthal field for model H4]{%
    Profiles of the regular radial and azimuthal field for model H4 at various
    times. The results of the simulation (grey lines) are compared to
    reconsructed fields (black lines), computed from $\EMF(z,t)$ via the
    mean-field induction equation (cf. Sec.~\ref{sec:mf}). At $t\simeq
    0.85\Gyr$, a field reversal with pronounced dipolar symmetry occurs.}
  \label{fig:B_of_z}
\end{figure}

In Figure~\ref{fig:B_of_z} we plot vertical profiles of the mean radial and
azimuthal magnetic field components for our model H4. With a scale height of
$\simeq80\pc$, we observe the field to be largely confined to the inner disk,
i.e., the molecular gas near the midplane. In contrast, the simulations of the
CR buoyancy instability by \citet*{2007ApJ...668..110O}, who do not include
radiative cooling, show a shallower and also more irregular vertical
structure. In particular, their profiles exhibit a dip in the azimuthal field
along with a zero in the radial field at $z=0$. In our simulations, the
predominant symmetry with respect to the midplane is found to be even (i.e.
quadrupolar); this base mode is interrupted by field reversals displaying odd
(i.e.  dipolar) symmetry. The distinct, oscillating behaviour has been
successfully reproduced in the 1D toy model (cf. Sec.~\ref{sec:transport}),
where the occurrence and frequency of the periodic field reversals depend
critically on the interplay of the diamagnetic transport and the mean vertical
velocity \citep[cf. also][]{2001A&A...370..635B}.

Observations support quadrupolar symmetry \citep[cf. Sec.~8.1
in][]{1996ARA&A..34..155B}, which is also the prevailing mode in most dynamo
simulations. Even for a fast dynamo mechanism, this poses the question of
suitable seed fields of quadrupolar type to produce equipartition fields
within the required time. As has been shown by \citet{2004A&A...424..565K},
such a seed field geometry can be provided by means of MRI.

Because of the tremendous timescale, the reversal phenomenon will, of course,
never be observed directly. This unexpected finding, nevertheless, poses an
interesting question for galactic dynamos: It has been found that four out of
five galaxies show a radial field directed towards their centre
\citep[cf.][]{1998A&A...335..789K}. Based on this small observational sample,
the hypothesis has been put forward that the radial magnetic field has a
distinct direction in all spiral galaxies.\footnote{%
  Because dynamo theory is invariant with respect to this property, the
  preffered direction is thought to be rooted in the underlying seed field
  mechanism \citep{1998A&A...335..789K}.} %
Such a prediction, of course, is incompatible with an oscillating dynamo mode
where both directions are expected to appear equally frequent. Ultimately,
three-dimensional, global dynamo models -- see discussion in Sec.~6.4 of
\citet{2004maun.book.....R} -- will have to show whether the reversals are an
artifact of the chosen box geometry.

\subsection{Pitch angles} 

\begin{figure}
  \begin{center}
    \includegraphics[width=0.32\columnwidth]{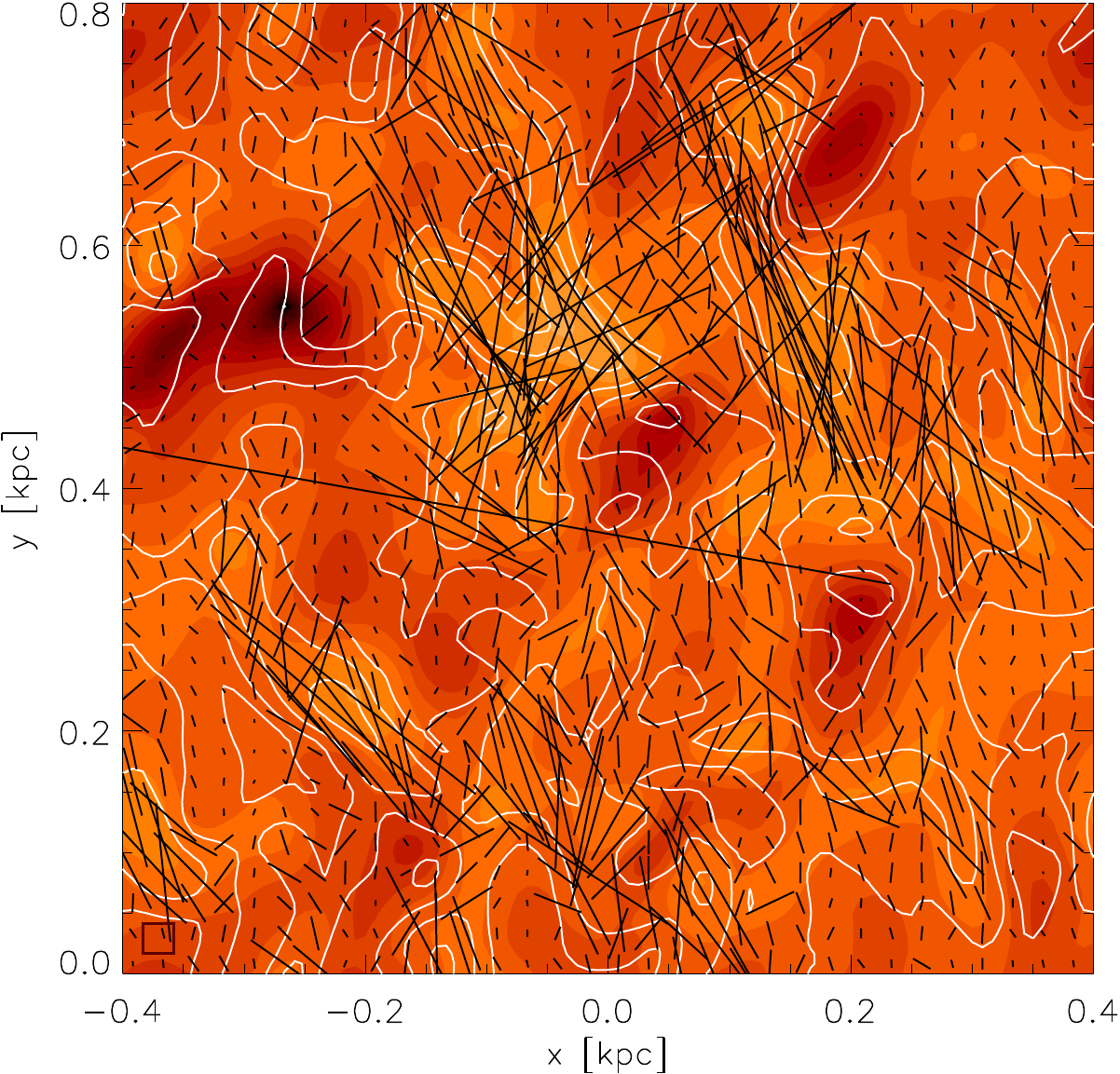}\hfill%
    \includegraphics[width=0.32\columnwidth]{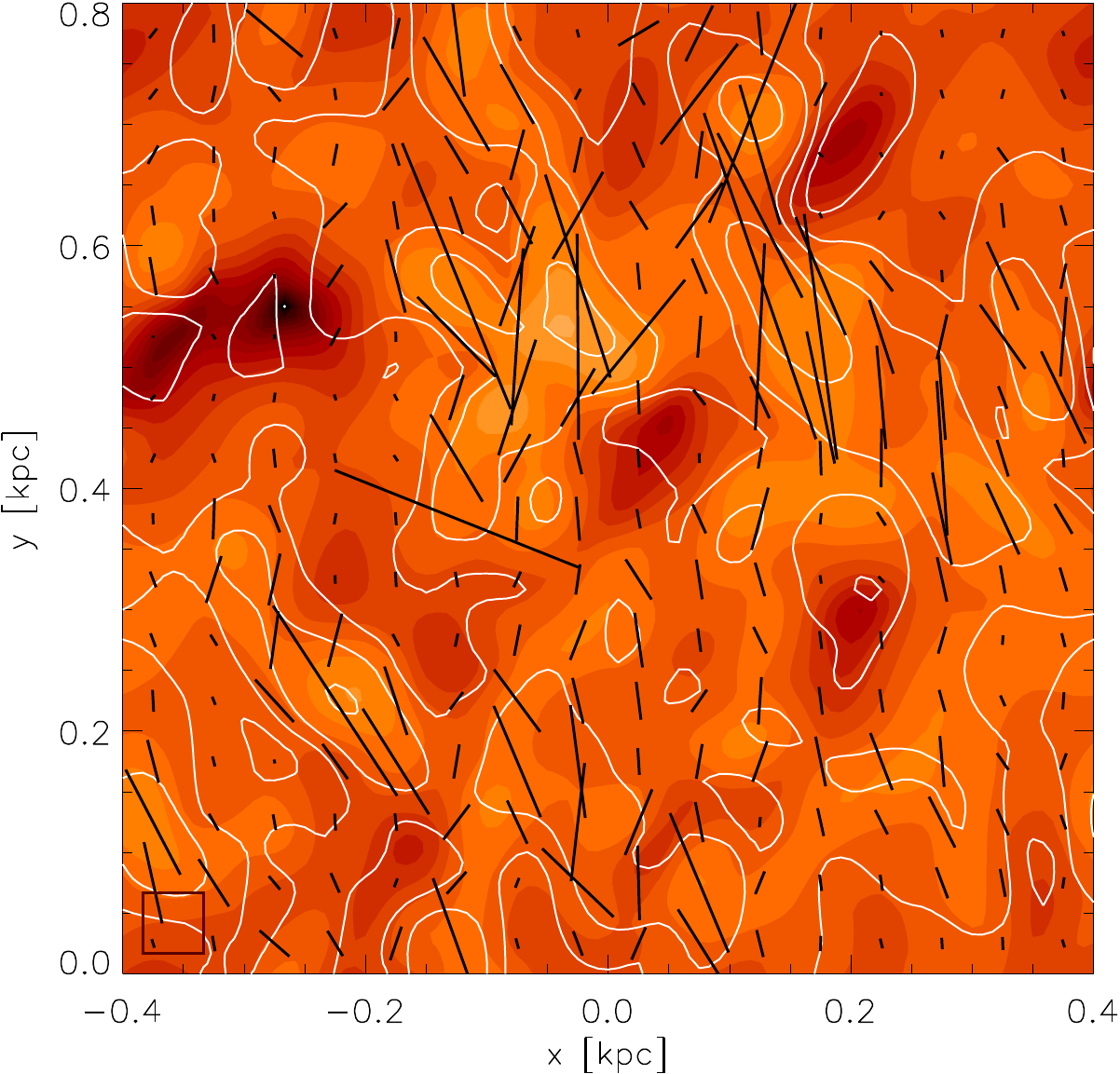}\hfill%
    \includegraphics[width=0.32\columnwidth]{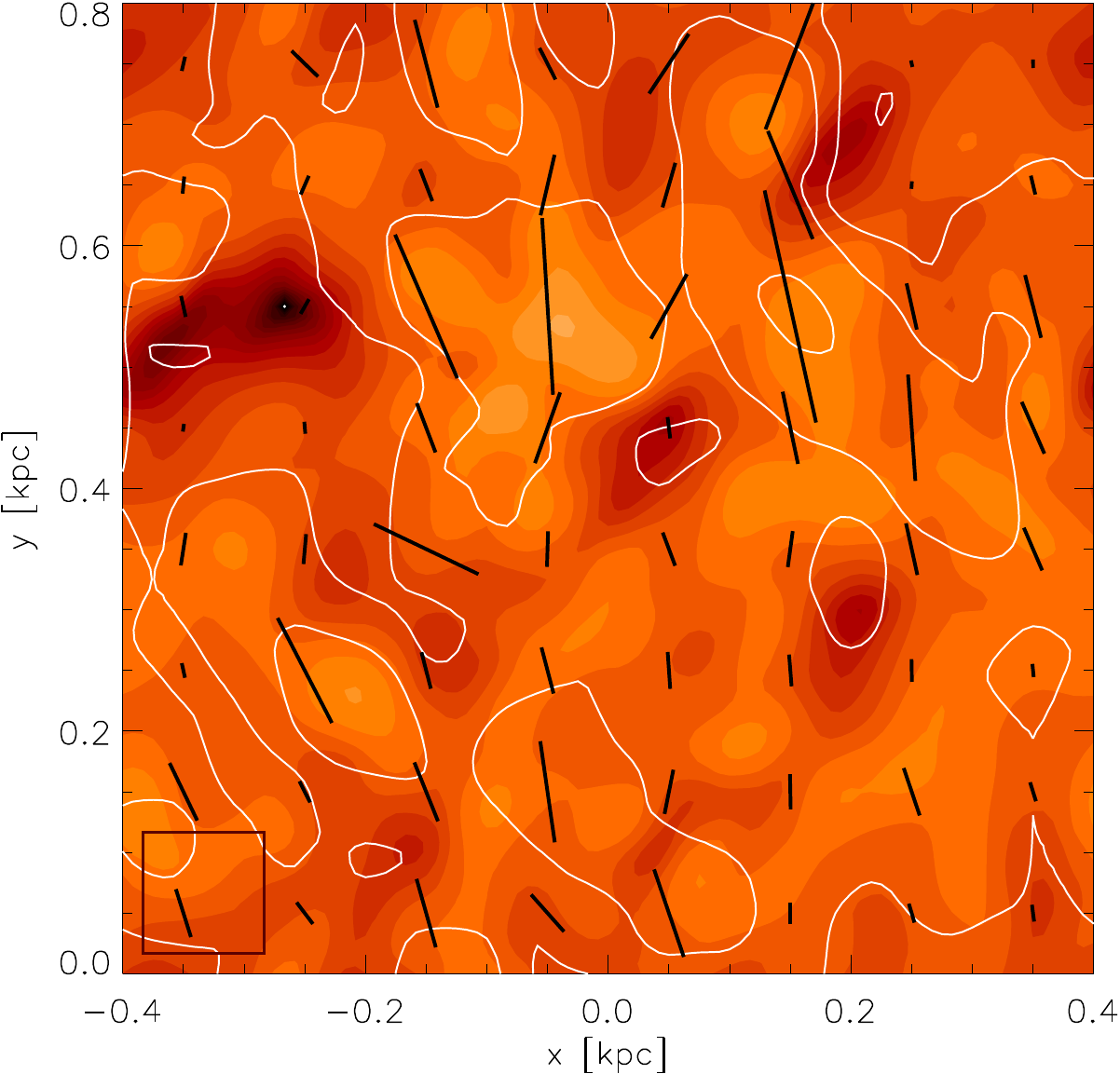}
  \end{center}
  \vskip-10pt%
  \caption[Synthetic radio observation]{%
    Synthetic observations of \HI\ column density (colour coding), total
    intensity (contour lines), and polarized intensity (vectors). The three
    panels demonstrate the effect of beam depolarisation when the maps are
    convolved with the kernel indicated by the square.}
  \label{fig:pol_map}
  \vskip20pt%
  \begin{center}
  \includegraphics[height=0.21\columnwidth]{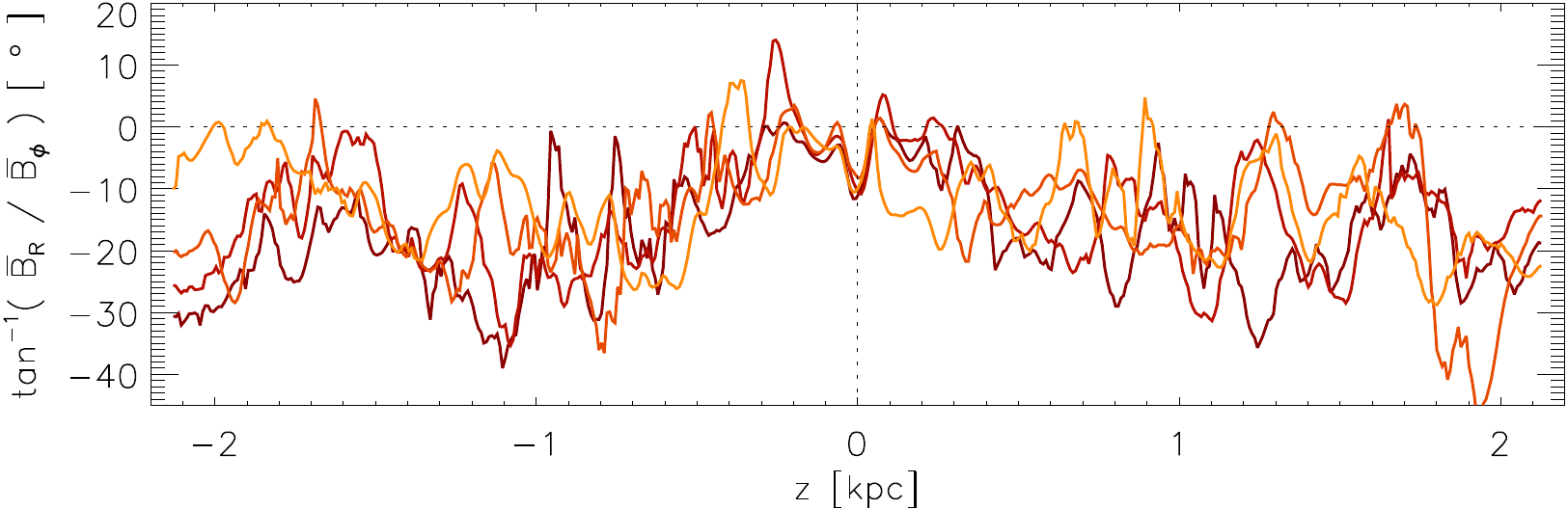}\hskip2ex%
  \includegraphics[height=0.21\columnwidth]{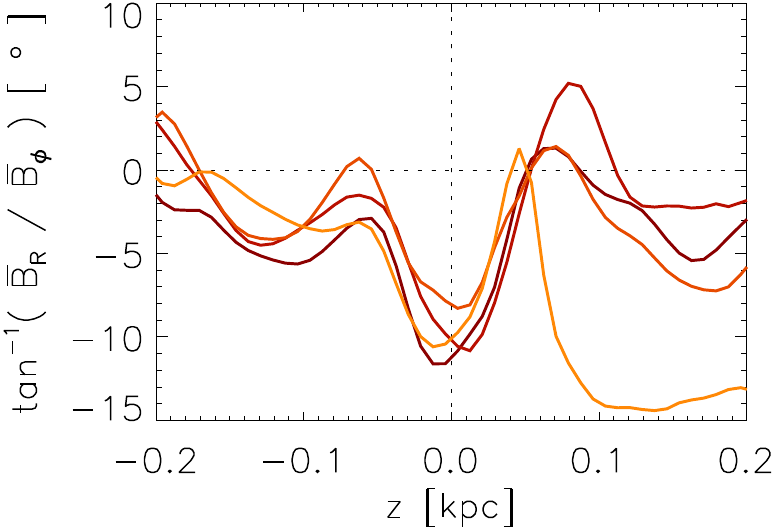}
  \end{center}
  \vskip-10pt%
  \caption[Radial pitch angle of the mean magnetic field]{%
    Radial pitch angle of the mean magnetic field averaged over four intervals
    of $\simeq 200\Myr$ at the beginning of model H4. The whole vertical range
    is displayed in the left panel while the right panel shows a close-up of
    the inner disk.}
  \label{fig:pitch}
\end{figure}

As we have already pointed out in the introductory section, the observed
radial pitch angles of the magnetic field lend strong support to the dynamo
paradigm. This means, in turn, that for any successful description of galactic
magnetic fields it is mandatory to explain the large pitch angles. The direct
comparison of simulation data with observations is somwhat complicated by the
fact that radio polarisation maps (i) only provide LoS integrated polarisation
vectors, and (ii) are affected by beam depolarisation, i.e., anti-parallel
components below the resolution given by the beam cross-section will cancel
each other out. In Figure~\ref{fig:pol_map}, we present synthetic polarisation
maps of our simulation results that have been obtained assuming a background
cyclotron emission of a relativistic electron gas with scaleheight $h_{\rm
  rel} \simeq 0.5\kpc$. Since in our simulations the field is largely confined
to the inner disk, the results are, however, rather insensitive to the
particular value of the scale height. The total intensity is computed as the
LoS integral of the synchrotron emissivity, which is assumed to be
proportional to the square of the perpendicular magnetic field component.

During their passage through the ISM, polarised radio waves are subject to
Faraday rotation. This is because, in the presence of a LoS component of the
magnetic field, left- and right-handed circularly polarized waves see a
different effective refraction index. For simplicity, we neglect the effects
of Faraday rotation in the integration of the Stokes Q- and U-parameters,
which reflect the local orientation of the field component orthogonal to the
line-of-sight. This is justified by the fact that we do not consider a mean
vertical field in our simulations. It shall ,however, be mentioned that the
turbulent component will cause some degree of Faraday depolarisation.

In the left panel of Figure~\ref{fig:pol_map}, we see that there exist regions
of strongly polarised emission showing a considerable pitch angle. As
expected, the orientation of the vectors corresponds to a trailing spiral,
i.e., a negative sign of the pitch angle. To mimic the effect of the finite
beam cross section, we apply convolutions with increasing kernel sizes of
$25$, $50$, and $100\pc$ to the obtained maps (indicated by the square in the
lower left corner of the maps). The vector scale is held fixed and we see that
the amplitude of the polarisation vectors is drastically reduced at a beam
size of $100\pc$. The contrast in the total intensity is less affected, which
implies that the overall polarisation level is generally underestimated by
observations of limited resolution.

Applying density and temperature thresholds (cf. Sec.~\ref{sec:dispersion}) to
select the \HI\ gas, we are able to compare the \HI\ column density to the
total intensity in the radio emission. High column densities correspond to
light colours in Figure~\ref{fig:pol_map}, and we see that the radio emission
is reasonably correlated with the atomic gas. This is little surprising
considering the strong correlation of the magnetic field with the gas density
in the midplane (cf. Sec.~\ref{sec:rhoB}). It remains to be studied whether
this correlation with density carries over to the larger scales, i.e. if the
coherent magnetic fields decouple from the neutral gas. Observations of the
ringed galaxy NGC4736 (see Fig.~\ref{fig:ngc4736}), e.g., show a pronounced
anti-correlation between the polarised intensity and the $H\alpha$ emission
\citep{2008ApJ...677L..17C}.

Independent of observational uncertainties, Figure~\ref{fig:pitch} provides a
more quantitative measure of the pitch angles present in our simulations. The
different curves show time averaged vertical profiles for the orientation of
the mean magnetic field in the horizontal direction. Despite the strong
fluctuations, we consistently observe negative values ranging up to $-40\degr$
throught the disk. With the exception of the midplane, the profiles closely
follow the shape of the $\alpha$~profiles that we will introduce in
Section~\ref{sec:dynamo_rot}. This is in accordance with mean-field theory,
which predicts that a strong $\alpha$~effect is necessary to produce a
substantial pitch angle. In the midplane, where the magnetic field is
strongest, we find a value of $-10\degr$ (see right panel of
Fig.~\ref{fig:pitch}). This value is found to be largely independent of the
rotation rate and supernova frequency.

\subsection{Regular versus turbulent component} 
\label{sec:reg_tur}

\begin{figure}
  \begin{minipage}[b]{0.66\columnwidth}
    \includegraphics[width=0.95\columnwidth]{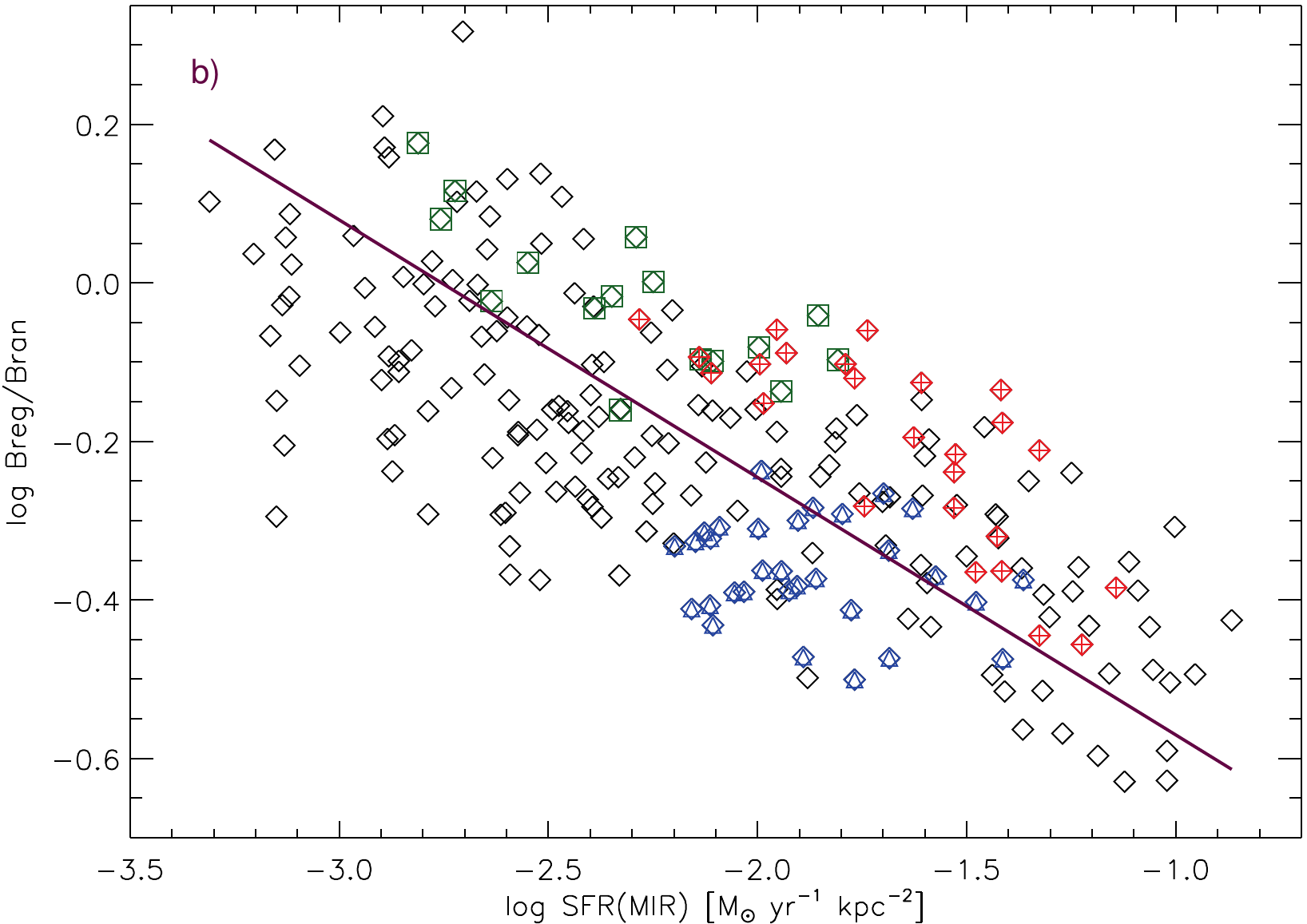}
  \end{minipage}
  \begin{minipage}[b]{0.33\columnwidth}
    \caption[Field regularity as a function of IR-based SFR within NGC4254]{%
      Field regularity as a function of IR-based SFR within NGC4254
      \citep{2008A&A...482..755C}. The symbols indicate the different magnetic
      arms. Figure courtesy of K.T. Chy{\.z}y.\vspace{10pt}}
    \label{fig:chyzy_7b}
  \end{minipage}
\end{figure}

To study integral properties as well as the temporal evolution of the arising
fields, we introduce vertically integrated rms~values $\rms{\bar{B}_R}$ and
$\rms{\bar{B}_{\phi}}$. In our simulations, the turbulent component dominates
over the regular by a factor of 2--3; we find values $\rms{\bar{B}}:\rms{B'}$
of $0.52\,(\pm0.02)$ for model Q4, $0.40\,(\pm0.03)$ for model H4, and
$0.31\,(\pm0.01)$ for model F4. This trend with $\sigma$ is consistent with
observations of strong regular fields in the inter-arm regions of spiral
galaxies \citep{2007A&A...470..539B}. From IR-based star formation rates,
\citet{2008A&A...482..755C} observes a correlation
\begin{equation}
  \log\ (B_{\rm reg}:B_{\rm tur}) =
  -0.32\,(\pm 0.01) \log {\rm SFR} - 0.90\,(\pm 0.03)\,.
\end{equation}
From our values cited above, we find a somewhat steeper slope of
$-0.38\,(\pm0.01)$. To obtain an absolute comparison, we assume a Salpeter
initial mass function (IMF) approximated by $\Psi(M)=k\,M^{-\gamma}$ with
$\gamma=2.35$ and integrate the local SFR/SNR via
\begin{eqnarray} 
  {\rm SFR} = & \int M\Psi(M)\,{\rm d}M & [\Msun\Myr^{-1}\kpc^{-2}]
  \,,\quad{\text and} \\
  {\rm SNR} = & \int  \Psi(M)\,{\rm d}M & [\Myr^{-1}\kpc^{-2}]\,.
\end{eqnarray}
Choosing appropriate mass limits for the two integrals, we can convert our
reference SN rate $\sigma_0$ to an equivalent of $\log\,{\rm SFR} = -2.4$.
This means that our values fall short of the observed values and lie to the
lower left of the scatter plot depicted in Figure~\ref{fig:chyzy_7b}.
Considering that our model is mainly based on parameters obtained in the
context of the Galaxy (e.g. the gravitational potential), we have to be
careful when comparing to different galaxies. It may be interesting to check,
whether the correlation can be matched more closely by a modified setup for
the case of NGC4254. Also further observations will be needed to proof the
universality of the observed correlation. Nevertheless, the quite good
agreement between our simulations and the observations indicates that this
relation is probably rather general.

\section{Mean-field dynamo} 
\label{sec:mf_dynamo}

Despite the early success of the mean-field models, until now there has been
no direct numerical verification of the dynamo process in the galactic
context. Although \citetalias{2005A&A...436..585D} considered the most
realistic model of the interstellar medium to date, they did not include the
galactic rotation and shear necessary for a mean-field dynamo to operate.
\citet{2004ApJ...617..339B} found a small-scale dynamo, but also neglected
rotation and even vertical stratification -- an even more important
prerequisite for large-scale dynamo action. Except for their neglect of
thermal instability,\footnote{%
  The possible influence of TI on the mean-field dynamo will be discussed in
  Sec.~\ref{sec:dynamo_TI}.} %
the simulations of \citetalias{1999ApJ...514L..99K} are very similar to ours.
In fact, their model served as a starting point for our investigations.
However, due to the limited computational resources available at the time, the
simulations suffered from a too small box size which prohibited a long term
evolution into developed turbulence.

In theory, a mean-field dynamo is already possible under the combined action
of Coriolis forces and stratification, i.e., without shear. However, for our
$q=0$ models we only observe a marginal amplification of the mean magnetic
field after the kinetic energy has reached a quasi-stationary state.
Estimations based on the derived $\alpha$~parameters indeed show that the
dynamo numbers for the $\alpha^2$~dynamo are slightly subcritical. At the
current point, we cannot conclusively state whether the reason for this is
indeed physical or merely a result of the limited magnetic Reynolds number of
the present simulations.

\begin{figure}
  \begin{minipage}[b]{0.66\columnwidth}
  \includegraphics[width=0.85\columnwidth]{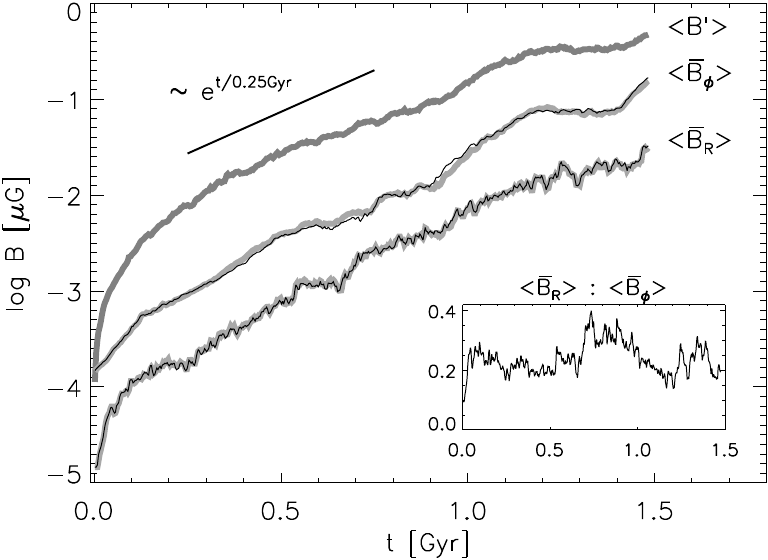}
  \end{minipage}
  \begin{minipage}[b]{0.33\columnwidth}
  \caption[Evolution of the turbulent and regular magnetic field for model H4]{%
    Evolution of the turbulent and regular magnetic field for model H4. For
    $\rms{\bar{B}_R}$ and $\rms{\bar{B}_{\phi}}$ we show the results from the
    direct simulation (grey lines) together with the reconstruction from
    $\EMF(z,t)$ (black lines). The inlay depicts the ratio of the mean radial
    versus azimuthal magnetic field corresponding to an average pitch angle of
    $\simeq 10\degr$.\vspace{10pt}}
  \label{fig:B_of_t}
  \end{minipage}
\end{figure}

If we include galactic differential rotation with $q=1$, our simulations
successfully produce a galactic dynamo, i.e., we observe an exponential
amplification of the mean magnetic field. The exponential growth of the
regular and fluctuating field in the simulation run H4 is depicted in
Figure~\ref{fig:B_of_t}. The e-folding time is on the order of $250\Myr$ and
varies with the reversals.

In the early phase of the evolution (but after the kinetic energy has reached
a quasi stationary state) we observe an e-folding time of $\simeq100\Myr$,
which is comparable to the values obtained in simulations of the cosmic ray
dynamo by \citet*{2007ApJ...668..110O} and about four times larger than the
expected growth time for the magneto-rotational instability
\citep*{2004A&A...423L..29D}. Such short growth times, which are rather
unexpected from classical theory, are beneficial in explaining strong magnetic
fields at high cosmological redshift \citep{2008Natur.454..302B}.

\subsection{Dependence on the main model parameters} 

After we have demonstrated that the galactic dynamo can indeed be operated by
the turbulence from supernova explosions we, of course, want to learn how the
growth rate of the dynamo is affected by the various parameters of our model.

\begin{figure}
  \center\includegraphics[width=0.9\columnwidth]{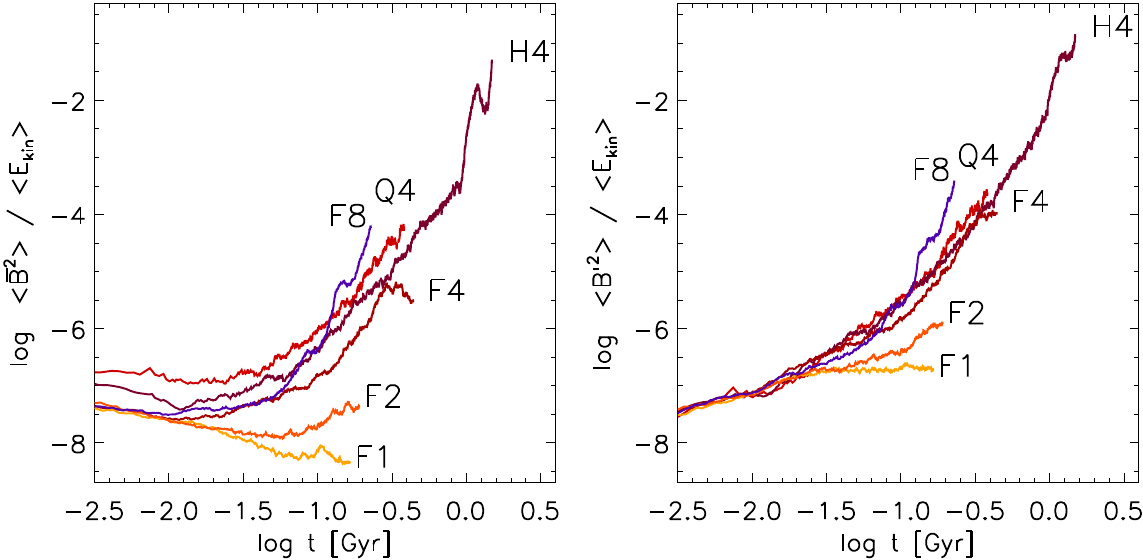}
  \caption[Comparison of evolution for the standard set of models]{%
    Comparison of the time evolution of the regular and fluctuating magnetic
    field strength over kinetic energy for the various setups as defined in
    Table~\ref{tab:models}.}
  \label{fig:B_of_t_comp}
\end{figure}

\begin{table}
\begin{minipage}[c]{0.66\columnwidth}
\begin{tabular}{llllllll}\heading{
growth time & & Q4 & H4 & F1 & F2 & F4 & F8}
of $\rms{B'}$ & $[\Myr]$ & 
$90$ & $92$ & $500$         & $140$ & $102$ & $54$\\
of $\rms{\bar{B}}$ & $[\Myr]$ & 
$90$ & $89$ & $500^\dagger$ & $147$ &  $99$ & $52$\\\hline
\end{tabular}
\end{minipage}
\begin{minipage}[c]{0.33\columnwidth}
  \caption[Exponential growth times for the standard set of models]{%
    Exponential growth times for the standard set of models. $\,^\dagger$) for
    model F1 we find the mean-field $\rms{\bar{B}}$ to decay in time.}
  \label{tab:growth_times}
\end{minipage}
\end{table}

Before we go into the discussion of our first basic parameter study
\citepalias{2008A&A...486L..35G}, we want to point out that, albeit many of
the chosen values are representative of our own galaxy, this is merely for
practical reasons. Another, more subtle issue is the inherent limitation of
the dynamic range of any numerical simulation. In view of the tremendous value
of the magnetic Prandtl number within the ISM, we especially have to keep in
mind that any result drawn from our simulations ideally has to be transformed
to this regime by means of a suitable scaling relation. Computations at even
moderate $\Pm$ are, however, extremely demanding. This is because both the
viscous and resistive length scales have to be resolved on the numerical grid.
For all practical purposes, this means that a strict proof of convergence is
currently beyond the available numerical capabilities.

We present the time evolution of the regular and fluctuating components of the
various models in the left panel of Figure~\ref{fig:B_of_t_comp} (cf. also
Tab.~\ref{tab:growth_times}). Rather surprisingly, even the absolute value of
the mean-field $\rms{\bar{B}}$ increases with decreasing SN rate (models F4,
H4, and Q4). This is consistent with the trend in the turbulent diffusivity
$\eta_{\rm t}$, for which we measure values of $2.0$, $1.7$, and
$1.4\kpc\kms$, respectively. Observations, on the other hand, suggest that
$\rms{\bar{B}}$ is independent of the star formation activity
\citep{2008A&A...482..755C}. 

For the range of parameters studied, we do not observe a significant
dependence of the growth rate on the supernova frequency $\sigma$, but only on
the rotation rate $\Omega$.\footnote{%
  Because we do not vary the shear parameter $q$ independently, we cannot yet
  determine whether the actual dependence is on $\Omega$, or rather on
  $q\Omega$. This will, however, be subject to future studies.} %
For the models F1--F8 we find e-folding times of $\simeq 500$, $140$, $102$,
and $54\Myr$ for the amplification of $\rms{B'}$. With exception of model F1,
that directly corresponds to the parameters used in
\citetalias{1999ApJ...514L..99K}, we find exponentially growing regular fields
$\rms{\bar{B}}$ with e-folding times of $147$, $99$, and $52\Myr$ for model
F2, F4, and F8, respectively. For model F1 the regular field decays at $\simeq
500\Myr$. The listed values have been obtained for a time frame of about
$100\Myr$ after the turbulence reaches a steady state. Due to the different
time base, these values are not directly comparable to the long-term growth
rate of model H4, where the field reversals become important. The initial
amplification time scale for the models H4 and Q4 is $\simeq90\Myr$, which is
consistent with the value observed in model F4.

In conclusion, it turns out that $\Omega \simgt 25\kms\kpc^{-1}$ is necessary
for dynamo action to occur, which coincides with the prediction of
\citet*{1994A&A...286...72S}. Nevertheless, this value may still depend on the
assumed gas density and the gravitational potential.

\subsection{The importance of rotation} 
\label{sec:results_shear_vs_rot}

As has been noted in the previous section, we did not observe a dynamo in our
simulations with rotation alone. Since, on the other hand, we observe a dynamo
in the case of combined rotation and shear, it occurs natural to ask whether
rotation then plays a role at all -- or whether it is simply the effect of the
shear that efficiently closes the dynamo loop. 

Ever since the origins of mean-field theory, rotation was considered the pivot
point in the generation of ``cyclonic turbulence''. As we already learned in
the introduction, mean-field dynamos critically depend on some sort of
anisotropy of the flow. The exception to this rule is the so-called
${\mathbf\Omega}\tms{\mathbf J}$-effect\footnote{%
  This effect is indeed present in our simulations as will be discussed in
  Section~\ref{sec:dynamo_tensor}.} %
\citep{1969MDAWB..11..272R}, which, in the presence of an inhomogeneous
magnetic field and differential rotation, already shows dynamo action for
homogeneous, isotropic turbulence \citep*{1994AN....315..157K}. Based on
recent shearing box simulations with peculiarly elongated aspect ratios and
$\Pm\simeq1$, \citet{2008PhRvL.100r4501Y} claim that a dynamo can already be
excited in the case of plain shear, i.e., in the absence of rotation. In
addition, the authors observe the generation of a large-scale vorticity in
their simulations. While the latter finding disagrees with the analytical
prediction of \citet{2006AN....327..298R}, the former is at least excluded for
order of unity magnetic Prandtl numbers by the same considerations.
\citet{2008ApJ...676..740B}, who infer dynamo coefficients from simulations,
conclude that ``the shear-current effect [without stratification] is
impossible''. Besides the possibility to design laboratory based dynamo
experiments with simplified geometry, the reason for the increased interest in
the particular case without rotation is mainly motivated by astrophysical
objects that do not show prominent rotation patterns such as irregular
galaxies and galaxy clusters.

\begin{figure}
  \center\includegraphics[width=0.90\columnwidth]{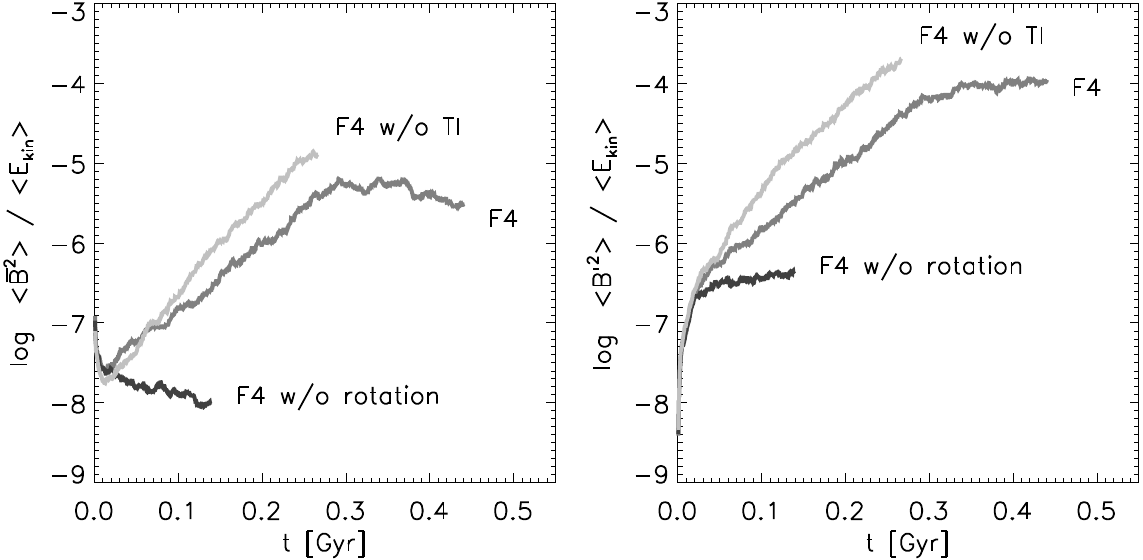}
  \caption[\same{for models F4--noTI and F4--SHR}]{%
    Evolution of the regular and fluctuating magnetic field strength over
    kinetic energy for model F4 compared to the cases of shear only (dark) and
    without thermal instability (light colour).}
  \label{fig:tmag2}
\end{figure}

With this recent controversy in mind, we now want to address the question, in
how far the galactic dynamo depends on rotation as a source of anisotropy. It
is known from theory that in the presence of a gradient in the turbulence
intensity there exists an $\alpha$~effect for a non-rotating sheared fluid
\citep{2006AN....327..298R}. Because our SN-driven galactic turbulence
exhibits a strong vertical stratification, we should hence be able to observe
this mechanism. In Figure~\ref{fig:tmag2} we plot the temporal evolution of
the mean and fluctuating magnetic field strength for our model F4, with and
without rotation.

We, first of all, notice that the irregular field is indeed amplified by the
combined action of turbulence and shear, although at a much lower rate as
compared to the case where the Coriolis force acts as an additional source of
helicity. Whereas the turbulent field grows steadily, we observe a decaying
solution for the mean magnetic field. As we will see from a detailed analysis
of the measured dynamo coefficients in Section~\ref{sec:dynamo_rot}, this is
not because there is no $\alpha$~effect, but because the diamagnetic pumping
is too weak to support the dynamo against the galactic wind. In conclusion, we
record that (differential) rotation is a necessary prerequisite for an
efficient galactic dynamo to operate.

\subsection{The effects of secondary parameters} 

Besides the two fundamental effects of rotation and shear, our model depends
on a number of assumptions regarding further physical input parameters. The
two most prominent features which might have an influence on the overall
outcome of our simulations are both related to the vertical structure of the
disk and, in the following, shall be briefly explored.

\subsubsection{The role of thermal instability}
\label{sec:results_TI}

As we mentioned before, the models of \citetalias{1999ApJ...514L..99K} served
as a starting point for our investigations. A major difference of our model,
compared to theirs, is the inclusion of the thermally unstable branch of the
radiative cooling function, leading to condensations via a cooling
instability. Because \citeauthor{1999ApJ...514L..99K} did not find a galactic
dynamo in their simulations, this poses the interesting question whether TI
plays an essential role in enhancing the inward diamagnetic pumping which
helps to improve the prospects of a super-critical dynamo. Recalling the
change of the disk morphology in the case without TI (see
Sec.~\ref{sec:gas_prof}) and the resulting change in the velocity structure
(see Sec.~\ref{sec:velocity_prof}), this seems a reasonably plausible
hypothesis. An alternative explanation, of course, might be given by the lower
rotation rate of $\Omega=25\kms\kpc^{-1}$ in their models. Because we find
this value to be marginal for the operation of a dynamo, the role of thermal
instability, however, remains ambiguous without a direct comparison run.

In Figure~\ref{fig:tmag2}, we display the regular and irregular components of
the magnetic field for the model F4--noTI. We measure e-folding times of $78$
and $87\Myr$ for the irregular and mean magnetic field, respectively. This is
somewhat smaller than the $100\Myr$ in the standard case. If we take the
gradient of the turbulent velocity as the defining criterion for the dynamo
efficiency, we should arrive at the conclusion that the weaker gradient in the
turbulent velocity will cause a smaller value for the inward pumping. Even
with the less pronounced wind this would mean that the conditions for our
dynamo should have become worse -- instead we find a higher growth rate. We
already see at this point, that even a qualitative analysis of the results
becomes tedious, based on kinematic quantities and intuition alone. The
interpretation of the results will, however, become easier if we include the
knowledge about the dynamo coefficients which will be derived in the following
chapter. While we have to postpone the explanation for the increased growth
rate of model F4--noTI to Section~\ref{sec:dynamo_TI}, we already want to
mention that the thermal instability does not seem to affect the
characteristic ratio $\hat{\gamma}$ of the diamagnetic pumping over the
$\alpha$~effect. We conclude that \citetalias{1999ApJ...514L..99K} simply did
not reach the critical rotation rate, and the inclusion of TI does not
significantly alter the picture.

\subsubsection{Varying the external potential}
\label{sec:results_grav}

\begin{figure}
  \center\includegraphics[width=0.90\columnwidth]{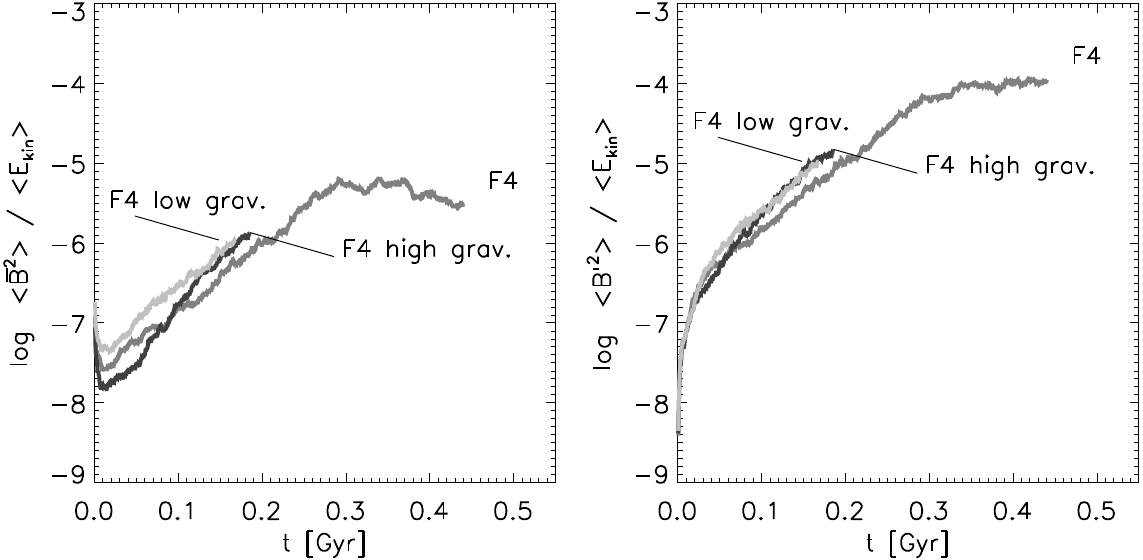}
  \caption[\same{for model F4 at varying gravity}]{%
    Time evolution of the regular and fluctuating magnetic field strength over
    kinetic energy for model F4 under low, standard, and high gravity.}
  \label{fig:tmag3}
\end{figure}

\begin{table}
\begin{minipage}[c]{0.66\columnwidth}
\begin{tabular}{llccc}\heading{growth time & & low & F4 & high}
  of $\rms{B'}$ & $[\Myr]$      & $94$ & $102$ & $83$ \\
  of $\rms{\bar{B}}$ & $[\Myr]$ & $97$ & $99$  & $76$ \\
\end{tabular}
\end{minipage}
\begin{minipage}[c]{0.33\columnwidth}
  \caption[\same{for model F4 at varying external gravity}]{%
    Measured exponential growth times for model F4 at varying external
    gravity.}
\end{minipage}
\label{tab:growth_grav}
\end{table}

As we can see in Figure~\ref{fig:tmag3}, the impact on the dynamo when varying
external gravity is less drastic than the effects of neglecting rotation or
thermal instability. Even though we alter the gravitational force by a factor
of two, we do not observe a significant change in the growth rate of the
dynamo. The e-folding times for the corresponding models are listed in
Table~\ref{tab:growth_grav}. With $94$ and $97\Myr$ for the irregular and
regular component, the amplification in the low gravity case is comparable to
the standard case, whereas the high gravity model with $83$ and $76\Myr$,
respectively, shows a slightly faster field amplification. Taken the
uncertainties in the modelling, e.g. the effects of numerical resolution
(which could not be studied, but might be of influence here) this is
considered to be within the range of fluctuations.

What dependence on the gravity would one have expected? As we learned in
Section~\ref{sec:velocity_prof}, the effective velocity profiles are very
similar for the three cases (see Fig.~\ref{fig:zdisp_grav}), and only the
vertical scale height of the structures is modified. Because we here consider
vertically integrated values, these differences do not seem to be reflected in
the overall growth rates. After all, the dynamo numbers relevant for the field
amplification are integral quantities of the dynamo active volume. In this
sense, the result of a universal growth rate with respect to the gravitational
potential seems plausible.

\subsection{Slow versus fast dynamo} 
\label{sec:convergence}

In laminar dynamos, the diffusion time, i.e., the relevant time scale for
magnetic field reconnection,\footnote{%
  Reconnection is important to change the topology of the fields. This becomes
  obvious in the schematic stretch-twist-fold picture of magnetic field
  amplification \citep[see e.g.][]{1995stf..book.....C}.} %
defines a lower limit for the growth time of the mean magnetic field. Because
the microscopic diffusivity is usually low, these dynamos are also referred to
as ``slow dynamos''. Within the ISM, the diffusion time $\tau_{\eta}=L^2/\eta$
related to the microscopic value $\eta\simeq10^8\vis$ of the magnetic
diffusivity by far exceeds the Hubble time. This implies that any efficient
galactic dynamo will have to be a ``fast dynamo'' in the sense that it works
on a time scale different than the diffusive one. If turbulence is involved,
one expects the dynamo to operate on time scales defined by $\eta_{\rm t}$,
accordingly. Although this still implies a dynamo of the ``slow'' type, the
associated time scale can be rather fast due to the much higher value of
$\eta_{\rm t}$.

While from theoretical considerations, this classification is rather
straightforward, matters get more intricate when numerical modelling is
involved. In view of the limited magnetic Reynolds numbers of numerical
simulations, the two regimes may not as easily be distinguished. The
requirement of a ``fast dynamo'', however, defines a necessary condition for
the robustness of the field amplification observed in simulations: To be able
to explain magnetic fields in the galactic context, the effect has to persist
for low $\eta$, or high $\Rm$, correspondingly.

\citet*{2007ApJ...668..110O}, in their models of the CR-driven buoyant
instability, observed that the CR dynamo crucially relies on the presence of
an atomic diffusivity $\eta$. In fact, the authors found the efficiency of the
field amplification to scale with this parameter. This can be seen as an
indication that the buoyant rise of the CR bubbles rather comprises a laminar
process,\footnote{%
  The authors, actually, state that no scale separation is manifest in
  the spectra of their simulations.} %
and only the high value of the cosmic ray diffusivity leads to the observed
fast amplification time scale of $\simeq 100\Myr$.

\begin{figure}
  \center\includegraphics[width=\columnwidth]{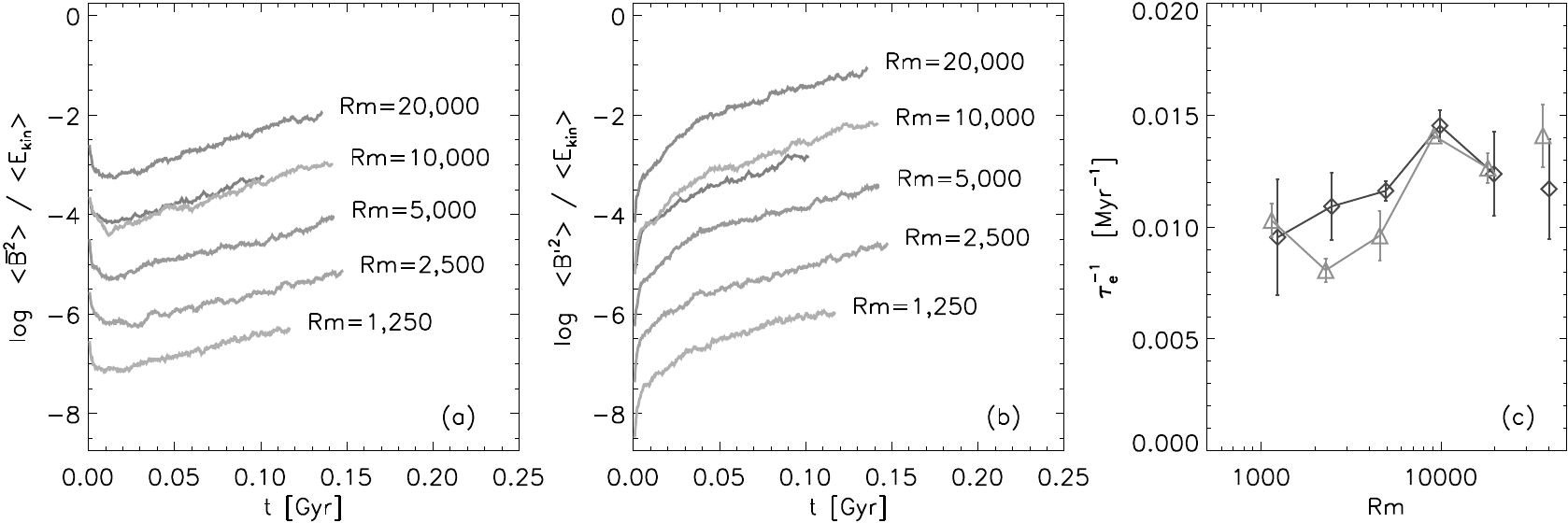}
  \caption[Field evolution at different magnetic Reynolds numbers]{%
    Regular (a) and fluctuating (b) magnetic field strength normalised to the
    kinetic energy for model H4 for different magnetic Reynolds numbers $\Rm$
    and constant $\Pm=2.5$. For clarity the ordinate of the different models
    has been offset by an order of magnitude each. For ${\Rm=10,000}$ we also
    show a comparison run at half grid resolution. In panel (c) we compare the
    growth rates for the turbulent (diamonds) and regular (triangles) magnetic
    field as obtained from a linear regression. The unconnected data points to
    the right correspond to a run with $\eta=0$, providing an indication for
    the level of numerical diffusivity.}
  \label{fig:tmag4}
\end{figure}

The described picture changes significantly as soon as the Reynolds number is
high enough to allow for developed turbulence. The integral length scale $L$
is now efficiently broken down to the Kolmogorov microscale where the atomic
diffusion takes over. Varying the microscopic value of $\eta$ does now only
change the extent of the inertial range towards higher wavenumbers. Because
the dynamics in the intermediate range of wavenumbers is only governed by the
non-linear term of the Navier-Stokes equation, the viscous dissipation is no
longer relevant for the large-scale flow. This implies that a turbulent dynamo
should be insensitive to variations in $\eta$, as soon as a critical value
$\Rm>\Rm_{\rm c}$ is exceeded. With a rotation frequency of $100\kms\kpc^{-1}$
and a box dimension of $L_x=0.8\kpc$, we yield a magnetic Reynolds number
$\Rm=L^2\Omega/\eta \simeq10,000$ for our model H4. This is marginally
sufficient to guarantee for developed turbulence (cf. Sec.~\ref{sec:spec}). In
Figure~\ref{fig:tmag4}, we present growth times for model H4 as a function of
magnetic Reynolds number. The different values for $\Rm$ are obtained by the
variation of $\eta$ at fixed rotation rate $\Omega$.

As can be seen from the comparison with the values for $\eta=0$ (unconnected
data points in panel (c) of Fig.\ref{fig:tmag4}), the highest value
$\Rm=20,000$ is probably already affected by the finite value of the inherent
numerical diffusivity of our code. This quantity, however, cannot easily be
assessed directly. Moreover, the interpretation of the results becomes
complicated by the fact that the assumed asymptotic growth rate in the limit
of high $\Rm$ can be explained in two contrary ways, i.e., the possible effect
due to a finite numerical diffusivity cannot be distinguished from the
expected turbulent behaviour without a proper convergence study. To estimate
the influence of the finite grid resolution, we have performed a fiducial run
at double the grid spacing for $\Rm=10,000$ (see panels (a) and (b) of
Fig.~\ref{fig:tmag4}). The obtained values are consistent with the higher
resolved run and thus provide an indication that the simulation results are
converged at this level of $\Rm$ and below.

Having confirmed the reliability of the numerical results, we can now try and
interprete the obtained growth times. Contrary to the laminar case, we observe
the growth rate $\tau_{\rm e}^{-1}$ to be an increasing function of the
magnetic Reynolds number $\Rm$ for both the regular and turbulent magnetic
field. This lends further evidence to the hypothesis that the SN-driven dynamo
operates in the turbulent regime and thus remains efficient in the limit of
high $\Rm$. As pointed out above, this finding is subject to verification by a
full-blown numerical convergence study (which is currently infeasible). We,
nevertheless, are confident that the observed trend will persist, and the
SN-driven dynamo is indeed capable of explaining galactic magnetic fields at
realistically high magnetic Reynolds numbers.



\cleardoublepage
\chapter{Understanding the Galactic Dynamo}
\label{ch:dynamo}

\section{Mean-field theory} 
\label{sec:mf}

In the previous chapter we have seen that supernova turbulence together with
rotation and stratification can drive a galactic dynamo with e-folding times
of $\simeq 100\Myr$. The numerical verification of the operability of such a
dynamo based on first principles marks an important step towards understanding
the process of field amplification in the interstellar medium. The results
also show that direct simulations constitute a valuable tool in studying the
dependence on certain parameters. However, because of the immense complexity
of the non-linear MHD equations at high Reynolds numbers, the outcome of the
simulations has to be interpreted according to an underlying theory that has
to be rooted at an intermediate level of complexity. As we have learned in the
introductory chapter, before the advent of powerful computers the exploration
of dynamo effects in turbulent flows was largely based on the mean-field
approach. In the following, we want to discuss how this analytical
description can be compared with direct simulations -- and how this will aid
us in understanding the simulation results.

In the framework of mean-field electrodynamics%
\footnote{sometimes also referred to as mean-field MHD} %
\citep{1980mfmd.book.....K,2004maun.book.....R}, one usually splits the fluid
variables into a mean part, denoted by over-bars, and a fluctuating part, here
indicated by a prime. The basic idea behind this is to treat the large-scale
evolution of the system independently from the underlying turbulence.
Accordingly, we split the fluid velocity $\U$ into $\mU+\U'$ and the vector of
the magnetic induction density $\B$ into $\mB+\B'$. The concept of averaged
equations goes back to the end of the nineteenth century when Osborne Reynolds
applied the formalism to the Euler equations and thereby derived an additional
turbulent stress. Similar to the Reynolds equation for a turbulent
hydrodynamic flow, one can derive an induction equation for the mean-field
$\mB$. For our case of a differentially rotating medium this mean-field
equation reads:
\begin{equation}
  \partial_t \mB = \nabla \times \left[\  (\mU+q\Omega x\yy)\tms\mB
    +\EMF - \eta\nabla\tms\mB\ \right]\,,
  \label{eq:MF_ind}
\end{equation}
where we included the term $q\Omega x\yy\tms\mB$, representing the background
shear. The effect of the turbulence on the mean flow, i.e., the creation of
the large-scale magnetic field from unordered motion, is described by the
correlation between the fluctuating velocity and magnetic field components,
more specifically the so-called mean electromotive force (EMF), defined by
$\EMF=\overline{\U'\tms \B'}$. The dissipative term in
Equation~(\ref{eq:MF_ind}) is due to the microscopic diffusivity $\eta$ which
shall not be confused with its counterpart $\eta_{\rm t}$, created by the
turbulence itself. As we will see in a moment, $\eta_{\rm t}$ is embodied in
the electromotive force $\EMF$ and will add up with the microscopic
diffusivity $\eta$ to yield an effective value of $\eta_{\rm T}=\eta_{\rm
  t}+\eta$. Usually the microscopic value can be neglected in comparison with
the turbulent one.

\subsection{The closure ansatz} 

As common to mean-field theories, the newly derived formula is not
self-contained with respect to the large-scale variables. This is because the
additional term $\EMF$, albeit only in an averaged fashion, still contains the
small scall fluctuations $\U'$ and $\B'$. To find a closure for the mean-field
equation, one therefore strives to parameterise the electromotive force with
respect to averaged quantities, i.e., $\EMF$ is regarded as a functional of
$\mU$, $\mB$, and statistical properties of $\U'$. For simplicity, we here
adopt the standard description, where $\EMF$ is supposed to depend on the
mean-field and its gradients:
\begin{equation}
  \EMF_i = \alpha_{ij} \bar{B}_j + \eta_{ijk}\partial_k \bar{B}_j.
  \label{eq:param1}
\end{equation}
Specifying the averaging procedure, one can compute the values of the
tensorial parameters $\alpha_{ij}$ and $\eta_{ijk}$ by means of direct
numerical simulations. In our case, we use spatial averages along horizontal
slabs which further simplifies the equations, as only vertical gradients
arise. \citet{2002GApFD..96..319B} showed that in this case the eddy
diffusivity tensor can be reduced and one yields:
\begin{equation}
  \EMF_i = \alpha_{ij} \bar{B}_j 
         - \tilde{\eta}_{ij}\varepsilon_{jkl}\partial_k \bar{B}_l\,,
  \qquad i,j \in \left\{R,\phi\right\}, k=z\,,
  \label{eq:param2}
\end{equation}
where there exists a unique mapping 
$\tilde{\eta}_{il}=\eta_{ijk}\varepsilon_{jkl}$ such that
\begin{equation}
\tilde{\eta}_{xx}=\eta_{xyz}\,,\quad \tilde{\eta}_{xy}=-\eta_{xxz}\,,\quad
\tilde{\eta}_{yx}=\eta_{yyz}\,,\quad \tilde{\eta}_{yy}=-\eta_{yxz}\,.
\end{equation}
We tacitly assume that the mean-field coefficients are constant in time (in a
quasi-stationary sense) and furthermore describe the instantaneous and local
influence of the turbulence on the mean-field. In the most general framework,
the $\talph$~tensor is regarded as a kernel including non-local and
retardation effects \citep*[see e.g.][]{2008A&A...482..739B}.

The assumption of constant dynamo coefficients is strictly only applicable in
the weak field limit. As soon as the magnetic field is strong enough to affect
the turbulence itself, the dynamo parameters will be inverse functions of the
magnetic field strength, i.e., the stronger the field becomes, the stronger it
inhibits its amplification. In this so-called quenching regime the field
amplification is slowed down until a stationary state is reached. We want to
point out that in the scope of the current work the dynamo coefficients are
measured in the unquenched regime where the magnetic field is well below
equipartition.

\subsection{The dynamo tensors}\label{sec:dynamo_tensor} 

In the following, we want to briefly describe, how the different coefficients
from Equation~(\ref{eq:param2}) affect the field amplification process they
parameterise. The role of the non-vanishing tensor components becomes more
obvious when we write them down in their matrix representation with respect to
a cylindrical coordinate system $[\ee_R,\ee_\phi,\ee_z]$. For simplicity, we
furthermore assume that the off-diagonal entries of both tensors are totally
antisymmetric. In this case, we can replace the according elements by the
components of vectors ${\boldsymbol \gamma}$ and ${\boldsymbol \delta}$,
respectively. Because, in our simulations, the axis of rotation is parallel to
the gravity force, this assumption is in accordance with the theoretical
prediction for $\talph$ by \citet*{1994AN....315..157K}. The question whether
$\teta$ possesses a significant symmetric contribution in its off-diagonal
elements will be addressed in Section~\ref{sec:dynamo_shear}.  The
parameterisation, in its simplified form, now reads:
\begin{equation}
  \EMF(z,t) = \left( \begin{array}{ccc}
      \alpha_R    & \!-\gamma_z   & \gamma_\phi \\
      \gamma_z    & \alpha_\phi & \!-\gamma_R    \\
     \!-\gamma_\phi & \gamma_R    & \alpha_z      \\
    \end{array} \right)\,\mB(z,t)
  - \left(\begin{array}{ccc}
      \tilde{\eta}_R & \delta_z          & \!-\delta_\phi \\
     \!-\delta_z       & \tilde{\eta}_\phi & \delta_R      \\
      \delta_\phi    & \!-\delta_R         & \tilde{\eta}_z \\
    \end{array} \right)\,\nabla\tms\mB(z,t)\,,
\end{equation}
where we make use of the convention $\alpha_R\equiv\alpha_{RR}$ etc. to
abbreviate the notation for the diagonal elements. Because we consider
mean-fields which only depend on the vertical coordinate $z$, the
anti-symmetric vectors ${\boldsymbol \gamma}$ and ${\boldsymbol \delta}$ can
be expressed by the scalars $\gamma_z$ and $\delta_z$. If we further assume
$\bar{B}_z=0$, we yield the following expression for the radial and azimuthal
components of the electromotive force:
\begin{equation}\label{eq:EMF_comp}
  \lft \begin{array}{c} \EMF_R \\ \EMF_\phi \end{array}\rgt
  = \left( \begin{array}{cc} 
      \alpha_R & -\gamma_z   \\
      \gamma_z & \alpha_\phi \end{array} \right)
  \lft \begin{array}{c} \bar{B}_R \\ \bar{B}_\phi \end{array}\rgt
  - \left(\begin{array}{cc}
      \tilde{\eta}_R & \delta_z          \\
      \!-\delta_z      & \tilde{\eta}_\phi \end{array}\right)
  \lft \begin{array}{c} \!-\bar{B}_{\phi,z} \\ 
                        \;\bar{B}_{R,z}    \end{array}\rgt\,.
\end{equation}
A brief look at the last term of this equation shows that the curl operator
swaps the radial and azimuthal components of the mean magnetic field and
introduces a change of sign in one of them. This mixing of the directions is
important to understand the role of the various coefficients. Since the
electromotive force itself enters the induction equation as $\nabla\tms\EMF$,
this effect is similarly introduced to the terms including $\talph$, whereas
it is cancelled again for the terms including $\teta$, leaving them with a
global change in sign. When we insert Equation~(\ref{eq:EMF_comp}) into the
mean-field induction equation (\ref{eq:MF_ind}), we thus yield
\begin{eqnarray}
  \bar{B}_{R,t} & = & 
    \left[\: -(\bar{u}_z+\gamma_z) \,\bar{B}_R  \nonumber
      \quad  -\alpha_\phi \,\bar{B}_\phi \right.   \\ & & \left.  
      \quad  +(\tilde{\eta}_\phi+\eta) \,\bar{B}_{R,z}
      \quad  +\delta_z \,\bar{B}_{\phi,z} \:\right]_{,z} 
    \label{eq:dynamo_R} \\
  \bar{B}_{\phi,t} & = &
    \left[\quad\alpha_R \,\bar{B}_R                       \nonumber
      \quad  -(\bar{u}_z+\gamma_z) \,\bar{B}_\phi \right. \\ & & \left. 
      \quad  -\delta_z \,\bar{B}_{R,z}
      \quad  +(\tilde{\eta}_R+\eta) \,\bar{B}_{\phi,z} \:\right]_{,z}
    + q\Omega\,\bar{B}_R\,.
    \label{eq:dynamo_phi}
\end{eqnarray}
Note that for the diagonal elements of the $\talph$~tensor the radial and
azimuthal components of $\mB$ are now exchanged -- this is where the main
feedback loop of the dynamo process is closed. As a note of caution, we want
to emphasis that the one-dimensional, simplified system of equations does, of
course, not comprise the full set of dynamo solutions. Particularly, it does
not reflect the thin disk geometry commonly considered for global mean-field
models. The ansatz, however, closely resembles the geometry of our vertically
elongated box, and many results of the direct simulations can in fact be
reproduced qualitatively in the 1D approach.

\subsubsection{Diagonal elements}

A closer look at the Equations~(\ref{eq:dynamo_R}) and (\ref{eq:dynamo_phi})
reveals that the diagonal elements of $\talph$ are not the only terms that
mutually couple the poloidal and toroidal field. Together with the
differential rotation $q\Omega$, they yield the dominant contribution to the
dynamo effect, however.

In the absence of other effects, one usually speaks of an $\alpha^2$-type
dynamo, referring to the intertwined effects of $\alpha_R$ and $\alpha_\phi$.
In the case of differential rotation, $\alpha_R$ becomes dispensable and a
feedback loop can be achieved by the mutual coupling via the $\alpha_\phi$ and
$q\Omega$ terms. This kind of amplification mechanism, which is also thought
to be responsible for the magnetic field creation inside the sun, is commonly
referred to as $\alpha\Omega$~type dynamo. Because the $\Omega$~mechanism is
usually dominating, this class of dynamos is characterised by very small pitch
angles. Since we observe galactic fields with large pitch angles, we
consequently have to seek for an $\alpha^2 \Omega$-type solution.

Because the sign of the $\alpha$~effect depends on ${\mathbf\Omega} \cdot
{\mathbf g}$, i.e., the orientation of the axis of rotation with respect to
gravitation, one expects $\alpha_R$ to have odd symmetry with respect to
the midplane.\footnote{%
  This is the very same effect that makes the trade winds in the Earth's
  atmosphere follow different directions in the northern and southern
  hemisphere and determines the sense of rotation in cyclones.} %
If we assume positive (negative) values in the top (bottom) half of the
simulation box, consequently, the gradient in $\alpha_R$ is positive near the
midplane. Given a mean radial field of quadrupolar symmetry, this property
carries over to $\partial_z\,(\alpha_R\,\bar{B}_R)$, which implies that this
term has the same sign as $\bar{B}_R$ itself. Since we introduced the shear
parameter $q$ to be negative, a positive value of $\alpha_R$ (in the northern
``hemisphere'') implies that the induction from the turbulence works against
the background shear $q\Omega$, increasing the pitch angle, whereas a negative
value would imply the opposite. If we, on the other hand, assume dipolar
symmetry for $\bar{B}_R$, the two terms will in any case work in accordance in
one half of the box and in discordance in the other half, respectively. In
this case one would expect oscillating solutions for the dynamo equation.

\subsubsection{Diamagnetic pumping}

In our derivation of the dynamo coefficients, we have replaced the
anti-symmetric part of the $\talph$~tensor by a vector ${\boldsymbol \gamma}$.
This is equivalent to writing a term ${\boldsymbol \gamma}\tms\mB$ in the EMF.
Taking a look at Equation~(\ref{eq:MF_ind}) we see that ${\boldsymbol
  \gamma}\tms\mB$ is formally identical to the term $\mU\tms\mB$. We hence
conclude that ${\boldsymbol \gamma}$ describes the transport of the mean-field
due to the turbulence. It is important to note that, although this term
formally looks like an advection term, it is not. This is because, unlike
$\mU$, which does also transport the fluctuating field $\B'$, the so-called
diamagnetic pumping ${\boldsymbol \gamma}$ per definition does only affect the
mean-fields.%
\footnote{As we shall see in Sec.~\ref{sec:transport}, this difference poses a
  nice solution to the helicity flux issue.}

The turbulent pumping can be understood in analogy to a diffusion process,
which follows a gradient in concentration. It can be shown that, similarly, a
gradient in the turbulence intensity $u'^2$ will lead to a turbulent transport
of the mean-field towards regions of lower turbulence amplitude. Since we
restrict ourselves to vertical gradients in the mean-fields and assume
$\bar{B}_z=0$, the only non-vanishing component of this vector is
\begin{equation}
  \gamma_z = \frac{1}{2}\,( \alpha_{\phi R} - \alpha_{R\phi} )\,,
\end{equation}
describing the turbulent pumping in the vertical direction. If we go back to
Equations~(\ref{eq:dynamo_R}) and (\ref{eq:dynamo_phi}), we see that
$\gamma_z$ simply adds up to the mean vertical velocity $\mn{u}_z$.

\subsubsection{Turbulent diffusivity}

By a similar argument as for the pumping velocity $\gamma_z$, we can derive
the meaning of the diagonal elements of $\teta$. Since they appear in the same
places as the molecular diffusivity $\eta$, it is self-evident to interprete
them as turbulent diffusivity
\begin{equation}
  \eta_{\rm t} = \frac{1}{2}\,( \tilde{\eta}_{R} + \tilde{\eta}_{\phi} )\,.
\end{equation}
Due to its origin in the turbulent nature of the flow, this quantity is
sometimes also referred to as ``eddy diffusivity''.

Because the magnetic field components have undergone two curl operations in
the diffusive part of the induction equation, the coefficients
$\tilde{\eta}_R$ and $\tilde{\eta}_\phi$ have switched places and
$\tilde{\eta}_R$ (somewhat confusingly) is now responsible for the diffusion
of $B\phi$ and vice versa. This can, however, be understood if one recalls
that the tensor index refers to the spatial direction rather than the
component of the magnetic field. Although the turbulent contribution to the
resistivity is found to be dominant by many orders of magnitude, the molecular
quantity still plays an important role in defining the magnetic Reynolds
number $\Rm$ of the flow.

\subsubsection{The R{\"a}dler effect}
\label{sec:Raedler}

As we have already noted, the diagonal elements of $\talph$ are not the only
coefficients that couple the radial and azimuthal component of the mean
magnetic field. Whereas $\alpha_R$ and $\alpha_\phi$ directly mix the
mean-field back into the electromotive force, the off-diagonal elements of
$\teta$ feed back the associated gradients. The possibility of driving a
so-called $\mathbf{\Omega}\tms{\mathbf J}$-type dynamo via this
effect\footnote{%
  Applying spectra of the mixing-length type, this term is found to vanish in
  the so-called $\tau$-approximation -- see Section~4.3 in
  \citet{2004maun.book.....R}.} %
was first discovered by \citet{1969MDAWB..11..272R}. Like for the pumping term
$\gamma_z$, the vector ${\boldsymbol \delta}$ reduces to
\begin{equation}
  \delta_z = \frac{1}{2}\,( \tilde{\eta}_{R\phi} - \tilde{\eta}_{\phi R} )\,.
\end{equation}
If we write down the coupled system only retaining the shear- and
$\delta$-terms and apply a Fourier decomposition, we yield
\begin{eqnarray}
  -i\omega \bar{B}_R    & = & -\delta_z\,k_z^2\,\bar{B}_\phi \nonumber\\
  -i\omega \bar{B}_\phi & = & (\delta_z\,k_z^2+q\Omega)\,\bar{B}_R\,,
\end{eqnarray}
and from evaluating the associated determinant we can derive a necessary
condition
\begin{equation}
  -\delta_z k_z^2\,(\delta_z k_z^2+q\Omega) > 0
\end{equation}
for dynamo action \citep{2005AN....326..787B}. Because of the (assumed)
antisymmetric nature of $\tilde{\eta}_{R\phi}$ and $\tilde{\eta}_{\phi R}$,
this is only possible for $q\Omega\ne 0$, i.e., in connection with
differential rotation. More specifically, since $q$ was defined negative, the
coefficient $\delta_z$ has to be positive to allow for a growing dynamo
solution. A dispersion relation including the dissipative terms can be found
in Appendix B of \citet{2005AN....326..787B}.

\section{The {SOCA} approach} 
\label{sec:SOCA}

Reynolds averaged equations draw their practical usability from the assumed
closure. In our case, we have chosen the closure given by
Equation~(\ref{eq:param1}). This parameterisation seems rather arbitrary at
first glance. In the following we want to briefly demonstrate, how this
approach can be justified by the neglect of third order correlations.

Generally speaking, the attempt to evaluate the highest order terms leads to a
cascade of equations including even higher order moments of the fluctuations
-- this is known as the closure problem. An example for a closure based on
higher order correlations is the so-called $\tau$-approximation
\citep{1983GApFD..24..273V} where triple correlations are approximated in
terms of quadratic moments via a relaxation time $\tau(k)$.

\subsection{Homogeneous turbulence} 

Following \citet{1980mfmd.book.....K}, we now focus on the second order
closure (SOCA) and write down the induction equation for the perturbed
magnetic field $\B'$:
\begin{equation}
  \partial_t \B' = 
  \nabla \times \left[\,
    \U'\tms\mB + (\mU+q\Omega x\yy)\tms\B' 
    - \overline{\U'\tms\B'} + \U'\tms\B' - \eta\nabla\tms\B' \,\right] \,.
  \label{eq:ind_Btur}
\end{equation}
We see that the turbulent EMF, i.e. $\overline{\U'\tms\B'}$, enters this
equation with a negative sign. This occurs naturally if we recall that this
term is responsible for the coupling between the mean and fluctuating field.
Besides the turbulent EMF and the dissipative term, there is an induction term
due to the mean-field, $\U'\tms\mB$, the advection/induction term
$(\mU+q\Omega x\yy)\tms\B'$, and the term $\U'\tms\B'$, quadratic in the
fluctuations.

\begin{figure}
  \center\includegraphics[width=\columnwidth]{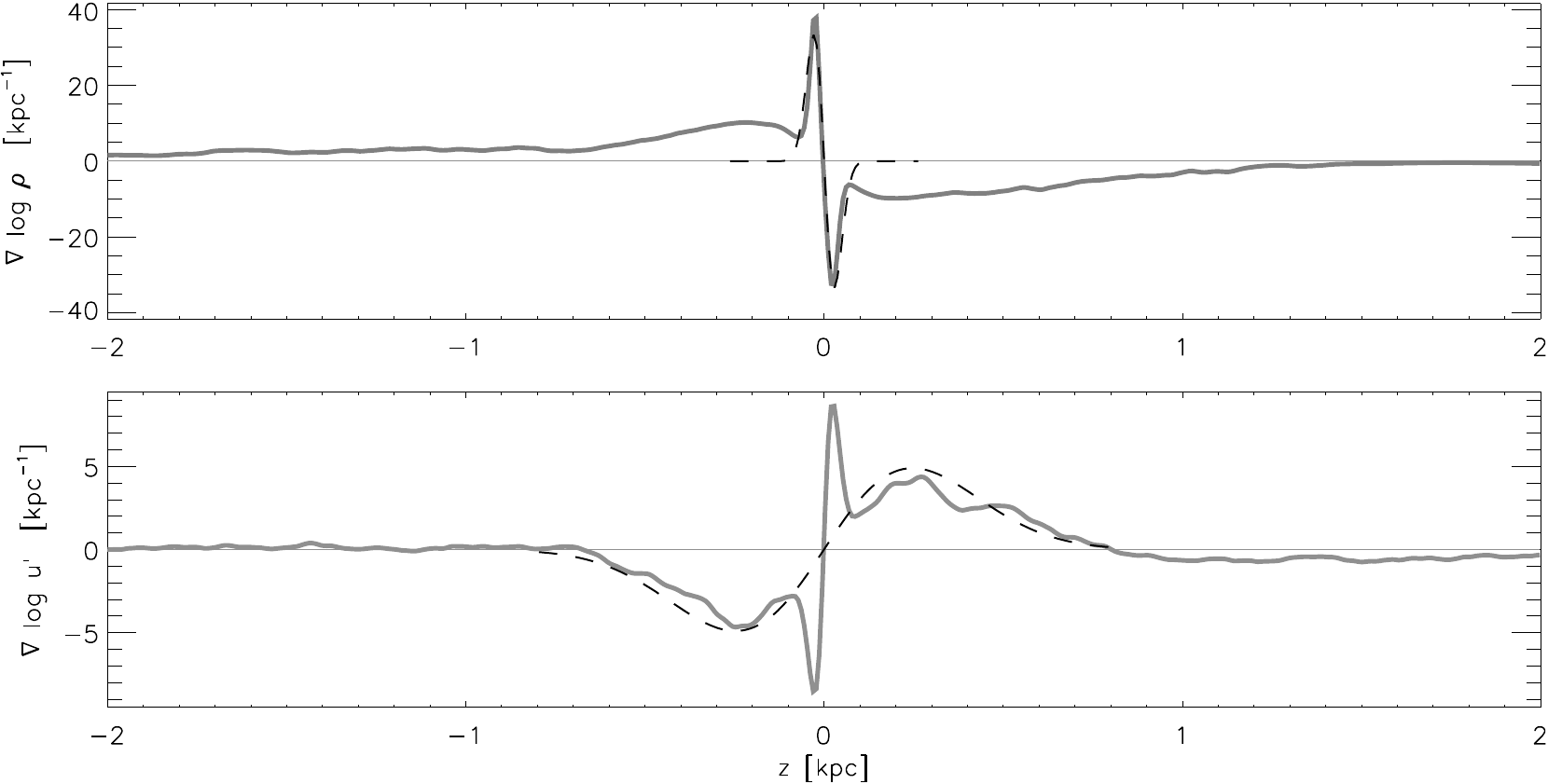}
  \caption[Logarithmic gradients of gas density and turbulent velocity]{%
    Time averaged logarithmic gradients of the gas density $\rho$ (upper
    panel) and the turbulent velocity $u'$ (lower panel) for model Q4. Two
    distinct disk components with scale heights of $40\pc$ and $350\pc$ are
    clearly visible (dashed lines).}
  \label{fig:dlog}
\end{figure}

The simplification introduced by SOCA now is to neglect the terms quadratic%
\footnote{These terms would enter the mean EMF as triple correlations, hence
  the name.} %
in the fluctuating quantities within the evolution
equation~(\ref{eq:ind_Btur}) for $\B'$, i.e., the turbulent EMF and the
correlation $\U'\tms\B'$. In the limit of high conductivity we thus yield
\begin{equation}
  \B'(\rr,t) = \int_{0}^{t}
  \nabla \times \left[\,\U'(\rr,t')\tms\mB(\rr,t')\,\right]\,{\rm d}t'\,,
\end{equation}
where we have simultaneously dropped the contribution by the mean flow
$(\mU+q\Omega x\yy)$ to focus on the effects caused by the turbulence. From
this formulation, we already see that the parameterisation via $\talph$ should
ideally be considered as some sort of Green's function. To yield an explicit
expression for the mean EMF we now evaluate the correlation 
\begin{equation}
  \U'(\rr,t)\tms\B'(\rr,t) = \int_{0}^{t}
  \U'(\rr,t) \times \left[\,\nabla \times \left[\,
      \U'(\rr,t')\tms\mB(\rr,t')\,\right]\,\right]\,{\rm d}t'\,.\label{eq:corr}
\end{equation}
To arrive at a closed expression with respect to $\mB$, we have to make
further assumptions -- this is where the Strouhal number $\St=\tau_{\rm c}\,
u'/l_{\rm c}$ comes into play. In the limit $\St\ll 1$, we can take the
mean-field to be constant over the time interval relevant for the integration.
If we accordingly assume $\mB(\rr,t')\simeq\mB(\rr,t)$ for $|t-t'|\ll
\tau_{\rm c}$, this implies that the right hand side of
Equation~(\ref{eq:corr}) can be expressed in terms of $\mB$ and $\nabla \tms
\mB$ alone, hence justifying the chosen parameterisation. After some tensor
algebra one arrives at the fundamental (scalar) SOCA expressions for $\talph$
and $\teta$ in the limit of high conductivity and homogeneous, isotropic
turbulence:
\begin{eqnarray}
  \alpha & = & -\frac{1}{3}\,\int_{0}^{\infty}\,\mn{
    \U'(\rr,t) \cdot \nabla \times \U'(\rr,t-\tau)
  }\ {\rm d}\tau\,, \nonumber\\
  \eta_{\rm t} & = & \frac{1}{3}\,\int_{0}^{\infty}\,\mn{
    \U'(\rr,t) \cdot \U'(\rr,t-\tau)
  }\ {\rm d}\tau\,.
\end{eqnarray}
If we further approximate the integrals, we finally yield
\begin{equation}
  \alpha = -\frac{1}{3}\,\mn{\U'\cdot\nabla\tms\U'}\ \tau_{\rm c}
  \qquad{\rm and}\qquad
  \eta_{\rm t} = \frac{1}{3}\,\mn{\U'^2}\ \tau_{\rm c}\,,
\end{equation}
where we infer that the $\alpha$~effect is proportional to the negative mean
value of the kinetic helicity density, whereas the turbulent diffusion is
proportional to the square of the fluctuating velocity. Note that both
expressions are in fact quadratic in $\U'$ as would be expected from the
second order approach. Of course, these expressions have to be modified in the
presence of inhomogeneities in the turbulent velocity field and density.

\subsection{The case of stratified turbulence} 

Applying the SOCA framework, \citet[hereafter
\citetalias{1993A&A...269..581R}]{1993A&A...269..581R} derived the turbulent
$\alpha$~effect resulting from gradients in the density $\rho$ and the
turbulent velocity $\U'$. The according horizontal components of the turbulent
$\alpha$~effect in this case are
\begin{equation}
  \alpha_{RR} = \alpha_{\phi\phi} = -\tau^2 \Omega\,\mn{u'^2}
  \left(\ \Psi^{\rho}\,\nabla\log\rho 
    \  +\ \Psi^{u}\,\nabla\log u'  \ \right)\,, \label{eq:alpha_strat}
\end{equation}
with weighting functions $\Psi^{\rho}$ and $\Psi^{u}$. The corresponding
gradients $\nabla\log\rho$ and $\nabla\log u'$ for our model Q4 are depicted
in Figure~\ref{fig:dlog}, where we, first of all, notice that the two
gradients show opposite signs. This implies that, at least in the central part
of the disk, the related effects work against each other. Note that, unlike
$\nabla\log\rho$, the gradient in $u'$ changes its sign in the halo. The
logarithmic slopes show distinct peaks near the midplane corresponding to the
dense inner disk. Beyond $|z| \simgt 100\pc$ a second peak with a scale height
of $h\simeq 350\pc$ becomes visible. As is illustrated\footnote{%
  While the components are visible in both quantities, we only plot one curve
  per panel for clarity.} %
by dashed lines in Figure~\ref{fig:dlog}, the shape of the profiles
approximately corresponds to a functional dependence $\sim \partial_z\, {\rm
  e}^{-(z/h)^2}$.

To estimate the relative contribution of the two effects, one has to consider
that the profiles are multiplied by an additional factor of $u'^2$ in the
final expression for $\alpha$. Because $u'$ has a minimum at $z=0$, the strong
gradient of the inner disk will only show up weakly in the $\alpha$~profile.
For the outer peak in the gradients, the density dominates by about a factor
of two. In the following, we want to infer how these amplitudes are modified
by the relative weights introduced by SOCA theory.

\begin{figure}
  \begin{minipage}[b]{0.75\columnwidth}
    \includegraphics[width=0.95\columnwidth]{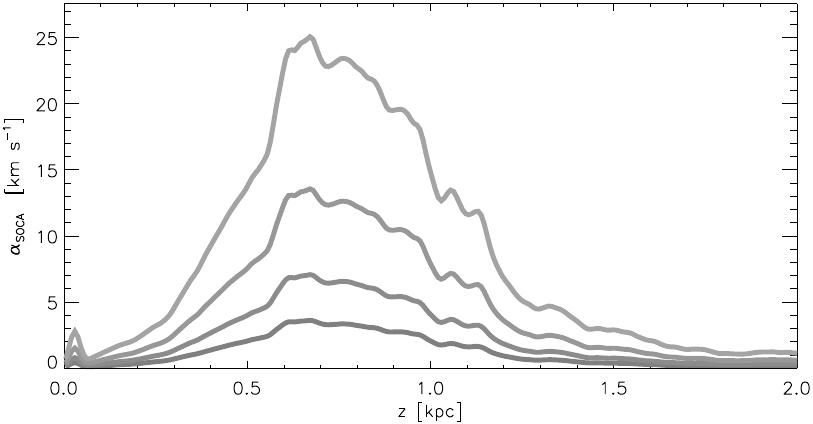}
  \end{minipage}
  \begin{minipage}[b]{0.24\columnwidth}
    \caption[$\alpha$~effect as computed from SOCA theory]{%
      Quasi\-lin\-ear $\alpha$~effect for model Q4, computed via
      Equation~(\ref{eq:alpha_strat}). The different curves correspond to the
      values of $\tau$ listed in Table~\ref{tab:weight}, below.\vspace{20pt}}
    \label{fig:alpha_soca}
  \end{minipage}
\end{figure}

\begin{table}\begin{minipage}[c]{0.50\columnwidth}
\begin{tabular}{ccccc}\heading{
 $\tau\ [\Myr]$& $\Co$& $\Psi^{\rho}$& $\Psi^{u}$& $s$}
 2.0 &  0.41 &  0.79 &  0.55 &  1.42 \\
 2.8 &  0.58 &  0.77 &  0.57 &  1.37 \\
 4.0 &  0.82 &  0.74 &  0.58 &  1.29 \\
 5.6 &  1.16 &  0.69 &  0.57 &  1.21 \\\hline
\end{tabular}\end{minipage}
\begin{minipage}[c]{0.49\columnwidth}
  \caption[SOCA weighting functions]{%
    SOCA weighting functions \citepalias{1993A&A...269..581R} for different
    values of the assumed correlation time $\tau$ applied to model Q4.
    \label{tab:weight}}
\end{minipage}
\end{table}

The dependence on the two weighting functions $\Psi^{\rho}$ and $\Psi^{u}$ can
alternatively be regarded as a weighted gradient $\nabla\log(\rho^s\,u')$,
where the newly introduced factor $s=\Psi^{\rho}/\Psi^{u}$ represents the
ratio of the two functions characterising the contributions of the density and
velocity stratification. The mixing functions are found to depend on the
rotation rate, expressed by the Coriolis number $\Co=2\tau\Omega$, as
\begin{eqnarray}
  \Psi^{\rho}(\Co) & = & \Co^{-2} + 6\Co^{-4} 
                     - \frac{6+3\Co^2-\Co^4}{\Co^5}\,\tan^{-1}\Co
                     \quad{\rm and}\nonumber\\
  \Psi^{u}(\Co)    & = &  \Co^{-2} + 9\Co^{-4} 
                     - \frac{9+4\Co^2-\Co^4}{\Co^5}\,\tan^{-1}\Co\,
\end{eqnarray}
\citepalias{1993A&A...269..581R}. In the limit of slow rotation ($\Co\simlt
1$) the arithmetic function can be expanded and we yield the approximations
\begin{equation}
  \Psi^{\rho}(\Co) \simeq \frac{4}{5}-\frac{8}{105}\Co^2
  \qquad{\rm and}\qquad  
  \Psi^{u}(\Co) \simeq \frac{8}{15}+\frac{16}{105}\Co^2\,,
\end{equation}
implying that the weighting factor approaches the value $s\rightarrow3/2$ in
the limit $\Co\rightarrow 0$. Note that the first order correction is rather
weak for $\Co\simlt 1$, which implies an $\alpha$~effect that (in the slow
rotation limit) increases linearly with $\Omega$. The exact values of the
functions for different assumed correlation times $\tau$ are listed in
Table~\ref{tab:weight} and we see that the turbulent $\alpha$~effect is
largely determined by the gradient stemming from the density stratification
rather than the turbulent velocity profile. 

Altogether, the two effects are not easily separable, however. This is because
the density profile is, at least partly, determined by the kinetic pressure
from the SNe and, vice versa, the turbulent velocity profile depends on the
momentum balance (cf. Sec.~\ref{sec:velocity_prof}) and thus on the density.
In Figure~\ref{fig:alpha_soca}, we present the $\alpha$~profiles computed from
the density and velocity distribution of model Q4. Because the positive
gradient in $u'$ is compensated by the strong decline in density, the
resulting $\alpha$~effect is found to be positive in the the northern
``hemisphere'' of our box. We also observe that the gradients are sufficiently
steep to already produce a strong dynamo effect for moderately low correlation
time $\tau$. This, of course, has to be seen in perspective with the
relatively high rotation rate of $\Omega=100\kms\kpc^{-1}$ in model Q4.
Nevertheless, even if we scale down the results to $25\kms\kpc^{-1}$, as
representative of our own Galaxy at $R_\odot$, the value needed to produce
$\alpha\simeq 10\kms$ is still somewhat shorter than the commonly assumed
$\tau=10\Myr$. This will become more obvious if we directly compare the
profiles measured from the simulations with the predictions made by SOCA
theory in the following section.

\section{Dynamo coefficients from solid body rotation} 
\label{sec:dynamo_rot}

For the practical purpose of our analysis of the simulation data, we restrict
ourselves to the parameterisation defined by Equation~(\ref{eq:param1}). To
efficiently perform the inversion of this tensor equation we apply the test
field approach proposed by \citet{2005AN....326..245S,2007GApFD.101...81S}
which has been adopted for the shearing box case by
\citep{2005AN....326..787B}.

Earlier attempts to extract the dynamo coefficients from simulations (with
non-trivial field geometry) were based on least square fit methods
\citep{2002GApFD..96..319B,2005mpge.conf..171K}. These approaches mainly
suffered from the fact that in regions where $\mB$ or $\nabla \mB$ becomes
small the inversion of the (inherently overdetermined) problem will become
inaccurate. To circumvent these difficulties, one can try and solve
Equation~(\ref{eq:param1}) for well behaved tracer fields, i.e., fields with
simple geometry and gradients which are fixed in time. The main advantage of
this method is that one can choose as many test-fields as are necessary to
make the problem well determined. The fact that the tracer fields $\mTf$
remain constant in time, however, does not imply that the electromotive forces
$\mn{\U'\tms\Tf'}$ that comprise the LHS of Equation~(\ref{eq:param1}) are
only depending on $\U'$. To properly include the non-linear evolution of the
test-field fluctuations $\Tf'$, one thus has to solve an extra set of passive
induction equations (see Appendix~\ref{ch:test_fields}). Although this adds a
considerable overhead to the computation, the direct determination of the
dynamo coefficients from simulations outweighs this additional demand in
computing resources by far.

\subsection{The \texorpdfstring{$\alpha$}{alpha} effect from rotation} 

As we already mentioned in the outline, we aim to understand the galactic
field amplification process in a bottom-up approach, i.e., starting from the
most basic setup. In the picture of cyclonic turbulence, there are two
indispensable prerequisites: rotation\footnote{%
  Or alternatively: shear, as we will see in Sec.~\ref{sec:dynamo_shear}
  below.} %
and stratification. While the Coriolis force creates the vorticity in the
expanding remnants, the positive (negative) vertical expansion velocity in the
top (bottom) half of the remnant will introduce opposite signs in the
helicity. With a homogeneous distribution of the remnants, this would imply a
cancellation of the contribution from the top and bottom half of the remnants
when taking the ensemble average. We thus see that in addition to the helicity
generation mechanism we need an additional source of inhomogeneity.

\begin{figure}
  \includegraphics[width=\columnwidth]{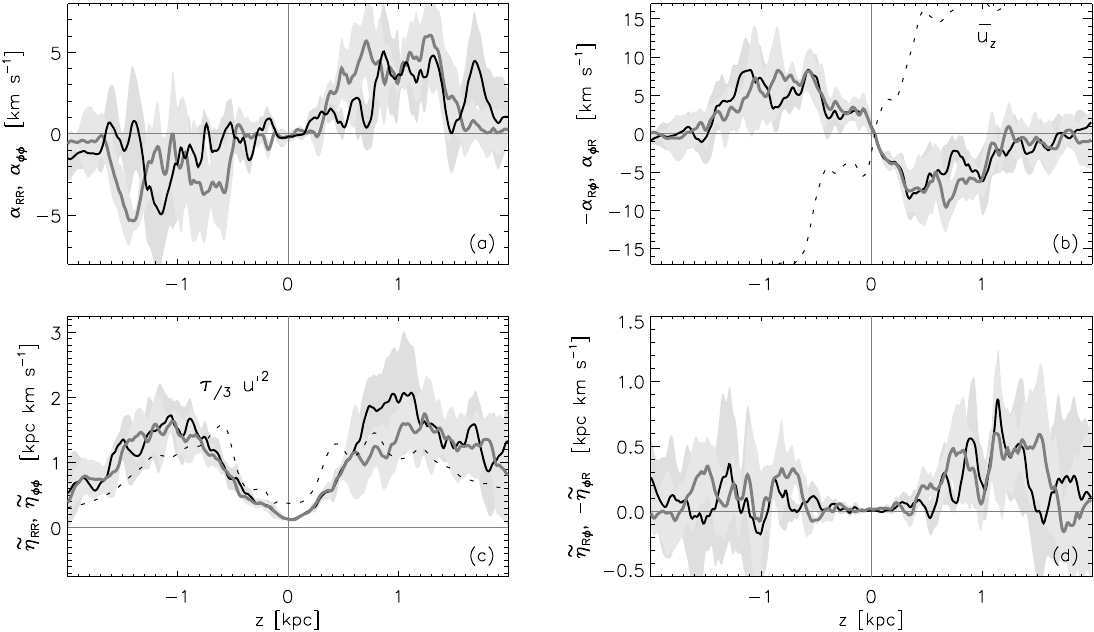}
   \caption[Dynamo $\alpha$ and $\tilde{\eta}$~profiles for model 
     F4--ROT]{%
     Dynamo $\talph$ and $\teta$~coefficients for model F4--ROT, i.e., for
     the case of solid body rotation. Quantities indicated by the ordinate
     labels are plotted in dark ($\alpha_{RR},\dots$) and light
     ($\alpha_{\phi\phi}, \dots$) colours, respectively. The panels show: (a)
     the diagonal elements of the dynamo tensor, (b) the transport
     coefficients -- compared with the profile of the mean flow (dashed line),
     (c) the turbulent diffusion which roughly follows the SOCA prediction
     (dashed line), and (d) the off-diagonal elements of $\teta$.}
  \label{fig:ae_F4_rot}
\end{figure}

From these considerations, our T4 series of models as well as model F4-ROT
should embody a non-vanishing $\alpha$~tensor. We indeed find this effect, as
is shown in Figure~\ref{fig:ae_F4_rot} where we plot the dynamo $\talph$ and
$\teta$~coefficients for model F4-ROT averaged over four consecutive time
intervals spanning a range of $t\simeq50\Myr$. The subintervals were chosen to
allow for the definition of some sort of standard deviation and get a feel for
the level of scatter present in the simulations.  Apart from the inherent
fluctuations due to the violent driving, the test-field method suffered from a
considerable amount of noise in our early models. This is why we could not
accurately determine the off-diagonal elements of $\teta$ for model T4 (cf.
Sec.~\ref{sec:noise}).

In accordance with SOCA theory, the (horizontal) diagonal elements of $\talph$
are positive (negative) in the top (bottom) half of our simulations box -- see
panel (a) of Fig.~\ref{fig:ae_F4_rot}. The amplitude of the observed
$\alpha$~effect is on the order of $5\kms$ with approximately equal values for
$\alpha_{RR}$ and $\alpha_{\phi\phi}$. The equal strength of the two
components is in accordance with the expected isotropy in the horizontal part
of the tensor. This occurs natural since, in the absence of differential
rotation, the radial and azimuthal direction are locally indistinguishable.
From the comparison of the amplitudes with the SOCA prediction (cf.
Sec.~\ref{sec:SOCA}), we infer a correlation time $\tau \simeq 3.6\Myr$
\citepalias{2008AN....329..619G} which is somewhat lower than the commonly
assumed value of $10\Myr$.

Our dynamo profiles can, in principle, be directly compared with the results
based on the uncorrelated-ensemble approach \citep[cf. Fig.~8 in][hereafter
\citetalias{1998A&A...335..488F}]{1998A&A...335..488F}. This model yields
peak values of $\alpha_{RR}\simeq6\kms$, and $\alpha_{\phi\phi}\simeq2.6\kms$
(see also Tab.~\ref{tab:ae_T4}). While we typically observe amplitudes of the
same order, the results of \citetalias{1998A&A...335..488F}, in general,
depend strongly on the variation of the input parameters with the
galactocentric radius $R$ -- a variation that is not explicitely considered in
our models, and hence, makes an exact comparison difficult. Whereas the
profiles derived from the semi-analytical approach in
\citetalias{1998A&A...335..488F} extend to galactic heights of up to $6\kpc$,
the $\alpha$~effect in our simulations vanishes at the box interfaces.
Although this is, most likely, an effect of the limited vertical box size, the
profiles of the model F4--noTI (see Fig.~\ref{fig:ae_F4_noTI} on
page~\pageref{fig:ae_F4_noTI}) are well contained within the vertical extent
of the box, giving rise to the assumption that the $\alpha$~effect might be
restricted to a much narrower vertical range than suggested by the profiles in
\citetalias{1998A&A...335..488F}. A distinct conclusion will, however, require
additional simulation runs covering greater galactic heights.

\subsection{Diamagnetic pumping} 

The coefficients describing the diamagnetic pumping are plotted in panel (b)
of Figure~\ref{fig:ae_F4_rot}. As already pointed out in connection with the
velocity profiles (see Fig.~\ref{fig:zdisp} on p.~\pageref{fig:zdisp}), the
structure of the turbulent velocity dispersion implies an inward transport of
the mean magnetic field. This inward pumping is indeed present in our
simulations, and the off-diagonal elements $\alpha_{\phi R}$ and
$\alpha_{R\phi}$ have negative and positive signs in the top half of our box,
respectively. The resulting transport coefficient $\gamma_z$ thus becomes
negative and has an amplitude of the order of $5$--$10\kms$. From the similar
shape of the two profiles, we conclude that the residual symmetric part of the
off-diagonal elements lies in the margins of fluctuations and is thus
negligible.

The sign of the transport coefficient $\gamma_z<0$ is in so far remarkable, as
it contradicts the ``escape velocity'' $V_{\rm esc}>0$ predicted by the models
of \citet{1992ApJ...391..188F}. These analytical models were based on
isolated, non-interacting remnants, and the finding of an outward diamagnetic
transport is, in fact, true for these single remnants. The assumption that the
ensemble average can be performed by a simple convolution, however, is wrong,
and non-linear simulations can yield important new insights.

Another important prediction of the said models was the relative strength
${\hat{\gamma}=|\alpha|:|\gamma_z|}$ of the dynamo process with respect to the
turbulent pumping: Early analytical calculations by
\citet{1992ApJ...391..188F} and semi-analytical models by
\citet*{1993A&A...274..757K} and \citet*{1996A&A...305..114Z} arrived at
prohibitively high values for $\hat{\gamma}$. After all, the last work in this
series of papers \citepalias[i.e.,][]{1998A&A...335..488F} added the effect of
stratification (resulting in pea- and peanut-shaped remnants) and thus arrived
at moderate values of $\hat{\gamma}\simeq 6$, which would, in principle, allow
for growing dynamo solutions \citep*{1994A&A...286...72S}. In contrast, second
order theory \citep{2004maun.book.....R} arrives at a considerably lower value
of $\hat{\gamma}\simeq 2.5$.

If we compare the amplitudes in the two upper panels of
Figure~\ref{fig:ae_F4_rot}, we, in accordance with SOCA theory, infer a ratio
of about two. Being a factor of three smaller than the most optimistic value
from analytical models, this greatly enhances the possible efficiency of the
dynamo process. Moreover, the negative sign of $\gamma_z$ allows for a
compensation of the diamagnetic transport by a galactic wind. This is indeed
observed in our simulations, as is demonstrated by the dashed line
representing the mean vertical velocity $\mn{u}_z$ in panel (b) of
Figure~\ref{fig:ae_F4_rot}. With the effective transport velocity determined
by the residuum of the two effects, the prospects for dynamo action are
further improved. Apart from the cancellation of the unhelpful vertical
transport, the combined action of the pumping and wind might pose a nifty
solution to the ``catastrophic quenching'' scenario.

\begin{table}\begin{center}\small
\begin{tabular}{lccr@{.}lcrlccc}\heading{
&$\alpha_{RR}$&$\alpha_{\phi\phi}$&\multicolumn{2}{c}{$\gamma_z$}&$\hat{\gamma}$
&\multicolumn{2}{c}{$\eta_{\rm t}$}&$\tau$&$\Omega^*$&$C_{\alpha}$ \\
& $[\kms]$ & $[\kms]$ &\multicolumn{2}{c}{$[\kms]$} 
& &\multicolumn{2}{c}{$[\kpc\kms]$}
& $[\Myr]$ & & }
T4       & 3.1 & 3.8 &$\quad$-7&8 & 2.1 &~~~0.9 & 2.7 & 3.6 & 0.7 & 3.3 \\
T4--NCL  & 7.3 & 4.3 &-24&7 & 5.9 & 2.4 & 2.2 & 2.8 & 0.6 & 1.4 \\
T4--SNII & 0.8 & 0.8 & -1&8 & 2.5 & 0.6 & 1.0 & 3.6 & 0.7 & 1.2 \\
T4--KIN  & 3.2 & 2.0 & -8&3 & 4.1 & 0.6 & 1.2 & 3.4 & 0.7 & 2.6 \\[6pt]
F98      & 6.0 & 2.6 & 16&0 & 6.2 & \multicolumn{2}{c}{18.} & 
 -- & -- & 0.2 \\
         & 0.7 & 0.5 &  4&0 & 6.6 & \multicolumn{2}{c}{2.} & 
 -- & -- & 0.3 \\\hline
\end{tabular}
\end{center} 
\caption[Dynamo coefficients for solid body rotation]{%
  Dynamo parameters for solid body rotation. Signs refer to the top
  half of the box, the values for $\eta_{\rm t}$ apply to the inner ($|z|\le
  0.8\,\kpc$) and outer region of the disk, respectively. Coherence time
  $\tau$ and Coriolis number $\Omega^*$ are estimated from a comparison with
  SOCA-profiles for $\alpha_{\phi\phi}$. For comparison we also list peak values
  from \citetalias{1998A&A...335..488F}, for $R=7\kpc$, and $R=8.5\kpc$,
  respectively.}
\label{tab:ae_T4}
\end{table}

\subsection{Turbulent diffusion} 
\label{sec:dynamo_diff}

The feedback loop via the diagonal elements of $\talph$ and the diamagnetic
transport via $\gamma_z$ are only one side of the coin. In the end the
operability (and growth rate) of a dynamo is determined by the dissipative
effects from the turbulent diffusion $\eta_{\rm t}$. In panel (c) of
Figure~\ref{fig:ae_F4_rot}, we plot the diagonal elements $\tilde{\eta}_{RR}$,
and $\tilde{\eta}_{\phi\phi}$, which show the same amplitude of $\simeq
2\kpc\kms$. Similar as with the diagonal elements of $\talph$, this can be
understood via the indistinguishability of the radial and azimuthal radial
direction in the case without shear. The profiles of the turbulent diffusion
are compared to the classic prediction $\tau/3\,\mn{u'^2}$ (with a constant
coherence time $\tau$), and a reasonable match is obtained. If we study the
curves more closely, we see that we have to decrease the value for $\tau$ near
the midplane, and increase it away from the midplane to improve the congruence
of the profiles. This trend of an increased coherence time $\tau$ with
galactic height $z$ agrees well with the assumed expansion of the remnants
near and away from the midplane.

The amplitude of the turbulent diffusivity $\beta_{\rm v}(z)$ in
\citetalias{1998A&A...335..488F} agrees well with the value we obtain for
$\tilde{\eta}_{RR}$, and $\tilde{\eta}_{\phi\phi}$ -- at least for
$z\simlt1.5\kpc$ and $R=8.5\kpc$. Whereas we observe a decline in $u'^2$
beyond $z\simeq1\kpc$,\footnote{%
  This result might, however, still be affected by the limited horizontal box
  size of our simulations.} %
the profiles in \citetalias{1998A&A...335..488F} peak at $z=2$--$4\kpc$ for
the various values of $R$. The source of this discrepancy is either rooted in
the different input parameters, or, more likely, in the contrary conceptual
approach: If we look at Figure~9 in \citetalias{1998A&A...335..488F}, we see
pea-shaped super bubbles extending over several kiloparsec; within the
semi-analytical framework, these bubbles are thought to expand coherently.
Accordingly, they give rise to the high peak values in the dynamo coefficients
far away from the midplane, where their lifetime is supposedly higher due to
the lower external gas pressure. In our simulations, we only see such
``bubbles'' at a very early stage, before the turbulence is fully developed.
Once the thermal instability has produced a clumpy, heterogenous ISM, the
bubbles easily break up into unordered turbulence and the coherent expansion
is suppressed. In our simulations, the turbulence is not only created by the
SBs but also acts back on them, i.e., unlike in the theoretical model,
we actually measure the turbulent diffusivity in a self-regulated regime.

The relative amplitudes of the processes relevant for dynamo action are
commonly measured by means of dimensionless numbers. The dynamo number
characteristic for the $\alpha$~effect is defined as
\begin{equation}
  C_{\alpha}=\alpha H/\eta_{\rm t}\,.
\end{equation}
We compute this number assuming a characteristic length $H=0.8\,\kpc$ and
applying the $\eta_{\rm t}$ value from the inner part of the disk (cf.
Tab.~\ref{tab:ae_T4}). In accordance with an analytical estimation based on
this number \citep*{1994A&A...286...72S}, the runs without shear are still
sub-critical, i.e., no field amplification is expected for the pure
$\alpha^2$~dynamo. We remark that this finding should in no way be regarded as
a general exclusion of the possibility of a SN-driven dynamo in the case of
solid rotation. Further simulations at higher magnetic Reynolds number yet
have to demonstrate whether differential rotation is indeed essential for the
dynamo to work, or, if it was only that the critical values for dynamo action
were not achieved in the rigid rotating case, given the assumed density
profile and gravitational potential used in the current simulations.

Finally, as illustrated in panel (d) of Figure~\ref{fig:ae_F4_rot}, the
off-diagonal elements of the diffusion tensor $\teta$ indeed possess the
correct sign to make the parameter $\delta_z$ positive. This is the case over
a wide range of the vertical domain and the term is negative only within a few
minor regions. According to the discussion in Section~\ref{sec:Raedler}, this
means that the antisymmetric off-diagonal entries can in fact contribute to
the overall dynamo effect. With an amplitude of $\simlt0.5\kpc\kms$, the
effect is somewhat smaller than the diffusion from the diagonal elements, but
by far not negligible. In the absence of any $\alpha$~effect or vertical
transport, these values, in the 1D toy model described in
Appendix~\ref{sec:toy}, produce an exponentially growing stationary mode with
quadrupolar symmetry.

\subsubsection*{Damping of MRI modes}

Apart from its importance for the dynamo, the turbulent diffusion caused by
the SNe might also have implications for other processes in the ISM.
\citet{1999ApJ...511..660S} have proposed that (mainly in regions of little
star formation activity) the observed uniform level of turbulence in the
neutral hydrogen may be explained by the magneto-rotational instability.
Simulations by \citet*{2004A&A...423L..29D} and
\citet{2005ApJ...629..849P,2007ApJ...663..183P} demonstrate that this is
indeed possible. Irrespective of the fact that MRI is not needed to explain
the turbulence in regions with strong SN feedback, it is nevertheless
interesting to ask whether it provides a contribution to the amplification of
magnetic fields in galactic disks. This is in so far worthwhile to be
considered, as MRI provides a very efficient way to extract kinetic energy
from the background shear flow. Based on the considerations in
\citet{1996ApJ...457..798J}, we crudely estimate that the inferred amount of
diffusion present in our simulations would in principle suffice to damp short
wavelength MRI modes for reasonably high $\beta_{\rm P}$. A definite
conclusion will, of course, require further investigations by means of
combined direct simulations. Such models will, however, have to properly
resolve the unstable MRI modes as well as the outer scale of the supernova
remnants -- which sets very high demands on the computational resources. Even
if MRI is found to be critically damped by the SNe, it will still have to be
considered as a possible mechanism to explain halo fields and, beyond that,
might serve as a pre-amplifier generating stronger seed fields.

\subsection{Comparison of models} 

In Table~\ref{tab:ae_T4}, we compile dynamo parameters from a set of
simulation runs based on the setup T4, discussed above. To study the effect of
coherent SNe within super bubbles, we have disabled the clustering
prescription for type~II SNe within the model T4--NCL. This means that all SNe
are now placed at random positions, and only the vertical distribution remains
constrained. As described in Section~\ref{sec:morphology} on
page~\pageref{sec:morphology}, the morphology is quite different in this case
and strong vertical streaming motions, commonly referred to as ``chimneys'',
are observed. These are also reflected in the high values for $\gamma_z$ and
$\mn{u}_z$. Also the turbulent diffusivity is high in the disk midplane, which
is not the case for the other runs, where $\eta_{\rm t}$ scales with the
galactic height. Although the $\alpha$~effect is somewhat stronger in this
case, the ratio $\hat{\gamma}$ is substantially increased due to the high
level of diamagnetic transport. Along with the high level of diffusion, the
dynamo number is decreased to a value close to unity, thus rendering dynamo
action rather improbable. The low value of $\tau=2.8\Myr$ in the model T4--NCL
indicates that a higher coherence time might be achieved by a more realistic
prescription for the modelling of SBs -- ultimately a self-consistent,
self-regulatory approach via a star formation criterion would be highly
desirable to properly mimic the spatial distribution of OB-associations
\citep{2005A&A...436..585D}.

In model T4--SNII, we neglect the field SNe and only consider (clustered)
type~II events. Here we see that the velocity dispersion, and related to it
the turbulent diffusion, predominantly arises from the more broadly
distributed type~I SNe -- despite their lower rate by a factor of eight.
Because theses events are mainly located in the dilute hot plasma, where the
cooling time is long compared to the dense plasma in the midplane, these
explosions, can develop a much higher velocity dispersion and more easily
break up into turbulence. From the drastic changes in the models T4--SNII and
T4--NCL, we reason that an accurate representation of the SN distribution is
of uttermost interest. In this respect, the current models will have to be
refined based on the growing knowledge from observations. The different level
of turbulence in the case of clustered SNe might particularly be important in
view of cosmological simulations, where the kinetic feedback from SNe may
change the efficiency of the structure formation.

The dynamo coefficients listed in Table~\ref{tab:ae_T4} are representative for
the unquenched regime of the dynamo, i.e., the field strength is well below
its equipartition value. It, nevertheless, seems valuable to check on the
influence of the small-scale magnetic field on the inferred mean-field
parameters. For example, \citet{1992A&A...260..494K} have proposed that the
anisotropy in the magnetic field created by a small-scale dynamo can lead to
an additional source of magnetic field transport. By means of the test fields
it is possible to obtain the dynamo coefficients even for the case of a
hydrodynamic simulation. This has been done for the model T4--KIN,
representing the kinematic case, with the back reaction of the magnetic fields
ignored.

In fact, the obtained values for model T4--KIN (see Tab.~\ref{tab:ae_T4}) show
some deviations from the standard case T4. Whereas the level of the turbulent
diffusion seems to be reduced, the amplitude of the vertical pumping is found
to be very similar. Because of the smaller value for $\alpha_{\phi\phi}$, the
ratio $\hat{\gamma}$ seems to be increased. Due to the inherent uncertainties
in these early models, it is, however, not clear whether this trend is indeed
significant. For more conclusive results, the kinematic case will have to be
compared to models with stronger magnetic fields and with the diagnostics
improved as described above.

\section{Dynamo coefficients from differential rotation} 

In the previous section, we have studied the $\alpha$~effect arising from the
combined action of stratification and solid body rotation. Although we indeed
found a non-vanishing contribution to the $\alpha$~effect, we did not observe
a dynamo. This is not the end of the world, of course, since the galactic
environment is characterised by strong differential rotation, anyway.
Proceeding with our bottom-up approach of understanding the galactic dynamo,
we now want to explore how the picture changes in the regime of differential
rotation. If we add shear to the rotation, we expect consequences in three
different ways: (i) the flow is now, at least in principle, unstable against
the magneto-rotational instability, (ii) the critical dynamo number is
considerably lowered due to the new induction term, and (iii) there is an
additional source of anisotropy, and potentially helicity, in the turbulence
itself.

As we have discussed above, MRI is probably overwhelmed by the high value of
the turbulent diffusion. Moreover, even in the absence of the SNe, the
magnetic field is still too weak in our simulations to be able to adequately
resolve the unstable MRI modes on the numerical grid. We hence conclude that
any effects due to (i) can be safely ignored at the current point -- which
does not imply, that this subject should not be studied at a later stage, of
course; approaching equipartition field strengths, MRI might very well become
important.

The striking simplicity and beauty of the MRI lies in the way it draws its
power from the vast reservoir of kinetic energy stored in the differential
rotation. The very same source of power can also be tapped by a more general
dynamo mechanism. As we already mentioned, in the presence of differential
rotation, the $\alpha$~effect merely has to serve in closing the feed-back
loop. The prospects of field amplification are thus drastically improved. To
quantify this effect, we define the dynamo number 
\begin{equation}
  C_{\Omega}=-q\Omega H^2/\eta_{\rm t}\,,
\end{equation}
which describes the relative strength of the mean induction over the
dissipation. For $\alpha\Omega$~dynamos the product $D=C_{\alpha}\,
C_{\Omega}\simeq 30$ now serves as a characteristic number determining whether
field amplification can be obtained. As we have seen in the previous chapter,
this type of dynamo is indeed super-critical in the galactic context and shows
growth times of $\simeq 100\Myr$. Apart from changing the very basic
conditions for the dynamo, linear shear, even in the absence of rotation, can
already alter the structure of the turbulence and provide a substantial source
of helicity \citep{2008PhRvL.100r4501Y}. Like vertical stratification and
rotation, it introduces a new preferred direction to the flow. How does this
term contribute to the overall $\alpha$~effect in our model?

\begin{table}\begin{center}\small
\begin{tabular}%
{lr@{$\pm$}lr@{$\pm$}lr@{$\pm$}lr@{$\pm$}lr@{$\pm$}lr@{$\pm$}lcc}
\heading{  & 
  \multicolumn{2}{c}{$\alpha_{RR}$} & 
  \multicolumn{2}{c}{$\alpha_{\phi\phi}$} & 
  \multicolumn{2}{c}{$\gamma_z$} &
  \multicolumn{2}{c}{$\tilde{\eta}_{RR}$} &  
  \multicolumn{2}{c}{$\tilde{\eta}_{\phi\phi}$} & 
  \multicolumn{2}{c}{$\delta_z$} &
  $C_{\alpha}$ & $C_{\Omega}$ \\ & 
  \multicolumn{2}{c}{$[\kmss]$} & 
  \multicolumn{2}{c}{$[\kmss]$} & 
  \multicolumn{2}{c}{$[\kmss]$} &
  \multicolumn{2}{c}{$\![\kpc\kmss]\!$} &  
  \multicolumn{2}{c}{$\![\kpc\kmss]\!$} & 
  \multicolumn{2}{c}{$\![\kpc\kmss]\!$} & &}
 Q4 & 
 $1.3$&$0.8$ &$~1.2$&$0.4$ &$ -3.1$&$0.8$ &
 $~~1.3$&$0.3$ &$~~1.6$&$0.4$ &$~~0.3$&$0.2$ &$0.7$&$43.$ \\
 H4 & 
 $  0.8$&$0.8$ &$  1.7$&$0.5$ &$ -3.5$&$0.8$ &
 $  1.5$&$0.2$ &$  1.8$&$0.1$ &$  0.3$&$0.1$ &$0.6$&$39.$ \\
 F4 &
 $  1.6$&$  0.7$ &$  2.0$&$  1.0$ &$ -4.1$&$  1.3$ &
 $  1.6$&$  0.2$ &$  2.0$&$  0.1$ &$  0.4$&$  0.3$ &$  0.8$&$  36.$ \\[6pt]
 F4--ROT &
 $  1.6$&$  0.5$ &$  2.1$&$  0.6$ &$ -3.3$&$  0.8$ &
 $  1.4$&$  0.2$ &$  1.2$&$  0.1$ &$  0.2$&$  0.2$ &$  1.1$& -- \\
 F4--SHR &
 $ -1.5$&$  0.6$ &$ -1.1$&$  0.3$ &$ -2.9$&$  0.4$ &
 $  1.3$&$  0.2$ &$  1.5$&$  0.2$ &$ (0.1$&$  0.2)$&$  0.7$&$  45.$ \\
 F4--noTI &
 $  0.8$&$  0.5$ &$  1.2$&$  0.1$ &$ -2.3$&$  0.5$ &
 $  1.2$&$  0.2$ &$  1.3$&$  0.2$ &$  0.2$&$  0.1$ &$  0.6$&$  50.$ \\
 \hline
\end{tabular}\end{center}
\caption[\same{for differential rotation}]{%
  Overview of the obtained dynamo coefficients for differential
  rotation. Signs for the coefficients $\alpha_{RR}$, $\alpha_{\phi\phi}$, and
  $\gamma_z$ apply to the northern ``hemisphere''. The numbers for these
  parameters are integral mean values, computed separately for the top and
  bottom half of the box. The remaining coefficients apply to $|z|>0.8\kpc$,
  except for model F4--noTI, where we integrate the mean values over
  $0.4\kpc<|z|<1.2\kpc$ to account for the smaller size of the dynamo-active
  region.}
\label{tab:ae}
\end{table}

\subsection{Cartesian shear} 
\label{sec:dynamo_shear}

Before we turn to the complete picture including shear \emph{and} rotation, we
want to follow a brief detour and study the $\alpha$~effect that linear shear
imprints on the interstellar turbulence. The more subtle question, why no
dynamo is observed in this case, will be discussed separately in
Section~\ref{sec:dynamo_shear_vs_rot}.

\begin{figure}
  \center\includegraphics[width=\columnwidth]{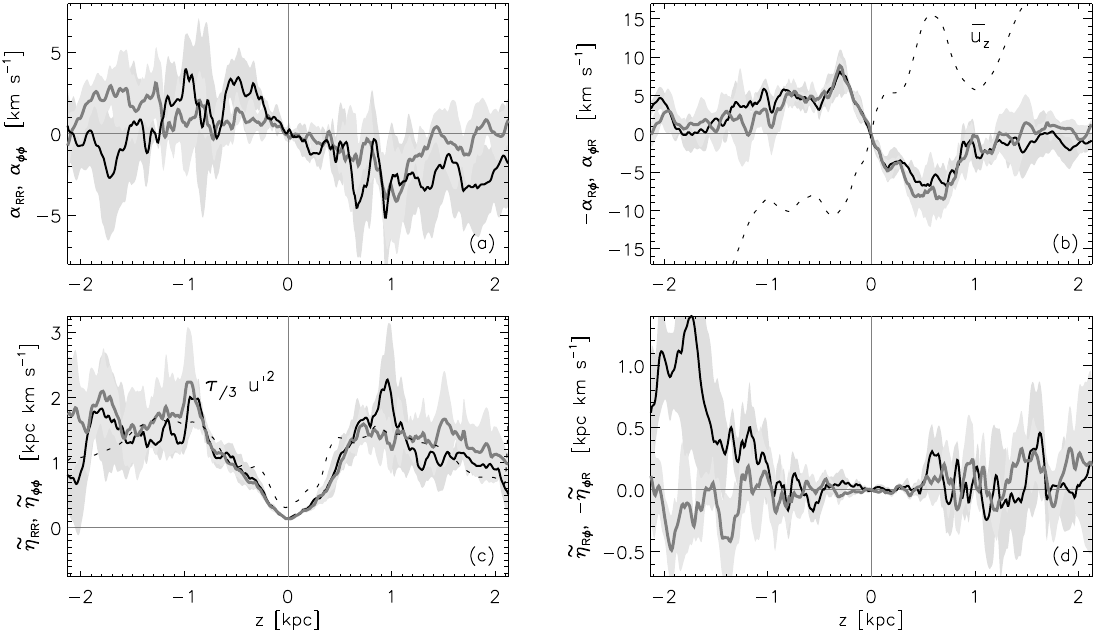}
  \caption[\same{for model F4--SHR}]{%
    Same as Fig.~\ref{fig:ae_F4_rot} but for model F4--SHR, i.e., for the case
    of plain Cartesian shear. Note the inverse sign (compared to F4--ROT) of
    the diagonal elements of $\talph$ in panel (a) and the odd parity of the
    coefficient $-\tilde{\eta}_{\phi R}$ in panel (d).}
  \label{fig:ae_F4_SHR}
\end{figure}

The dynamo parameters are again obtained via the test-field method, and we
present the corresponding results in Figure~\ref{fig:ae_F4_SHR}. Integral
amplitudes of the various profiles are listed in Table~\ref{tab:ae}, where the
sign of the coefficients is representative of the top half of the simulation
box. As can be seen in panel (a) of Figure~\ref{fig:ae_F4_SHR}, the
$\alpha$~effect from Cartesian shear has the opposite sign as in the case of
solid body rotation \citep[cf. Sec.~2.1 in][]{2006AN....327..298R}. This can
intuitively be understood in terms of the opposite radial velocity gradients
in the cases of solid body rotation and galactic shear. Unlike the Coriolis
force, which equally acts on the radial and azimuthal component of the
velocity field, the shear only affects the former. Consequently, we expect the
$\alpha$~effect to be anisotropic, and we indeed find the radial component
$\alpha_{RR}$ to be prevailing. In sight of the substantial fluctuations (see
Fig.~\ref{fig:ae_F4_SHR}), we cannot yet decide whether this trend is in fact
significant. A stronger radial contribution would, however, be consistent with
the results of \citet{1995PhDT........90Z}, who studied the evolution of
isolated remnants and found $\alpha_{RR}$ to be significantly reduced in the
case of differential rotation, while $\alpha_{\phi\phi}$ was less affected by
the additional shear.

If we compare the transport coefficient $\gamma_z$ for the cases of solid body
rotation and Cartesian shear, we see that the values are consistent within the
error margins. This is in excellent agreement with the prediction by SOCA
theory, which states that this term is solely determined by the combined
vertical gradient $\nabla\log(\rho\,u')$ and is independent of the rotation
rate $\Omega$ \citepalias{1993A&A...269..581R}. Whereas the tensor component
$\tilde{\eta}_{RR}$ related to the dissipation of $\bar{B}_{\phi}$ is also
consistent for the two cases, the component $\tilde{\eta}_{\phi\phi}$ is
somewhat enhanced in the case of linear shear. This anisotropy seems plausible
since the shearing of the radial field $\bar{B}_{R}$ enhances the curl of the
field and hence the susceptibility to dissipation. A similar increase of
$\eta_{\rm t}$ has been found by \citet{2008arXiv0806.1608M}, who consider
helically forced turbulence under the influence of external shear.

With respect to the R\"adler effect, we find a possibly interesting difference
between the effects of rotation and shear: In the model F4-ROT
$\tilde{\eta}_{R\phi}$ and $-\tilde{\eta}_{\phi R}$ are positive in the whole
domain corresponding to the totally antisymmetric case. Also both terms
display an even parity with respect to the midplane. Contrary to this, in the
model F4-SHR, the $\tilde{\eta}_{\phi R}$ term shows a trend towards odd
parity. Because the sign of $\tilde{\eta}_{R\phi}$ remains unclear for $z>0$,
we cannot draw a definite conclusion yet. There is some indication, however,
that the off-diagonal part of the $\teta$~tensor, at least for $z<0$, might
possess a non-vanishing symmetric contribution.

\subsection{Differential rotation} 

\begin{figure}
  \center\includegraphics[width=\columnwidth]{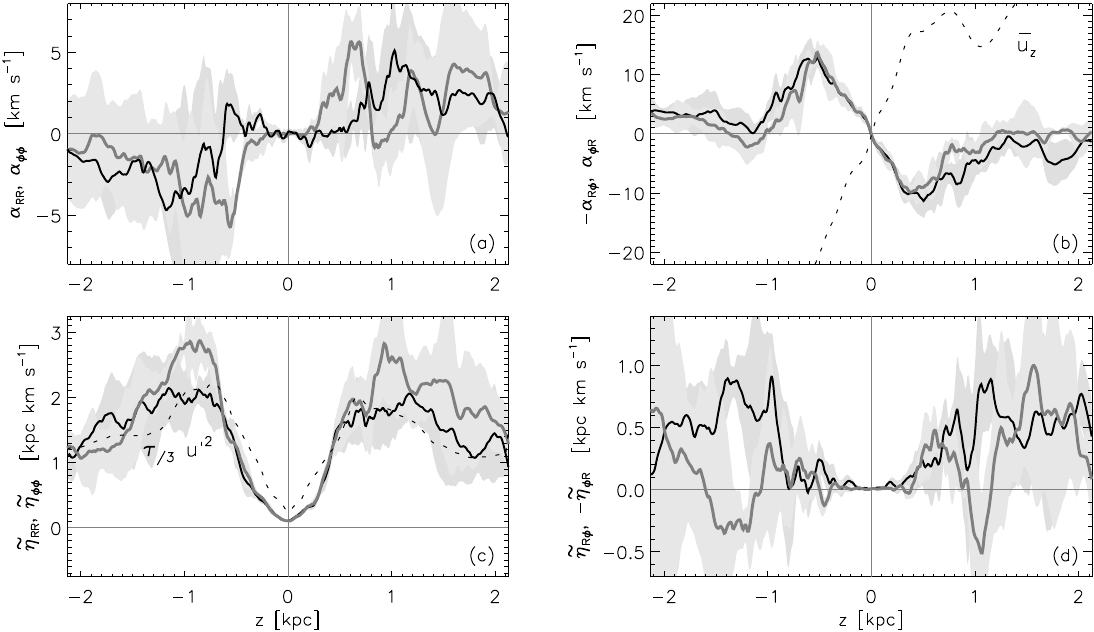}
  \caption[\same{for model F4}]{%
    Same as Fig.~\ref{fig:ae_F4_rot} but for model F4, i.e., for the case of
    differential rotation.}
  \label{fig:ae_F4}
\end{figure}

\begin{figure}
  \center\includegraphics[width=\columnwidth]{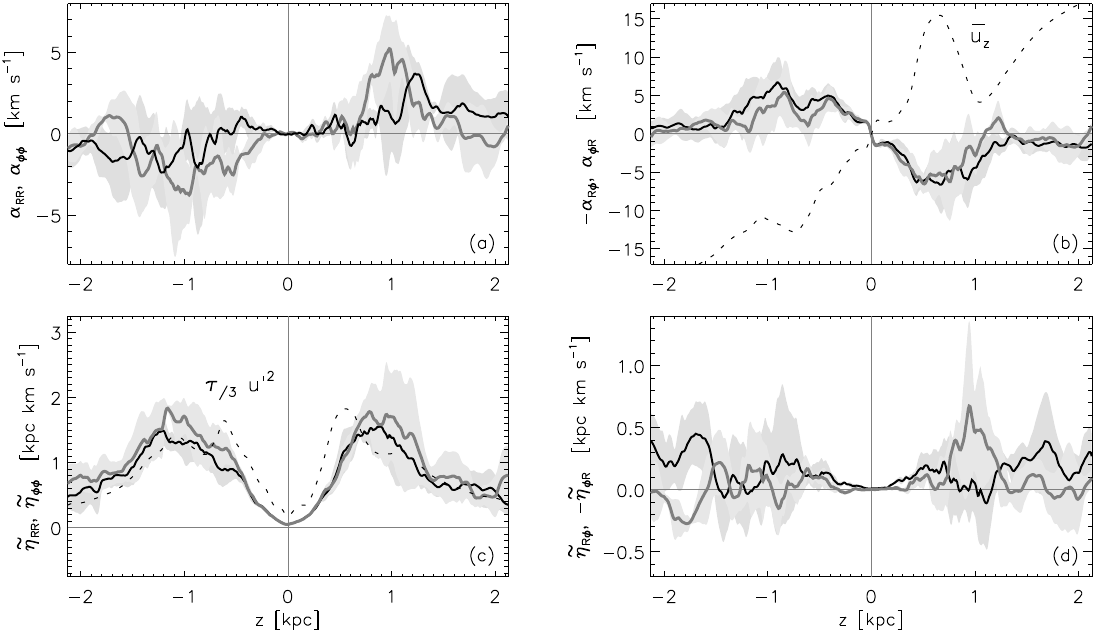}
  \caption[\same{for model H4}]{%
    Same as Fig.~\ref{fig:ae_F4_rot} but for model H4.}
  \label{fig:ae_H4}
\end{figure}

\begin{figure}
  \center\includegraphics[width=\columnwidth]{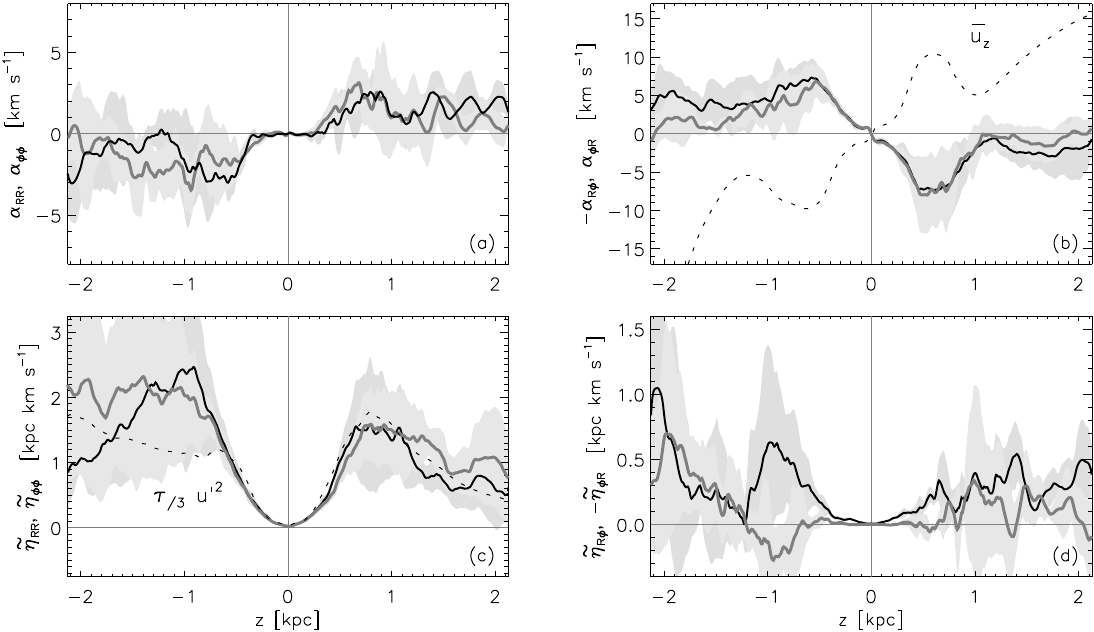}
  \caption[\same{for model Q4}]{%
    Same as Fig.~\ref{fig:ae_F4_rot} but for model Q4.}
  \label{fig:ae_Q4}
\end{figure}

Considering the non-linearity of the underlying equations, one may reason
whether differential rotation can simply be regarded as a superposition of
``rotation plus shear''. Although this is certainly not the whole truth, we in
fact find some evidence, that the effects we observed in the models F4--ROT
and F4--SHR are also reflected in model F4, which represents the case of
differential rotation.

Compared to the case of solid body rotation, the coefficient $\alpha_R$, in
the very inner part of the disk, seems to pick-up the negative slope
$\partial_z \alpha|_{z=0}$ present in model F4--SHR. A possible explanation
for this is the smaller asymptotic radius of the supernova shells in the thick
inner disk. With the limited expansion phase, the effect of the Coriolis force
is relatively weaker as for the larger shells away from the midplane. This, on
the other hand, means that the effects due to the shear might be more
important for $|z|<100\pc$. The negative slope at $z=0$ becomes particularly
evident for model F4--noTI, where we neglect thermal instability (see
Fig~\ref{fig:ae_F4_noTI}). Because of the different thermal structure of the
disk without TI, the effect seems to be more pronounced in this case.

As we see in Figure~\ref{fig:ae_Q4}, for model Q4, the $\alpha$~effect
vanishes for $z<400\pc$ -- a trend that is already visible in model H4 (cf.
Fig.~\ref{fig:ae_H4}). The $\alpha$~effect from rotation seems to be
``quenched'' by the additional negative contribution from the shear.
\footnote{%
  Since the inner region does not significantly contribute to the overall
  amplitude, this effect is not visible in the values of Table~\ref{tab:ae}.} %
A similar result of $\alpha$~quenching via shear has recently been obtained by
\citet{2008arXiv0806.1608M} for simulations of helically forced turbulence
\citep[see also][for a theoretical account]{2008PhRvL.100n4502L}. Although
our supernova-driven turbulence, under the influence of the Coriolis force,
might be comparable to the helical forcing, at the current point, it can only
be speculated whether the two observed effects have a common origin.

As discussed at the end of Section~\ref{sec:dynamo_shear}, the second major
difference between the models F4--ROT and F4--SHR concerned the symmetry of
the tensor element $\tilde{\eta}_{\phi R}$ (represented by a grey line in the
corresponding figures). While its counterpart $\tilde{\eta}_{R\phi}$ (black
line) is invariably positive for differential rotation (models Q4,H4, and F4),
$\tilde{\eta}_{\phi R}$ does not show a distinct sign, in this case. As one
would suspect, this might be inherited from model F4--SHR.

\subsubsection{Dependence on supernova rate}

In the following, we want to briefly study how the dynamo coefficients depend
on the applied supernova rate. Since the models Q4, H4, and F4 only span a
factor of four in $\sigma$, we are, however, restricted to a rather limited
region in parameter space. Moreover, the three different models are, more or
less, all in the saturated regime with respect to the driving. This means that
the time between two distinct explosions is short compared to the relaxation
time of the turbulence. As we have seen from the horizontal slices in
Figures~\ref{fig:slc_Q4} and \ref{fig:slc_F4}, there are, nevertheless,
pronounced differences in the sheared out structure of the remnants. To this
respect, the main effects of the variaton in the SN frequency are (i) the
overall level of the turbulence (see Fig.~\ref{fig:zdisp} on
p.~\pageref{fig:zdisp}) and (ii) the relative strength of the driving compared
to the shear (see lower panels in Figs.~\ref{fig:slc_Q4} and \ref{fig:slc_F4}
on p.~\pageref{fig:slc_F4}).

If we compare the first three rows of Table~\ref{tab:ae}, we find a slight
trend towards increasing amplitudes for higher SN rates in all the parameters.
This trend is rather marginal for $\alpha_{RR}$, $\alpha_{\phi\phi}$ and
$\gamma_z$, which are thought to depend on the gradient of $u'$, whereas it is
more pronounced in the diffusive coefficients $\tilde{\eta}_{RR}$ and
$\tilde{\eta}_{\phi\phi}$, which, according to SOCA theory, scale directly
with $u'^2$. For the parameter $\delta_z$, we do not observe a significant
dependence on $\sigma$. Overall, the obtained coefficients are in reasonable
agreement with theoretical considerations. Because the diagonal $\alpha$ and
$\tilde{\eta}$~coefficients show a comparable dependence on the supernova
rate, the corresponding dimensionless number $C_{\alpha}$ has a constant value
of $0.6\pm0.1$ for all the models. The trend towards higher turbulent
diffusivities $\eta_{\rm t}$ is reflected in decreasing numbers $C_{\Omega}$,
which are also consistent with the anti-correlation of the field regularity
with the star formation activity (see Sec.~\ref{sec:reg_tur}).

\section{The effect of vertical transport processes} 
\label{sec:transport}

The kinetic energy deposited into the interstellar medium by supernovae is
tremendous; nobody doubted that this kind of driving would create an
$\alpha$~effect of sufficient strength to power a galactic dynamo. It was
perceived from the very beginning, however, that the vigorous driving from the
SNe, at the same time, would create a strong diamagnetic transport. Both
inward and outward pumping were found to drastically diminish the prospects of
dynamo action -- \citet*{1994A&A...286...72S}, in fact, found a rather narrow
range of permissible values for $\hat{\gamma}$. The scepticism was finally
affirmed by the dominant pumping found under the assumption of an ensemble of
uncorrelated explosions (cf. Sec.~\ref{sec:isolated}). In view of the
unquestionable importance of the vertical pumping, we now want to try and
understand the simulation results presented in Section~\ref{sec:mf_dynamo}.
For this, we shall be aided by a simple 1D mean-field model based on the
dynamo coeffcients derived above.

\subsection{Symmetry considerations} 

\begin{figure}
  \center\includegraphics[width=\columnwidth]{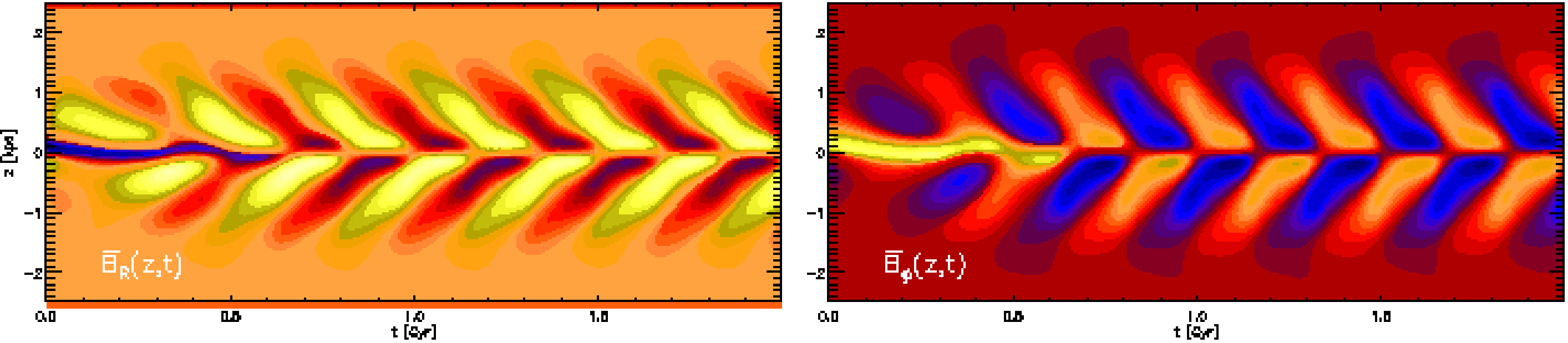}
  \caption[``Butterfly'' diagram of the mean radial and azimuthal field]{%
    ``Butterfly'' diagram of the mean radial and azimuthal field as computed
    from the 1D toy model with parameters $\alpha_R$, $\alpha_{\phi}$, and
    $\eta_{\rm t}$ taken from model H4. Vertical transport processes
    $\gamma_z$ and $\bar{u}_z$ are neglected. The colour coding is corrected
    for the exponential growth of the rms~values $\rms{\bar{B}_{R}}$ and
    $\rms{\bar{B}_{\phi}}$, respectively.}
  \label{fig:bf_H4}
  \vskip4ex%
  \center\includegraphics[width=\columnwidth]{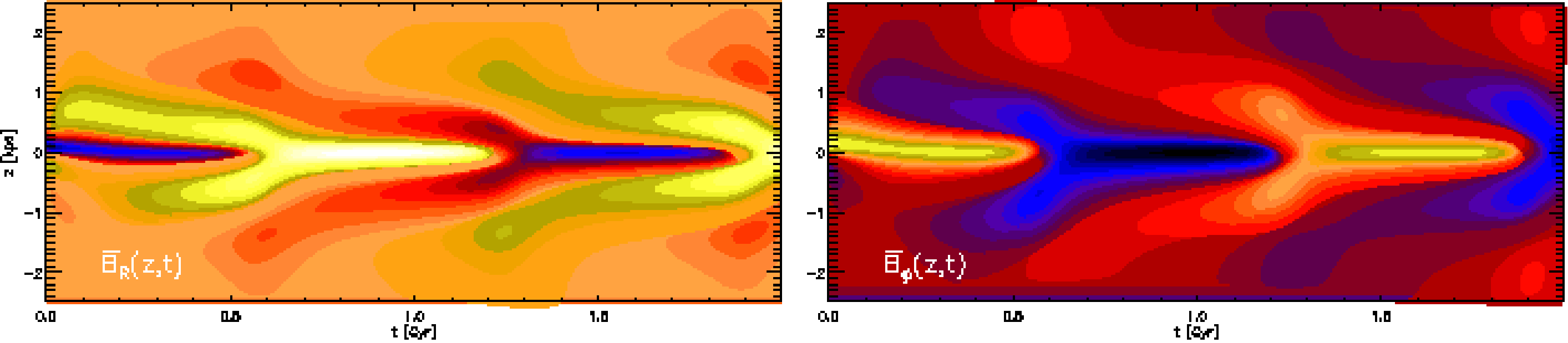}
  \caption[\same{including diamagnetic pumping and wind}]{%
    Same as above, but now including the contribution of the diamagnetic
    pumping $\gamma_z$ and the mean flow $\bar{u}_z$. The vertical asymmetry
    at the first reversal is a relic of the mixed inital conditions and
    disappears already for the second reversal.}
  \label{fig:bf_H4_v}
\end{figure}

With $C_{\alpha}\,C_{\Omega}\simeq 35$--$40$, the models Q4, H4, and F4 are
well super-critical under the $\alpha\Omega$~mechanism. In the absence of
vertical transport, we, accordingly, obtain an oscillatory solution with
dipolar symmetry, i.e., odd parity with respect to $z=0$. The corresponding
dynamo pattern is depicted in Figure~\ref{fig:bf_H4}, where we draw the
``butterfly'' diagram of $\bar{B}_R(z,t)$ and $\bar{B}_{\phi}(z,t)$. This
result, obviously, does not match the quadrupolar symmetry found in our direct
simulation runs. The picture is, however, drastically changed if we add the
diamagnetic transport and wind, which we have neglected so far.

\begin{figure}
  \center\includegraphics[width=\columnwidth]{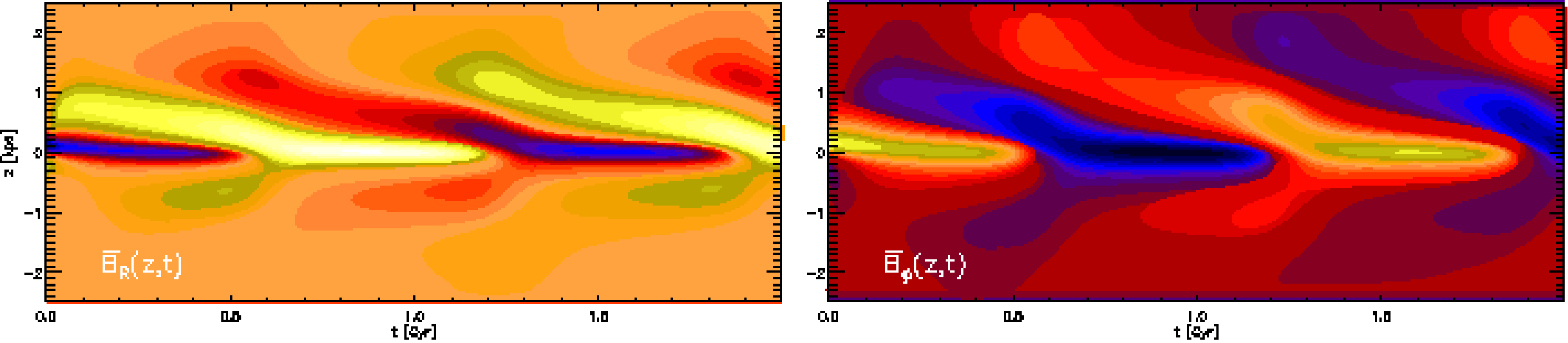}
\vskip4ex%
  \center\includegraphics[width=\columnwidth]{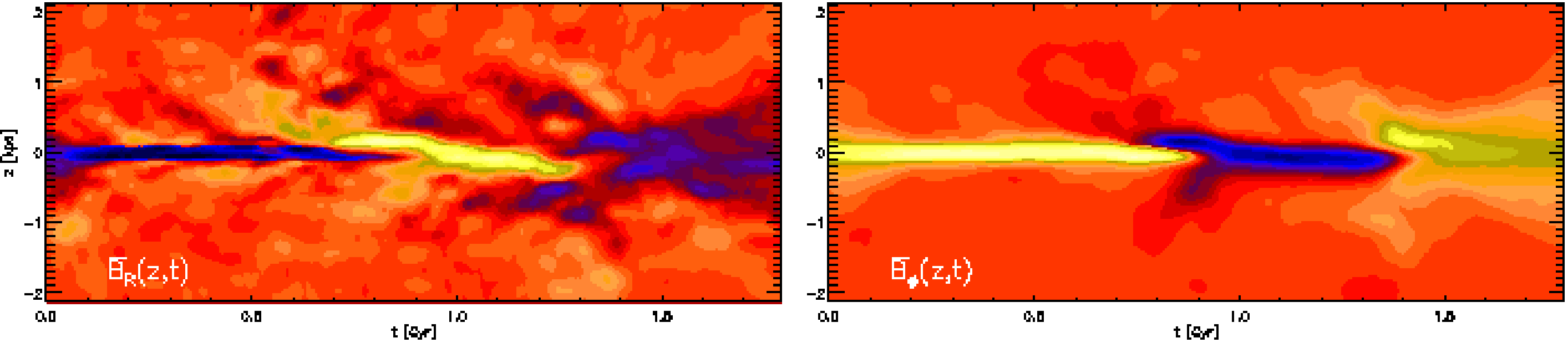}
  \caption[\same{including an off-diagonal term in $\tilde{\eta}$}]{%
    Same as Fig.~\ref{fig:bf_H4_v}, but additionally including a mixed
    (anti-)symmetric contribution in the off-diagonal elements of $\teta$
    (upper panels). Now the lopsided dipolar symmetry in the field reversals
    persists and closely resembles the features seen in the \emph{direct}
    simulation H4 (lower panels).}
  \label{fig:bf_H4_vR}
\end{figure}

Due to the inclusion of the vertical transport processes, the effective dynamo
number is somewhat reduced and we enter a new regime, where now the
quadrupolar mode is dominant \citep[cf. Fig.~2 in][]{2001A&A...370..635B}; the
resulting pattern is shown in Figure~\ref{fig:bf_H4_v}. While the dynamo
solution still shows an oscillatory character, the period between the
reversals is now considerably longer. Varying the ampitude of the pumping, we
infer that the corresponding time scale is related to the effective wave speed
of the field patterns, i.e., it depends critically on the residual transport
velocity.

The first field reversal in Figure~\ref{fig:bf_H4_v} already seems to exhibit
the characteristic skewness we observe in the direct simulations. The effect,
however, already disappears again at the second reversal and we conclude, that
it is merely a relic of the initial conditions. Because we expect our
numerical discretisation to exactly preserve the overall symmetry, we have to
be careful not to exclude certain dynamo modes ab initio. We therefore
initialise our models with a mixed dipolar and quadrupolar initial field
geometry. Although the dipolar mode quickly decays away, it is still visible
at $t=0.5\Gyr$, resulting in the slightly asymmetric field pattern.

If we now include the effects due to the off-diagonal part of the
$\teta$~tensor, the picture is again refined (see upper panels of
Fig.~\ref{fig:bf_H4_vR}). As we have discussed in
Section~\ref{sec:dynamo_shear}, there is some indication of a possible mixed
symmetric and antisymmetric contribution in these tensor elements.
Particularly in the case of Cartesian shear, we observed a significant
symmetric part -- but only in the lower half of our simulation box. If we add
this symmetric term to our mean-field model, the vertical symmetry is broken
and we recover the lopsidedness present in the direct simulations (see lower
panels of Fig.~\ref{fig:bf_H4_vR}). Considering the simplicity of the
parameterisation, we regard this as a remarkable level of resemblance. The
similarity of the results might even seem more striking if we compare the
computational effort necessary to produce the respective figures: whereas the
direct simulation took several weeks on a cluster with 128 CPUs, the toy model
ran in less than five seconds on the authors work station -- overall this
makes a difference of a factor of $\sim10^9$ in computing time. The example
illustrates the possible ``gain'' inherent in the mean-field approach, albeit
realistic global models, of course, will have to be performed in 3D. In the
end, one should, however, not be misguided by the impressive speedup of
mean-field models -- we have to keep in mind that these models are merely a
toy without the knowledge from direct simulations, and that results based on
first principles cannot simply be replaced by intuition.

\subsection{The role of thermal instability} 
\label{sec:dynamo_TI}

In Section~\ref{sec:results_TI}, we have seen that the overall changes
introduced by the neglect of thermal instability enhance the growth rate of
the galactic dynamo by about twenty five percent. From the examination of the
main kinematic characteristics, we could not derive a conclusive explanation
for this. The main differences compared to the standard case were a slightly
lower overall level of the turbulent velocity dispersion along with a
drastically reduced inner gradient (see Fig.~\ref{fig:zdisp} on
p.~\pageref{fig:zdisp}). Whereas the former finding implies a higher value for
$C_{\alpha}$ and thus favours a more efficient dynamo, the latter implies a
weaker diamagnetic pumping $\gamma_z$, which poses a threat due to the
possibly enhanced outward transport by a dominating galactic wind.

\begin{figure}
  \center\includegraphics[width=\columnwidth]{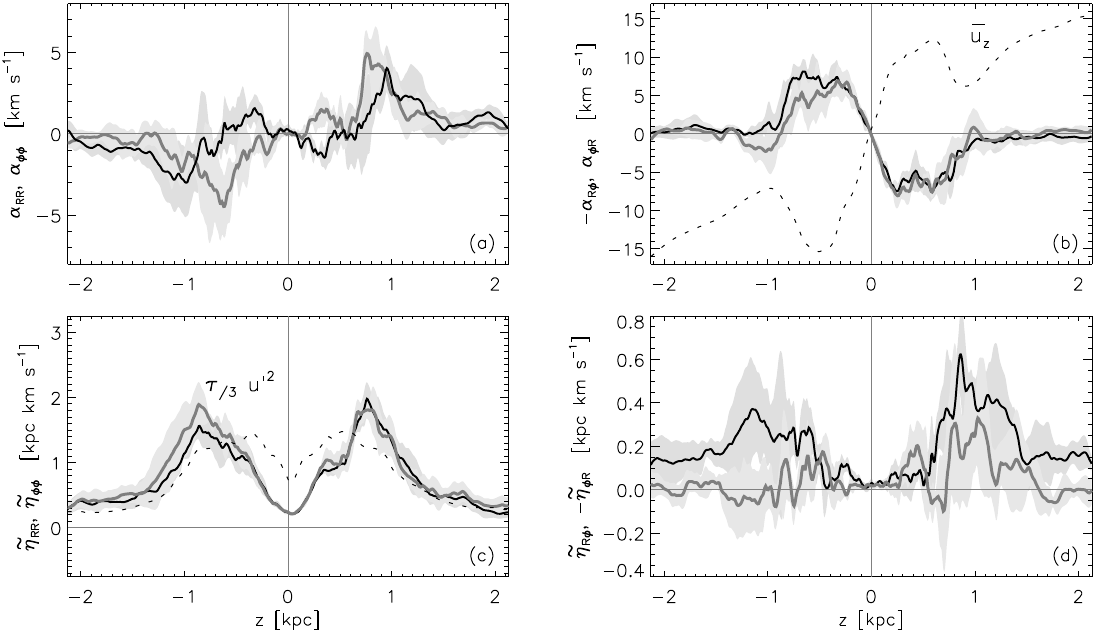}
  \caption[Dynamo $\alpha$ and $\tilde{\eta}$~coefficients for 
    model F4--noTI]{%
    Same as Fig.~\ref{fig:ae_F4_rot} but for model F4--noTI, i.e., without
    thermal instability.}
  \label{fig:ae_F4_noTI}
\end{figure}

With the diagnostics of the test-field method at hand, we are now in the
position to directly query these hypotheses. In Figure~\ref{fig:ae_F4_noTI},
we present the eight tensor components for the model neglecting thermal
instability. First of all, we notice that the radial $\alpha$~effect is
reversed in the inner part of the disk. As has been discussed above, this
might be a consequence of a dominating contribution from the shear effect in
the absence of a strong density gradient. The particular shape of the
resulting $\alpha$~profile is found to slightly alter the field geometry. The
growth rate of the dynamo, however, remains largely unaffected by this
additional feature. As expected from the weaker gradient in both $\rho$ and
$u'$, the overall amplitude of the dynamo coefficient $\alpha_{\phi}$ is
reduced in the inner part of the disk. Further away from the midplane, the
influence of the stratification becomes weaker and the contribution from
$u'^2$ becomes more important (cf. Eq.~\ref{eq:alpha_strat}) -- consequently
the peak amplitudes are comparable to model F4.

The curves for model F4--noTI are well contained within the vertical extent of
the simulation box. This is particularly evident for the pumping terms in
panel (b) of Figure~\ref{fig:ae_F4_noTI}. Despite the apparent differences in
$\nabla\log u'$, the amplitude of $\gamma_z$ is only moderately lower than in
model F4. In accordance with the standard model, the mean flow $\bar{u}_z$ is
opposite to the pumping and has a slightly higher amplitude compared to
$\gamma_z$. In view of the inward travelling dynamo pattern in the case of a
positive $\alpha$~effect, this implies near ideal conditions for the dynamo.
Furthermore, the characteristic ratio $\hat{\gamma}\simeq2$ is unaffected by
the neglect of TI -- a finding that points into the direction that this ratio
is ultimately governed by a very fundamental mechanism. Much as we expected
from the lower overall level of $u'$, the value for the turbulent diffusion is
somewhat reduced in model F4--noTI.

In conclusion we can say that all relevant processes are slightly attenuated
in the case neglecting TI -- or, to use a positive formulation -- they are
enhanced under the effect of the thermal condensation. If we put together all
the pieces, the effective dynamo number $C_{\alpha}\,C_{\Omega}$ is only
marginally higher compared to the standard case. Because the effect of the
vertical transport processes is not reflected in these numbers, they
ultimately have little significance for the real case. Experiments based on
the 1D toy model, however, indicate that the different growth rates are
consistent with the varying level of transport present in the models. As we
will see in the next section, the approximate balance of the diamagnetic
pumping and the mean flow has profund implications for the overall efficiency
of the dynamo process.

\subsection{The importance of rotation} 
\label{sec:dynamo_shear_vs_rot}

From the analysis of our simulation runs presented in
Section~\ref{sec:results_shear_vs_rot}, we have demonstrated that no
amplification of the mean-field was obtained under the influence of plain
shear. As we want to point out, the only difference of the models lies in the
fact that the Coriolis force is disabled for model F4--SHR. Because curvature
terms are neglected in the shearingbox approximation, the linear radial
profile of the background velocity is identical in both cases and the
amplitude of the shear is still given by the term $q\Omega$ -- which is why we
kept the ``F4'' in the nomenclature of the model. The ``negative'' outcome of
the run F4--SHR marks an important result in support of the picture of
cyclonic turbulence. Since the particular setup only represents a single
region in parameter space, the general implication of the finding may,
however, be challenged. To broaden the scope of its applicability, we have to
study the underlying mechanism. To do so, we conduct a parameter study based
on the afforementioned 1D toy model. To closely resemble the simulation runs,
the basic parameter sets will be taken from the models F4--SHR and F4,
respectively.

\begin{figure}
  \center\includegraphics[width=\columnwidth]{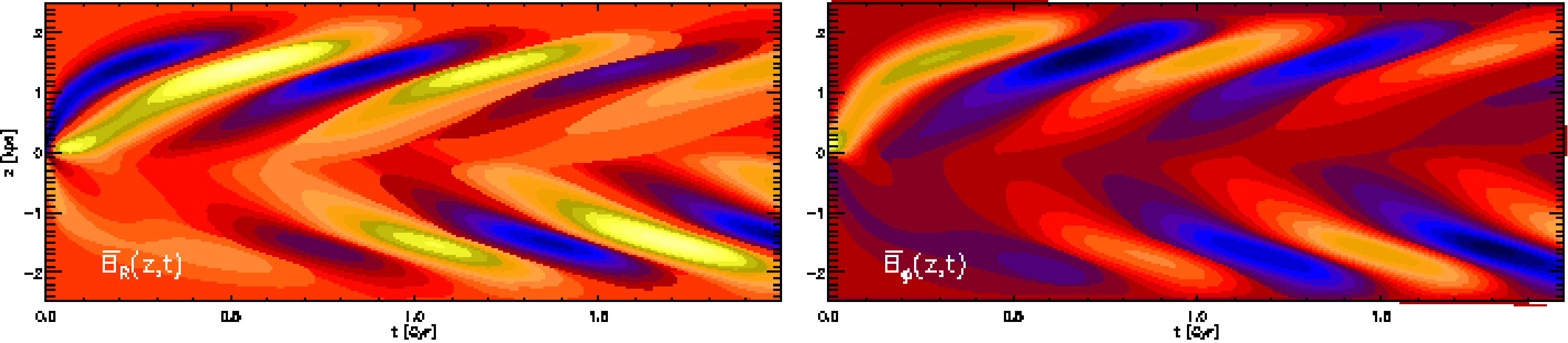}
  \caption[\same{for model F4--SHR}]{%
    Same as Fig.~\ref{fig:bf_H4_v} but for model F4--SHR, where the
    $\alpha$~effect possesses the opposite sign compared to model F4. In
    contrast to the case of rotation, the dynamo pattern (in the absence of
    vertical transport) now travels outwards.}
  \label{fig:bf_F4_SHR}
\end{figure}

The paramount difference between the cases of differential rotation and
cartesian shear was the reverse sign in the $\alpha$~effect. Because the shear
parameter independently defines a sense of orientation, this overall sign is
indeed significant.\footnote{%
  This is contrary to the case of an $\alpha^2$~dynamo, where the overall sign
  of the $\alpha$~coefficients is unimportant.} %
As we can see in Figure~\ref{fig:bf_F4_SHR}, the dynamo solution corresponding
to negative values of $\alpha_R$ and $\alpha_{\phi}$, in fact, exhibits a very
different wave pattern. It shall be noted that, like in the upper panel of
Figure~\ref{fig:bf_H4_v}, the depicted solution corresponds to the case
neglecting vertical transport. Whereas the actual modes are very distinct in
the two cases, the overall growth times are very similar -- which particular
implies that we observe field amplification in both cases. To explain the even
qualitatively different behaviour found in the direct simulations we have to
consider the additional effects caused by the vertical transport.

\begin{figure}
  \center
    \includegraphics[height=0.33\columnwidth]{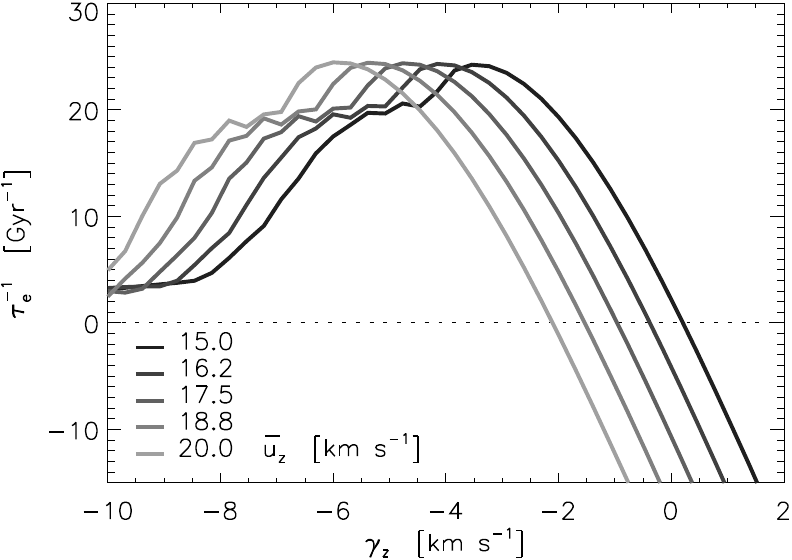}
    \hskip2ex%
    \includegraphics[height=0.33\columnwidth]{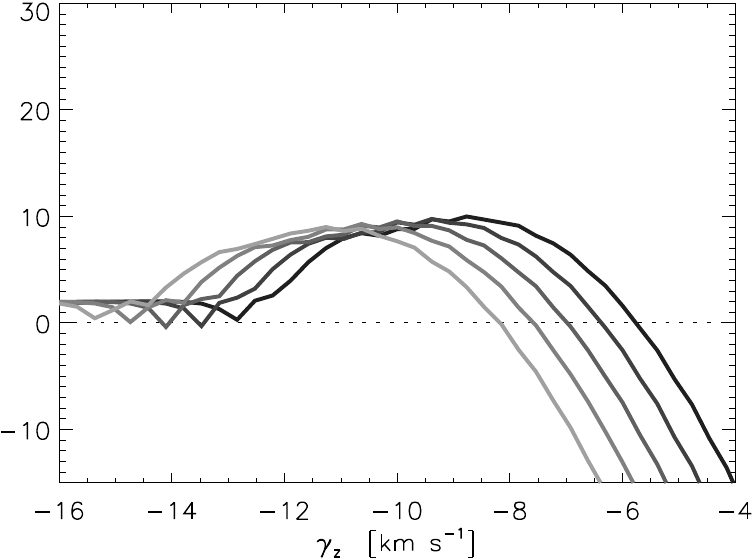}
  \caption[Growth rate as a function of the diamagnetic pumping velocity]{%
    Exponential growth rate of $\rms{\bar{B}_R}$ as a function of the
    diamagnetic pumping velocity $\gamma_z$ for different peak values
    $\bar{u}_z$ of the galactic wind. The curves are computed from the 1D
    toy model applying parameters from setup F4 (left panel) and F4--SHR (righ
    panel). Note the different intervals of the abscissae. Negative values for
    $\tau_{\rm e}$ correspond to decaying solutions.}
    \label{fig:gmzw}
\end{figure}

In Figure~\ref{fig:gmzw}, we present growth rates of dynamo solutions obtained
from a small parameter study varying the amount of the diamagnetic pumping
$\gamma_z$ and the amplitude of the mean flow $\bar{u}_z$. The results shown
in the left hand panel correspond to parameters $\alpha_R$, $\alpha_{\phi}$,
and $\eta_{\rm t}$ inferred from run F4. We want to remark that the two
velocities cannot simply be subtracted to compute the residual velocity. This
is due to the different shape of the two contributions (cf.
Sec.~\ref{sec:toy_prof}). Whereas the definition of $\gamma_z$, for better
comparison, is adopted from the integral mean values listed in the tables, the
numbers for $\bar{u}_z$ mark the peak values at $z=2\kpc$. As a rule of thumb,
the peak values for $\alpha_{R\phi}$ and $\alpha_{\phi R}$ can be estimated by
doubling the numbers given for $\gamma_z$. Following this definition, the
fastest growing solutions are reached at an outward ``residual'' velocity of
$\simeq 7\kms$, which agrees well with the findings of
\citet*{1994A&A...286...72S} and \citet{2001A&A...370..635B}. The e-folding
time of these models can be as low as $40\Myr$, which is already comparable to
the timescales obtained from MRI simulations.

For strong inward pumping (i.e., for $\gamma_z<0$) we observe a slow growing
asymptotic mode with an e-folding time of $\simeq 0.3\Gyr$. In this limit, the
growth rate is probably determined by the midplane value of the diffusivity
$\eta_{\rm t}$. If, on the other hand, the amplitude of the diamagnetic
transport is reduced, the wind can efficiently remove the created field and
the dynamo mechanism is drastically quenched. In the case of differential
rotation, we find a limit of $\simeq15\kms$ for the critical residual
velocity, i.e., even a weak inward pumping allows for growing dynamo
solutions. For $\gamma_z\simeq-3.5\kms$ and $\bar{u}_z=20\kms$ we
approximately recover the growth rate of the direct simulation.

In the case of plain shear, the situation is dramatically changed. Because of
the adverse sign of the dynamo coefficients, the dependence on the diamagnetic
pumping is found to be more critical (see right panel of Fig.~\ref{fig:gmzw}).
Since the basic solution already constitutes of outward travelling dynamo
waves, a much stronger inward pumping becomes necessary to balance already a
moderate wind. Although the configuration, at least in principle, would allow
for growing solutions with e-folding times of $\simeq0.1\Gyr$, the critical
threshold is well below the level present in our simulations. Without an
additional effect enhancing the rate of turbulent transport, the possibility
of a supernova-driven galactic dynamo based on shear alone is thus rendered
unlikely.

Because our direct simulations are based on very fundamental assumptions, this
result has to be seen as a rather general one. The main restriction to the
given argumentation lies in the neglect of the vertical field component and
the corresponding tensor element $\alpha_{zz}$. Further investigations will
have to show whether the result remains valid if this term is included.
Although the 1D model closely resembles the elongated shearing box of our
direct simulations, it has to be checked carefully, in how far this 1D
approach puts constraints on the admissible dynamo solutions. With respect to
the simulations we have already explored the validity of the results with
respect to the magnetic Reynolds number (see Sec.~\ref{sec:convergence}).
There, however, remains the possibility that the relevant effects also depend
non-trivially on the magnetic Prandtl number $\Pm$.



\cleardoublepage
\chapter{Conclusions}
\label{ch:conclusions}

In this thesis, I present local box simulations of a differentially rotating,
vertically stratified, turbulent interstellar medium threaded by weak magnetic
fields. The two main aspects of this work are (i) the verification of a
helicity-based field amplification mechanism driven by SNe, and (ii) the study
of this effect via mean-field coefficients.

Starting from the perceived theoretical requirements for a disk galaxy dynamo,
a ``minimal'' model for the interstellar medium has been developed. As a
central ingredient, we apply optically thin radiative cooling and heating to
account for the heterogeneous, multi-phase nature of the ISM. The vertical
stratification of the galactic disk is approximated by a static external
potential, representing the contributions of a stellar disk population and a
dark matter halo. Improving over existing models, I compute a radiatively
stable initial solution to avoid the transient collapse observed in other
models. The central feature of our simulations is the driving of turbulence
via (several thousand) localised injections of thermal energy, which closely
resemble the kinetics of the vigorous supernova feedback present in the ISM.
Unlike for artificial forcing, the energy and distribution of the SNe are
determined by observable physical parameters. This approach, which was first
pursued by \citet{1999ApJ...514L..99K}, marks an important step towards the
understanding of field amplification in spiral galaxies based on first
principles.

With respect to the more general morphological features (i.e., volume filling
factors, mass fractions, velocity dispersions, energy spectra, and structure
function scalings) our results agree reasonably well with similar findings by
\citeauthor{2006ApJ...653.1266J} and \citeauthor{2004Ap&SS.292..207D}. If
details are concerned, there is also a certain amount of tension between the
models. The reason for this lies in the choice of input parameters, as well as
in differences in numerical resolution. As has been demonstrated by
\citet{2004A&A...425..899D}, the demands for modelling a numerically converged
cold ISM phase are tremendous.

The focus of the presented thesis lies in the investigation of the generation
and evolution of galactic magnetic fields. Therefore, it becomes mandatory to
include differential rotation -- an effect which is not taken into account in
the models of \citeauthor{2004Ap&SS.292..207D} and
\citeauthor{2006ApJ...653.1266J}. From a technical point of view, the
inclusion of shear means a major complication in connection with adaptive mesh
techniques. The aforementioned models, however, draw their particular strength
from this strategy, which allows to follow the evolution of the turbulent ISM
to greater galactic heights without sacrificing numerical resolution near the
midplane. In comparison, our simulations come up with grid spacings coarser by
a factor of eight, and a vertical box size of $4\kpc$ as compared to $20\kpc$.
Despite the smaller vertical extent of our model, we demonstrate that we
capture the fundamental process of a disk-halo circulation. By means of a
fiducial run at half the grid resolution (compared to our standard model) we,
furthermore, provide evidence that the findings relevant for dynamo excitation
are reasonably converged and robust with respect to the numerical
approximation.

In addition to the general, ISM-related findings, we show that, in the case of
differential rotation, the turbulence created by SNe does, in fact,
exponentially amplify the mean magnetic field
\citepalias{2008A&A...486L..35G}. The timescale of the associated mechanism is
on the order of hundred million years and, thus, can easily explain the
observed strong magnetic fields in young galaxies \citep{2008Natur.454..302B}.
The very possibility of such a dynamo has long been disputed, specifically,
after semi-analytical models predicted prohibitively high values for the
related diamagnetic transport. We highlight the importance of rotation in the
generation of helicity by showing that a similar mechanism, based on Cartesian
shear alone, does not lead to an amplification of the mean field. This finding
impressively confirms the classical picture of a dynamo based on cyclonic
turbulence \citep*{1966ZNatA..21..369S,1971ApJ...163..255P}.

The symmetry of the mean magnetic field with respect to the midplane is found
to be quadrupolar. This predominant mode is interleaved with field reversals
of dipolar symmetry. The observation of this distinct oscillating behaviour
required the simulation of several hundred turnover times and could only be
achieved on massively parallel computers. Observationally, the quadrupolar
mode is favoured \citep[cf. Sec.~8.1 in][]{1996ARA&A..34..155B}, and suitable
seed fields can be provided by MRI \citep{2004A&A...424..565K}. Although the
reversal phenomenon will, of course, never be observed directly, this
unexpected finding, nevertheless, poses an interesting result for dynamo
theory in general.

The numerical representation \citep{2004ApJ...605L..33H} of the cosmic ray
driven dynamo proposed by \citet{1992ApJ...401..137P} depends critically on a
finite level of the microscopic diffusivity $\eta$. In contrast, we
demonstrate that the $\alpha$~effect resulting from the direct driving via SNe
persists at moderately high magnetic Reynolds numbers -- a finding which
strongly supports the relevance of the effect for the real ISM. Coinciding
with the value predicted by \citet*{1994A&A...286...72S}, we find a lower
threshold of $\Omega \simeq 25\kms \kpc^{-1}$ for magnetic field
amplification.  Furthermore, the observational correlation of field regularity
with star formation activity \citep{2008A&A...482..755C} is approximately
recovered within our simulations. The same holds for the observed high pitch
angles of up to $35\degr$, with the restriction that the dominating region of
the field only exhibits values of $10\degr$. Because it is not obvious which
region of the disk actually contributes to the observed angles, this does,
however, not necessarily imply a discrepancy. Finally, the influence of
thermal instability as well as the strength of the external potential are
found to be of minor importance for the overall field amplification.

To aid the elementary understanding of the underlying dynamo mechanism, we
interpret our results in the framework of mean-field magnetohydrodynamics,
i.e., we determine dynamo tensors $\talph$ and $\teta$, which parameterise the
effects of turbulence in the large-scale induction equation in terms of the
mean-field and its gradient, respectively.

The dynamo profiles $\alpha_{RR}(z)$ and $\alpha_{\phi\phi}(z)$ obtained with
this method agree well with the theoretically predicted $\alpha$~effect
\citep{1993A&A...269..581R}. As expected for the case of solid body rotation,
both coefficients have approximately the same amplitude and show a positive
(negative) sign in the top (bottom) half of the simulation box. From a
comparison with the SOCA prediction for stratified, rotating turbulence, we
infer a correlation time $\tau_{\rm c}\simeq3.5\Myr$, which is about a factor
of three lower than has been commonly assumed. Moreover, we find even shorter
coherence times for the models that neglect the clustering of type~II SNe.
This trend is consistent with the assumption of a higher level of coherency
for correlated super bubbles. At the current level of approximation, we apply
a prescribed clustering algorithm as a proxy for the distribution of stars
within OB associations. Hence, we conjecture that a higher value for the
coherence time might be obtained if a more accurate representation of the
clustering is achieved. It shall be noted that, within the paradigm of
mean-field modelling, the coherence time is a free parameter and little is
known about its true value within the interstellar medium.

The inferred coherence time $\tau_{\rm c}$ roughly corresponds to the
classical evolution time of a single remnant, which is defined by the
condition that the expansion velocity of the shell reaches the level of the
velocity dispersion within the ambient medium. This should, however, not be
taken too literally since the picture of a supernova remnant as a well defined
entity is certainly misleading in the context of the heterogeneous ISM. From
the computed structure functions, we estimate a correlation length of $l_{\rm
  c}=80$--$100\pc$. This agrees well with the findings of
\citet{2007ApJ...665L..35D}, who measure this quantity by means of two-point
correlation functions. While the inferred value of $l_{\rm c}$ is similar to
the asymptotic diameter of a single remnant near the midplane, it is
considerably smaller than the commonly assumed ``average size'' of evolved
super bubbles.

With these two characteristic flow parameters, we can compute the
dimensionless Strouhal number $\St = \tau_{\rm c}\,u'\,l_{\rm c}^{-1}$, for
which we obtain values of $0.1$, $0.5$, $0.7$, and $2.3$ for the cold,
unstable, warm, and hot phase, respectively. As we recall, the limit of small
Strouhal numbers, $\St\ll 1$, was a formal requirement of the SOCA
approximation. While the stated numbers are certainly not ``small'', they are
of the order of unity -- which implies that SOCA expressions can be regarded
as reliable order of magnitude estimates. This result is of considerable
importance since general numerical simulations are inherently limited to the
study of magnetic Prandtl numbers close to unity. To obtain robust scaling
relations with respect to $\Pm$, SOCA theory (and closures based on higher
order moments) will remain indispensable.

The dynamo profiles obtained for the case of differential rotation can, in
principle, be directly compared with the results based on the
uncorrelated-ensemble approach \citep{1998A&A...335..488F}. In general, we
measure similar amplitudes, but find the $\alpha$~profiles to be contained
within a smaller vertical range than predicted by this model. For the
turbulent diffusivity we find a similar trend. As has been discussed, a
dependence of our results on the box size cannot yet be excluded. We,
nevertheless, presume this disagreement to be rooted in the assumption of
``coherently expanding'' super bubbles, as opposed to the chaotic turbulence
represented in our direct simulations.

Concerning the diamagnetic pumping term $\gamma_z$, we encounter even more
severe discrepancies between the ``uncorrelated ensemble'' and our
simulations. In accordance with SOCA theory, which predicts that this term has
its origin in the spatial variation of the turbulence intensity, we find
negative (positive) values for $\gamma_z$ in the top (bottom) half of our
simulation box. In contrast, \citeauthor{1998A&A...335..488F} arrives at an
escape velocity $V_{\rm esc}>0$. While this is indeed correct for a single
explosion, this property, obviously, does not carry over to the ensemble,
which evidently shows that the interactions between the remnants must not be
ignored.

Within the scope of this thesis, we restrict ourselves to the unsaturated
regime of the dynamo mechanism. The very particular nature of the obtained
vertical transport processes, however, suggests an interesting quenching
scenario which shall be briefly discussed in the following: In the classical
notion, the dynamo is saturated via an inverse dependence of the diagonal
elements $\alpha_{RR}$ and $\alpha_{\phi\phi}$ on the magnetic field strength
approaching its equipartition value. A characteristic property of this
so-called $\alpha$~quenching is the suppression of the pitch angle in the
quenched regime. This occurs because the differential rotation $q\Omega$,
which is unaffected by the quenching, becomes the dominant effect.

If, on the other hand, the saturation process is controlled by the quenching
of the diamagnetic advection, the dynamo could be brought into saturation
without destroying the pitch angle. The strong inward transport which we
observe in our simulations might be opposed by a buoyant pumping\footnote{%
  This term, similarly to $\gamma_z$, describes an effective turbulent
  transport, and is thus rather distinct from the normal buoyancy due to the
  magnetic pressure.} %
\citep{1992A&A...260..494K}. For weak fields, this term scales with the square
of the Alfv{\'e}n velocity $v_{\rm A}$ and is directed towards lower
densities, i.e., upward. This, in turn, implies that the effective inward
pumping may be reduced at increasing field strength. Because the role of the
galactic wind will be reinforced if the turbulent transport is attenuated, the
dynamo may already saturate before the actual $\alpha$~quenching comes into
play. With an unquenched $\alpha$~effect, it becomes possible to explain the
high observed pitch angles at equipartition field strength. Preliminary
mean-field models (Elstner, private communication) suggest that such a ``wind
quenching'' can indeed produce higher pitch angles in the saturated regime.

Beyond this quenching scenario, we want to shortly address an ongoing
discussion on the efficiency of mean-field dynamos: Based on very general
considerations, \citet{1992ApJ...393..165V} have raised serious criticism
against dynamo theory in the limit of high magnetic Reynolds numbers $\Rm$.
\citeauthor{1992ApJ...393..165V} argue that, because magnetic helicity is
ideally conserved in this case, the creation of large-scale helicity by a
dynamo process has to be accompanied by an accumulation of small-scale
helicity of the opposite sign.  Consequently, the large-scale dynamo is
catastrophically quenched, where the term ``catastrophic'' refers to the fact
that the quenching function is proportional to $\Rm$. With an estimated
$\Rm=10^{18}$ for the interstellar medium, this quenching would indeed be
catastrophic. With an effective magnetic Reynolds number of $\Rm=10^4$, the
effect is most likely much less pronounced in our simulations -- which has to
be kept in mind, when interpreting the results.

A possible solution to the described issue may be given by the existence of
sufficiently strong helicity fluxes, as discussed in
\citet*{2007MNRAS.377..874S}. If the dynamo-active region can be efficiently
cleaned from the small-scale helicity, the quenching effect might be
annihilated. In their zero-dimensional mean-field model, the authors, in
addition to the mean induction equation, dynamically evolve a magnetic
contribution $\alpha_m$ to the dynamo effect. This term is based on the
current-helicity of the flow, and is supposed to represent the quenching due
to the small-scale fields. The related non-linear system shows an interesting
behaviour: A non-vanishing amplitude of the saturated field can only be
obtained under the contribution of an outward mean velocity. Such a wind will,
however, also remove the created mean magnetic field.

Alternatively, the authors consider a contribution of the helicity flux
discovered by \citet{2001ApJ...550..752V}, which similarly improves the
operability of the dynamo. Yet, the models do not yield saturation levels of
the field that are comparable with observations. As a complement to these
studies, we suggest that the vertical transport which we observe in our direct
simulations might naturally improve the situation: On one hand, the strong
wind efficiently removes the small-scale helicity while, on the other hand,
the mean field is subject only to the, much lower, residual velocity given by
the additional inward pumping. With an approximate balance of these two very
distinct transport processes, the dynamo operates in an optimal state, while
the catastrophic quenching is inherently circumvented. Although our models
are currently only representative of the regime of moderate $\Rm$ (where the
quenching is probably still weak), the described scenario might already be
embodied in our simulations -- a hypothesis which is further supported by the
sustained high growth rates for low values of the microscopic diffusivity
$\eta$. First preliminary simulations, moreover, show that our current setup
is indeed capable of obtaining equipartition field strengths -- which strongly
suggests that catastrophic quenching is not in effect.

Having demonstrated that direct driving via supernovae is a powerful mechanism
for amplifying galactic magnetic fields, the current work opens new
perspectives for global mean-field models. With simulation-based dynamo
parameters replacing analytical derivations, which were primarily based on
intuition, the mean-field approach can finally be put on a solid foundation.
As many dependencies remain to be explored, the current simulations can only
be regarded as a starting point for more comprehensive parameter studies.
Moreover, the highly important aspect of quenching has completely been
neglected within the current thesis. Ultimately, future simulations at
enhanced numerical resolution will have to show how the supernova driven
turbulence is affected in the regime of saturated fields.



\appendix

\cleardoublepage
\chapter{Code Validation}
\label{ch:validation}

\section{Flux-matching at sheared interfaces} 

To account for the large-scale galactic shear, our model makes use of the
so-called shearing box formalism, where the radial boundary conditions for the
fluid variables $\rho$, $\mathbf{m}$, $\epsilon$, and $e$, i.e., the mass-,
momentum-, thermal energy-, and total energy-density, respectively, take a
time-dependent shifted-periodic form:
\begin{eqnarray}
  f(x,y,z) & \mapsto & f(\tilde{x},\tilde{y},z), \qquad\qquad\qquad
  f \in \left\lbrace \rho, m_x, m_z, \epsilon \right\rbrace \nonumber\\
  m_y(x,y,z) & \mapsto & m_y(\tilde{x},\tilde{y},z)\, \mp\rho\,w, \nonumber\\
  e(x,y,z) & \mapsto & e(\tilde{x},\tilde{y},z)\,
                       \mp m_y\,w + \nicefrac{1}{2}\,\rho\,w^2,
\end{eqnarray}
where $w=q\Omega L_x$, $\tilde{x} = x \pm L_x$, and $\tilde{y} = y\mp w t$.
Since the $y$ coordinate of above mappings varies continously in time, there
is some kind of interpolation necessary to map the ghost-zone values onto the
finite grid. For our implementation, the same piecewise linear reconstruction
is used as for the numerical scheme.

\subsection{The conservation of hydrodynamic fluxes} 

As the numerical fluxes are nonlinear functions of the primitive variables,
any form of interpolation will lead to a small inconsistency. We avoid this by
matching the computed $x$ fluxes at the sheared domain boundaries.  It is
straightforward to map fluxes not containing $m_y$. For the remaining
quantities, we derived the following expressions for the numerical fluxes
\citep{2007CoPhC.176..652G}:
\begin{eqnarray}\label{eq:flux_map}
  F^x_{i\phlf,j,k}(m_y) &\, = \,& \widehat{F}^x_{\itl\phlf,\jtl,k}(m_y)
      \;\mp\;\widehat{F}^x_{\itl\phlf,\jtl,k}(\rho)\, w\,,      \nonumber\\
  F^x_{i\phlf,j,k}(e) &\, = \,&\widehat{F}^x_{\itl\phlf,\jtl,k}(e) 
      \;\mp\;\widehat{F}^x_{\itl\phlf,\jtl,k}(m_y)\, w
      \;+\;\nicefrac{1}{2}\,\widehat{F}^x_{\itl\phlf,\jtl,k}(\rho)\,w^2\,,
\end{eqnarray}
where the hat stands for the piecewise linear interpolation procedure used, and
tilde marks the corresponding indices of the zones to map from.

\subsection{The conservation of magnetic fluxes} 

Applying Gau{\ss}' theorem to the integral form of the induction equation, one
can show that the azimuthal field, in the case of ideal MHD, grows linearly
with the net radial magnetic flux:
\begin{equation}
  \partial_t \left<{\mathbf B}\right> =
  -\frac{w}{V}\; \yy \int_{\partial X}{\mathrm{dy\,dz} B_x}\,,
\end{equation}
that is, for zero net radial field, the mean magnetic flux through the
shearing box is conserved. Our implementation satisfies this condition to
machine accuracy for the $x$ and $z$ component of the magnetic field and to
truncation error for the $y$ component. To achieve this, we apply additional
boundary conditions to the electric field fluxes. This is necessary because
the velocity offset $w$ enters the fluxes via the electromotive force:
\begin{equation}
  {\mathbf E}(x,y,z) \mapsto {\mathbf E}(\tilde{x})\,
  \pm w\;\hat{{\mathbf y}}\times {\mathbf B} \label{eq:map_E}\,.
\end{equation}
Similar to the above expressions for the hydrodynamic fluxes, we derived
mappings for the electromotive forces \citep{2007CoPhC.176..652G}:
\begin{eqnarray}
  {\mathbf G}^x_{i\phlf,j,k} & = & 
    \widehat{\mathbf G}^x_{\itl\phlf,\jtl,k} \;\mp\; 
    w\;\hat{{\mathbf y}} B_{x\vert\itl\phlf,\jtl,k}\,, \nonumber\\
  {\mathbf G}^z_{i,j,k\phlf} & = & 
    \widehat{\mathbf G}^z_{\itl,\jtl,k\phlf} \;\mp\; 
    w\;\hat{{\mathbf y}} B_{z\vert\itl,\jtl,k\phlf}\,, \nonumber\\
  {\mathbf G}^y_{i,j\phlf,k} & = & 
    \widehat{\mathbf G}^y_{\itl,\jtl\phlf,k} \pm w\;\frac{
      b^{+}(B_x \xx + B_z \hat{\mathbf z})^N_{\itl,\jtl,k}
      -b^{-}(B_x \xx + B_z \hat{\mathbf z})^S_{\itl,\jtl,k}
    }{b^{+}-b^{-}}\,,
\end{eqnarray}
with the same notation as in Equation~(\ref{eq:flux_map}) above. In addition,
$b^{+}$ ($b^{-}$) is the the maximal (minimal) wave-propagation
direction-sensitive speed at $y_{j\phlf}$, and 'N' ('S') indicates piecewise
linear reconstruction at the northern (southern) cell interface.

\section{Momentum source terms} 

For our local Cartesian coordinate frame, we apply the so called Hill system.
This approximation is based on the local expansion of the equations of motion
resulting in a tidal force $2\rho\, q\Omega^2x\xx$. Together with the Coriolis
force $-2\rho\Omega \hat{\mathbf z} \times {\mathbf v}$, the momentum source
terms can be formally written as an effective Coriolis force $-2\rho\Omega
\hat{\mathbf z} \times ( {\mathbf v} + q\Omega x \yy )$ acting on the
perturbed velocity $\delta v_y=v_y+q\Omega x$. Although this formulation would
in principle allow for an exact Coriolis update (via an analytic rotation of
the velocity vector) numerical experiments indicate that such an update is not
compatible with the multi-stage integration scheme. Thus, we decide to
implement the source terms unsplit, i.e., as explicit forces within the
Runge-Kutta time integration.

\citet{2005AIPC..784..475G} point out the importance of conserving the energy
contained in the epicyclic mode. This ideally conserved quantity can be
derived from the energy budget in the limit of inviscid flow and reads
\begin{equation}
  E_{\rm epi} = \nicefrac{1}{2}\,\rho \left\langle u_R \right\rangle^2
              + \nicefrac{1}{2}\,\rho \left\langle u_\phi\right\rangle^2\;
              \frac{(2\Omega)^2}{\kappa^2}\,,
\end{equation}
with $\kappa$ the epicyclic frequency. While this formally looks like a
kinetic energy, it also includes the potential energy with respect to the
epicyclic displacement. We have found that it is important to implement the
source terms in an unsplit fashion to avoid oscillations in this energy that
would otherwise arise from systematic splitting errors.

\begin{figure}
  \begin{minipage}[b]{0.64\columnwidth}
    \includegraphics[width=0.9\columnwidth]{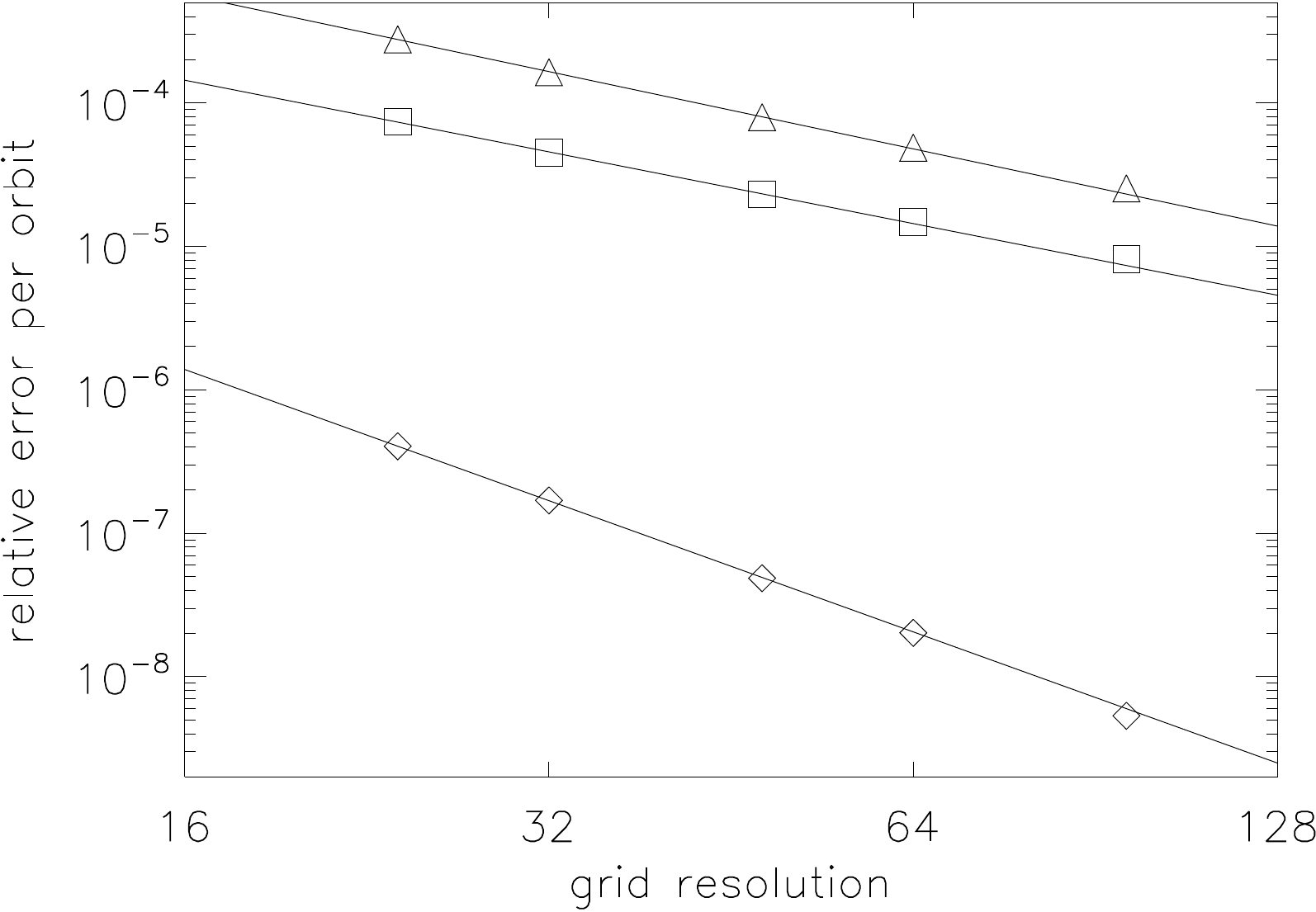}
  \end{minipage}
  \begin{minipage}[b]{0.35\columnwidth}
    \caption[Convergence of the relative errors in the epicyclic energy]{%
      Convergence of the relative errors in the epicyclic energy for A
      (triangles), B (squares), and the un\-split method (diamonds).
      Least-square fits with (logarithmic) slopes of -1.79, -1.66, and -3.04
      are indicated by the respective lines.\vspace{10pt}}
    \label{fig:epi}
  \end{minipage}
\end{figure}

From 2D shearing sheet simulations, the error in the epicyclic energy is found
to be independent of the mode amplitude. For constant excitation amplitude,
the error is growing linearly in time. The measured relative error per orbit
decreases with resolution, reflecting the third-order convergence of the
underlying time integration scheme. For comparison, we have also tested two
conventional (operator split) methods for the source terms: method (A)
forward-Euler integrates the Coriolis forces while method (B) treats the
Coriolis term via rotation of the momentum vector. By expanding the
trigonometric functions, one can show that method A gives a first order
approximation to method B. Both methods (in contrast to the unsplit one) lead
to oscillations, with frequency $2\Omega$, in the epicyclic energy.
Figure~\ref{fig:epi} compares the residual errors as a function of resolution,
clearly favouring the unsplit approach.

\section{Isolated remnants} 
\label{sec:isolated_SNR}
\begin{figure}
  \center\includegraphics[width=0.9\columnwidth]{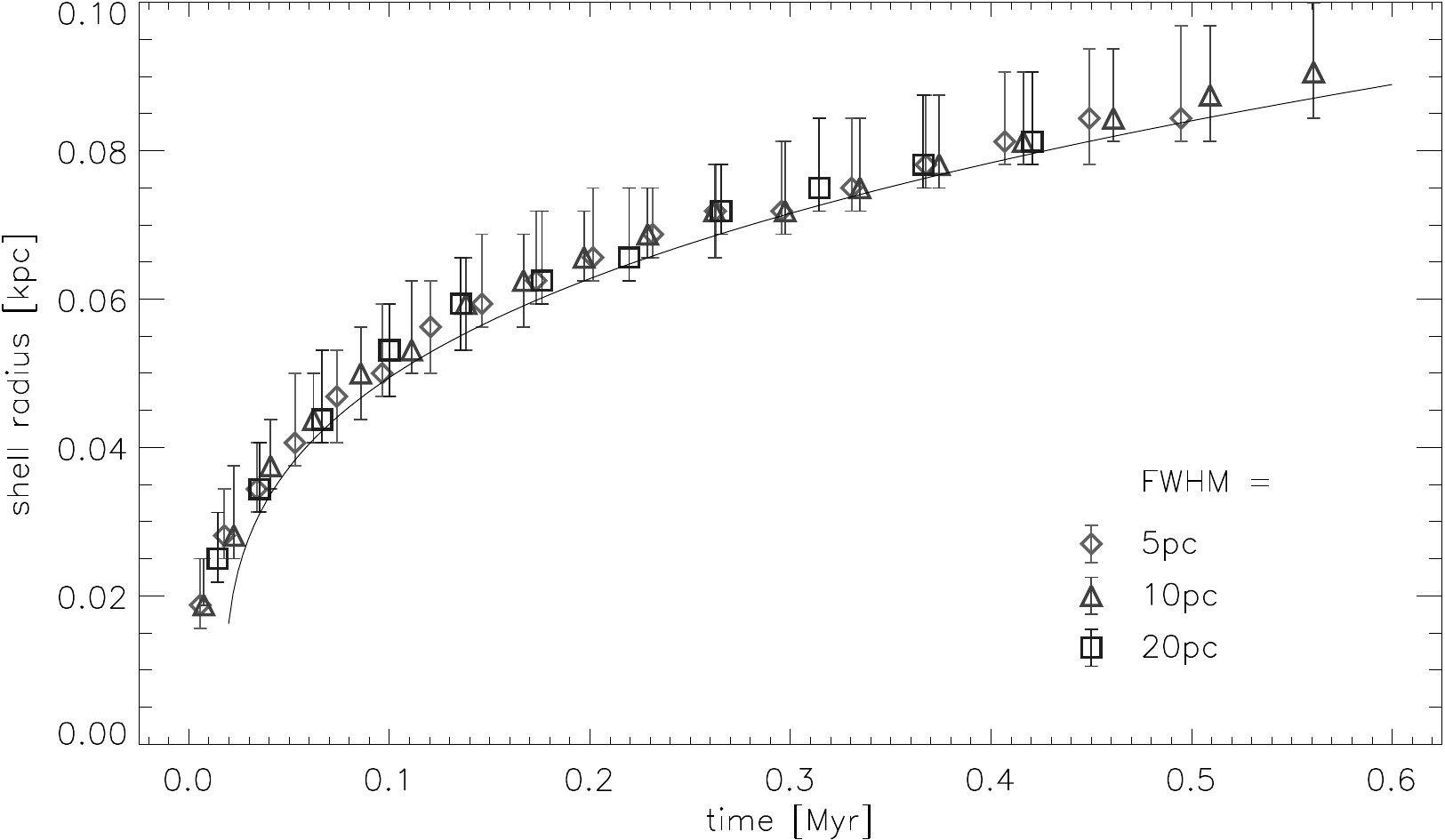}
  \caption[Parameter study for single SNR varying the kernel diameter]{%
    Parameter study for single SNR. Symbols mark the radius of the traced
    density peak for three different smoothing lengths of the initial energy
    profile. Error bars indicate the range where the density is above the
    ambient value. The solid line indicates the analytical solution by
    \citet*{1988ApJ...334..252C}.}
  \label{fig:fwhm}
  \vspace{10pt}
  \center\includegraphics[width=0.9\columnwidth]{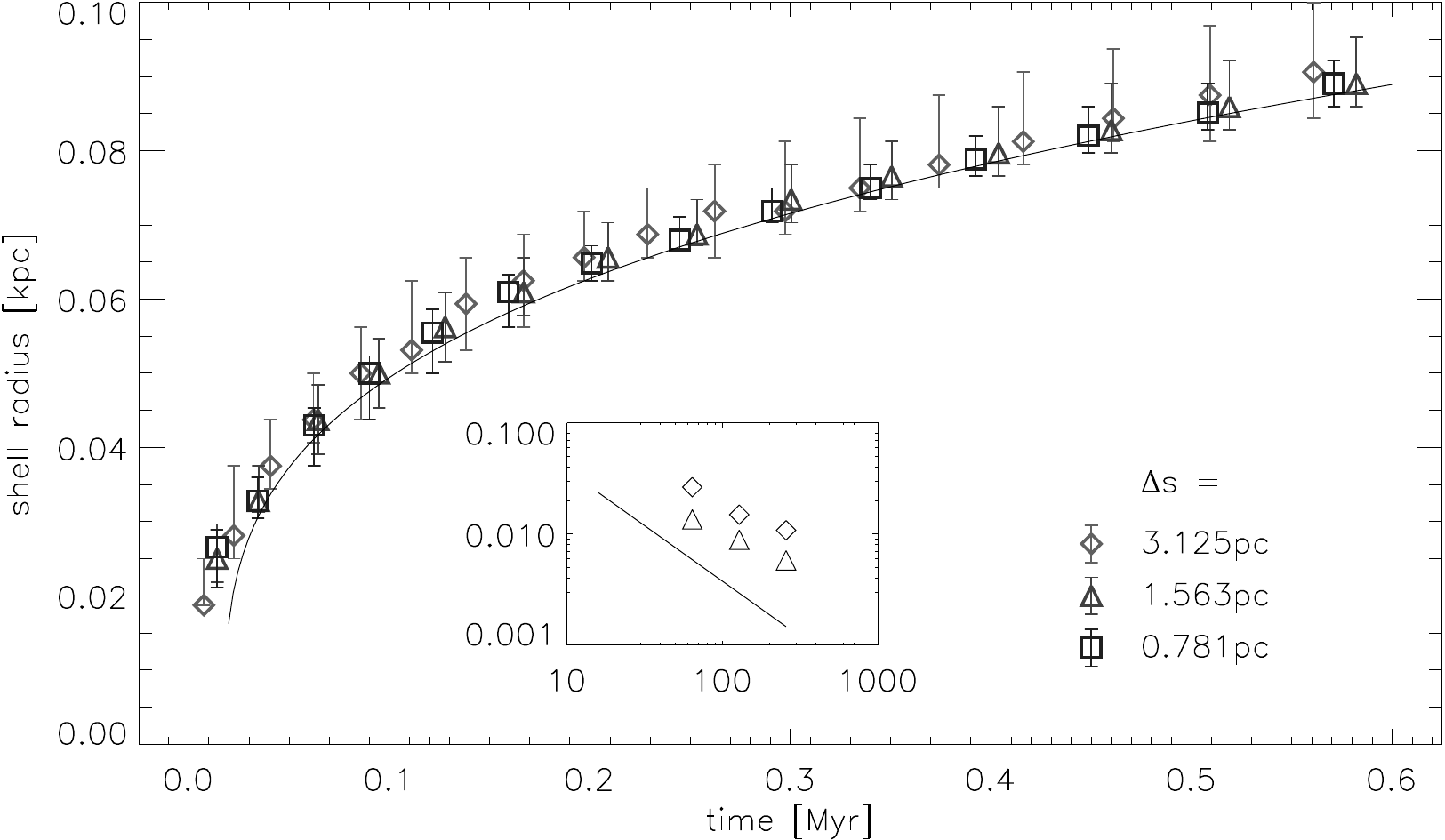}
  \caption[Convergence study for single SNR]{%
    Same as in figure~\ref{fig:fwhm}, but for three different grid spacings
    $\Delta s$, indicating convergence of the numerical solution (see inlay).}
  \label{fig:cioffi}
\end{figure}

In our ISM-model, turbulence is driven via supernova explosions, which are
modelled as local injections of thermal energy. The initial energy
distribution is smeared over three standard deviations of a Gaussian support.
To practically determine a feasible dimension for the kernel, we conducted
test simulations of single SNRs with initially Gaussian shape of varying width
as depicted in Figure~\ref{fig:fwhm}. Due to the self-similarity in the
adiabatic expansion phase, it is sufficient to apply a FWHM of $20\pc$ to
adequately represent the energy injection without excessively suppressing the
numerical time step. Numerical solutions with increasing spatial resolution
$\Delta s$ have been tested against the analytical description by
\citet*{1988ApJ...334..252C} as illustrated in Figure~\ref{fig:cioffi}. The
results agree well, and we see that convergence can be obtained for grid
spacings below $3.1\pc$. This corresponds to the findings reported in
\citet{2005ApJ...626..864M}.

\cleardoublepage
\chapter{The Mean-field Approach} 
\label{ch:test_fields}

\section{The test-field method} 

To avoid complications with the inversion of the tensorial
equation~(\ref{eq:param1}), we apply the test-field approach proposed by
\citet{2005AN....326..245S,2007GApFD.101...81S}. The method has also recently
been adopted to the shearing box case by \citet{2005AN....326..787B}. Earlier
approaches to the computation of dynamo coefficients from simulations
\citep{2002GApFD..96..319B,2005mpge.conf..171K} were based on least square fit
methods. The major drawback with these was that, in regions where $\mB$ or
$\nabla \mB$ vanishes, the inversion will become singular. One can circumvent
these difficulties by solving Equation~(\ref{eq:param1}) for properly defined
test-fields, i.e., fields with simple, well behaved geometry and gradients.
The price for this is that one has to evolve an extra (passive) induction
equation for each field. In reality we do not evolve the test-field
$\Tf_{(\nu)}$ itself but its associated fluctuations, i.e., we integrate
\begin{equation}
\begin{array}[pos]{rccl}
  \partial_t \Tf'_{(\nu)} & = & \nabla \times & \left[\,
    \U'\tms\mTf_{(\nu)} 
    + (\mU\!+\!q\Omega x\yy)\tms\Tf'_{(\nu)} \right. \\  & \, &
    & - \overline{\U'\tms\Tf'}_{(\nu)} + \left. \U'\tms\Tf'_{(\nu)} 
    - \eta\nabla\tms\Tf'_{(\nu)} \,\right] \,,\\
    \nabla\cdt\Tf'_{(\nu)} & = & 0\,, & \label{eq:testfields}
\end{array}
\end{equation}
with the velocity ${\mathbf u}$ taken from the direct simulations and for
constant $\mTf_{(\nu)}$, with $\nu=0\dots3$ for the case of $\bar{B}_z=0$ and
horizontal averages. Note that $\eta$ here denotes the physical, i.e.,
microscopic diffusivity.

We implemented these additional equations within NIRVANA employing the
constrained transport paradigm to exactly satisfy the solenoidal constraint.
The actual method uses up-winding to guarantee stability, while second order
in space is attained via piecewise linear reconstruction. For this, we apply
the same slope limiter as in the actual code. Our procedure is very similar to
the methods described in \citet*{2006JCoPh.218...44T}. The time integration is
finally performed with a second order Runge-Kutta method to minimise
synchronisation overhead compared to the full third order update of the MHD
equations. This is important if the number of test-fields is further
increased, especially, since the computation of $\mU$ and
$\overline{\U'\tms{\mathcal B}'}$ requires collective communication. The exact
discretisation of the system~(\ref{eq:testfields}) is described in detail in
section~\ref{sec:testfields}.
\begin{figure}
  \center\includegraphics[width=\columnwidth]{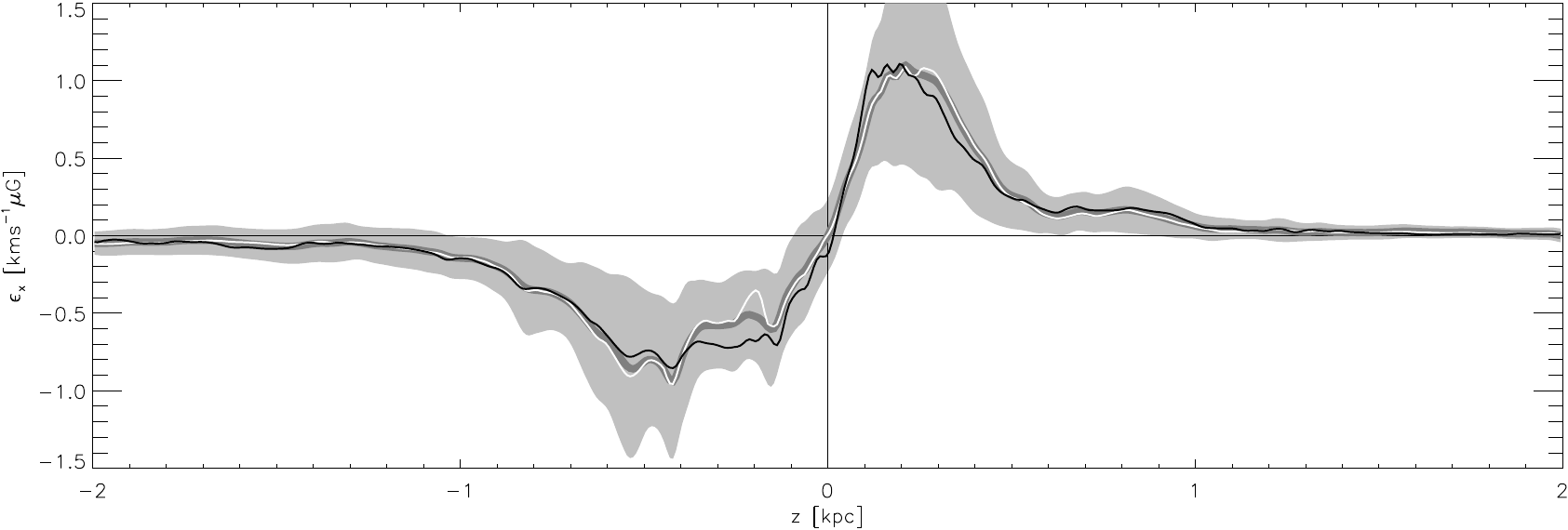}
  \caption[Electromotive force reconstructed from dynamo tensors]{%
    Radial component $\EMF_x(z)$ of the mean electromotive force. The thick
    gray line shows the time averaged value over a period of $50\Myr$, while
    the shaded area gives the rms~fluctuations. Thin lines are the mean EMFs
    as reconstructed from Equation~(\ref{eq:param1}) with dynamo parameters
    $(\talph,\teta)$ computed via the least square fit (white line) and
    test-field (black line) approach.}
  \label{fig:emf_test}
\end{figure}

For the particular choice of the four test-fields $\mTf_{(\nu)}$, we use the
ones from \citet{2005AN....326..787B}, which are
\begin{eqnarray}
  \bar{\mathcal B}_{(0)} = \cos(k_1 z)\,\xx\,, &\quad 
  \bar{\mathcal B}_{(1)} = \sin(k_1 z)\,\xx\,, &\nonumber\\
  \bar{\mathcal B}_{(2)} = \cos(k_1 z)\,\yy\,, &\quad
  \bar{\mathcal B}_{(3)} = \sin(k_1 z)\,\yy\,, &
\end{eqnarray}
with $k_1$=$\pi/L_z$ and $L_z$ the vertical extent of the box. For each of the
test-fields, we compute the corresponding mean electromotive force
\begin{equation}
  \bar{\EMF}^{(\nu)} = \overline{\U' \times \Tf'}\,.
\end{equation}
The dynamo coefficients can then be computed via Equation~(\ref{eq:param1}),
which, for the $x$ component, takes the form
\begin{eqnarray}
  \bar{\EMF}_x^{(0)}\Eq \alpha_{xx} c \Mn \eta_{xxz}\,k_1 s\,, & \ \ &
  \bar{\EMF}_x^{(1)}\Eq \alpha_{xx} s \Pl \eta_{xxz}\,k_1 c\,,\nonumber\\
  \bar{\EMF}_x^{(2)}\Eq \alpha_{xy} c \Mn \eta_{xyz}\,k_1 s\,, & \ \ &
  \bar{\EMF}_x^{(3)}\Eq \alpha_{xy} s \Pl \eta_{xyz}\,k_1 c\,,
\end{eqnarray}
with similar expressions for $\bar{\EMF}_y^{(\nu)}$ and 
$\bar{\EMF}_z^{(\nu)}$, and where we have abbreviated $\sin(k_1 z)$ and 
$\cos(k_1 z)$ with ``$s$'' and ``$c$'', respectively. By identifying the 
coordinates $x$, $y$, and $z$ with the indices 1, 2, and 3 the solution to
this set of equations can be compactly written as
\begin{equation}
  \left(\begin{array}{c}
      \alpha_{ij}\\
      k_1 \eta_{ij3}
    \end{array}\right) = 
  \left(\begin{array}{cc}
      \,c & s \\ \!-s & c
    \end{array}\right)
  \left(\begin{array}{c}
      \bar{\EMF}_i^{(2j\Mn2)} \\
      \bar{\EMF}_i^{(2j\Mn1)} 
    \end{array}\right)\,,\label{eq:tf_solution}
\end{equation}
with $i\in\{1,2,3\}$, and $j\in\{1,2\}$ in the case of $\bar{B}_z$=0. In
contrast to the least square fit method, Equation~(\ref{eq:tf_solution}) can
be directly computed for each $z$, yielding vertical profiles for the dynamo
parameters.

We have implemented and tested both the least square fit and the test-field
method. Figure~\ref{fig:emf_test} shows a comparison for the most basic check
one can apply, i.e., reconstructing the EMF from the computed coefficients
according to Equation~(\ref{eq:param1}). On can see that both methods perform
well on reproducing the simulated data. While for the fit method this seems
obvious, we want to point out that the test-field method in no way relies on
the actual magnetic field from the simulation (as long as there is no
significant quenching involved).

\section{Implementation details} 
\label{sec:testfields}

For the discretisation of the test-field induction equation, we follow the
concept of constrained transport (CT), in which the solenoidal constraint is
automatically satisfied to roundoff error. The guiding idea is to apply a
discrete version of Stokes' theorem. In the resulting finite-surface approach,
the magnetic field components $\fv{B}{x}$, $\fv{B}{y}$, and $\fv{B}{z}$ are
regarded as face averages and centred at the faces (of the corresponding
coordinate direction), while the electric fields $\fv{E}{x}$, $\fv{E}{y}$, and
$\fv{E}{z}$ are thought of as line integrals along the edges of a particular
face.

\subsection{Discretisation} 

We follow standard notation and index the centre of a cubic grid cell with
$(i,j,k)$. If we combine all the terms of the rhs of
Equation~(\ref{eq:testfields}) into one electromotive force
$\EMF$=$\U\tms\Tf'-\overline{\U'\tms\Tf'}+\U\tms\mTf-\eta\nabla\tms\Tf'$, the
basic update from time $t^n$ to time $t^{n+1}$ of the induction equation
$\partial_t \Tf'$=$\nabla\tms\EMF$ reads:
\begin{eqnarray}
\fv{B}{x}^{n+1}_{i\mhlf,j,k} = & \fv{B}{x}^n_{i\mhlf,j,k} &
  + \frac{\delta t}{\delta y} \left( 
    \fv{E}{z}^n_{i\mhlf,j\phlf,k} - \fv{E}{z}^n_{i\mhlf,j\mhlf,k} \right) 
    \nonumber \\ &&
  - \frac{\delta t}{\delta z} \left( 
    \fv{E}{y}^n_{i\mhlf,j,k\phlf} - \fv{E}{y}^n_{i\mhlf,j,k\mhlf} \right)\,,\\
\fv{B}{y}^{n+1}_{i,j\mhlf,k} = & \fv{B}{y}^n_{i,j\mhlf,k} &
  + \frac{\delta t}{\delta z} \left( 
    \fv{E}{x}^n_{i,j\mhlf,k\phlf} - \fv{E}{x}^n_{i,j\mhlf,k\mhlf} \right)
    \nonumber \\ &&
  - \frac{\delta t}{\delta x} \left( 
    \fv{E}{z}^n_{i\phlf,j\mhlf,k} - \fv{E}{z}^n_{i\mhlf,j\mhlf,k} \right)\,,\\
\fv{B}{z}^{n+1}_{i,j,k\mhlf} = & \fv{B}{z}^n_{i,j,k\mhlf} &
  + \frac{\delta t}{\delta x} \left( 
    \fv{E}{y}^n_{i\phlf,j,k\mhlf} - \fv{E}{y}^n_{i\mhlf,j,k\mhlf} \right)
    \nonumber \\ &&
  - \frac{\delta t}{\delta y} \left( 
    \fv{E}{x}^n_{i,j\phlf,k\mhlf} - \fv{E}{x}^n_{i,j\mhlf,k\mhlf} \right)\,.
\end{eqnarray}
The next task in constructing a numerical scheme is to specify how the
edge-centred electric fields are obtained: $-\overline{\U'\tms\Tf'}$ and
$\U\tms\mTf$ can be regarded as source terms and are discretised in a
straightforward fashion. The treatment of the diffusive part
$-\eta\nabla\tms\Tf'$ has been closely adopted from the original NIRVANA code.
The crucial term here is $\U\tms\Tf'$, describing the advection of $\Tf'$ with
the given flow $\U$. It can be shown that simple central averages of the form
\begin{eqnarray}
\fv{E}{z}_{i\mhlf,j\mhlf,k} & = &
  + \hlf u_x \left( \fv{B}{y}_{i\mhlf,j,k} + \fv{B}{y}_{i\mhlf,j-1,k} \right)
    \nonumber \\ &&
  - \hlf u_y \left( \fv{B}{x}_{i,j\mhlf,k} + \fv{B}{x}_{i-1,j\mhlf,k} \right)
\end{eqnarray}
would lead to an unconditionally unstable numerical scheme. This has also been
pointed out by \citet*{2006JCoPh.218...44T}. A strict proof for a
corresponding advection type equation can be found in
\citet{1988nuco.book.....H}. To obtain a numerically stable solution, we
decide to use the upwind states of the magnetic field components, i.e.,
\begin{eqnarray}
\fv{E}{x}_{i,j\mhlf,k\mhlf} & = & + u_y \begin{cases}
   \fv{B}{z}^{\rm S}_{i,j,k\mhlf}   & (u_y \le 0) \\
   \fv{B}{z}^{\rm N}_{i,j-1,k\mhlf} & (u_y > 0)\end{cases}
    \nonumber \\ &&
  - u_z  \begin{cases}
   \fv{B}{y}^{\rm B}_{i,j\mhlf,k}   & (u_z \le 0) \\
   \fv{B}{y}^{\rm T}_{i,j\mhlf,k-1} & (u_z > 0) \end{cases}\,,
\end{eqnarray}
\begin{eqnarray}
\fv{E}{y}_{i\mhlf,j,k\mhlf} & = & + u_z \begin{cases}
   \fv{B}{x}^{\rm B}_{i\mhlf,j,k}   & (u_z \le 0) \\
   \fv{B}{x}^{\rm T}_{i\mhlf,j,k-1} & (u_z > 0) \end{cases}
    \nonumber \\ &&
  - u_x  \begin{cases}
   \fv{B}{z}^{\rm W}_{i,j,k\mhlf}   & (u_x \le 0) \\
   \fv{B}{z}^{\rm E}_{i-1,j,k\mhlf} & (u_x > 0) \end{cases}\,,
\end{eqnarray}
\begin{eqnarray}
\fv{E}{z}_{i\mhlf,j\mhlf,k} & = & + u_x \begin{cases}
   \fv{B}{y}^{\rm W}_{i,j\mhlf,k}   & (u_x \le 0) \\
   \fv{B}{y}^{\rm E}_{i-1,j\mhlf,k} & (u_x > 0) \end{cases}
    \nonumber \\ &&
  - u_y  \begin{cases}
   \fv{B}{x}^{\rm S}_{i\mhlf,j,k}   & (u_y \le 0) \\
   \fv{B}{x}^{\rm N}_{i\mhlf,j-1,k} & (u_y > 0) \end{cases}\,,
\end{eqnarray}
instead. This comes at the price of reducing the spatial order of the scheme.
Second order can, however, be recovered by piecewise linear reconstruction of
the magnetic field components. This is indicated by the labels W, E, S, N, T,
and B, denoting reconstruction towards the west/east, south/north, and
top/bottom direction. Non-oscillatory behaviour is achieved by the use of a
slope limiter. For consistency, we chose the same limiter that is applied in
the general code \citep{2004JCoPh.196..393Z}. Finally, the time integration is
performed via a second-order Runge-Kutta scheme:
\begin{eqnarray}
  \Tf'^{(*)} & = & \Tf'^{(n)} 
      + \delta t\ \nabla\times\EMF(\Tf'^{(n)},\U^{(n)},\mTf)\,,\nonumber\\
  \Tf'^{(n+1)} & = & \frac{1}{2} \left[ \Tf'^{(n)} +\Tf'^{(*)} 
      + \delta t\ \nabla\times\EMF(\Tf'^{(*)},\U^{(n+1)},\mTf)\right]\,,
\end{eqnarray}
where we use the time step $\delta t$ computed from the (more restrictive)
Courant condition of the actual MHD-scheme. The test-fields are not updated
with the full third order Runge-Kutta method to minimise overhead. This is
important because the averaging procedure includes additional collective
MPI-communication. Application of boundary conditions, synchronisation, and
(in the case of AMR) mesh fix-up, however, are still aligned with the main
scheme since the predictor step is the same in both cases.

\subsection{Improvements} 
\label{sec:noise}

Within the test-field concept, the electromotive force $\mn{\U'\tms\Tf'}$,
related to the tracer fields, is supposed to pick up the effects of the
turbulent flow on the magnetic field. This tracing of the turbulence is
supposed to be sensitive for fluctuations over timescales comparable to the
turbulent turnover time. This is also consistent with the picture of scale
separation, i.e., slowly varying mean-fields and uncorrelated fluctuations on
intermediate time scales.

Since our models are evolved over time scales that are long compared to
$\tau_{\rm c}$, the fluctuations $\Tf'$ will, however, keep their memory for
much longer than is desired. As we can see from Equation~(\ref{eq:ind_Btur}),
the ``fluctuation'' of the EMF term, i.e. $\U'\tms\Tf'-\mn{\U'\tms\Tf'}$,
enters the induction equation for $\Tf'$ as a source term. These high order
correlations, in fact, introduce a considerable level of noise into the method
and drastically degrade the signal to noise ratio of the derived coefficients.
Matters become worse because there is no efficient damping term in the
evolution equation for $\Tf'$. Whereas turbulent diffusion per definition only
affects the mean-field, the microscopic diffusivity $\eta$ entering
Equation~(\ref{eq:ind_Btur}) results in a damping time scale of
$\tau_{\eta}=L^2/\eta\simeq 1.5\Gyr$ for $L=100\pc$.

To avoid the effects of the accumulation of long term fluctuations, we
periodically reset the test-field fluctuations $\Tf'$ to zero. This
(artificially) introduces a finite temporal domain of dependence, which is in
accordance with the assumptions of the mean-field concept and drastically
improves the signal to noise ratio of the measured dynamo coefficients. In an
alternative approach, we suppose an artificial damping time scale by
introducing an additional sink term of the form $-\Tf'/\tau_{\rm art}$. For
sufficiently low $\tau_{\rm art}\simgt 4\tau_{\rm c}$, this approach will also
result in a diminished level of noise.

As a note of caution, we do not want to conceal that one has to be careful
about choosing the periodic ``refresh rate'' or the artificial damping time
scale, respectively, as too low values will markedly decrease the amplitude of
the measured coefficients. However, since $\Tf'$ enters the EMF linearly, all
coefficients will be affected in the same manner and one could in principle
correct for this attenuation of the signal. Albeit the suggested ``cleaning
procedures'' might seem rather ad-hoc at first glance, we have to recall that
we can always check the consistency of the derived parameters via the
reconstruction of the mean EMF. Since we independently compute
$\mn{\U'\tms\B'}$ from the actual fields $\B'$, the mean-field coefficients
derived by means of the test-field approach have to fulfil $\EMF(z,t)=
\talph\,\mB(z,t)+\teta\,\nabla\tms\mB(z,t)$, with the mean electromotive force
$\EMF(z,t)$ and the mean-field $\mB(z,t)$ taken from the direct simulations.

\section{One-dimensional toy model} 
\label{sec:toy}

Based on the coefficients obtained from the tracer fields, we want to explore
more closely the effects due to the various tensor elements. If one can
reproduce the results of the direct simulations in the mean-field approach,
this drastically aids the understanding of the underlying fundamental
mechanisms. For this purpose, we have implemented a simple one-dimensional
``toy'' model representing the 1D mean-field induction equation. The basic
design idea was to stay as close as possible to the actual data from the
simulations. Like in the direct simulations, we neglect vertical fields and a
possible contribution of the term $\alpha_{zz}$. Because of these
restrictions, the resulting model can in no way claim to be representative of
the general case. In principle, there is, however, no objection against
performing full 3D mean-field models based on the obtained coefficients.

\subsection{Model profiles} 
\label{sec:toy_prof}

\begin{figure}
  \center\includegraphics[width=\columnwidth]{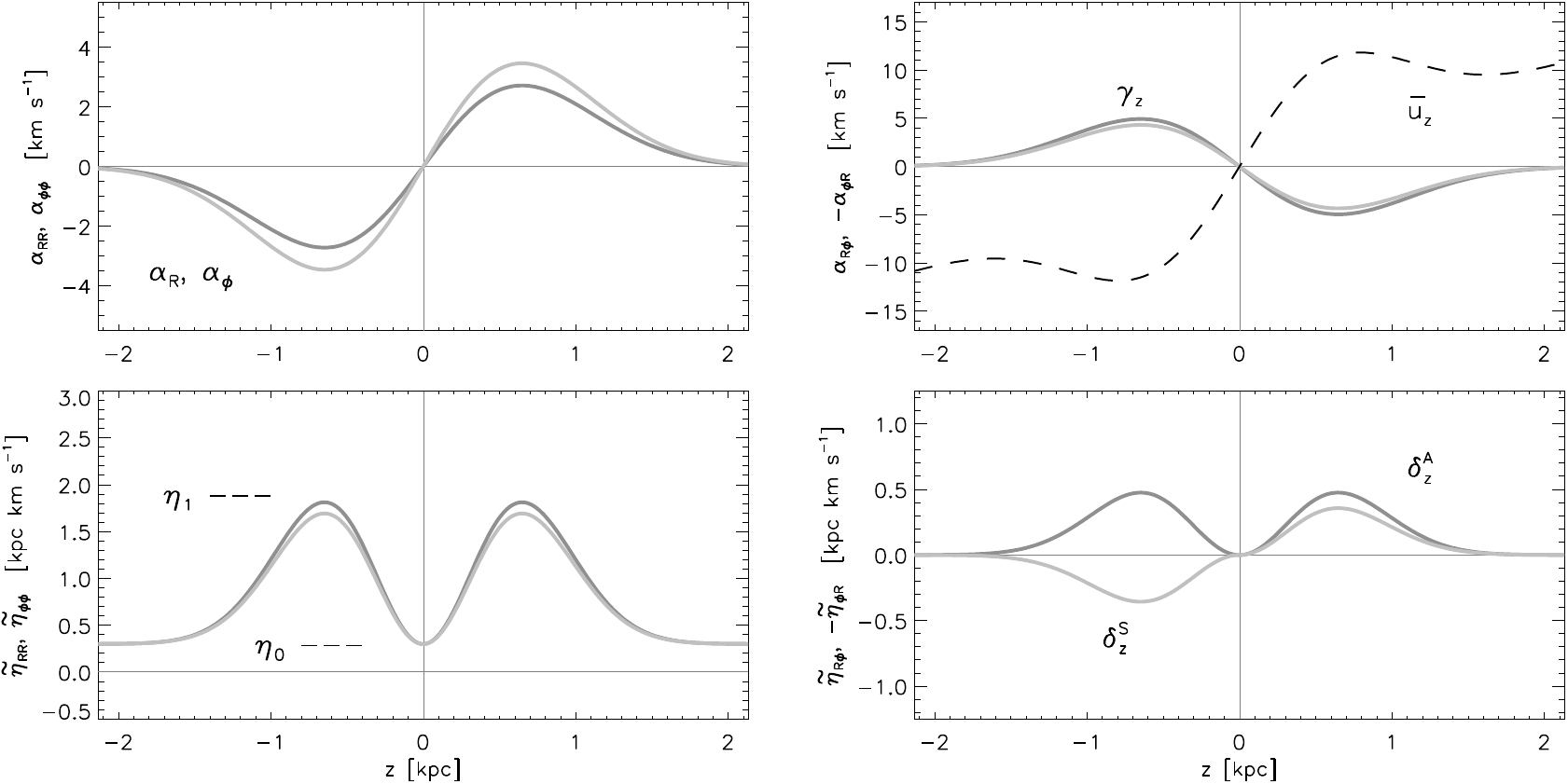}
  \caption[Vertical profiles for the 1D toy model]{%
    Assumed vertical profiles of the dynamo coefficients for the 1D toy model.
    The labels indicate the free parameters of the model related to the
    profiles: radial and azimuthal $\alpha$~effect, vertical pumping and wind,
    lower and upper bounds for the diffusivity $\eta_{\rm t}$, and
    (anti-)symmetric $\delta$-effect, respectively. Additional free parameters
    include the rotation rate $\Omega$ and the shear rate $q$.}
  \label{fig:dyna}
\end{figure}

It has been found that already the particular shape of the $\alpha$~profiles
can have profound implications on the excited dynamo modes \citep[see
e.g.][for a drastic case]{2005AN....326..693G}. We, therefore, try to mimic
the observed profiles as closely as possible. Figure~\ref{fig:dyna} depicts a
schematic representation of the assumed model profiles along with the
corresponding amplitudes which enter the model as free parameters. The
diagonal elements of the $\talph$ and $\teta$ tensors can be adjusted
independently to account for anisotropies. Because, in our simulations, we do
not find a clear trend with respect to the off-diagonal elements of $\teta$,
we explicitly allow for a possible symmetric contribution. Contrary to this,
the off-diagonal elements of $\talph$ are assumed to be totally antisymmetric
such that $\alpha_{\phi R}$ and $\alpha_{R\phi}$ are reduced to a common
parameter $\gamma_z$. The depicted model profiles are essentially governed by
the function
\begin{equation}
  \mathcal{P}(z) = \sin(\pi\,\frac{z}{2\,h_1})\ e^{-z^2/h_2}\,,
\end{equation}
with $h_1=1.5\kpc$ and $h_2=1.0\kpc$, respectively. Whereas $\alpha_R$,
$\alpha_\phi$, and $\gamma_z$ are directly proportional to $\mathcal{P}(z)$,
the diffusivity tensor $\teta$ is assumed to have the shape according to
$\pm\mathcal{P}(z)^2$. For the diagonal elements we, furthermore, add a floor
value $eta_0$. The wind profile $\bar{u}_z$, which also enters the mean-field
induction equation, comprises the same sinusoidal modulation on top of a
linear ramp. Although there is, certainly, space for fine-tuning of the
profiles, we want to keep the number of free parameters as low as possible. In
principle, one could even apply smoothed $\alpha$~profiles directly from the
simulations, or alternatively, take the SOCA expressions as a starting point
for the shape of the curves. We, however, believe that the chosen approach
strikes a reasonable balance between removing essential physics and getting
lost in too many free parameters.

\subsection{Discretisation of the equations} 

For the implementation of the 1D toy model we follow the same approach as for
the test-field fluctuations $\Tf'$, i.e., we apply the staggering given by
constrained-transport\footnote{%
The solenoidal constraint is, of course, trivially fulfilled in the case of
$\bar{B}_z=0$ and gradients in $z$, only.} %
in combination with up-winding to guarantee the numerical stability of the
resulting scheme. The implemented equations read:
\begin{eqnarray}
  \bar{B}_{R,t} & = & 
    \left[\: -(\bar{u}_z+\gamma_z) \,\bar{B}_R  \nonumber
      \quad  -\alpha_\phi \,\bar{B}_\phi \right.   \\ & & \left.   
      \quad  +(\tilde{\eta}_{\phi\phi}+\eta) \,\bar{B}_{R,z}
      \quad  -\tilde{\eta}_{\phi R} \,\bar{B}_{\phi,z} \:\right]_{,z} \\
  \bar{B}_{\phi,t} & = &
    \left[\quad\alpha_R \,\bar{B}_R                       \nonumber
      \quad  -(\bar{u}_z+\gamma_z) \,\bar{B}_\phi \right. \\ & & \left. 
      \quad  +\tilde{\eta}_{R\phi} \,\bar{B}_{R,z}
      \quad  +(\tilde{\eta}_{RR}+\eta) \,\bar{B}_{\phi,z} \:\right]_{,z}
    + q\Omega\,\bar{B}_R\,,
\end{eqnarray}
with $\eta$ the molecular value of the diffusivity, $\Omega$ the rotation
frequency, and $q$ the shear parameter. The off-diagonal elements of the
$\teta$~tensor are obtained as a superposition
\begin{equation}
  \tilde{\eta}_{R\phi} = \delta_z^{\rm S} + \delta_z^{\rm A} 
  \quad\text{and}\quad
  \tilde{\eta}_{\phi R}= \delta_z^{\rm S} - \delta_z^{\rm A}
\end{equation}
of the symmetric and antisymmetric contributions $\delta_z^{\rm S}$ and
$\delta_z^{\rm A}$, respectively. The system of equations is explicitly
evolved in time by means of a second-order Runge-Kutta scheme.





\cleardoublepage\addcontentsline{toc}{chapter}{\bibname}

\clearpage
\chapter*{Danksagung}

\center
\parbox[t]{0.815\columnwidth}{
  
  $\quad$Dank gebührt an erster Stelle Dr.~Udo Ziegler, ohne den dieses
  Projekt in vielerlei Hinsicht undenkbar gewesen wäre. Die Ursprünge des
  verwendeten NIRVANA-Codes, der die höchst anspruchsvollen numerischen
  Simulationen dieser Arbeit erst möglich gemacht hat, liegen schließlich in
  seiner eigenen Doktorarbeit, die sich mit dem Dynamoeffekt einzelner
  Supernovaüberreste beschäftigte. Besonders möchte ich Udo für die
  uneingeschränkte Be\-reit\-schaft bei der Betreuung dieser Arbeit danken --
  es gab für meine Fragen nie einen falschen Zeitpunkt und feature-requests
  zwecks Integration der erforderlichen Änderungen in NIRVANA fanden stets ein
  offenes Ohr. \vspace{10pt}
  
  $\quad$Prof.~Dr.~Günther Rüdiger möchte ich für die gewährten Freiräume und
  die bestimmten Impulse zur rechten Zeit danken. Von ihm und von Dr.~Detlef
  Elstner habe ich in meiner Zeit am AIP Grundlegendes über die an diesem
  Institut begründete Dynamotheorie gelernt. In diesem Zusammenhang muss auch
  Dr.~Leonid Kitchatinov erwähnt werden, der die bemerkenswerte Gabe besitzt,
  auf klare Fragen klare Antworten zu geben. Dr.~Krzysztof Chy\.zy danke ich
  für die freundliche Bereitstellung von Abbildungen zu Beobachtungsdaten.
  Prof.~Dr.~Axel Brandenburg gebührt Anerkennung für die Anregung zur
  Verwendung der Testfeldmethode, die eine erhebliche Verbesserung bei der
  Messung des $\alpha$-Effekts lieferte. \vspace{10pt}
  
  $\quad$Ein Lob geht an die Systemadministratoren Manfred Schultz, Mario
  Dionies und Dr.~Karl-Heinz Böning für die tadellose Betreuung des
  Arbeitsgeräts und die Unterstützung in technischen Angelegenheiten. Meinen
  Kollegen Dr.~Rainer Arlt und Dr.~Jacek Szklarski möchte ich für die
  zahlreichen anregenden Diskussionen danken. Gleiches gilt für Dr.~Simon
  Glover, Dr.~Robert Piontek und besonders Michael Zatloukal und Oliver Hahn,
  denen ich darüberhinaus für die kritische Durchsicht zahlreicher Entwürfe
  und Manuskripte danke. \vspace{10pt}
  
  $\quad$Zu guter Letzt danke ich meinen Eltern Katharina und Lothar Gressel,
  die mich stets vorbehaltlos und mit vollstem Vertrauen unterstützt haben,
  und denen ich diese Arbeit widmen möchte.

}

\end{document}